\documentclass[12pt,preprint]{aastex}   
\usepackage{longtable}
\usepackage{color}
\usepackage{natbib}
\usepackage{graphicx}
\usepackage{rotating}

\slugcomment{Submitted to {\it The Astrophysical Journal
    Supplement}}
\shortauthors{N. Giammichele et al.}

\begin{document}

\title{Toward High Precision Seismic Studies of White Dwarf Stars:
Parametrization of the Core and Tests of Accuracy}

\author{N. Giammichele\altaffilmark{1,2,3}, S. Charpinet\altaffilmark{2,3},
  G. Fontaine\altaffilmark{1}, and P. Brassard\altaffilmark{1}}

\altaffiltext{1}{D\'epartement de Physique, Universit\'e
  de Montr\'eal, Montr\'eal, QC H3C 3J7, Canada}
\altaffiltext{2}{Universit\'e de Toulouse, UPS-OMP, IRAP, Toulouse
  F-31400, France}
\altaffiltext{3}{CNRS, IRAP, 14 avenue Edouard Belin, F-31400
  Toulouse, France}

\begin{abstract}

We present a prescription for parametrizing the chemical profile in the
core of white dwarfs in the light of the recent discovery that pulsation
modes may sometimes be deeply confined in some cool pulsating white dwarfs.
Such modes may be used as unique probes of the complicated chemical
stratification that results from several processes that occurred in
previous evolutionary phases of intermediate-mass stars. This effort is
part of our ongoing quest for more credible and realistic seismic models
of white dwarfs using static, parametrized equilibrium structures.
Inspired from successful techniques developed in design optimization
fields (such as aerodynamics), we exploit Akima splines for the tracing of
the chemical profile of oxygen (carbon) in the core of a white dwarf
model. A series of tests are then presented to better seize the
precision and significance of the results that can be obtained in an
asteroseismological context. We also show that the new parametrization
passes an essential basic test, as it successfully reproduces the
chemical stratification of a full evolutionary model.
\end{abstract}


\section{INTRODUCTION}

In the era of space missions such as $Kepler$ and $Kepler$-2 (K2) which have
been providing asteroseismic data of unprecedented quality for several
types of pulsating stars (including white dwarfs), we decided to revisit
the problem of the seismic modeling of this latter type of pulsators
using current techniques, typical of those developed quite successfully in
recent years for pulsating hot B subdwarfs (see, e.g., Charpinet et
al. 2013). To test these techniques in a white dwarf context, we first
carried out a detailed analysis of the pulsation properties of two
classical hot ZZ Ceti stars -- GD 165 and Ross 548 -- using the best
available data from the ground. Those targets were chosen on the basis
of the simplicity of their light curves (associated with their locations
near the blue edge of the ZZ Ceti instability strip), their near
spectroscopic twin nature, and the availability of time-series data sets
of exceptional quality. The results of that analysis were presented by
Giammichele et al. (2015; 2016) who found a credible seismic model for
each star able to reproduce simultaneously the six observed periods well
within ~$\sim0.3\%$ on the average, which is comparable to the best
results achieved so far in asteroseismology. The models provided robust
mode identification and were found to be perfectly compatible with the
expectations of linear pulsation theory in both its adiabatic and
nonadiabatic versions. In addition, these seismic models were shown to
be consistent with all the available external independent constraints
such as estimates of the atmospheric parameters derived from
time-averaged spectroscopy, estimates of distances provided by parallax
measurements or spectrocopy, and even the measured rate of period change
for one mode in Ross 548. The study of Giammichele et al. (2015; 2016)
thus firmly established that our approach to quantitative seismology --
based on static, parametrized models of stars -- can be extended
reliably to the white dwarf domain.

An unexpected result of the study of Giammichele et al. (2016) is the
finding that the pulsation periods detected in GD 165 bear only a weak
dependence on the core composition, while those observed in Ross 548
are, on the contrary, quite sensitive to a variation of the core
composition. This a priori puzzling result finds a natural explanation
in that all six modes in the GD 165 model have amplitudes and weight
functions that do not extend into the deep core and, consequently, their
periods are not sensitive to a variation of the core composition. In
comparison, three of the six modes of interest in the Ross 548 model are
partly confined below its thin envelope and, therefore, bear a
strong sensitivity on the core composition. This discovery of the
existence of deeply confined modes in Ross 548 and, potentially, in many
other pulsating white dwarfs, opens up the most interesting possibility
of using those modes as probes of the internal composition profile in
the C-O core. In the Giammichele et al. (2016) study, to keep things as
simple as possible, it was assumed that the core composition is
homogeneous, which is a crude approximation in the light of the rather
complicated C-O stratification expected in the cores of former AGB stars
as revealed by detailed evolutionary calculations (see, e.g., Salaris et
al. 2010 or Romero et al. 2012). In retrospect, such an approximation
was amply sufficient in the case of GD 165, but it is now clear that for
stars such as Ross 548, with deeply confined modes, our modeling could be
perfected by including a parametrized description of the chemical
layering in the C-O core. Presumably, this would lead to still more
realistic seismic models of pulsating white dwarfs, with theoretical
periods approaching the ``Holy Grail'' of asteroseismology, i.e.,
perhaps able to reproduce the observed periods at the accuracy of the
observations. Currently, as indicated above, the best one can do is to
obtain average dispersions of less than $\sim 0.3\%$ in periods, still a
long way from the measurement uncertainties as indicated, for example, in
Tables 3 and 4 of Giammichele et al. (2015).
The advent of ultra-high precision photometry from space, in particular
with the {\sl Kepler} mission, further enlarges this gap as some
pulsation frequencies can be measured with outstanding precision of
the order of nanohertz. Six pulsating white dwarfs were extensively monitored
with {\sl Kepler} (see, e.g., {\O}stensen et al. 2011;
Greiss et al. 2014, 2015) and high precision data for more objects have
been (or will be) delivered by the still ongoing K2 mission
(Hermes et al. 2014; Howell et al. 2014). Thus, there are exceptionally
high-quality data available for a handful of pulsating white dwarfs that
should provide particularly stringent testbeds for asteroseismology.
We present, in this paper, a prescription for parametrizing the C-O core
of a white dwarf, with the hope of improving future seismic models of
stars of this type. We also discuss series of tests that evaluate the
precision that can potentially be achieved on stellar and structural
parameters with our technique.

\section{PARAMETRIZATION OF THE CORE OF A WHITE DWARF}

\subsection{The Need for a New Parametrization}

Parametrizing the core composition is not by itself  a new idea as
several attempts have been made in the past to improve white dwarf
modeling for the purposes of asteroseismic studies. Different
parametrizations for the distribution of chemical species in the core
were employed, the simplest one being a fixed homogeneous core. For
instance, \cite{1992ApJS...80..369B} explored the potential of white dwarf
seismology, with the study of the adiabatic properties of pulsating DA
white dwarfs, by using a homogeneous carbon core in their calculations
(see also Fontaine \& Brassard 2002). More recently, Castanheira \&
Kepler (2008; 2009) also used a fixed homogeneous C/O core (50\% each in
mass fraction) through their analyses of 83 ZZ Ceti stars. The main
argument that was used to justify this crude approach was simplicity, in
view of the usual paucity of observed periods and the perceived need to
limit the number of free parameters that would characterize a model. In
addition, on general grounds, it was expected that the g-modes detected
in the more evolved, more degenerate pulsating white dwarfs, the ZZ Ceti
stars in particular, would not be very sensitive to the details of the
core structure because of the outward migration of g-mode amplitudes and
weight functions associated with the cooling process. This can indeed be
true when the detected modes do not propagate deep inside the star, but
this is not the case in presence of modes confined below the helium
mantle as discussed by Giammichele et al. (2016). This
simple parametrization has also the benefit of not requiring the use of
any additional parameter in the search for a best-fit solution to the
observed pulsation periods. Nevertheless, it remains very far from even
approaching expected chemical profiles evolved from the zero age main
sequence (ZAMS), according to calculations from different groups
over the years (e.g., Salaris et al 1997, Althaus et al. 2002, and
Straniero et al. 2003, just to mention a few). According to these
evolutionary calculations, the oxygen profile inside the core of a white
dwarf follows a rather complicated descent from an inner central
homogeneous part, while carbon is the complementary element to it in
that region. Moreover, it is found that the point where the mass
fraction of oxygen is about to vanish is located above an interface
where three elements, carbon, oxygen and helium, can coexist in roughly
comparable proportions.

A first step in dealing more rigorously with core C/O profiles in models
of white dwarfs built for asteroseismology was performed by
\cite{1996ApJ...468..350B} who used a ramp-like shape as described by
\cite{1992ApJ...386..539W}, i.e., a simplified version of the
oxygen-rich core profiles computed by Mazzitelli \& D'Antona (1986a;
1986b). The core of this model is set to a homogeneous central part with
80 $\%$ (in mass) of oxygen out to 0.75 $M_{*}$ and it linearly changes
to pure carbon by 0.90 $M_{*}$. The central mass fraction of oxygen and
the location of the C/O transition zone has been allowed to vary for a
few possible configurations  in the study of  mode trapping in ZZ Ceti
stars in \cite{1996ApJ...468..350B}, as well as in the study of the two
pulsators G117-B15A and Ross 548 \citep{1998ApJS..116..307B}, and later
in a similar exercise concerning L19-2 and GD 165
\citep{2001ApJ...552..326B}. However, a major caveat of the optimization
technique performed at the time was the non-simultaneous fitting of the
various parameters and the use of a coarse grid of models as imposed by
the evolutionary approach. Even if the descent in the transition zone
turns out to be more realistic when compared to a basic homogeneous
core, this approach remains relatively crude. In particular, the triple
transition of carbon, oxygen and helium is still not a possible
configuration of this parametrization, as the limit of 0.90 $M_{*}$
where the pure carbon layer exists is fixed, and the helium mantle is
too thin to be connected to the oxygen reservoir. Moreover, the oxygen
descent is not closely reproduced when compared to what is found from
detailed evolutionary models with a ZAMS connection.

A more sophisticated core treatment came from \cite{2005MNRAS.363L..86M}
who introduced for the first time a parametrization that tries to mimic
the features of an evolutionary-based chemical profile for the C-O white
dwarf core, using Salaris et al. (1997) evolutionary tracks. This
parametrization employs two free parameters defining the constant
central abundance of oxygen, $X$(O), and the fractional mass, $q$, at
which the constant oxygen mass fraction starts dropping. The shape of
the drop is simply taken to be a scaled version of an evolutionary track
from the work of Salaris et al. (1997) at the specific white dwarf mass
of $M_{*}$ = 0.61 $M_{\rm \odot}$. Moreover, the drop ends at the fixed
value of 0.95 $M_{*}$. While doing a better job in reproducing a
specific core composition profile than the other proposed recipes
(relative to what is expected from stellar evolution calculations),
these physically motivated C/O profiles suffer from a major lack of
flexibility. The pre-determined parameters and shape do not allow a
thorough search in the stellar model space (there are only few
configurations possible with two parameters) and this lack of
flexibility does not allow a proper testing of the core chemical
profiles. Quite notably, this parametrization fails at simply
reproducing the core composition profiles from detailed evolutionary
models at other masses. They also fail, like other methods discussed
before, at reproducing the triple chemical transition at the edge of the
core, where oxygen, carbon, and helium are expected to coexist,
according to current evolutionary sequences computed to produce white
dwarfs star (see, e.g., Althaus et al. 2010). Here again, this is an
important limitation as the pulsations properties can be influenced by
those major predicted features.

For completeness, we also mention the efforts developed these past ten
years to tackle asteroseismology of white dwarf stars from a different
angle by using evolutionary models computed from the ZAMS (e.g., Romero et
al. 2012), or parametrized white dwarf evolutionary models with
pre-computed C/O profiles from grids of evolutionary models from the
ZAMS (e.g., Chen 2016). However, this approach, by definition, leaves no
flexibility at all at the level of the core structure and does not
constitute an option for our purposes, i.e., testing white dwarf stellar
structure with asteroseismology and, in particular, the predictions of
stellar evolution calculations. For this, we need an independent tool, a
new method to model and parametrize the core composition that does not
incorporate parts of pre-calculated profiles. We need a tool that is
capable of reproducing closely enough the stratification obtained in
evolutionary models, but which at the same time leaves open the
possibility that the composition profiles in the core of real white
dwarfs may be different from what is currently predicted. Relying
heavily on pre-calculated profiles would clearly not allow such tests.

In this context, we are thus left with a need for a new parametrization
for the core composition which would be more robust and flexible than
previous recipes proposed so far. We also need this new parametrization
to be coupled with our modern tools for asteroseismology based on a set
of powerful optimizations codes (see, e.g., the Introduction of
Giammichele et al. 2016, or Charpinet et al. 2015 and references
therein). We describe in the following subsection how these developments
are implemented.

\subsection{Implementation of the New Parametrization}

In our seismic studies, the problem revolves around minimizing a merit function
defined by the sum of the squared differences between theoretical and
observed periods obtained for a specific star (see, e.g., Giammichele
et al. 2016). In our previous efforts on GD 165 and Ross 548, we fixed
the convective efficiency to a standard version, we also imposed fixed
composition profiles at the H-He and He-core interfaces as obtained from
detailed evolutionary calculations of GD 165 taking into account
diffusion between the atomic species. We were thus left with five
parameters to define a full, static model of a ZZ Ceti star : (1) the
surface gravity, (2) the effective temperature, (3) the mass contained
in the He mantle, (4) the mass contained in the H outermost layer, and
(5) the core composition. Given the small number of modes available (six
for both GD 165 and Ross 548), we strove to keep the number of free
parameters to a minimum value, so we assumed crudely that the
composition in the core would be homogeneous and specified by a single
quantity, the oxygen mass fraction $X$(O). We now consider a more
realistic depth-dependent core composition as obtained from detailed
evolutionary calculations.

Inferring the most realistic composition profile given an
ensemble of observed periods sensitive to the details of the core can
be defined as a shape optimization problem. As
generally stated, shape optimization, or optimal design, is the set of
methods that gives the best possible form in order to fulfill the
desired requirements. The typical problem is to find the shape which is
optimal in that it minimizes a certain cost, or merit, function while
satisfying given constraints. These methods are widely encountered in
various domains such as aerodynamics, hydrodynamics, acoustics,
electromagnetism, and many more. Our situation, the ``drawing" of the
chemical profile to best match what is occurring in stars, quantified by
the minimization of our merit function, is, in all particulars,
analogous to airfoil shape optimization, just to name one.

In general, the ideal airfoil shape, entirely defined by a set of
parameters, is determined by minimizing a simple merit function. In the
case of aerodynamic problems, the merit function to minimize is usually
defined to be the inverse of the lift-to-drag ratio. It is important to
realize that the number of parameters used to define the airfoil shape
has nothing to do with the number of quantities involved in the merit
function. Every single airfoil shape give a lift-to-drag ratio,
independently of the number of parameters that are needed to define
the shape of the airfoil. We adopt the exact same approach for
parametrizing the chemical profile of the core. Every single shape of
the chemical profile leads to a single merit function value,
independent of the adopted parametrization of the shape and no matter
how many periods (that enter into the merit value computation) are
observed.

If the parametrization technique is not flexible enough, which
translates into not being able to represent all possible shapes, then a
true optimal shape cannot generally be attained. On the other hand, if the
number of design parameters is too large, the optimization problem becomes
unfeasible in term of excessive computational time. Different shape
parametrizations have been used in the aerodynamic field, among others:
analytical, discrete, polynomial and spline representations
(Gallart 2002). Every method has its advantages and drawbacks. In the
analytic representation, given an original shape, a set of functions
deform that shape. The design variables control the deformations added to
the original shape in order to create the new
shape. This method has the advantage of reducing the necessary number
of design variables to a small set, while obtaining a smooth surface. On
the other hand, it is only applicable to simple geometries and the
deformations are dependent on the shape functions used. The discrete
representation is more flexible, easy to implement, and can be used with any
geometry. However, it requires a large number of design variables and
the final shape can have high frequency oscillations. With polynomial
methods, the design variables are the coefficients of a polynomial. The main
advantage in this method is that a small set of design variables can be
used, but as for analytical methods, if only low order polynomials are
considered, some shapes become impossible to represent. The
most commonly used representations in areas of expertise such as
aerodynamics or automobile design are the spline representations, being
the sum of weighted polynomials. In this case, the set of weighting
parameters, called control points, are used as the design
variables. There exist several types of splines: Bezier curves,
B-splines and non-uniform rational B-spline (NURBS), just to name a
few. There are several reasons for using splines: it allows for a
reduction in the number of control points (or design variables); the
perturbation of one control point has only local effects on the design
shape; it produces curves with C$^{2}$ continuity, therefore guaranteeing
a shape that is smooth enough without high frequency oscillations.

In our context, we adopt spline representations to reproduce the
oxygen chemical profile in a white dwarf core. The left panel of Figure 1
introduces the different design parameters that constitute the new
parametrization. These parameters define a set of control points through
which interpolation is done (right panel of the same figure). After
testing various options, we opt for an interpolation scheme using Akima
splines (Akima 1970). The Akima interpolation is a continuously
differentiable sub-spline interpolation. It is built from piecewise
third order polynomials. Only data from the next neighbor points is used
to determine the coefficients of the interpolation polynomial. The
disadvantage of other schemes, such as cubic splines, is that they can
oscillate in the neighborhood of an outlier or when gradients change
abruptly (as illustrated in Figure 1). This is also known as ``Gibbs
noise''. The Akima spline is a special spline which is stable against
such points and provides a robust and smooth representation of the chemical
stratification in the core. The shape parameters that control the Akima
splines are specifically chosen in order to best imitate the main
structures in an evolved white dwarf chemical profile (see, again,
Salaris et al. 2010 or Romero et al. 2012). Eight design parameters are
necessary to fully define a two-transition chemical profile in the core:
core O, t1, $\Delta$t1, t1(O), t2, $\Delta$t2, t2(O), and envl O. We
point out that the last parameter, envl O, is here to account for
specific configurations involving DB white dwarfs for which it is
believed that the oxygen mass fraction does not drop to zero at the edge
of the C/O core. In a DA white dwarf context, which will be the main
focus in this paper, the envl O parameter is always set to zero and
the problem technically reduces to a seven parameter shape optimization.
The proposed parametrization is flexible enough that it allows us to
define a simpler one-transition profile by only using a subset of three
shape parameters, core O, t1 and $\Delta$t1. In that case and by
construction, in order to have a smooth profile, t1(O) and t2(O) are set
to zero, while the value of t2 is set far enough not to interfere and
$\Delta$t2 is given a non-zero value.

	The main improvement of this new parametrization relies on the
  definition of the oxygen profile as just described. The rest of the
  structure is defined using our standard procedure for building static
  parametrized models as described in Giammichele et al. (2016). For the
  benefit of the reader, however, we briefly recall here the approach
  that we have adopted for parametrizing the envelope as well. We
  address the definition of the envelope by using four different
  parameters: $D$(H), $D$(He), $Pf_{\rm H}$, and $Pf_{\rm He}$. In order to
  specify the quantity of hydrogen and helium in the star, the two
  parameters $D$(H) and $D$(He) are necessary and are directly
  proportional to the total (logarithmic) mass fraction of hydrogen and
  helium. $D$(H) and $D$(He) indicate the middle of the H/He and He/C/O
  transitions, respectively. The shape of the transition at the border
  of the H/He and the He/C/O layers is based on equations (26) and (28)
  of Tassoul et al. (1990) that define the profile for the tail of the
  complementary trace element when diffusive equilibrium is assumed. The
  two parameters $Pf_{\rm H}$ and $Pf_{\rm He}$ are introduced to be scaling
  factors that allow to compensate for the potentially unreached diffusive
  equilibrium and allow to explore a large range of possible shapes from
  sharp to broad. The implicit equation governing the chemical profile
  of two species in a transition zone is given by: 
\begin{equation}
\frac{X_2^{A_1}}{X_1^{A_2}}
\Bigl(\frac{A_1X_2+A_2X_1}{A_1+A_2}\Bigr)^{A_2-A_1}=
\Bigl(\frac{q}{D}\Bigr)^{P_f[A_2(Z_1+1)-A_1(Z_2+1)]},
\end{equation}
where $A_{i}(Z_{i})$ corresponds to the atomic weight (average charge
evaluated at the middle of the transition zone) of element i, $D$ is
the value of $q$ where $X_1=X_2=0.5$ and $P_f$ is the scaling factor
($1$ when diffusive equilibrium is reached). Such a functional form
belongs to the so-called family of sygmoid functions. In the case where
three elements overlap, such as the extension of the He tail into the
outer regions of the C/O core for example, we still use the above
equation with element 1 representing He, but ``element 2'' being the
weighted sum of the contributions of C and O to the local values of
$X_2$, $A_2$, and $Z_2$. Of course, in the innermost part of the core
where $X$(He) and $X$(H) fall to zero, the carbon mass fraction is
simply given by $X(C) = 1 - X(O)$.
 
Figure 2 presents a comparison of three helium profiles culled from a
representative DA evolutionary sequence incorporating diffusion (the
solid black curves) with the results of our fitting procedure obtained
by varying the four envelope parameters (the red dotted curves), in
particular $Pf_{\rm H}$ and $Pf_{\rm He}$. Our computed evolutionary
sequence is characterized by a total mass of $M_{*}$ = 0.6 $M_{\rm
\odot}$ and an envelope layering specified by log $M(He)/M_{*}= -3.0 $
and log $M(H)/M_{*}= -5.0 $. We can observe that the overall shape and,
specifically, the asymmetry in the evolving profile of helium is well
matched by the parametrized chemical profile in both the case of the H/He
transition and of the He/C/O transition. Note that in the rest of this
paper the argumentation is focussed on the parametrization of the core,
not the envelope. Consequently, for the sake of simplicity and for the
unique purpose of this paper, $Pf_{\rm H}$ and $Pf_{\rm He}$ were not 
varied and left to their calibrated values obtained from evolutionary
calculations performed for the pulsating white dwarf star GD 165, as
mentioned in Giammichele et al. (2016). 

With the above prescription, we now have in hands a new parametrization
that can imitate smoothly enough one or two steep "drops" in the oxygen
profile (and complementarily the carbon profile) in the core, without
directly relying on a specific evolutionary model. Quite importantly,
the flexibility of this parametrization also allows, as one of its many
configurations, the triple transition at the edge of a rather thick
helium mantle, between helium, carbon and oxygen, as predicted by
evolutionary calculations. This triple transition was never a
possibility in previous works using parametrized models, as mentioned in
details in the previous subsection. Figure 3 indeed illustrates how well
this new parametrization can mimick generic profiles obtained from
evolutionary calculations for a typical DA white dwarf (left panel)
adjusted to match the model shown in \cite{2010ApJ...717..897A} (their
Figure 6), and for a typical DB white dwarf (right panel) adjusted to
match the evolutionary model presented in \cite{2012A&A...541A..42C}
(their Figure 2). 

In the following section, we validate this new parametrization through a
series of tests more generally aimed at evaluating the theoretical
precision that can be achieved with white dwarf asteroseismology in
various contexts. 

\section{TESTS ON THE PRECISION ACHIEVABLE WITH ASTEROSEISMOLOGY}

The measurements of white dwarf stellar parameters using asteroseismology
are quite precise by current standards, if we refer to the latest
analysis of Ross 548 and GD 165 as alluded to above. However, periods are
still not fitted at the precision of the observations, indicating that
uncertainties are dominated by shortcomings in the models and suggesting
that the potential of this technique is not yet fully exploited.
In this section, we are interested in testing the degree of precision we
can expect on the measured stellar parameters from the results of a seismic
analysis if the uncertainties were only coming from observations,
either ground- or space-based.

To test this, we start by creating an artificial reference star from a
pulsating white dwarf model with given global and structural (shape)
parameters. The structure and pulsation properties of this reference star
are thus entirely known. We then select a subset of 10 $\ell=1$ and $2$
$g$-modes out of the pulsation spectrum to simulate a typical outcome
(e.g., in terms of number of modes) from a campaign of dedicated observations.
The artificial data set is finalized by adding to the frequency of each
selected mode a random perturbation that follows a normal (Gaussian)
distribution. The mean value, $\mu$, of the normal deviates is set to zero
in all our tests (implying that no bias is introduced in the perturbations),
while the standard deviation, $\sigma$, is set to the desired value to
mimic, e.g., random noise fluctuations in the frequency measurements.
Three representative cases corresponding to different levels of precision are
investigated. The first case is made comparable to the typical period fit
precision achieved currently in the field, which is obtained by setting the
standard deviation to $\sigma = 10$ $\mu$Hz. The second test case
corresponds to a precision that
would be limited by the accuracy of standard ground-based data, which is
of the order of 0.1 $\mu$Hz on the measured frequencies (hence,
$\sigma = 0.1$ $\mu$Hz).
The third and last case represents the ultimate situation where the
precision is only limited by the accuracy offered by {\sl Kepler} data,
which can reach down to  $\sim 1$ $n$Hz on the measured frequencies
for the longest runs and most stable pulsators (therefore,
$\sigma = 0.001$ $\mu$Hz).

From these modified subsets of periods, the objective is then to retrieve the
global and shape parameters of the artificial reference star with our
optimization tools. The latter have been described in some detail, in a
white dwarf context, by Giammichele et al. (2016) who used them
extensively for the analysis of the two ZZ Ceti stars R548 et GD 165.
In a nutshell, our approach is a double optimization procedure that both
matches the observed periods to computed periods assuming no a priori
knowledge of the mode identification and conducts a multimodal global
search of the best fit solution(s) in the vast model parameter space.
The global optimization is carried out by the code LUCY, a robust hybrid
genetic algorithm capable of identifying and exploring simultaneously,
if needed, several minima of the merit function in case the problem turns
out to be ill-posed with no uniquely defined solution. Among other
benefits, these tools ensure that the search is {\sl objective} in the
sense that the whole parameter space is thoroughly searched, the global
minimum of the merit function is robustly found, and the uniqueness (or not)
of the solution is assessed.

Along with these precision tests, three parametrizations are also
experimented. The first set of tests is performed with a simple variable
homogeneous core, while the second set uses the new one-transition
parametrization, and the third set the two-transition parametrization.

A last test is performed highlighting the flexibility of the two-transition parametrization, by reproducing a reference star with a triple transition of He/C/O in its core with a rather thick helium layer this time, to emphasize that this parametrization is capable of mimicking models from evolutionary calculations.

\subsection{Parametrization with a Varying Homogeneous Core}

The different sets of ``observed" periods are initially computed from a
reference model with the parameters given in Table \ref{tab3-1}. This
model represents a standard ZZ Ceti star, with a rather thick hydrogen
envelope. As mentioned before, once the pulsation periods are calculated from
the reference model, a subset of modes is selected, and their periods
are perturbed with normally distributed (Gaussian) random fluctuations
to the level of the chosen precision. This process is detailed in Table
\ref{tab3-2} which provides the values of the periods at all stages. The
first column shows the periods of the reference star, while the
following columns list the modified periods for each degree of
precision, now and hereafter dubbed ``current fit" ($\sigma=10$ $\mu$Hz),
``ground-based data" ($\sigma=0.1$ $\mu$Hz), and ``$Kepler$ data"
($\sigma=0.001$ $\mu$Hz) precision. We acknowledge that these sets of
modified periods represent only one realization over a somewhat small
number of modes of the random perturbations applied to the original model
periods. Ideally, each experiment would have to be repeated multiple times
with different realizations for a better statistical assessment of error
propagations, but this is not feasible practically speaking due to the
large computational resources needed by the optimization process for
each asteroseismic analysis. For our purpose, however, which is to
obtain an order-of-magnitude estimate of the uncertainties on the
derived model parameters, our approach is sufficient. The global search
range where the optimization is done is also indicated in Table
\ref{tab3-1} for each parameter. The explored domain is vast and
virtually allows for any model configurations potentially encountered
for ZZ Ceti stars.

Table \ref{tab3-3} presents the stellar parameters derived from the three
optimizations based on the modified subsets of periods with various
degrees of precision. Each entry has errors statistically calculated from
the likelihood function (linked to the $\chi^2$-type merit function $S^2$)
that was sampled by the optimization code during the search for the best-fit
models (see Giammichele et al. 2016 for more details on this procedure;
see also Fig. 3).
In test case 1 (current fit precision), all parameters are well retrieved
with a precision that is generally between $0.1\%$ and $2\%$, for an unweighted
$S^2$-value of 9.1. The optimization is less sensitive to chemical profiles
deeper than the envelope, in particular the central homogeneous oxygen value
is only determined with a precision of 6\%.
The achieved level of precision for the derived stellar
parameters is found comparable, as one could have anticipated, to the precision
claimed by former analyses of white dwarf stars. This shows a quantitative
consistency between the level of precision that is achieved on the stellar
parameters and the overall quality of the period fit.
This connection is further illustrated
with the other tests. With ground-based data precision, the $S^2$-value of the
best-fit solution reaches four orders of magnitude less, for a general
improvement of a factor $\sim 100$ in precision for most of the global
stellar parameters. Another four orders of magnitude are gained on $S^2$ by
switching to {\sl Kepler} data precision. The parameters of the reference model
are retrieved to an impressive precision of $\sim$ 0.0002 \% in that case.
The homogeneous core composition is still less precisely determined by one
order of magnitude compared to the other stellar parameters.

As a complementary view, Figures 4, 5 and 6
depict the maps of
the projected merit function $S^2$ (on a logarithmic scale) onto the $T_{\rm
  eff}$-log $g$ and $D$(He)-$D$(H) planes, as well as the derived probability
density function of all the retrieved stellar parameters when considering
the current fit precision ($\sim$ 10 $\mu$Hz), the typical ground-based
data ($\sim$ 0.1 $\mu$Hz), and the $Kepler$ precision ($\sim$ 0.001 $\mu$Hz),
respectively. White contours show regions where the period fits have
$S^2$-values within the 1-$\sigma$, 2-$\sigma$,and 3-$\sigma$ confidence
levels relative to the best-fit solution. The gain in precision for the
derived stellar parameters, from current fit to $Kepler$ data precision, is
clearly illustrated from the inspection of the 2-D maps. With current fit
precision (Fig. 4), the best-fit solution within 1-$\sigma$ is partly
diluted in the background noise. With increasing observational precision,
we get much narrower confidence level regions in the 2-D maps from Figures
5 and 6, as well as much narrower probability
distributions for all stellar parameters, in line with the values given in
Table \ref{tab3-3} derived from these distributions.

\subsection{Core Parametrization with One Transition Zone}

The same exercise is conducted bearing in mind the use of the new
parametrization. The reference star is this time computed with the
adjustable core composition using a single transition. The global and
shape parameters for that model are given in Table \ref{tab3-4}.
The pulsation periods for $\ell=1$ and 2 $g$-modes for this reference star
are then computed and perturbed using the same technique as described in
the prior section. The modified subsets of periods corresponding to
the three test cases, which allow investigating different levels of
precision on the ``measured" periods, are provided in Table \ref{tab3-5}.

In test case 1, where we lean toward the precision of current fits, we
obtain after the optimization procedure a best fit solution with a
$S^2$-value of 3.1, i.e., the same order of magnitude as previously found
from our test with the homogeneous core. The estimated errors on the retrieved
global and shape parameters given in Table \ref{tab3-6} (second column,
for test case 1) are slightly larger than the previous parametrization
with the homogeneous core, but remain of the same
order of magnitude. The optimization appears less sensitive to the shape
parameter $\Delta t_{1}$, as the error on the value of this specific
parameter reaches 20 $\%$. We note from the tabulated values compared to
the original model parameters given in Table \ref{tab3-4} that the
optimal solution uncovered in this case is slightly offsetted (by up to
$\sim 2\sigma$) relative to
the true solution. We briefly discuss this feature below. As we move to a
higher achievable period fit precision with test case 2, corresponding
to a standard deviation from the original periods set of $\sigma = 0.1$
$\mu$Hz, the increase in precision for the derived model parameters is
remarkable. The best-fit solution now matches the periods with an
unweighted $S^2$-value of 3.0 x 10$^{-4}$ and gives error on the stellar
parameters of the order of $\sim 0.01\%$. If seismic analyses were only
limited by the precision of the frequencies measured from typical
ground-based observations, we would reach an impressive internal
precision of 0.05 $\%$ for the determination of the mass, for example.
Shape parameters are found to be less precisely determined, with errors of
the order of 0.1-1$\%$. Quite interestingly, the offset pointed out in the
previous experiment has now disappeared, all seismically derived values
matching the true values within $1\sigma$.
With the typical {\sl Kepler} precision, we gain a
factor of a hundred relative to the achieved precision with ground-based
data. The best $S^2$-value obtained in that particular test case 3 is
6.1 x 10$^{-7}$ and the errors on the global parameters reduce further
to $\sim$ 0.0002$\%$, as for the seismic mass measurement. As before, the shape
parameters, the core oxygen central value and $\Delta t_{1}$, are less
precisely measured than the global parameters, while the transition
$t_{1}$ is well retrieved. This most certainly reflects the fact that
the probing modes bear a lower sensitivity to this particular region of
the star.

Figures 7, 8 and 9 show as before
the maps of the projected merit function $S^2$ (in logarithmic scale)
onto the $T_{\rm eff}$-log $g$ and $D$(He)-$D$(H) plane as well as
the probability density functions of all the retrieved stellar parameters
for the current fit ($\sim$ 10 $\mu$Hz), typical ground-based data
($\sim$ 0.1 $\mu$Hz), and $Kepler$ data ($\sim$ 0.001 $\mu$Hz) precisions,
respectively. The 2-D maps of Figure 7 show a slightly
shifted solution, if we only consider the 1-$\sigma$ region, but fall
well within the 2-$\sigma$ region. This is related to the small shift
already pointed out from the values given in Table \ref{tab3-6}. The
probability distributions for each measured parameters also illustrate
this shift, providing additional insight on its origin. Indeed, for several
parameters, the distribution appears bi-modal, with a main peak
corresponding to the optimal (but slightly offsetted relative to the true
model) solution, and smaller secondary peaks (a local minimum) that fall
almost exactly on the true values. We have here an example of two close
minima of the merit function, where the dominating solution (global minimum)
is not necessarily the true solution (although not very different in the
present case). Interestingly, this test case offers an illustration of the
ambiguity and biases that can potentially arise in the seismic solution
when the reached precision on the period fit is not sufficient.
We also point out that a slight bias in the solution may also be generated
by the fluctuations in the random numbers applied to the relatively few
selected periods (small number statistics), which can tip the balance
towards a specific region. This effect could eventually
be compensated by repeating many times the same experiment (for better
statistics), a refinement that we do not however implement in our present
tests because of the very high computational burden it would imply.
Hence, if we consider only the precision of the current fits, we can see
that the determination of the solution may be affected by the background
noise and potential secondary optima. We note also, based on the histograms,
that the seismic optimization is clearly not sensitive enough to the
parameter $\Delta t_{1}$ at this level of precision.
When increasing the precision on the measured frequencies, from
typical ground-based data to $Kepler$ data precision, we obtain much narrower
confidence level regions in the 2-D maps as shown in Figures 8 and
9, as well as much narrower probability distributions for all
stellar parameters, including the shape parameters of the core. This is in
line with the values provided in Table \ref{tab3-6} that derive from these
distributions. We point out again that the small offsets relative to the
true values have disappeared with the increased precision in the period fit,
and secondary structures are no longer visible in the histograms. The latter
shows that the best cure against potential ambiguities in the solution is an
overall increase in the quality of the period fit.

Overall, we conclude this section by emphasizing that the new adjustable
core parametrization used in the single-transition setup behaves well under
the seismic analysis and allow to retrieve with a remarkable precision
both the stellar and shape parameters of the reference star.

\subsection{Core Parametrization with Two Transition Zones}

For the last test of this series, we use the full potential and flexibility
of the new core parametrization with the two-transition setup, both for the
reference star and the global search. The parameters defining the standard reference
star are provided in Table \ref{tab3-7}, along with the search range
considered for the optimizations. The pulsation periods for dipole and
quadrupole $g$-modes are again computed from this reference model, and
then modified using the same technique described earlier. The perturbed
subsets of periods used for the three test cases corresponding to the
different levels of precision are given in Table \ref{tab3-8}.

We find, as summarized in Table \ref{tab3-9}, that the seismic measurements
of the global and shape parameters follow the same pattern as in the two
previous experiments involving a homogeneous core of variable composition
and a stratified core with the one-transition parametrization.
The internal precision
on the seismically derived model parameters improves by an important factor
of $\sim 100$ at each step from the current fit precision to typical
ground based data and {\sl Kepler} data precisions.
If we examine more closely Table \ref{tab3-9}, we notice that shape
parameters are less precisely determined, especially $\Delta t_{1}$ and
$\Delta t_{2}$ for which the error on the obtained value reaches up to
$\sim$ 30$\%$, in the least favorable test case based on current fit
precision. This situation improves as we achieve higher precision for
the period fits. In particular, we find that to accurately constrain
the shape parameters defining the core structure, the overall precision
of the seismic fit must be able to reproduce the observed periods at least
to the precision of $\sim 0.1$ $\mu$Hz (test case 2). In retrospect, this
should not be very surprising considering that the modes are usually less
sensitive to structures in the core, meaning that precise fits are needed
to capture correctly their signature in the period spectrum of the star.
Otherwise, these signatures are simply blurred or hidden by the noise.

Figures 10, 11, and 12 again illustrate the
solutions with maps of the projected merit function $S^2$ onto the
$T_{\rm eff}$-log $g$ and $D$(He)-$D$(H) plane, as well as probability
density functions derived for all the retrieved stellar parameters for
the current fit ($\sim$ 10 $\mu$Hz), typical ground-based data ($\sim$ 0.1 $\mu$Hz)
and $Kepler$ ($\sim$ 0.001 $\mu$Hz) precision levels, respectively. The 2-D
maps in Figure 10 (current fit precision test case) present again
a slightly shifted minimum compared to the true values, if we stay strictly
within the 1-$\sigma$ region, but consistency with the true model is achieved
well within the 2-$\sigma$ region. The probability distribution functions
(histograms) show that the seismic optimization is clearly not sensitive
enough to the parameter $\Delta t_{1}$ and $\Delta t_{2}$ at this level of
precision. Increasing the fit precision to the level of typical ground-based
data and $Kepler$ data leads to much narrower confidence level regions in
the 2-D maps, as illustrated in Figures 11 and 12, as
well as much narrower probability distributions for all stellar
parameters. We also observe, in Figure 11 (typical ground-based
data precision), the presence of a weak secondary solution (visible on some
histograms, but not at the scale of the  2-D maps) slightly offsetted from the
main peaks and the true values. It disappears when considering fits that
can reproduce the observed periods with a greater accuracy
(see Figure 12).
If all the parameters are now found to be well constrained, we
nevertheless note that the values of the majority of the shape parameters
from the reference star do not fall within the 1-$\sigma$ limits of the
probability distribution of the recovered parameters.
This bias is most notably visible in Figure 12 (typical {\sl Kepler}
data precision). We trace down this problem to an under-sampling of the
merit function in the vicinity of the solution by the optimization code.
The parameter space volume to explore indeed increases considerably with
the addition of more parameters to define a 2-transition core. This can be
balanced by scaling the optimizer ressources to the size of the problem
considered (at the cost of an increase of the needed computational ressources)
but, for simplicity and to save on our limited (in terms of available CPU
time) computer ressources, the genetic algorithm was used with the same
initial population of 1,000 evolved solutions for all the tests with
different parametrizations.
Therefore, the error estimates in the most demanding cases (and the most
demanding experiment is the one shown in Figure 12) may be
underestimated for some parameters. This being said, a quick look at the
scales involved in each histogram shows that the bias is extremely small
and the issue is minor.

\subsection{The Special Case When a Triple He/C/O Transition Exists}

The choice of a relatively thin layer of helium to compute the reference models for our testing purposes
is one of the many possible configurations allowed by the new parametrization. However, it does not allow to see the triple transition of He/C/O noticed in models computed from evolutionary means that have rather thick helium mantles. To show that the parametrization is nonetheless capable of mimicking this feature and to assess that the precision obtained is comparable to the case previously presented with a thinner layer of helium with the two-transition parametrization, we deal with a last test case. We use again the new core parametrization with the two-transition zones, but this time having the precise setup of a triple transition of He/C/O at the edge of the core for the reference star. The parameters defining the standard reference star are provided in Table \ref{tab3-10-1}. No restrictions whatsoever are applied to the search parameters, e.g. the search range is wide enough to accommodate thin or thick helium layers with or without a triple transition of He/C/O, as can be seen from the search range exposed in Table \ref{tab3-10-1}. For this last test, we only look at the most stringent case, by using the level of precision given by the $Kepler$ data. Pulsations periods for dipole and quadrupole $g$-modes are again computed from the reference model, and then modified using the same technique as described earlier. The perturbed subset of periods used for this last test case corresponding to the $Kepler$ level of precision is given in Table \ref{tab3-10-2}.

The seismic measurements of the global and shape parameters follow
exactly the same pattern as previously presented in the experiment
involving a stratified core with the two-transition parametrization and
a thin helium layer, as summarized in Table \ref{tab3-10-3}. If we look
more into details, the internal precision reached for the seismically
derived parameters is similar in all points to what was previously found
for the stratified core with the two-transition parametrization and the
thin helium layer, if we compare to the test case 3 in Table
\ref{tab3-9}. Periods, global and shape parameters are reproduced to the
same precision. We still observe that shape parameters are less
accurately determined than global parameters, an effect of the lesser
sensitivity of the probing modes onto the structures in the core.

Figure 13 illustrates the internal chemical profiles (top panel) of the reference star (red curves) with a thick helium layer and a triple transition of He/C/O, superimposed with the chemical profiles retrieved from the optimization (black curves). The bottom panel presents the run of the Brunt-V\"ais\"al\"a frequency as well as the Lamb frequencies for both the reference and the retrieved model. We can only notice the perfect superposition of the two models on both panels, demonstrating that the parametrization and our optimization tools are equally capable of retrieving models with thick helium layer that involves a triple transition of He/C/O at the core boundary.

In summary, we have demonstrated that our optimization procedure
is perfectly capable of retrieving the stellar and shape parameters from
a reference star for different configurations and for different levels of precision in fitting the periods
measured from seismic observations. If the limiting factor was purely
observational, that is governed only by the actual precision on the
measured frequencies (or periods), we find that stellar and shape parameters
could be recovered to an outstanding internal precision, much higher than
currently obtained. This result illustrates the very strong potential offered
by asteroseismology which still has to be fully exploited.
Although our asteroseismic method gives robust, precise, and accurate
constraints on either stellar or shape parameters, the current limiting factor
is usually not the precision of the observations but the uncertainties associated
with the constitutive physics included in the models. The latter need to be
addressed in order to go beyond current achievements in white dwarf asteroseismology,
especially if we aim at setting better constraints on the core chemical
profiles which is only possible if we significantly improve the overall
quality of the obtained seismic fits compared to current standards.

\section{TESTS WITH RANDOM PERIOD SEQUENCES}

In this section, we present additional tests using randomly generated period
sets. When we perform a seismic analysis, the optimization process is
governed by the minimization of a merit function. During the
optimization, the merit function will take on different values as we
search in parameter space. The point here is that there will always
be a minimum value found at the end of the process, but this minimum
value is not necessarily statistically significant. The proposed tests allow
us to roughly estimate a threshold on the best-fit value of the
merit function over which a random set of periods could likely be fitted
equally well. Hence, this threshold indicates when a seismic analysis is
credible or not, based on the overall quality of the obtained period fit.
Obviously, such tests should be performed a large number of times and for
different observed mode configurations in order to get significant
statistics and a truly global overview of the problem.
However, because we are mostly interested here in getting a rough idea of
this threshold in the context of the experiments conducted in this paper and
because calculating power and time was limited for this
project, we only perform this test five times for each core parametrization
presented earlier. We point out that, ideally, this kind of experiment
should be done (and tuned) for each specific seismic analysis of a pulsating
white dwarf star in order to assess the reliability of the results.

In the following experiments, ten periods are randomly computed with the only
restriction that they fall between 100 and 500 seconds in order to remain
in the same range as the previous tests and calculations. Five random tests
are performed, and the pulsation spectra used in each case are given
in Table \ref{tab3-10}.

The results of the five test cases conducted for each core
parametrization are presented in Table \ref{tab3-11}. The merit function
$S^2$, an unweighted $\chi^2$, as well as $< \Delta \nu >$, the average
frequency dispersion, and $<\Delta P>$, the average period
dispersion of the best-fit solution are disclosed.
The first observation that can be made out of this test, when we go
through the columns, is that the values of either $S^2$, $< \Delta \nu >$
and $<\Delta P>$ are greatly varying.
But, again, since we are just interested in an estimation of the threshold
for a credible seismic analysis, these five test cases are sufficient for
having a rough idea of the value of $S^2$ above which the fit must be
considered as not statistically significant.
The second observation is that the increase in flexibility of the core
definition from the varying homogeneous core to the one- and two-transition
parametrization leads to an overall decrease of the values
obtained from these random tests. Therefore, not surprisingly, the threshold
is not the same if we use an homogeneous core or a more complex core
chemical profile parametrization, being more stringent in the latter case.

With the varying homogeneous core, the best-fit merit value went down to as
low as $\sim 15$ (Case 5; see Table \ref{tab3-11}), even though the
average value of the 5 tests is quite higher. The corresponding average
period dispersion is around $<\Delta P> \sim 2$ s.
With the one- and two-transition parametrizations, the dispersion between
the values obtained for each random test is less important. The average
value for $S^2$ is of the order of ten. In these cases, the average
period difference of the best-fit solution becomes less than 1 s. Hence,
depending on the parametrization used, we can define a conservative
limit of about 1 s on the average period dispersion between the optimal
model and the observations above which a seismic solution becomes
essentially meaningless, considering that it could have reproduced a
random period spectrum equally well. This finding, although needing
to be tuned to each specific analysis for a more precise quantitative
statement, is of importance to judge the reliability of past and future
asteroseismic inferences of white dwarf stellar properties.

\section{TESTING THE CORE PARAMETRIZATION WITH A STRATIFICATION FROM EVOLUTIONARY MODELS}

In this section, we investigate the flexibility of the new
parametrization and its capability to reproduce the chemical stratification
derived from standard evolutionary models. For this test, we start with a
static model using the chemical profiles in the core derived from
Maurizio Salaris' evolutionary models. The core profile is obtained by
interpolating between the available profiles derived for different
masses by Salaris et al. (2008). The resulting core profile is then
smoothed out to remove as much as possible the numerical noise still
unfortunately present in the resulting curve. This reference model
has the global parameters listed in Table \ref{tab3-12}.

The resulting chemical profiles as well as the Brunt-V$\ddot{\rm
 a}$is$\ddot{\rm a}$l$\ddot{\rm a}$ and Lamb frequencies for $\ell=1$ and
 $\ell=2$ are illustrated in Figure 14. We can still observe some
 noisy features in the Brunt-V\"ais\"al\"a frequency which are most
 probably residuals from the numerical noise that was propagated during
 the evolution in the pre-white dwarf phases.

Table \ref{tab3-13} (first column) presents the ``observed" periods and mode
identification chosen from the reference model. A set of ten periods were
selected: five dipole modes with consecutive values of the radial order $k$
from 2 to 6, and a similar group of five quadrupole modes with $k$ ranging
from 3 to 7. These modes were chosen in order to best represent ZZ Ceti
stars close to the blue edge of the instability strip. Also, we made sure
that our choice would include modes that probe the deep interior, a property
that is confirmed in Figure 15 showing the weight functions of the
selected modes. These modes are the equivalent of the deeply confined
modes discovered in Ross 548 (Giammichele et al. 2016), and their
presence in the ``observational'' data considered here is of course
necessary for testing our different core parametrizations. The ultimate
test is then to try to recover the global parameters and structure of the
reference model (the white dwarf defined with a core composition
derived from evolution calculations) from the pulsations periods, with
chemical profiles in the core defined by the different levels of
sophistication of our parametrization. It is important to realize that this
experiment differs from the tests presented in prior sections in that
the structure to recover cannot be exactly reproduced even by our most
flexible 2-transition parametrization of the core. Only the main features
of the chemical profiles can be approximated. This test is therefore
more stringent for evaluating the overall robustness of the method.

The first experiment consists of using the simplest parametrized
static model we have, the varying homogeneous core, to search for a
best-fit solution matching the ``observed" periods.
The search range used in the optimization process for every parameter
of interest is indicated in Table \ref{tab3-12}.
The resulting global parameters for this first test are given in Table
\ref{tab3-14}, and the associated chemical profiles are illustrated in
Figure 16 (black curves) relative to the reference model
(red curves). The $S^{2}$-value of 17.5 obtained for the best solution
is somewhat high and has to be considered carefully as it reaches some
of the values obtained in the random tests described in the previous
section. A close examination of Table \ref{tab3-13} reveals, however,
that a substantial part of this large value of $S^{2}$ is due to the
particularly poor mismatch between the theoretical and ``observed''
period for the mode with $\ell = 1$ and $k = 6$. Otherwise, the period fits
for the nine other modes are quite acceptable. We point out that four
out of the ten modes of interest have not been identified correctly in
terms of their defining indices $\ell$ and $k$. And, of course, the
detailed composition profile of our reference model cannot be reproduced
in the context of this first test with homogeneous cores. This is well
illustrated in Figure 16. Despite these shortcomings, and
perhaps somewhat surprisingly, the hypothesis of an homogeneous core
still leads to quite reasonable estimates of the global structural
parameters of the reference model as can be observed in Table
\ref{tab3-14}. We recall in
this context that the hypothesis of homogeneous cores has proven itself
a good approximation in the cases of GD 165 -- in particular -- and Ross
548 for which the six observed periods were well fitted for each star
(Giammichele et al. 2016). We also remind the reader here that the current
set of ten periods was specifically chosen to probe deep into the
reference star, since we were mainly concerned in showing the
improvement brought up by the new parametrization.

A substantial gain on all fronts is obtained by using parametrized cores
defined by three parameters and corresponding to a chemical profile
with a single transition zone. This parametrization, a particular
case of the complete two-transition model, is obtained by setting $t1_o$ and
$t2_o$ to zero, $\Delta t2$ to a non-zero value and $t2$ fixed at $-$2.0,
far enough that it does not interfere within the range of $t1$. Table
\ref{tab3-14} reveals a clear improvement of the merit
function, from $S^2$ = 17.5 to 10.4, and the average dispersion in
period, $<\Delta P>$, has dropped below 1 s. In both cases, much of the
deviation is dominated by a particularly poor mismatch, this time
associated with the $\ell = 2$ and $k = 7$ mode (see Table \ref{tab3-13}).
The period fits are quite acceptable for the nine other modes. In
addition, the mode identification is this time perfectly recovered and the
global parameters of the solution are noticeably closer to the ones
from the reference model. A look at Figure 17 also reveals a
clear improvement in the resemblance of the Brunt-V\"ais\"al\"a
frequency with the one of the reference model. The
first core feature is almost perfectly positioned at the first
transition, along with the two chemical transitions between the core and
the helium-rich mantle, and between the helium layers and the hydrogen
envelope. The central values of carbon and oxygen are somehow inverted,
a rather striking result, but since this inversion does not reflect
strongly on the Brunt-V\"ais\"al\"a frequency (inverting the values of $X(C)$
and $X(O)$ leads to a very similar profile), the sensitivity of the
period-fitting optimization process to this parameter is certainly weak.
Therefore it most probably cannot be well constrained in view of the still
relatively high $S^2$-value of the best-fit solution. The latter is almost
entirely due to the absence of the second feature in the core corresponding
to the second drop in the oxygen profile, but the chemical stratification
is overall quite well reproduced with the one-transition parametrization.

At this stage, the need for a two-transition parametrization should be
clear. The Salaris profile presents two major drops in the oxygen
stratification, so it is most natural to try to recover the reference
model with a  two-transition core parametrization. The advantage of a
more flexible parametrization at this level is expected to be visible
through an improved value of $S^2$ and, therefore, a better match
between the ``observed" periods of the reference model and the computed
theoretical periods. The question remaining is how accurately the
periods and the chemical profiles in the core can be reproduced.

As above, the results of this third test are presented in Table
\ref{tab3-14}. The $S^2$-value of 0.1 and the average
$<\Delta P>$ of 0.09 s obtained for the best-fit model are,
respectively, two and one orders of magnitude smaller than previously
achieved with the variable homogeneous core and the one-transition
parametrization. This corresponds to an impressive improvement, which
demonstrates that our prescription can indeed mimic quite well, although
not perfectly, the complex chemical stratification expected in the cores
of white dwarf stars. By inspecting the theoretical periods matched to
the periods of the reference model in Table \ref{tab3-13}, we note that
the period 376.01 s is wrongly identified. However, a closer inspection
of the model shows that the correct identification would have only very
slightly increased the merit function $S^2$, because the value of the
$\ell=1, k=6$ mode period is almost identical to the $\ell=2, k=12$ mode
period. Beside this very minor problem, the global atmospheric parameters
of the reference star are significantly better constrained than before,
save for the helium layer mass.

Finally, inspecting Fig. 18 shows that the two-transition
prescription for the oxygen profile matches rather well the original
oxygen profile taken from Salaris et al. (2008). The central homogeneous
values of the oxygen and carbon mass fractions are correctly recovered, and
the two transitions are perfectly placed. The layering of the helium and
the hydrogen envelope is well superimposed on the chemical profiles of
the reference model. The exact shapes of the first and, mostly, the second
feature of the Brunt-V\"ais\"al\"a frequency
corresponding to the reference model are not so well matched, but
this does not seem a major problem as the periods and the overall model
are finely reproduced. The shape of these inner transitions most likely
have a smaller impact than other structures on the period spectrum
and would require higher precision period fits to be more accurately
reproduced.

In summary, with this test, the flexibility of the new two-transition core
parametrization was explored and proven to be sufficient, as well as
necessary, to recover sufficiently accurately the details of the complex
chemical stratification expected from evolutionary calculations. Global
parameters can be estimated fairly reliably from the approximation of
homogeneous cores, but this has to be handled with care, especially when
the modes probe deeper than the envelope of
the star. To best reproduce the core chemical profile, the two-transition
parametrization is necessary and cannot be replaced by the simpler
one-transition setup.

\section{CONCLUSION}

In this paper, we introduced and tested a new parametrization that can
replicate smoothly one or two steep "drops" in the chemical profile of
oxygen in the core of a white dwarf model in order to better reproduce the
shape of the stratification usually predicted from evolutionary models.
The flexibility of the new two-transition core
parametrization was explored and proven to be both necessary and sufficient
for recovering the structure expected from evolutionary calculations. To be clear, the evolutionary structures are not reproduced to perfection, but we showed that the method is amply sufficient for testing the most important features of the chemical profile of a white dwarf through asteroseismology. And if need be, in special cases, the flexibility of our approach could also be improved.
This new parametrization will be of high interest in upcoming seismic
studies of pulsating white dwarfs, particularly those with available
{\sl Kepler} and {\sl K2} data, which we plan to analyze thoroughly
with our optimization tools developed for asteroseismology.

For the future, we retain the following lessons from the current work.
Based on tests generated with random sequences of periods, we can roughly
define a conservative threshold of $\sim 1$ s for the average period
dispersion of the best-fit solution above which results from seismic analyses
should not be considered as very robust (i.e., statistically significant).
We showed that our seismic method, based on parametrized models in
hydrostatic equilibrium, is perfectly capable of retrieving stellar and
shape parameters from a reference star with different levels of precision
in the "observational" data. If the limiting factor was purely
observational, i.e., only governed by the errors in measuring the pulsation
frequencies (periods), stellar and shape parameters could be recovered to a
remarkable precision. However, although our asteroseismic method gives
robust, precise, and accurate constraints on both stellar or shape
parameters, the limiting factor is usually not the precision of the
observations themselves, but the uncertainties associated with the
constitutive physics involved in current stellar models which still
dominate the error budget. The latter need to be addressed in order to go
beyond current achievements and improve overall the quality of the seismic
fits toward solutions that approach the very stringent limit of observational
measurement errors.

\acknowledgments
This work was supported in part by the NSERC Canada through a doctoral
fellowship awarded to Noemi Giammichele, and through a research grant
awarded to Gilles Fontaine. The latter also acknowledges the
contribution of the Canada Research Chair Program. St\'ephane Charpinet
acknowledges financial support from ``Programme National de Physique
Stellaire'' (PNPS) of CNRS/INSU, France, and from the Centre National
d'\'Etudes Spatiales (CNES, France). This work was granted access to the
HPC resources of CALMIP under allocation number 2015-p0205.

\clearpage
\begin{table*}[!h]
\begin{center}
\caption{Parameters of the reference star as well as the search range
  for the optimization with a variable homogeneous core.}\label{tab3-1}
\begin{tabular}{clc}
  \hline
  Parameter & Value & Search range \\
  \hline \hline
  $T_{\rm eff}$ (K) & 12,000 & 11,000 - 13,000 \\
  Log $g$ & 8.00 & 7.80 - 8.20 \\
  $D$(H) & $-$5.0 & $-$9.0 - $-$4.0 \\
  $D$(He) & $-$3.0 & $-$4.0 - $-$1.5 \\
  Total mass ($M_{\rm \odot}$) & 5.9479x10$^{-1}$ & ... \\
  Radius ($R_{\rm \odot}$) & 1.2765x10$^{-2}$ & ... \\ \hline
  Core oxygen fraction ($\%$) & 50 & 0 - 100 \\ \hline

\end{tabular}
\end{center}
\end{table*}

\clearpage
\begin{table*}[!h]
\begin{center}
\caption{Selected and modified periods for the reference star with a variable homogeneous core.}\label{tab3-2}
\begin{tabular}{c|c|c|c|c|c}
  \hline
  $\ell$ & $k$ & Period (s) & $\sigma$ = 10 $\mu$Hz & $\sigma$ = 0.1 $\mu$Hz & $\sigma$ = 0.001 $\mu$Hz\\
  \hline\hline
  1 & 1 & 151.53791917 & ... & ... & ...\\
  \textbf{1} & \textbf{2} & \textbf{227.016706244} & \textbf{226.98946604} & \textbf{227.01643381} & \textbf{227.01670352}\\
  \textbf{1} & \textbf{3} & \textbf{313.807911837} & \textbf{313.66203636} & \textbf{313.80645241} & \textbf{313.80789724}\\
  \textbf{1} & \textbf{4} & \textbf{325.278884448} & \textbf{325.58422098} & \textbf{325.28193498} & \textbf{325.27891495}\\
  \textbf{1} & \textbf{5} & \textbf{379.026095735} & \textbf{380.55088833} & \textbf{379.04128317} & \textbf{379.02624760}\\
  \textbf{1} & \textbf{6} & \textbf{436.100785261} & \textbf{433.16916032} & \textbf{436.07127260} & \textbf{436.10049011}\\
  1 & 7 & 459.45372755 &...&...& ...\\\hline \hline
  2 & 2 & 131.65045899 &...&...& ...\\
  \textbf{2} & \textbf{3} & \textbf{181.920837778} & \textbf{181.85326019} & \textbf{181.92016175} & \textbf{181.92083102}\\
  \textbf{2} & \textbf{4} & \textbf{212.206268924} & \textbf{212.35274848} & \textbf{212.20773272} & \textbf{212.20628356}\\
  \textbf{2} & \textbf{5} & \textbf{224.636765373} & \textbf{224.25083274} & \textbf{224.63289947} & \textbf{224.63672671}\\
  \textbf{2} & \textbf{6} & \textbf{260.460240785} & \textbf{259.40908188} & \textbf{260.44968703} & \textbf{260.46013524}\\
  \textbf{2} & \textbf{7} & \textbf{282.114308484} & \textbf{283.05460137} & \textbf{282.12368049} & \textbf{282.11440220}\\
  2 & 8 & 306.14572754 &...&...& ...\\
  2 & 9 & 336.77469004 &...&...& ...\\
  2 & 10& 357.86963676 &...&...& ...\\
  2 & 11& 387.57588290 &...&...& ...\\
  2 & 12& 414.11284901 &...&...& ...\\
  2 & 13& 437.72710109 &...&...& ...\\
  2 & 14& 467.36842954 &...&...& ...\\
  2 & 15& 491.40529857 &...&...& ...\\
  \hline
\end{tabular}
\end{center}
\end{table*}

\clearpage
\begin{sidewaystable}
\begin{center}
\caption{Results from the three optimizations searching for the reference star
  with a variable homogeneous core and with different levels of
  precision: current fit precision, ground-based data and $Kepler$ data
  precision.}\label{tab3-3}
\begin{tabular}{lccc}
\hline
   & Test case 1 & Test case 2 & Test case 3 \\
   & $\mu = 0$ , $\sigma = 10$ $\mu$Hz & $\mu = 0$ , $\sigma = 0.1$ $\mu$Hz & $\mu = 0$ , $\sigma = 0.001$ $\mu$Hz \\ \hline \hline
  $S^{2}$ & 9.1 & 9.1 x 10$^{-4}$ & 9.2 x 10$^{-8}$ \\
  $<\Delta\nu>$ ($\mu$Hz) & 8.29 & 0.083 & 0.0083 \\
  $<\Delta X/X>$ (\%) & 0.24 & 2.35 x 10$^{-3}$ & 2.37 x 10$^{-5}$ \\ \hline
  $T_{\rm eff}$ (K) & 11,970 $\pm$ 140 (0.3 $\%$) & 12,000.5 $\pm$ 4.0 (0.0042 $\%$) & 11,999.99 $\pm$ 0.06 (0.0001 $\%$) \\
  Log g & 8.008 $\pm$ 0.030 (0.1 $\%$) & 7.9999 $\pm$ 0.0008 (0.0012 $\%$) & 8.00000 $\pm$ 0.00001 (0.0001 $\%$) \\
  $D$(H) & $-$5.03 $\pm$ 0.15 (0.6 $\%$) & $-$5.000 $\pm$ 0.002 (0.02
   $\%$) & $-$5.00000 $\pm$ 0.00004 (0.0002 $\%$)\\
  $D$(He) & $-$3.07 $\pm$ 0.10 (2.3 $\%$) & $-$3.000 $\pm$ 0.003 (0.03 $\%$) & $-$3.00001 $\pm$ 0.00004 (0.0003 $\%$) \\
  Total mass ($M_{\rm \odot}$) & 0.60 $\pm$ 0.02 (0.9 $\%$) & 0.5948 $\pm$ 0.0005 (0.002 $\%$) & 0.594791 $\pm$ 0.000006 (0.0002 $\%$) \\
  Radius ($R_{\rm \odot}$) & 0.0127 $\pm$ 0.0003 (0.5 $\%$) & 0.01277 $\pm$ 0.00001 (0.04 $\%$) & 0.01276541 $\pm$ 0.00000008 (0.003 $\%$) \\ \hline
  Core oxygen fraction ($\%$) & 53 $\pm$ 8 (6 $\%$) & 50.0 $\pm$ 0.3 (0.2 $\%$) & 50.001 $\pm$ 0.003 (0.002 $\%$) \\ \hline
\end{tabular}
\end{center}
\end{sidewaystable}

\clearpage
\begin{table*}[!h]
\begin{center}
\caption{Parameters of the reference star and search range for the optimization
procedure using models with a one-transition core.}\label{tab3-4}
\begin{tabular}{clc}
  \hline
  Parameter & Value  & Search range \\
  \hline \hline
  $T_{\rm eff}$ (K) & 12,000 & 11,000 - 13,000 \\
  Log $g$ & 8.00 & 7.80 - 8.20 \\
  $D$(H) & $-$5.0 & $-$9.0 - $-$4.0 \\
  $D$(He) & $-$3.0 & $-$4.0 - $-$1.5 \\
  Total mass ($M_{\rm \odot}$) & 5.9593x10$^{-1}$ & ... \\
  Radius ($R_{\rm \odot}$) & 1.2778x10$^{-2}$ & ... \\ \hline
  Core oxygen fraction ($\%$) & 70 & 0 - 100 \\
  $t_{1}$ & $-$0.3 & $-$0.50  -  $-$0.15\\
  $\Delta t_1$ & 0.06  &  0.010  -  0.10\\
  \hline

\end{tabular}
\end{center}
\end{table*}

\clearpage
\begin{table*}[!h]
\begin{center}
\caption{Selected and perturbed periods for the reference star with the one-transition core.}\label{tab3-5}
\begin{tabular}{c|c|c|c|c|c}
  \hline
  l & k & Period (s) & $\sigma$ = 10 $\mu$Hz & $\sigma$ = 0.1 $\mu$Hz & $\sigma$ = 0.001 $\mu$Hz\\
  \hline\hline
  1 & 1 & 151.75446102 & ... & ... & ... \\
  \textbf{1} & \textbf{2} & \textbf{216.81536572} & \textbf{216.79051854} & \textbf{216.81511722} & \textbf{216.81536323}\\
  \textbf{1} & \textbf{3} & \textbf{233.92848154} & \textbf{233.08021901} & \textbf{233.91996835} & \textbf{233.92839640}\\
  \textbf{1} & \textbf{4} & \textbf{315.62545841} & \textbf{315.91293268} & \textbf{315.62833057} & \textbf{315.62548713}\\
  \textbf{1} & \textbf{5} & \textbf{372.35739163} & \textbf{373.82889673} & \textbf{372.37204934} & \textbf{372.35753820}\\
  \textbf{1} & \textbf{6} & \textbf{407.89150141} & \textbf{405.32576007} & \textbf{407.86568322} & \textbf{407.89124321}\\
  1 & 7 & 456.09585372 & ... & ... & ...\\\hline \hline
  2 & 2 & 133.59921893 & ... & ... & ...\\
  \textbf{2} & \textbf{3} & \textbf{151.31237295} & \textbf{151.26561948} & \textbf{151.31190527} & \textbf{151.31236828}\\
  \textbf{2} & \textbf{4} & \textbf{182.34310127} & \textbf{182.45124394} & \textbf{182.34418206} & \textbf{182.34311208}\\
  \textbf{2} & \textbf{5} & \textbf{216.56591828} & \textbf{216.20719723} & \textbf{216.56232518} & \textbf{216.56588235}\\
  \textbf{2} & \textbf{6} & \textbf{241.19856963} & \textbf{241.88556292} & \textbf{241.20542025} & \textbf{241.19863814}\\
  \textbf{2} & \textbf{7} & \textbf{264.34846623} & \textbf{264.24494250} & \textbf{264.34743059} & \textbf{264.34845587}\\
  2 & 8 & 299.27669790 & ... & ... & ...\\
  2 & 9 & 328.49590688 & ... & ... & ...\\
  2 & 10& 350.35377602 & ... & ... & ...\\
  2 & 11& 385.35347602 & ... & ... & ...\\
  2 & 12& 407.54276181 & ... & ... & ...\\
  2 & 13& 422.40023701 & ... & ... & ...\\
  2 & 14& 447.50516837 & ... & ... & ...\\
  2 & 15& 475.14880300 & ... & ... & ...\\
  \hline
\end{tabular}
\end{center}
\end{table*}

\clearpage
\begin{sidewaystable}
\begin{center}
\caption{Results from the three optimizations from the reference star
  with a one-transition core and with different levels of precision:
  current fit precision, ground-based data and $Kepler$ data
  precision.}\label{tab3-6}
\begin{tabular}{cccc}
\hline
   & Test case 1 & Test case 2 & Test case 3 \\
   & $\mu = 0$ , $\sigma = 10$ $\mu$Hz & $\mu = 0$ , $\sigma = 0.1$ $\mu$Hz & $\mu = 0$ , $\sigma = 0.001$ $\mu$Hz \\ \hline\hline
  $S^{2}$ & 3.1 & 3.0 x 10$^{-4}$ & 6.1 x 10$^{-7}$ \\
  $<\Delta\nu>$ ($\mu$Hz) & 7.29 & 0.083 & 0.0043 \\
  $<\Delta X/X>$ (\%) & 0.17 & 1.90 x 10$^{-3}$ & 9.52 x 10$^{-5}$ \\ \hline
  $T_{\rm eff}$ (K) & 11,770 $\pm$ 120 ( 1.9 $\%$) & 11,998.5 $\pm$ 3.0 ( 0.01$\%$) & 11999.96 $\pm$ 0.05 ( 0.0003$\%$) \\
  Log $g$ & 8.045 $\pm$ 0.022 ( 0.6 $\%$) & 8.0006 $\pm$ 0.0005 ( 0.008$\%$) & 8.000005 $\pm$ 0.000008 ( 0.0001$\%$) \\
  $D$(H) & $-$5.17 $\pm$ 0.10 ( 3.4 $\%$) & $-$5.002 $\pm$ 0.02 ( 0.04$\%$) & $-$5.00002 $\pm$ 0.0000025 ( 0.0004$\%$)\\
  $D$(He) & $-$3.15 $\pm$ 0.06 ( 5.0 $\%$) & $-$3.002 $\pm$ 0.0012 ( 0.07$\%$) & $-$3.00002 $\pm$ 0.00002 ( 0.0007$\%$) \\
  Total mass ($M_{\rm \odot}$) & 0.62 $\pm$ 0.01 ( 4.0 $\%$) & 0.5962 $\pm$ 0.0003 ( 0.05$\%$) & 0.595930 $\pm$ 0.000003 ( 0.0002$\%$) \\
  Radius ($R_{\rm \odot}$) & 0.0124 $\pm$ 0.0002 ( 3.0 $\%$) & 0.012772 $\pm$ 0.000004 ( 0.05$\%$) & 0.0127775 $\pm$ 0.0000001 ( 0.004$\%$) \\ \hline
  Core oxygen fraction ($\%$) & 66 $\pm$ 3 ( 5.7 $\%$) & 0.6998 $\pm$ 0.0015 ( 0.03$\%$) & 0.70000 $\pm$ 0.00002 ( 0.001$\%$) \\
  $t_{1}$ & $-$0.300 $\pm$ 0.018 ( 0.3 $\%$) & $-$0.3003 $\pm$ 0.0006 ( 0.1 $\%$) & $-$3.00000 $\pm$ 0.000008 ( 0.0003$\%$) \\
  $\Delta t_{1}$ & 0.072 $\pm$ 0.015 ( 20 $\%$) & 0.0607 $\pm$ 0.0015 ( 1.2 $\%$) & 0.060015 $\pm$ 0.000013( 0.03$\%$) \\ \hline
\end{tabular}
\end{center}
\end{sidewaystable}

\clearpage
\begin{table*}[!h]
\begin{center}
\caption{Parameters of the reference star and search range
  for the optimization using a two-transition core.}\label{tab3-7}
\begin{tabular}{clc}
  \hline
  Parameter & Value & Search range  \\
  \hline \hline
  $T_{\rm eff}$ (K) & 12,000 & 11,000 - 13,000\\
  Log $g$ & 8.00 & 7.80 - 8.20 \\
  $D$(H) & $-$5.0 & $-$9.0 - $-$4.0 \\
  $D$(He) & $-$3.0 & $-$4.0 - $-$1.5 \\
  Total mass ($M_{\rm \odot}$) & 5.9521x10$^{-1}$ & ... \\
  Radius ($R_{\rm \odot}$) & 1.2770x10$^{-2}$ & ... \\ \hline
  Core oxygen fraction ($\%$) & 70 & 0 - 100 \\
  $t_{1}$ & $-$0.3 & $-$0.50  -  $-$0.15\\
  $\Delta t_1$ & 0.06 & 0.010  -  0.10 \\
  $t_{1}$ oxygen ($\%$) & 40 & 0 - 100  \\
  $t_{2}$ & $-$1.5 & $-$2.00 -   $-$0.60 \\
  $\Delta t_2$ & 0.08 & 0.010  -  0.10 \\
  $t_{2}$ oxygen ($\%$) & 20 & 0 - 100  \\
  \hline
\end{tabular}
\end{center}
\end{table*}

\clearpage
\begin{table*}[!h]
\begin{center}
\caption{Selected and modified periods for the reference star with the
  two-transition core.}\label{tab3-8}
\begin{tabular}{c|c|c|c|c|c}
  \hline
  $\ell$ & $k$ & Period (s) & $\sigma$ = 10 $\mu$Hz & $\sigma$ = 0.1 $\mu$Hz & $\sigma$ = 0.001 $\mu$Hz\\
  \hline\hline
  1 & 1 & 151.653409370 & ... & ... & ... \\
  \textbf{1} & \textbf{2} & \textbf{228.787742059} & \textbf{228.76007521} & \textbf{228.78746536} & \textbf{228.78773929}\\
  \textbf{1} & \textbf{3} & \textbf{264.030696374} & \textbf{263.92742134} & \textbf{264.02966322} & \textbf{264.03068604}\\
  \textbf{1} & \textbf{4} & \textbf{315.524781925} & \textbf{315.81207274} & \textbf{315.52765225} & \textbf{315.52481063}\\
  \textbf{1} & \textbf{5} & \textbf{369.012046557} & \textbf{370.45717843} & \textbf{369.02644206} & \textbf{369.01219051}\\
  \textbf{1} & \textbf{6} & \textbf{399.888135738} & \textbf{397.42178861} & \textbf{399.86332075} & \textbf{399.88788757}\\
  1 & 7 & 455.034684809 & ... & ... & ...\\\hline \hline
  2 & 2 & 133.361713913 & ... & ... & ...\\
  \textbf{2} & \textbf{3} & \textbf{178.011613603} & \textbf{177.94690859} & \textbf{178.01096632} & \textbf{178.01160713}\\
  \textbf{2} & \textbf{4} & \textbf{183.545746313} & \textbf{183.65532062} & \textbf{183.54684141} & \textbf{183.54575726}\\
  \textbf{2} & \textbf{5} & \textbf{216.944553720} & \textbf{216.58457827} & \textbf{216.94094804} & \textbf{216.94451766}\\
  \textbf{2} & \textbf{6} & \textbf{238.597497023} & \textbf{237.71509923} & \textbf{238.58864062} & \textbf{238.59740846}\\
  \textbf{2} & \textbf{7} & \textbf{263.710718836} & \textbf{264.53215547} & \textbf{263.71890795} & \textbf{263.71080072}\\
  2 & 8 & 298.974939191 & ... & ... & ...\\
  2 & 9 & 327.792644523 & ... & ... & ...\\
  2 & 10& 348.111904002 & ... & ... & ...\\
  2 & 11& 380.831879546 & ... & ... & ...\\
  2 & 12& 405.577093095 & ... & ... & ...\\
  2 & 13& 427.351916311 & ... & ... & ...\\
  2 & 14& 445.672564712 & ... & ... & ...\\
  2 & 15& 473.357290231 & ... & ... & ...\\
  \hline
\end{tabular}
\end{center}
\end{table*}

\clearpage
\begin{sidewaystable}
\begin{center}
\caption{Results of the three optimizations for a reference star
  with a two-transition core and assuming different levels of precision:
  current fit, ground-based data and $Kepler$ data precision.}\label{tab3-9}
\begin{tabular}{cccc}
\hline
   & Test case 1 & Test case 2 & Test case 3 \\
   & $\mu = 0$ $\sigma = 10$ $\mu$Hz & $\mu = 0$ $\sigma = 0.1$ $\mu$Hz & $\mu = 0$ $\sigma = 0.001$ $\mu$Hz \\ \hline\hline
  $S^{2}$ & 2.8 & 2.7 x 10$^{-4}$ & 2.9 x 10$^{-8}$ \\
  $<\Delta\nu>$ ($\mu$Hz) & 5.41 & 0.061 & 0.00072 \\
  $<\Delta X/X>$ (\%) & 0.14 & 1.50 x 10$^{-3}$ & 1.60 x 10$^{-5}$ \\ \hline
  $T_{\rm eff}$ (K) & 11880 $\pm$ 90 ( 1 $\%$) & 11999.0 $\pm$ 1.5 ( 0.008 $\%$) & 11999.98 $\pm$ 0.05 ( 0.0002$\%$) \\
  Log $g$ & 8.037 $\pm$ 0.020 ( 0.5 $\%$) & 8.00052 $\pm$ 0.00015 ( 0.007 $\%$) & 8.000005 $\pm$ 0.000013 ( 0.001$\%$) \\
  $D$(H) & $-$5.15 $\pm$ 0.08 ( 3 $\%$) & $-$5.002 $\pm$ 0.0012 ( 0.04 $\%$) & $-$5.00002 $\pm$ 0.00004 ( 0.0004$\%$)\\
  $D$(He) & $-$3.15 $\pm$ 0.06 ( 5 $\%$) & $-$3.002 $\pm$ 0.001 ( 0.07  $\%$) & $-$3.00005 $\pm$ 0.00003 ( 0.002 $\%$) \\
  Total mass ($M_{\rm \odot}$) & 0.62 $\pm$ 0.01 ( 4 $\%$) & 0.5955 $\pm$ 0.0002 ( 0.05 $\%$) & 0.595209 $\pm$ 0.000006 ( 0.0002 $\%$) \\
  Radius ($R_{\rm \odot}$) & 0.0124 $\pm$ 0.0002 ( 3 $\%$) & 0.012765 $\pm$ 0.000002 ( 0.04 $\%$) & 0.0127698 $\pm$ 0.0000001 ( 0.002$\%$) \\ \hline
  Core oxygen fraction ($\%$) & 70 $\pm$ 5 ( 1 $\%$) & 70.0 $\pm$ 0.1 ( 0.1 $\%$) & 70.005 $\pm$ 0.002 ( 0.007 $\%$) \\
  $t_{1}$ & $-$0.310 $\pm$ 0.013 (   3 $\%$) & $-$0.3004 $\pm$ 0.0003 ( 0.1 $\%$) & $-$0.29989 $\pm$ 0.00001 ( 0.04 $\%$) \\
  $\Delta t_{1}$ & 0.044 $\pm$ 0.023 (  27 $\%$) & 0.0607 $\pm$ 0.0010 ( 1.2 $\%$) & 0.05956 $\pm$ 0.00004 ( 0.07 $\%$) \\
  $t_{1}$ oxygen ($\%$) & 45 $\pm$ 5 (  13 $\%$) & 40 $\pm$ 0.2 ( 0.3 $\%$) & 40.038 $\pm$ 0.003 ( 0.1 $\%$) \\
  $t_{2}$ & $-$1.61 $\pm$ 0.10 (  7 $\%$) & $-$1.502 $\pm$ 0.002 ( 0.1 $\%$) & $-$1.5009 $\pm$ 0.0001 ( 0.06 $\%$) \\
  $\Delta t_{2}$ & 0.057 $\pm$ 0.028 (  29 $\%$) & 0.0803 $\pm$ 0.0015 ( 0.4 $\%$) & 0.08008 $\pm$ 0.00047 ( 0.1 $\%$) \\
  $t_{2}$ oxygen ($\%$) & 15 $\pm$ 4 (  25 $\%$) & 20.1 $\pm$ 0.1 ( 0.5 $\%$) & 19.96 $\pm$ 0.01 ( 0.2 $\%$) \\ \hline
\end{tabular}
\end{center}
\end{sidewaystable}

\clearpage
\begin{table*}[!h]
\begin{center}
\caption{Parameters of the reference star and search range
  for the optimization using a two-transition core with the presence of the evolutionary predicted triple transition of helium, carbon and oxygen.}\label{tab3-10-1}
\begin{tabular}{clc}
  \hline
  Parameter & Value & Search range  \\
  \hline \hline
  $T_{\rm eff}$ (K) & 12,000 & 11,000 - 13,000\\
  Log $g$ & 8.00 & 7.80 - 8.20 \\
  $D$(H) & $-$4.2 & $-$9.0 - $-$4.0 \\
  $D$(He) & $-$1.8 & $-$4.0 - $-$1.5 \\
  Total mass ($M_{\rm \odot}$) & 6.0541x10$^{-1}$ & ... \\
  Radius ($R_{\rm \odot}$) & 1.2879x10$^{-2}$ & ... \\ \hline
  Core oxygen fraction ($\%$) & 72 & 0 - 100 \\
  $t_{1}$ & $-$0.5 & $-$0.80  -  $-$0.15\\
  $\Delta t_1$ & 0.2 & 0.010  -  0.30 \\
  $t_{1}$ oxygen ($\%$) & 40 & 0 - 100  \\
  $t_{2}$ & $-$1.6 & $-$2.00 -   $-$0.60 \\
  $\Delta t_2$ & 0.25 & 0.010  -  0.30 \\
  $t_{2}$ oxygen ($\%$) & 40 & 0 - 100  \\
  \hline
\end{tabular}
\end{center}
\end{table*}

\clearpage
\begin{table*}[!h]
\begin{center}
\caption{Selected and modified periods for the reference star with the
  two-transition core with the presence of the evolutionary predicted triple transition of helium, carbon and oxygen.}\label{tab3-10-2}
\begin{tabular}{c|c|c|c}
  \hline
  $\ell$ & $k$ & Period (s) & $\sigma$ = 0.001 $\mu$Hz\\
  \hline\hline
  \textbf{1} & \textbf{1} & \textbf{120.452983850} & \textbf{120.45298857} \\
  \textbf{1} & \textbf{2} & \textbf{189.003923848} & \textbf{189.00386827} \\
  \textbf{1} & \textbf{3} & \textbf{251.065310769} & \textbf{251.06532894} \\
  \textbf{1} & \textbf{4} & \textbf{288.146522125} & \textbf{288.14660990} \\
  \textbf{1} & \textbf{5} & \textbf{308.669247080} & \textbf{308.66909922} \\
  1 & 6 & 362.527251742 & ... \\
  1 & 7 & 407.584693069 & ... \\
  1 & 8 & 465.047565605 & ... \\
  1 & 9 & 495.871289508 & ... \\
  \hline \hline
  \textbf{2} & \textbf{2} & \textbf{111.787448206} & \textbf{111.78744565} \\
  \textbf{2} & \textbf{3} & \textbf{145.458275720} & \textbf{145.45825951} \\
  \textbf{2} & \textbf{4} & \textbf{168.756963139} & \textbf{168.75696163} \\
  \textbf{2} & \textbf{5} & \textbf{200.954894937} & \textbf{200.95494249} \\
  \textbf{2} & \textbf{6} & \textbf{211.835583311} & \textbf{211.83557666} \\
  2 & 7 & 236.460839397 & ... \\
  2 & 8 & 270.690939915 & ... \\
  2 & 9 & 294.553272794 & ... \\
  2 & 10& 320.151012113 & ... \\
  2 & 11& 341.454056864 & ... \\
  2 & 12& 364.828511100 & ... \\
  2 & 13& 383.707614573 & ... \\
  2 & 14& 406.414434402 & ... \\
  2 & 15& 434.387667041 & ... \\
  2 & 16& 456.565643881 & ... \\
  2 & 17& 478.331486497 & ... \\
  \hline
\end{tabular}
\end{center}
\end{table*}

\clearpage
\begin{sidewaystable}
\begin{center}
\caption{Results of the optimization for a reference star
  with a two-transition core with the presence of the evolutionary predicted triple transition of helium, carbon and oxygen and assuming $Kepler$ data precision.}\label{tab3-10-3}
\begin{tabular}{cc}
\hline
   & Test case 3  \\
   & $\mu = 0$ $\sigma = 0.001$ $\mu$Hz \\ \hline\hline
  $S^{2}$ &  4.0 x 10$^{-9}$ \\
  $<\Delta\nu>$ ($\mu$Hz) & 0.001 \\
  $<\Delta X/X>$ (\%)  & 1.6 x 10$^{-5}$ \\ \hline
  $T_{\rm eff}$ (K) & 11,999.69 $\pm$ 0.02 ( 0.003 $\%$) \\
  Log $g$ & 8.000028 $\pm$ 0.000003  ( 0.0004 $\%$) \\
  $D$(H)  & $-$4.20008 $\pm$ 0.00001 ( 0.002 $\%$)\\
  $D$(He) & $-$1.80005 $\pm$ 0.00001 ( 0.003 $\%$) \\
  Total mass ($M_{\rm \odot}$) & 0.605420 $\pm$ 0.000001 ( 0.002 $\%$) \\
  Radius ($R_{\rm \odot}$) & 0.01287854 $\pm$ 0.00000003 ( 0.002 $\%$) \\ \hline
  Core oxygen fraction ($\%$) & 72.0325 $\pm$ 0.0001 ( 0.05 $\%$) \\
  $t_{1}$ & $-$0.499775 $\pm$ 0.000002 ( 0.05 $\%$) \\
  $\Delta t_{1}$ & 0.19972 $\pm$ 0.00001 (  0.1 $\%$) \\
  $t_{1}$ oxygen ($\%$) & 40.0512 $\pm$ 0.0005 ( 0.1 $\%$) \\
  $t_{2}$ & $-$1.60010 $\pm$ 0.00002 ( 0.06 $\%$) \\
  $\Delta t_{2}$ & 0.2497 $\pm$ 0.0001 ( 0.1 $\%$) \\
  $t_{2}$ oxygen ($\%$) & 39.999 $\pm$ 0.003 ( 0.003 $\%$) \\ \hline
\end{tabular}
\end{center}
\end{sidewaystable}

\clearpage
\begin{table*}[!h]
\begin{center}
\caption{Pulsation spectra for the five test cases.}\label{tab3-10}
\begin{tabular}{ccccc}
  \hline
 Case 1 & Case 2 & Case 3 & Case 4 & Case 5 \\ \hline \hline
 109 s& 128 s& 111 s& 118 s& 113 s\\
 125 s& 173 s& 113 s& 148 s& 143 s\\
 148 s& 179 s& 136 s& 199 s& 193 s\\
 228 s& 183 s& 146 s& 207 s& 234 s\\
 282 s& 196 s& 281 s& 362 s& 247 s\\
 285 s& 209 s& 283 s& 363 s& 301 s\\
 329 s& 236 s& 323 s& 367 s& 326 s\\
 374 s& 377 s& 338 s& 389 s& 367 s\\
 400 s& 388 s& 399 s& 430 s& 450 s\\
 423 s& 486 s& 490 s& 455 s& 460 s\\ \hline
\end{tabular}
\end{center}
\end{table*}

\clearpage
\begin{sidewaystable}
\scriptsize
\begin{center}

\caption{Global fit properties obtained for each random period sequence and
each core parametrization.}\label{tab3-11}
\begin{tabular}{cccccccc}
\hline
 & Case 1 & Case 2 & Case 3 & Case 4 & Case 5 & Mean & Median \\ \hline \hline
  Hom. core:    & S$^{2}$ = 23.9& S$^{2}$ = 209.0 & S$^{2}$ = 212.0 & S$^{2}$ = 57.6 & S$^{2}$ = 15.2 & 103.5 & 57.6 \\
                & $< \Delta \nu >$ = 35.1 $\mu$Hz & $< \Delta \nu>$ = 90.6 $\mu$Hz & $<\Delta \nu>$ = 159.1 $\mu$Hz & $<\Delta \nu>$ = 41.6 $\mu$Hz & $<\Delta \nu>$ = 16.1 $\mu$Hz &  68.5  &  41.6  \\
                & $<\Delta P>$ = 1.6 s          & $<\Delta P>$ = 3.8 s          & $<\Delta P>$ = 3.9 s           & $<\Delta P>$ = 2.0 s          & $<\Delta P>$ = 0.9 s & 2.44     & 2.0  \\ \hline
  1 transition: & S$^{2}$ = 11.3 & S$^{2}$ = 22.0  & S$^{2}$ = 4.0   & S$^{2}$ = 31.1 & S$^{2}$ = 6.6  & 15.0  & 11.3  \\
                & $< \Delta \nu >$ = 10.5 $\mu$Hz & $< \Delta \nu>$ = 34.2 $\mu$Hz & $<\Delta \nu>$ = 12.7 $\mu$Hz & $<\Delta \nu>$ = 17.8 $\mu$Hz & $<\Delta \nu>$ = 7.6 $\mu$Hz &  16.6  &  12.7  \\
                & $<\Delta P>$ = 0.7 s          & $<\Delta P>$ = 1.2 s          & $<\Delta P>$ =  0.5s           & $<\Delta P>$ =  1.3 s          & $<\Delta P>$ = 0.6 s &  0.9    & 0.7  \\ \hline
  2 transitions:& S$^{2}$ = 6.6 & S$^{2}$ = 9.0   & S$^{2}$ = 1.9   & S$^{2}$ = 10.3 & S$^{2}$ = 3.7  & 6.3   & 6.6  \\
                & $< \Delta \nu >$ =  21.6   $\mu$Hz & $< \Delta \nu>$ = 18.0    $\mu$Hz & $<\Delta \nu>$ =  12.0   $\mu$Hz & $<\Delta \nu>$ =  10.9    $\mu$Hz & $<\Delta \nu>$ = 13.0 $\mu$Hz & 15.1 & 13.0 \\
                & $<\Delta P>$ = 0.5   s          & $<\Delta P>$ = 0.7   s          & $<\Delta P>$ =  0.3  s           & $<\Delta P>$ =  0.9  s          & $<\Delta P>$ =  0.5 s & 0.6  & 0.5  \\
  \hline
\end{tabular}
\end{center}
\end{sidewaystable}

\normalsize

\clearpage
\begin{table*}[!h]
\begin{center}
\caption{Parameters of the reference model.}\label{tab3-12}
\begin{tabular}{clc}
  \hline
  Parameter & Value & Search range  \\
  \hline \hline
  $T_{\rm eff}$ (K) & 12,000 & 11,000 - 13,000 \\
  Log $g$ & 8.00 & 7.80 - 8.20 \\
  $q$(H) & $-$4.04 & [$D$(H)] $-$9.0 - $-$4.0 \\
  $q$(He) & $-$2.10 & [$D$(He)] $-$4.0 - $-$1.5 \\ \hline
\end{tabular}
\end{center}
\end{table*}

\clearpage
\begin{sidewaystable}
\begin{center}
\caption{Periods from the reference model and from the resulting optimizations.}\label{tab3-13}
\begin{tabular}{|c|c|c||c|c|c||c|c|c||c|c|c|}
  \hline
  $\ell$ & $k$ & Chosen periods (s) & $\ell$ & $k$ & Var. hom. core (s) & $\ell$ & $k$ & 1-transition (s) & $\ell$ & $k$ & 2-transition (s) \\
  \hline \hline
  1 & 2 &206.429533263 &2 & 5 &207.80387574 &1 & 2 &205.91667143 &1 & 2 &206.30355349 \\
  1 & 3 &259.558241480 &1 & 3 &261.26271139 &1 & 3 &260.35665876 &1 & 3 &259.72976629 \\
  1 & 4 &285.787098098 &2 & 9 &285.94654263 &1 & 4 &285.73410324 &1 & 4 &285.88885831 \\
  1 & 5 &316.227894252 &2 &10 &315.89655226 &1 & 5 &314.83482318 &1 & 5 &316.31323695 \\
  1 & 6 &375.973051700 &1 & 6 &373.22825616 &1 & 6 &375.51411398 &2 & 12&376.01101225 \\ \hline
  2 & 3 &150.725138295 &2 & 3 &151.01496289 &2 & 3 &150.89406262 &2 & 3 &150.78465417 \\
  2 & 4 &177.098921306 &2 & 4 &178.19931882 &2 & 4 &178.35162669 &2 & 4 &177.05147091 \\
  2 & 5 &195.326732918 &1 & 2 &196.91534658 &2 & 5 &195.37498858 &2 & 5 &195.34102115 \\
  2 & 6 &218.924655751 &2 & 6 &217.81162674 &2 & 6 &217.73166428 &2 & 6 &218.75794933 \\
  2 & 7 &245.287135903 &2 & 7 &245.41225685 &2 & 7 &247.37133055 &2 & 7 &245.34295988 \\ \hline
\end{tabular}
\end{center}
\end{sidewaystable}

\clearpage
\begin{table*}
\begin{center}
\caption{Retrieved stellar parameters from the three
  optimizations.}\label{tab3-14}
\begin{tabular}{c|c|c|c}
\hline
Parameter & Var. hom. core & 1-transition & 2-transition \\
\hline \hline
$S^2$ & 17.5 & 10.4 & 1.0 x 10$^{-1}$\\
$< \Delta \nu >$ ($\mu$Hz) & 19.6 & 15.0 & 1.7 \\
$<\Delta P>$ (s) & 1.1 & 0.8 & 0.09 \\ \hline
$T_{\rm eff}$ (K) &  12,820 (6.8\%) & 12,540 (4.5\%) & 11,770 (1.9\%)\\
Log $g$ & 8.11 (1.4\%) & 8.04 (0.5\%) & 8.02 (0.3\%)\\
$q$(H)  & $-$4.60 (13.9\%) & $-$4.35 (7.7\%) & $-$4.15 (2.7\%) \\
$q$(He) & $-$2.53 (20.5\%)& $-$2.25 (7.1\%) & $-$2.35 (11.9\%) \\ \hline
\end{tabular}
\end{center}
\end{table*}

\clearpage
\centerline{\bf{FIGURE CAPTIONS}}

\noindent Fig. 1 --- Proposed parametrization of $X$(O) using a
  two-transition model white dwarf core. All the parameters are labeled
  in the schematic view shown in left panel. As illustrated in right
  panel, the parametrization defines control points (red dots) through
  which the profile is interpolated using Akima splines (the thick blue
  curve). Other interpolation schemes are possible, e.g., linear (thin
  curve) or cubic splines (dotted curve), but either lead to less smooth
  or less stable profiles. 
	
\noindent Fig. 2 --- Calculated helium profiles extracted from a
 representative evolutionary sequence that includes diffusion (black
 solid curves) with corresponding $M_{*}$ = 0.6 $M_{\rm
\odot}$, log $M(He)/M_{*}=
 -3.0 $ and log $M(H)/M_{*}= -5.0 $, at three different effective
 temperatures around 30,000K (top panel), 20,000K (middle panel) and
 12,000K (bottom panel), compared to static parametrized models (red
 dotted curves) obtained with optimized values of $Pf_{\rm H}$ and
 $Pf_{\rm He}$.  

\noindent Fig. 3 --- Top panel: Chemical abundance profiles of a generic
  model using our two-transition model white dwarf core mimicking the
  general structures of a typical model of a DA white dwarf star (left
  panel) taken from evolutionary calculations connecting with the ZAMS
  (from Althaus et al. 2010, Figure 6), and of a typical model of a DB
  white dwarf star (right panel) taken as well from detailed
  evolutionary calculations (from C\'{o}rsico et al. 2012, Figure
  2). Oxygen (long-dashed curve), carbon (dashed curve), helium (dotted
 curve), and hydrogen (solid curve) are depicted on the Figure. The
  abscissa shows the fractional mass depth (with $\log q = 0$
  corresponding to the center of the star). Bottom panel: run of the
  square of the Brunt-V$\ddot{\rm a}$is$\ddot{\rm a}$l$\ddot{\rm a}$
  frequency (solid curves) and of the Lamb frequency for $l=1$ (lower
  dotted curve) and $l=2$ (upper dotted curve). The left part of both
  panels emphasizes a zoomed-in view of the deep interior of the star.

\noindent Fig. 4 --- Map of the projected merit function $S^2$ (on a log scale)
  onto the $T_{\rm eff}$-log $g$ and $D$(He)-$D$(H) plane as well as the
  probability density functions for all the retrieved stellar parameters
  from the reference star with a variable homogeneous core. The degree
  of precision of the ``observations" is set to the current fit precision
  level. The red-hatched region between two vertical solid red lines
  defines the 1$\sigma$ range, containing 68.3\% of the
  distribution. The blue vertical dashed line indicates the value from
  the reference model.

\noindent Fig. 5 --- Map of the projected merit function $S^2$ (on a log scale)
  onto the $T_{\rm eff}$-log $g$ and $D$(He)-$D$(H) plane as well as the
  probability density functions for all the retrieved stellar parameters
  from the reference star with a variable homogeneous core. The degree
  of precision of the ``observations" is set to the ground-based data
  level. The red-hatched region between two vertical solid red lines
  defines the 1$\sigma$ range, containing 68.3\% of the
  distribution. The blue vertical dashed line indicates the value from
  the reference model.

\noindent Fig. 6 --- Map of the projected merit function $S^2$ (on a log scale)
  onto the $T_{\rm eff}$-log $g$ and $D$(He)-$D$(H) plane as well as the
  probability density functions for all the retrieved stellar parameters
  from the reference star with a variable homogeneous core. The degree
  of precision of the ``observations" is set to the $Kepler$ data
  level. The red-hatched region between two vertical solid red lines
  defines the 1$\sigma$ range, containing 68.3\% of the
  distribution. The blue vertical dashed line indicates the value from
  the reference model.

\noindent Fig. 7 ---  Map of the projected merit function $S^2$ (on a log scale)
  onto the $T_{\rm eff}$-log $g$ and $D$(He)-$D$(H) plane as well as the
  probability density functions for all the retrieved stellar parameters
  from the reference star with one-transition core. The degree of
  precision of the ``observations" is set to the current fit precision
  level. The red-hatched region between two vertical solid red lines
  defines the 1$\sigma$ range, containing 68.3\% of the
  distribution. The blue vertical dashed line indicates the value from
  the reference model.

\noindent Fig. 7 --- Continued.

\noindent Fig. 8 --- Map of the projected merit function $S^2$ (on a log scale)
  onto the $T_{\rm eff}$-log $g$ and $D$(He)-$D$(H) plane as well as the
  probability density functions for all the retrieved stellar parameters
  from the reference star with one-transition core. The degree of
  precision of the ``observations" is set to the ground-based data
  precision level. The red-hatched region between two vertical solid red
  lines defines the 1$\sigma$ range, containing 68.3\% of the
  distribution. The blue vertical dashed line indicates the value from
  the reference model.

\noindent Fig. 8 --- Continued.

\noindent Fig. 9 --- Map of the projected merit function $S^2$ (on a log scale)
  onto the $T_{\rm eff}$-log $g$ and $D$(He)-$D$(H) plane as well as the
  probability density functions for all the retrieved stellar parameters
  from the reference star with one-transition core. The degree of
  precision of the ``observations" is set to the $Kepler$ data precision
  level. The red-hatched region between two vertical solid red lines
  defines the 1$\sigma$ range, containing 68.3\% of the
  distribution. The blue vertical dashed line indicates the value from
  the reference model.

\noindent Fig. 9 --- Continued.

\noindent Fig. 10 --- Map of the projected merit function $S^2$ (on a logarithmic scale)
  onto the $T_{\rm eff}$-log $g$ and $D$(He)-$D$(H) plane as well as the
  probability density functions for all the retrieved stellar parameters
  from the reference star with two-transition core. The degree of
  precision of the ``observations" is set to the current fit precision
  level. The red-hatched region between two vertical solid red lines
  defines the 1$\sigma$ range, containing 68.3\% of the
  distribution. The blue vertical dashed line indicates the value from
  the reference model.

\noindent Fig. 10 --- Continued.

\noindent Fig. 11 --- Map of the projected merit function $S^2$ (on a log scale)
  onto the $T_{\rm eff}$-log $g$ and $D$(He)-$D$(H) plane as well as the
  probability density functions for all the retrieved stellar parameters
  from the reference star with two-transition core. The degree of
  precision of the ``observations" is set to the ground-based data
  precision level. The red-hatched region between two vertical solid red
  lines defines the 1$\sigma$ range, containing 68.3\% of the
  distribution. The blue vertical dashed line indicates the value from
  the reference model.

\noindent Fig. 11 --- Continued.

\noindent Fig. 12 --- Map of the projected merit function $S^2$ (on a log scale)
  onto the $T_{\rm eff}$-log $g$ and $D$(He)-$D$(H) plane as well as the
  probability density functions for all the retrieved stellar parameters
  from the reference star with two-transition core. The degree of
  precision of the ``observations" is set to the $Kepler$ data precision
  level. The red-hatched region between two vertical solid red lines
  defines the 1$\sigma$ range, containing 68.3\% of the
  distribution. The blue vertical dashed line indicates the value from
  the reference model.

\noindent Fig. 12 --- Continued.

\noindent Fig. 13 --- Top panel: Chemical abundance profiles of the
 reference model with a triple transition He/C/O (red curves) and the retrieved model (black curves),
 oxygen (long-dashed curve), carbon (dashed curve), helium (dotted
 curve), and hydrogen (solid curve). The abscissa shows the fractional
 mass depth (with $\log q = 0$ corresponding to the center of the
 star). Bottom  panel: run of the square of the Brunt-V$\ddot{\rm
 a}$is$\ddot{\rm  a}$l$\ddot{\rm a}$ frequency (solid curves) and of the
 Lamb frequency for $l=1$ (lower dotted curve) and $l=2$ (upper dotted
 curve). The left part of both panels emphasizes a zoomed-in view of the
 deep interior of the star.

\noindent Fig. 14 --- Top panel: Chemical abundance profiles of the
 reference model,
 oxygen (long-dashed curve), carbon (dashed curve), helium (dotted
 curve), and hydrogen (solid curve). The abscissa shows the fractional
 mass depth (with $\log q = 0$ corresponding to the center of the
 star). Bottom  panel: run of the square of the Brunt-V$\ddot{\rm
 a}$is$\ddot{\rm  a}$l$\ddot{\rm a}$ frequency (solid curves) and of the
 Lamb frequency for $l=1$ (lower dotted curve) and $l=2$ (upper dotted
 curve). The left part of both panels emphasizes a zoomed-in view of the
 deep interior of the star.

\noindent Fig. 15 --- Weight functions of the selected modes from the reference
 model.

\noindent Fig. 16 --- Similar to Fig. 14,
 but showing the stratifications of the optimal model calculated with an
 homogeneous core (black curves) compared to those of the reference
 model (red curves).

\noindent Fig. 17 --- Similar to Fig. 14, but showing the stratifications of the
optimal model calculated with the adjustable core parametrization with
one transition (black curves) compared to those of the reference model
(red curves).

\noindent Fig. 18 --- Similar to Fig. 14, but showing the stratifications of the
optimal model calculated with the adjustable core parametrization with
two transitions (black curves) compared to those of the reference model
(red curves).


\clearpage
\begin{figure}
\includegraphics[width=.35\textwidth]{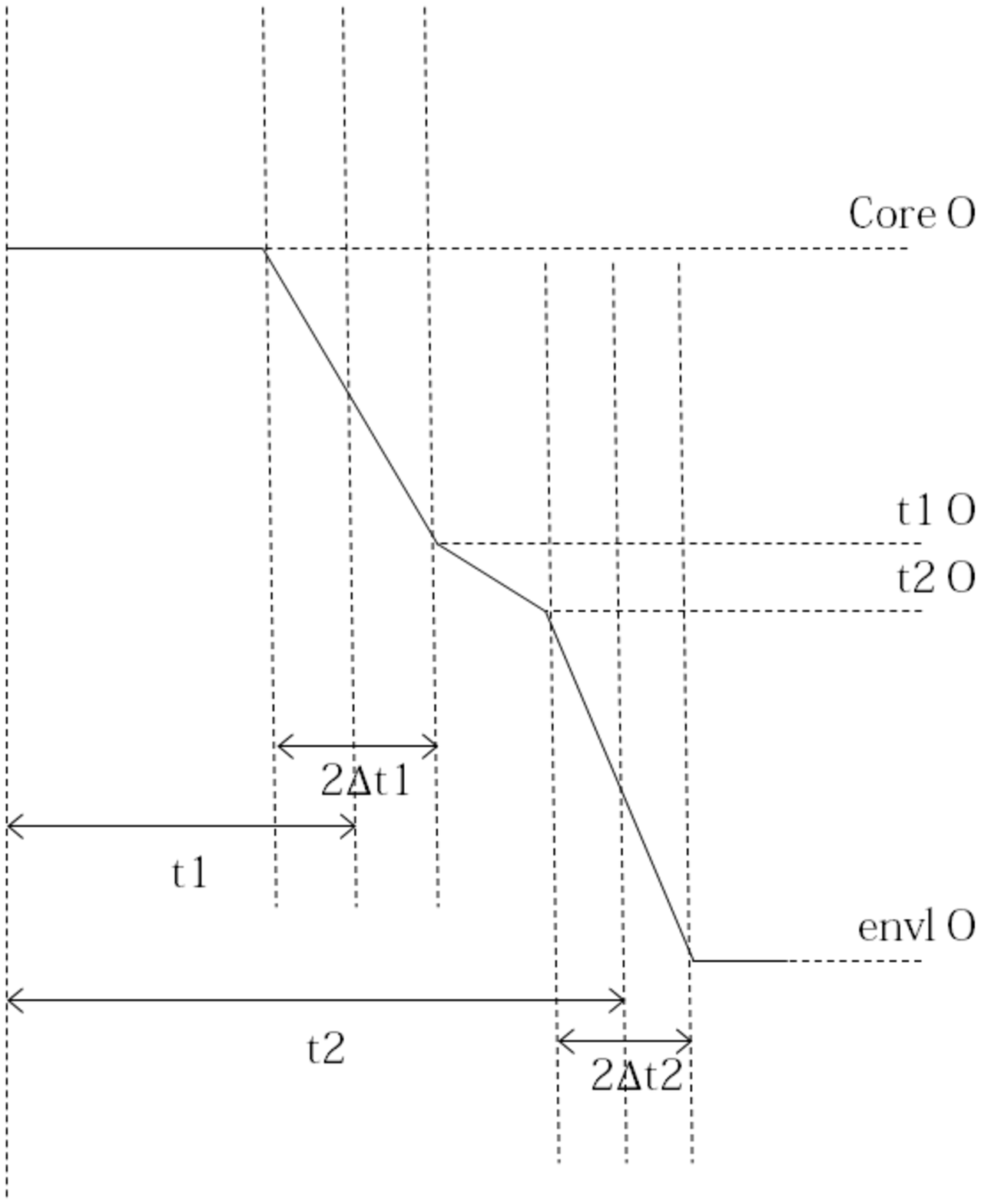}
\includegraphics[width=.65\textwidth]{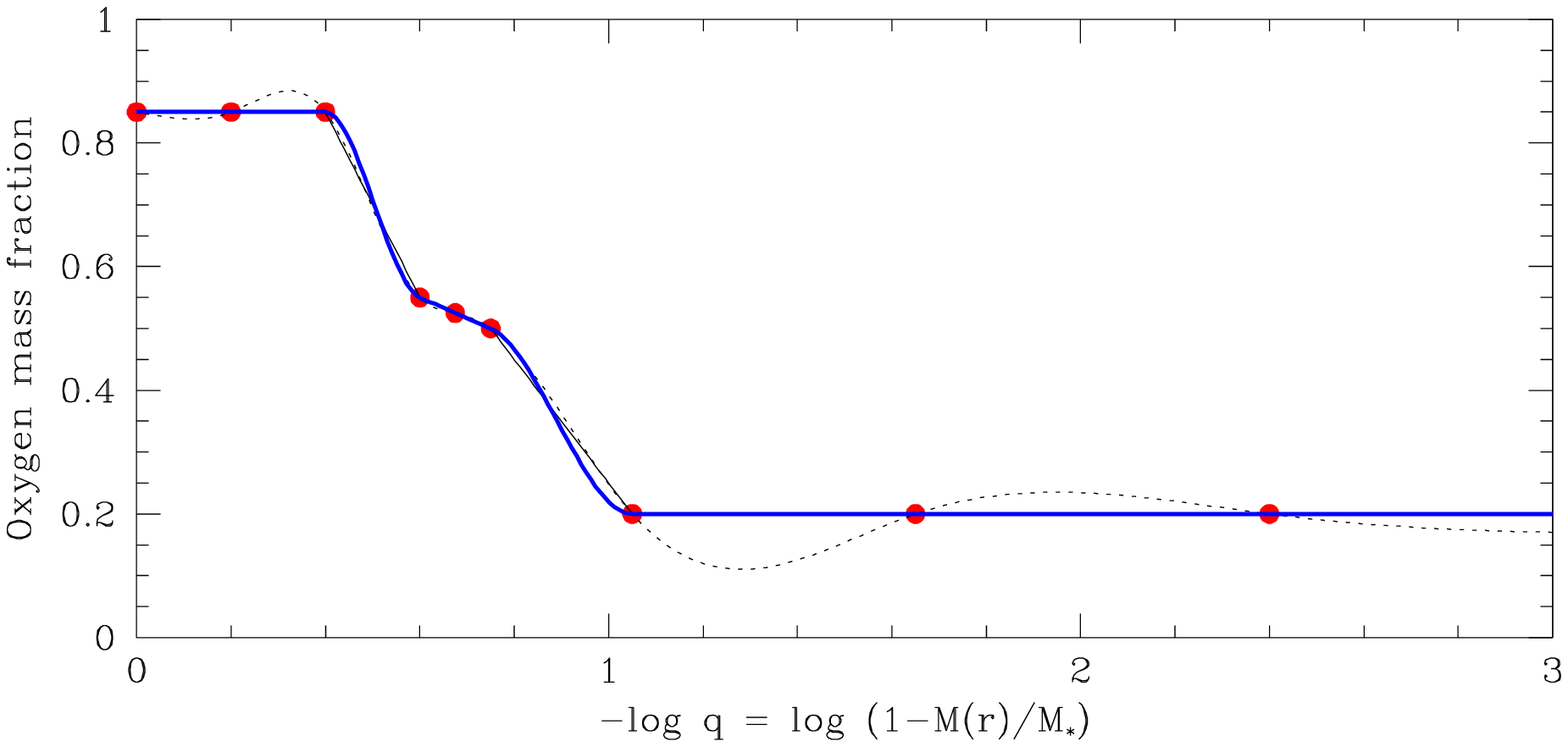}
  \begin{flushright}
Figure~1
\end{flushright}
\end{figure}

\clearpage
\begin{figure}
\includegraphics[width=.90\textwidth]{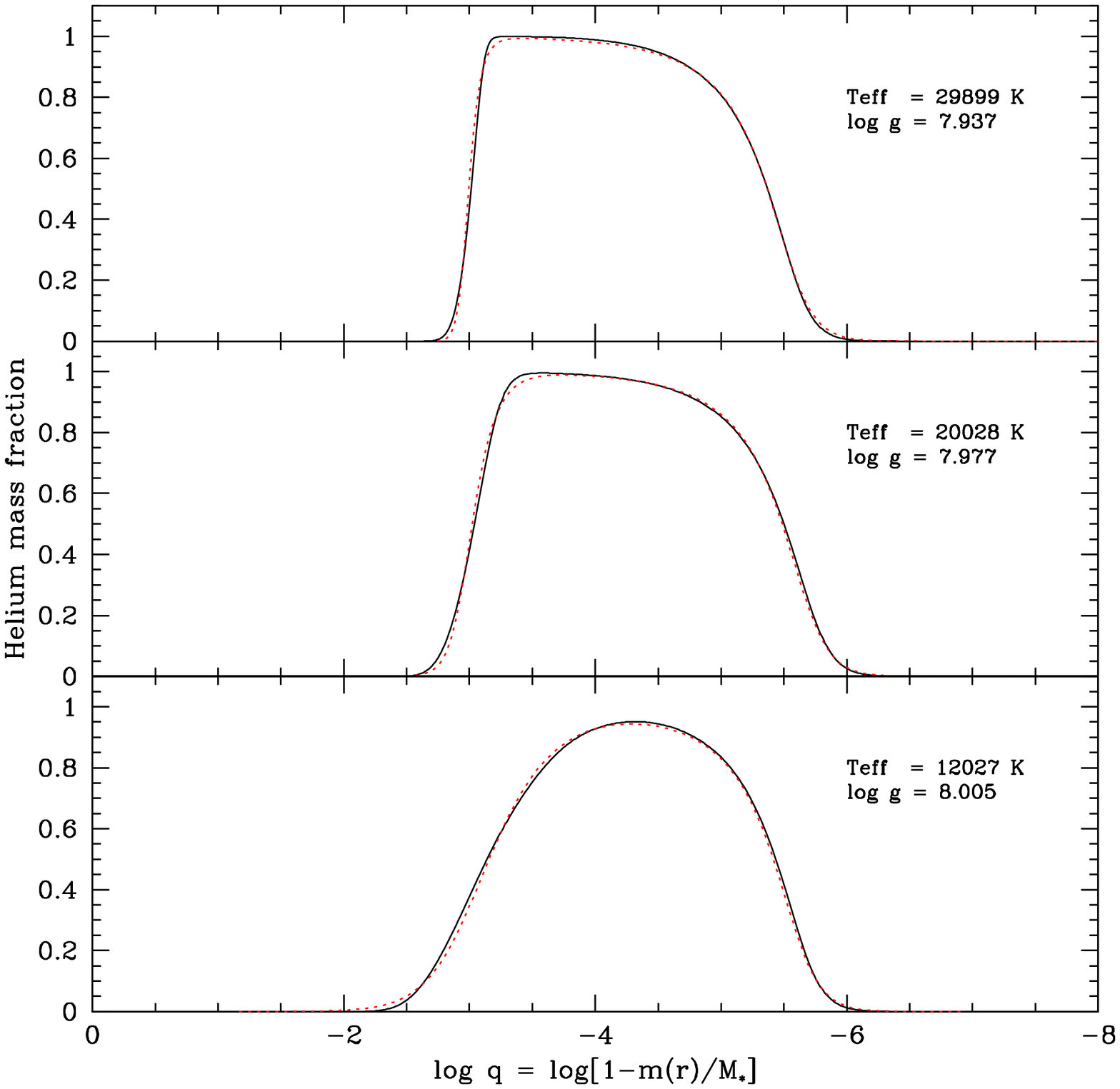}
  \begin{flushright}
Figure~2
\end{flushright}
\end{figure}

\clearpage

\clearpage
\begin{figure}
\includegraphics[width=.50\textwidth]{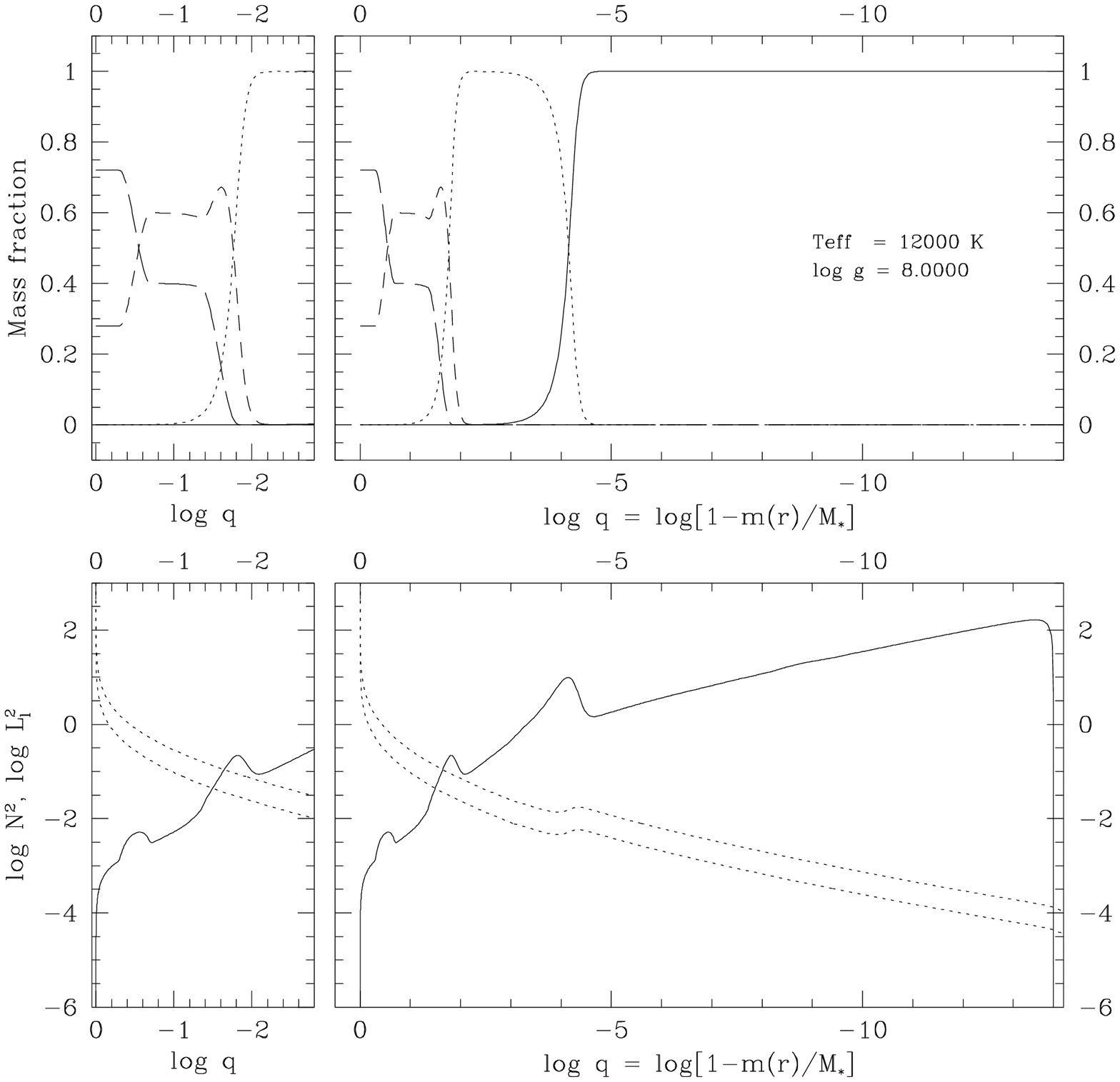}
\includegraphics[width=.50\textwidth]{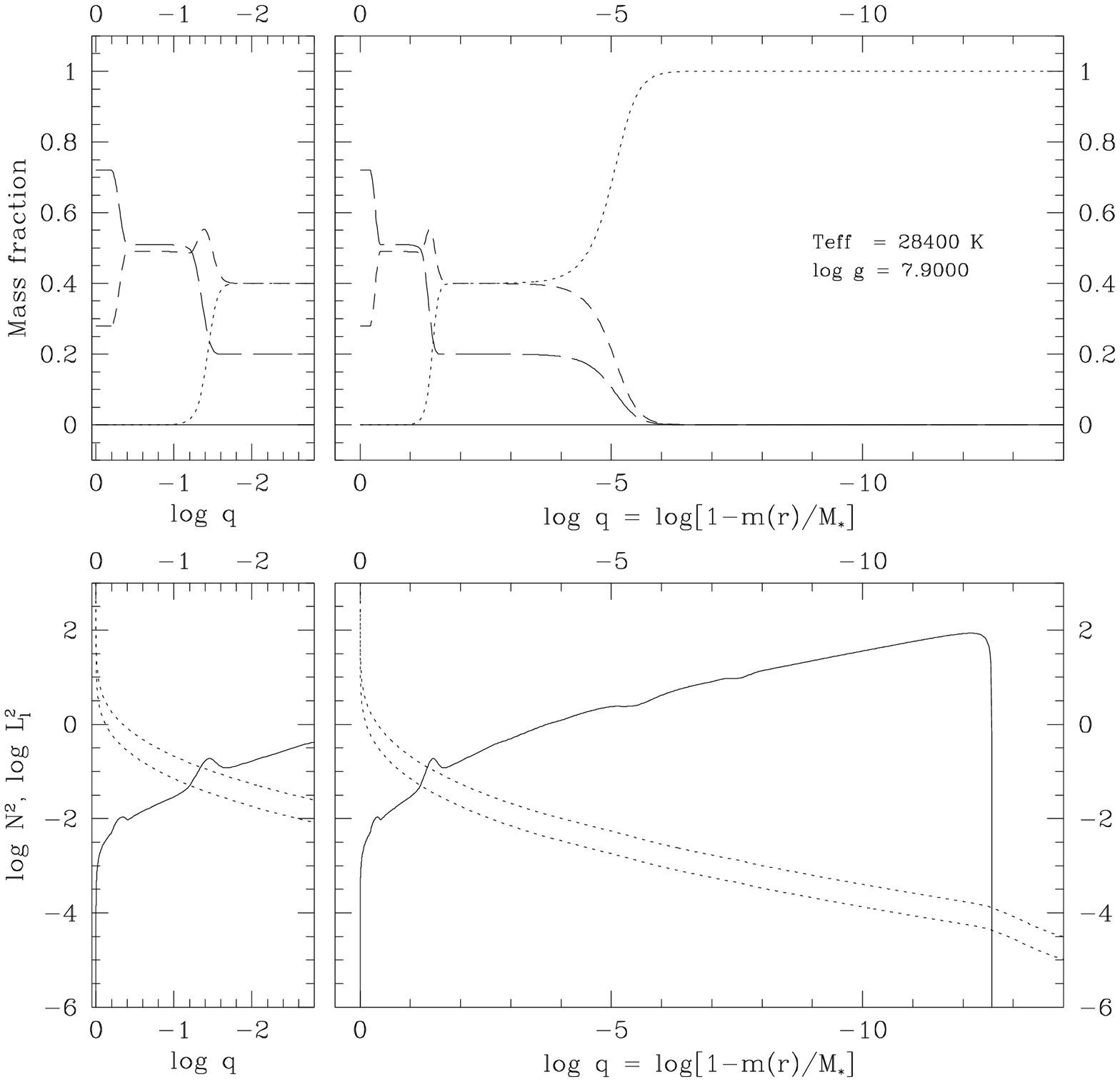}
  \begin{flushright}
Figure~3
\end{flushright}
\end{figure}

\clearpage

\begin{figure}[!h]
\centering
  \begin{tabular}{@{}ccc@{}}
    \includegraphics[width=.35\textwidth]{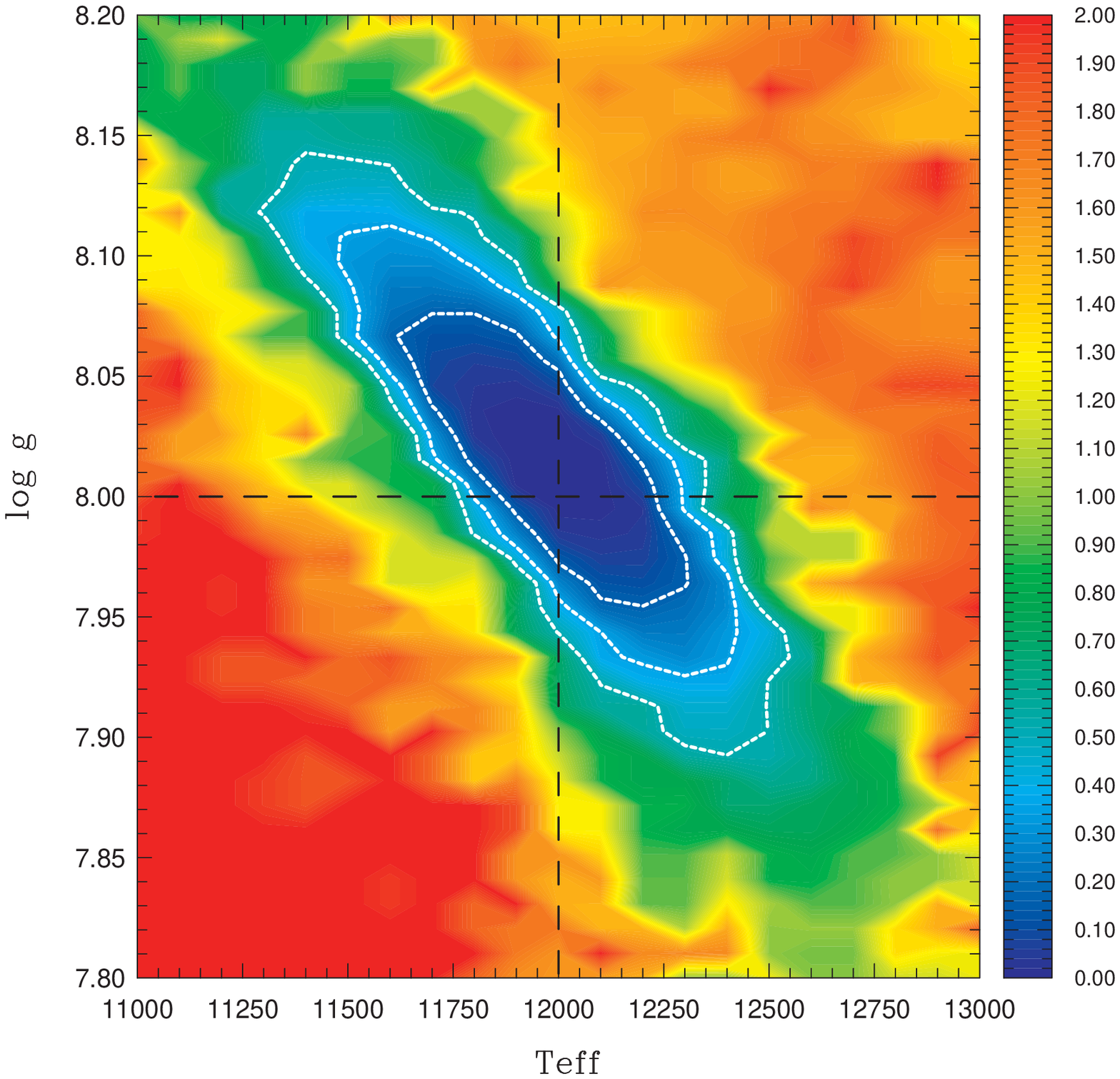} &
    \includegraphics[width=.35\textwidth]{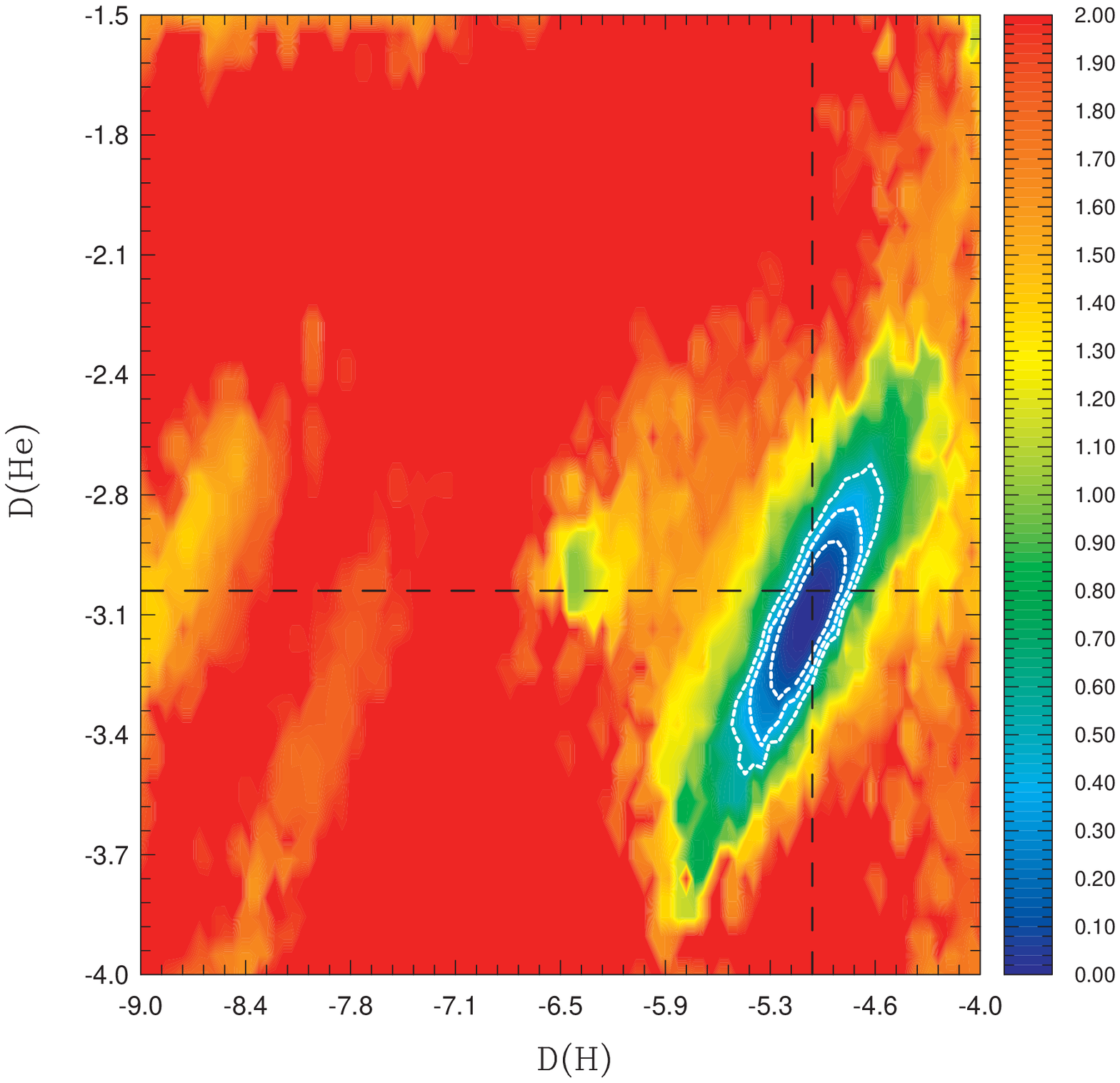} &
    \includegraphics[width=.35\textwidth]{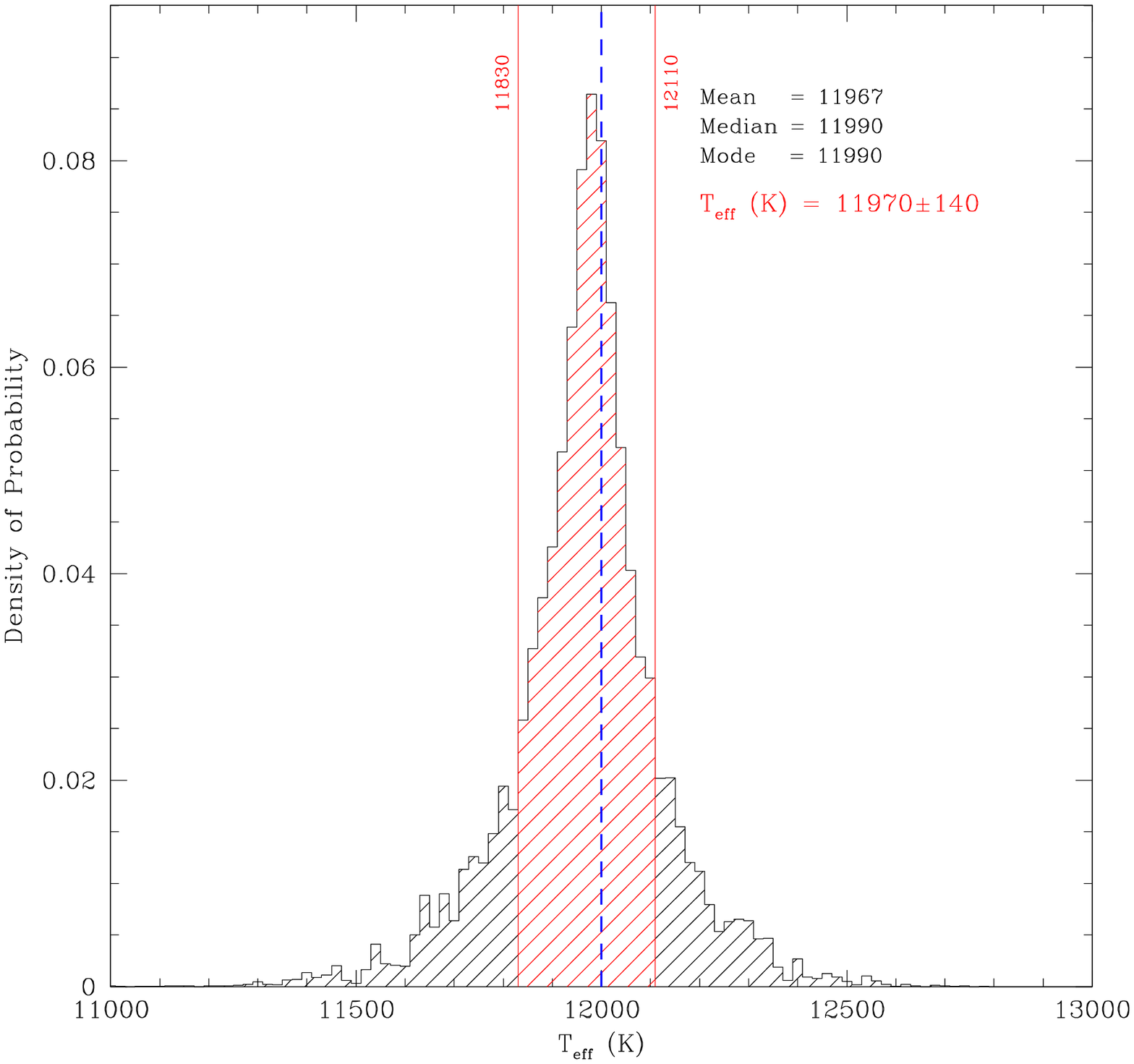} \\
    \includegraphics[width=.35\textwidth]{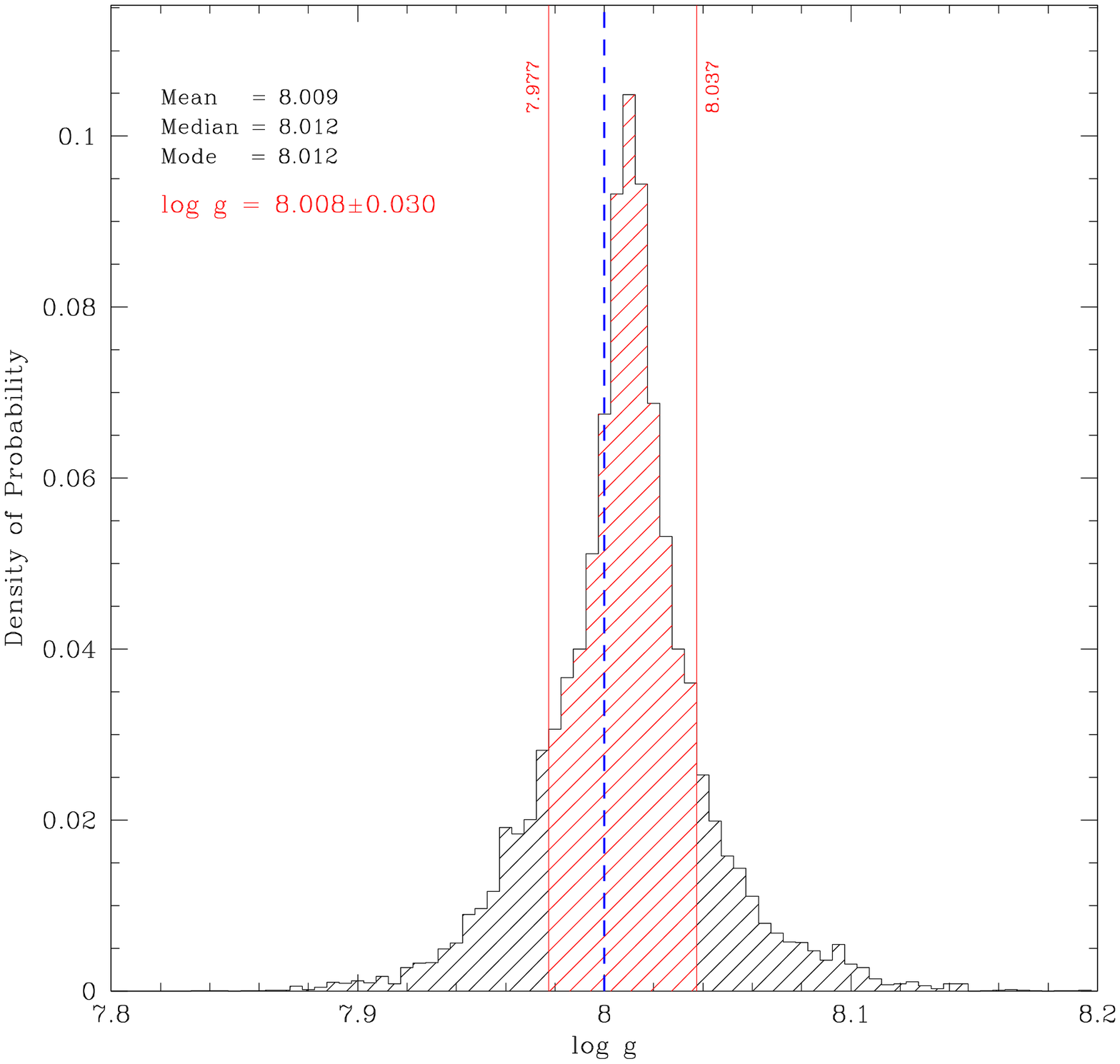}  &
    \includegraphics[width=.35\textwidth]{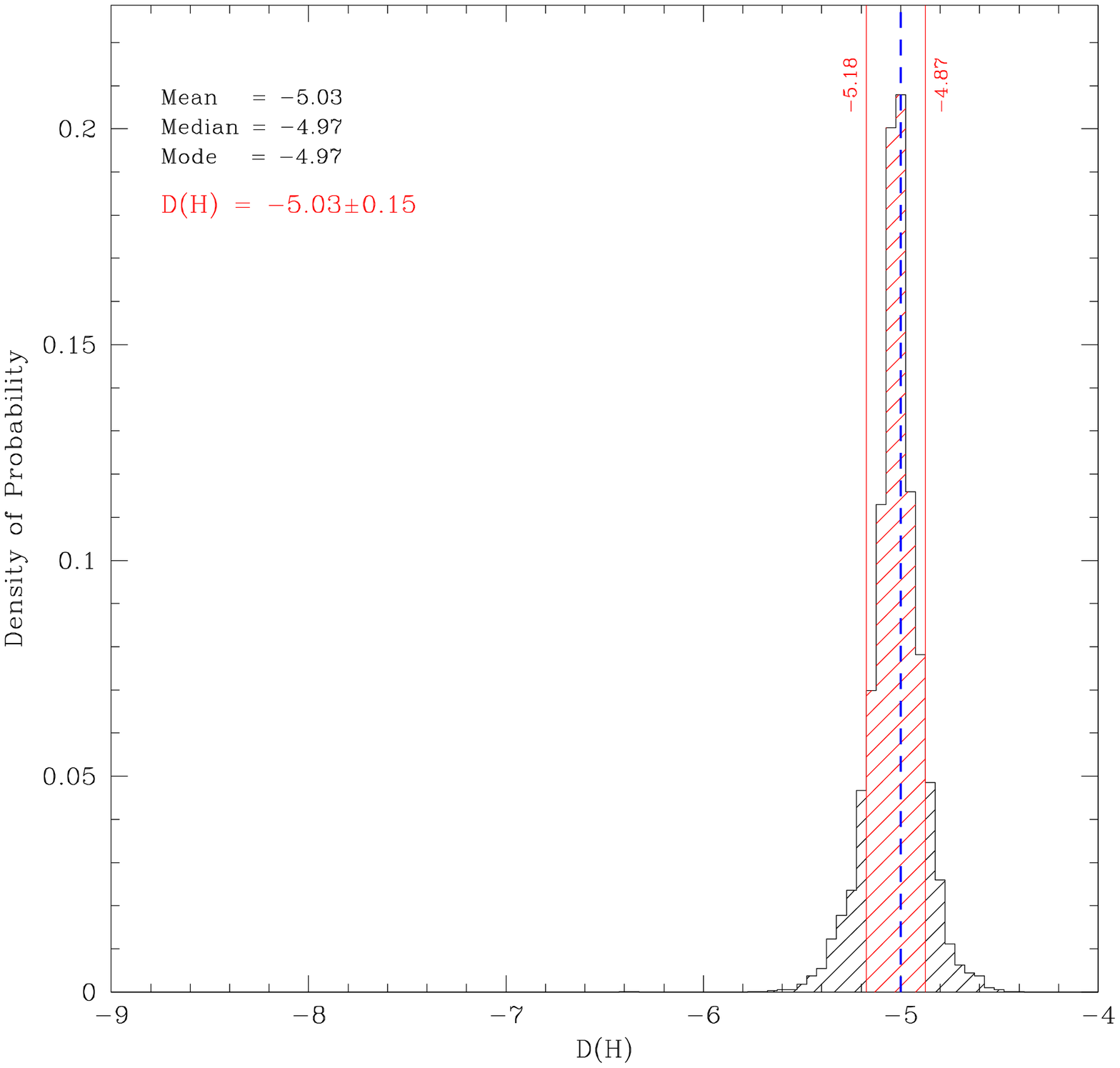} &
    \includegraphics[width=.35\textwidth]{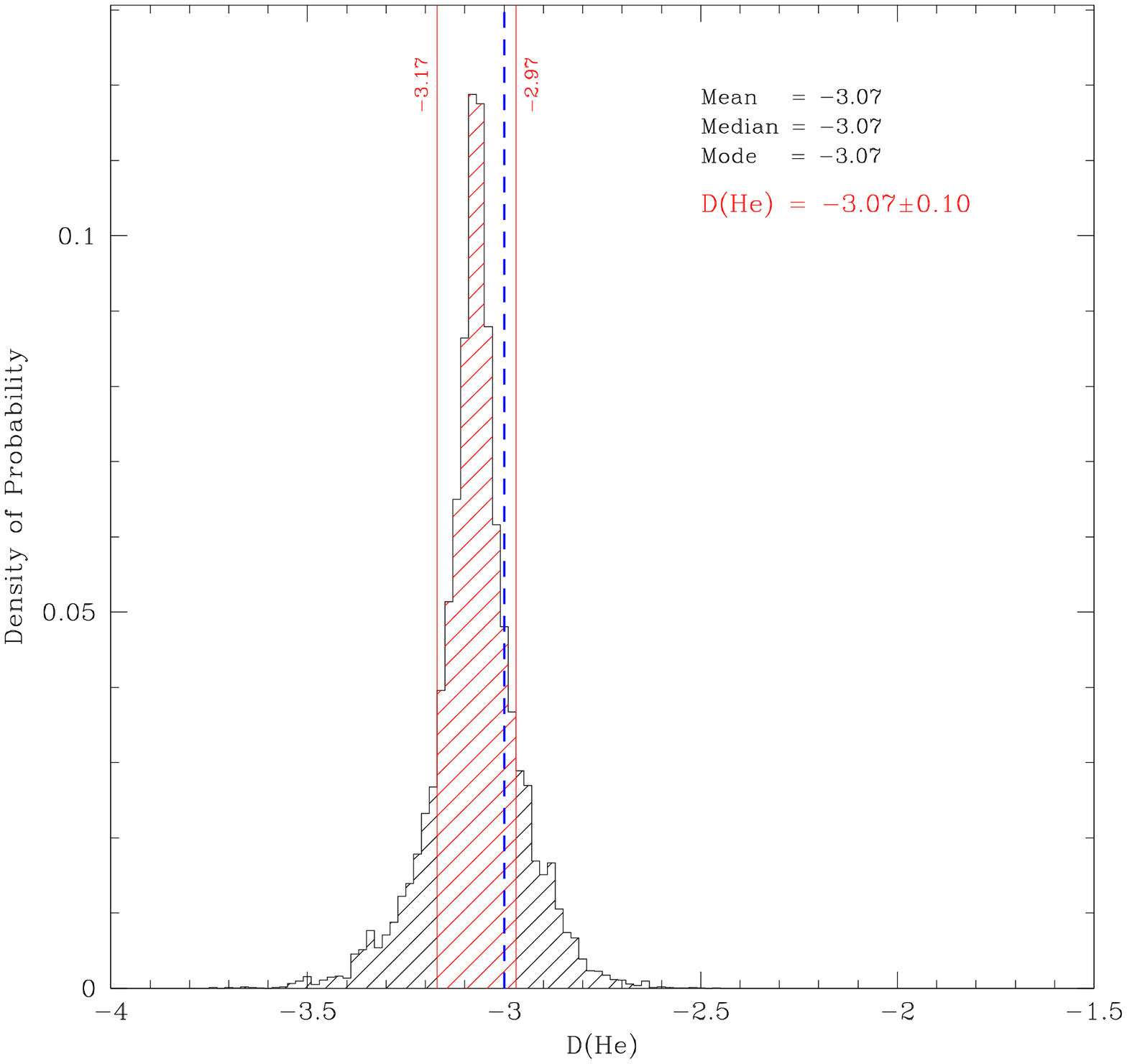} \\
    \includegraphics[width=.35\textwidth]{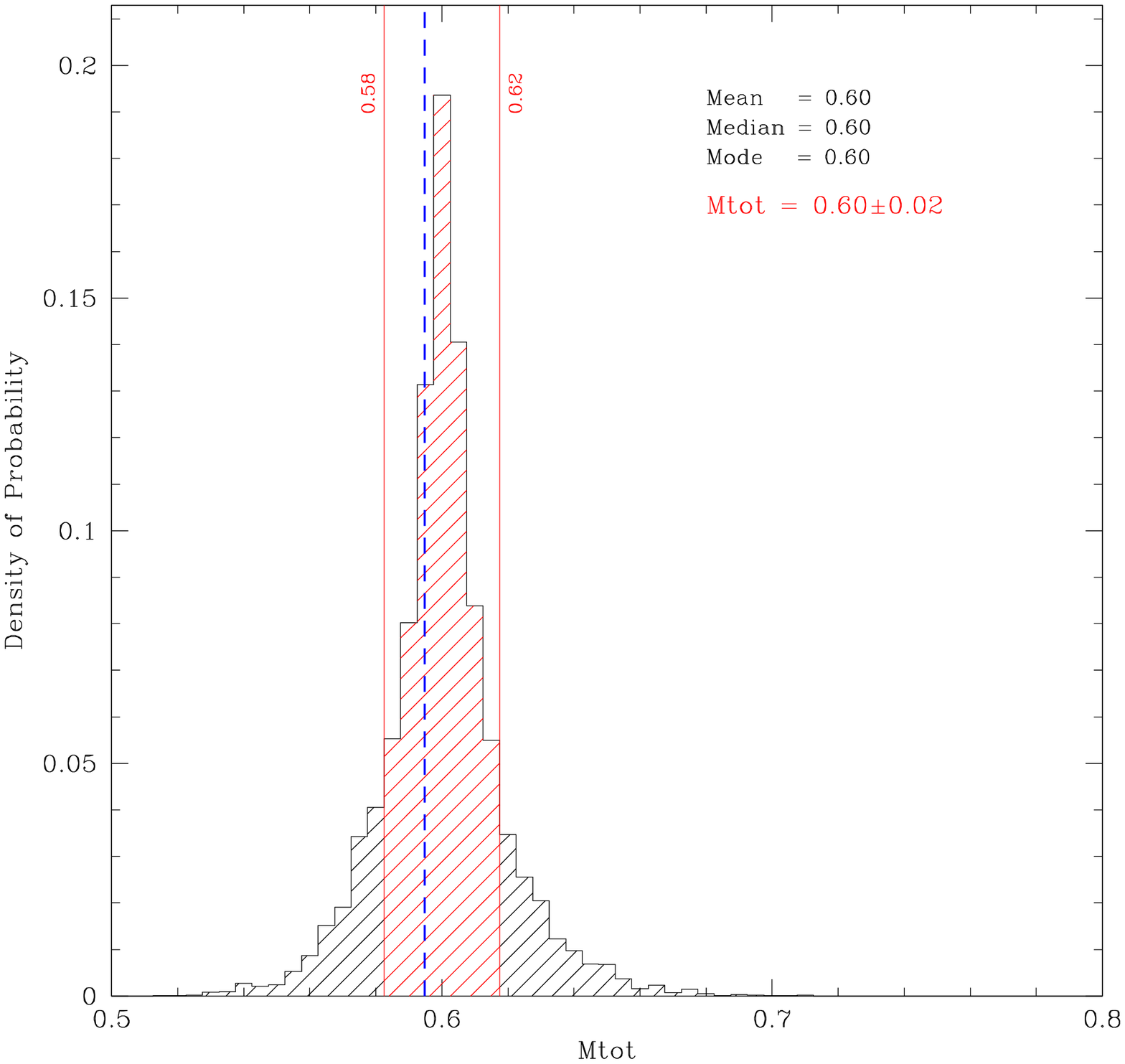} &
    \includegraphics[width=.35\textwidth]{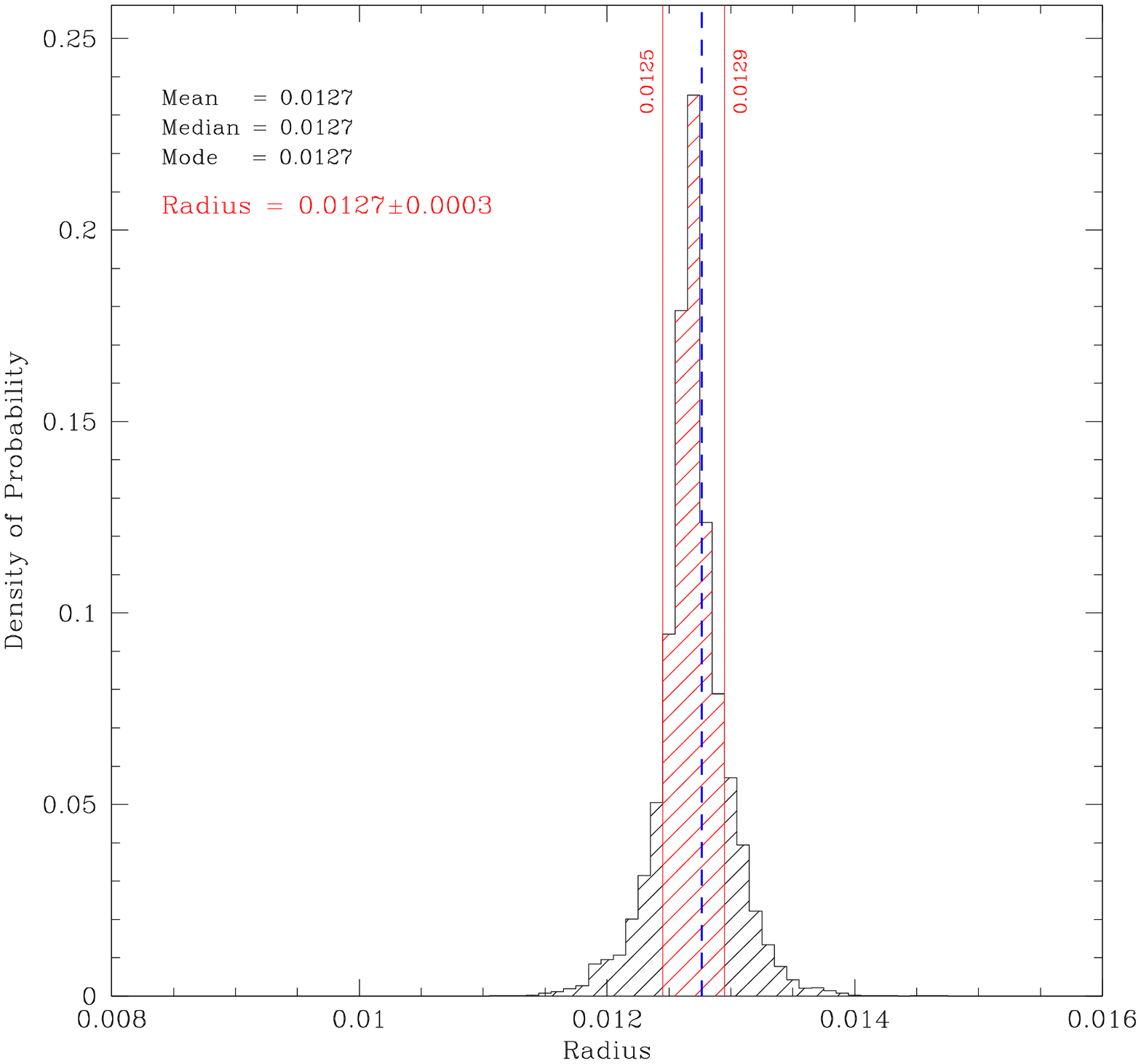} &
    \includegraphics[width=.35\textwidth]{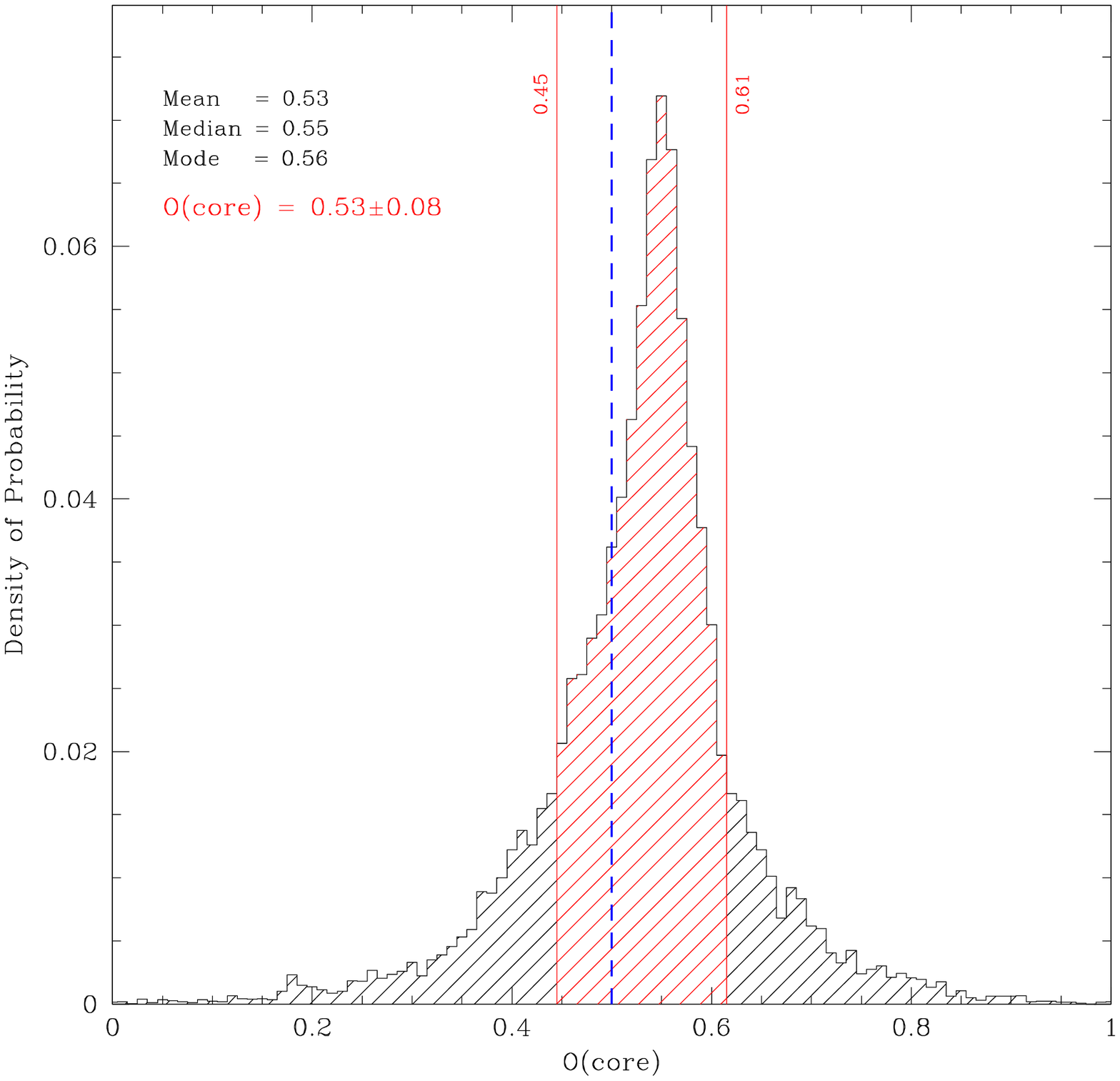}
    \end{tabular}
    \begin{flushright}
Figure~4
\end{flushright}
\end{figure}

\clearpage

\begin{figure}[!h]
\centering
  \begin{tabular}{@{}ccc@{}}
    \includegraphics[width=.35\textwidth]{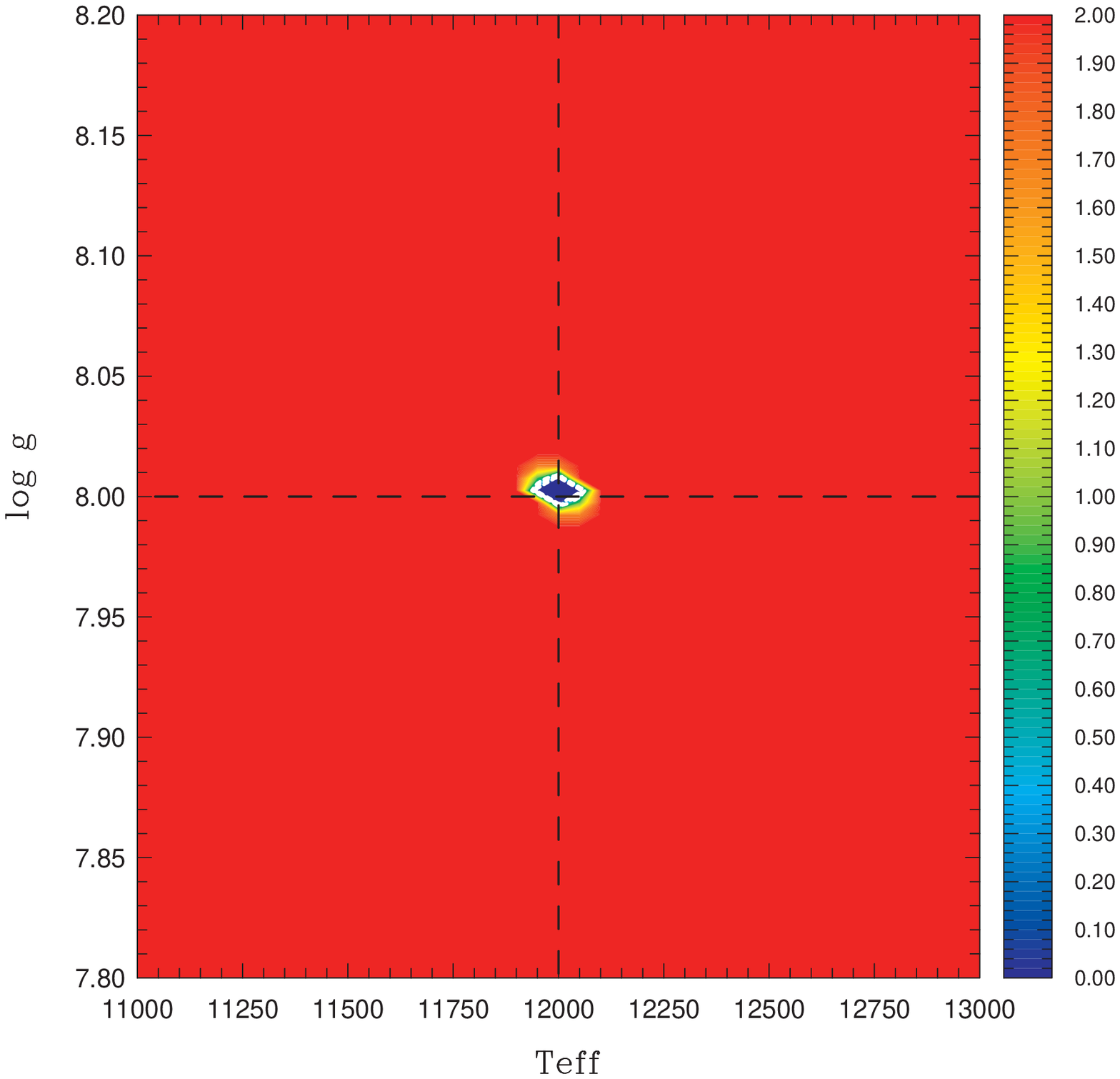} &
    \includegraphics[width=.35\textwidth]{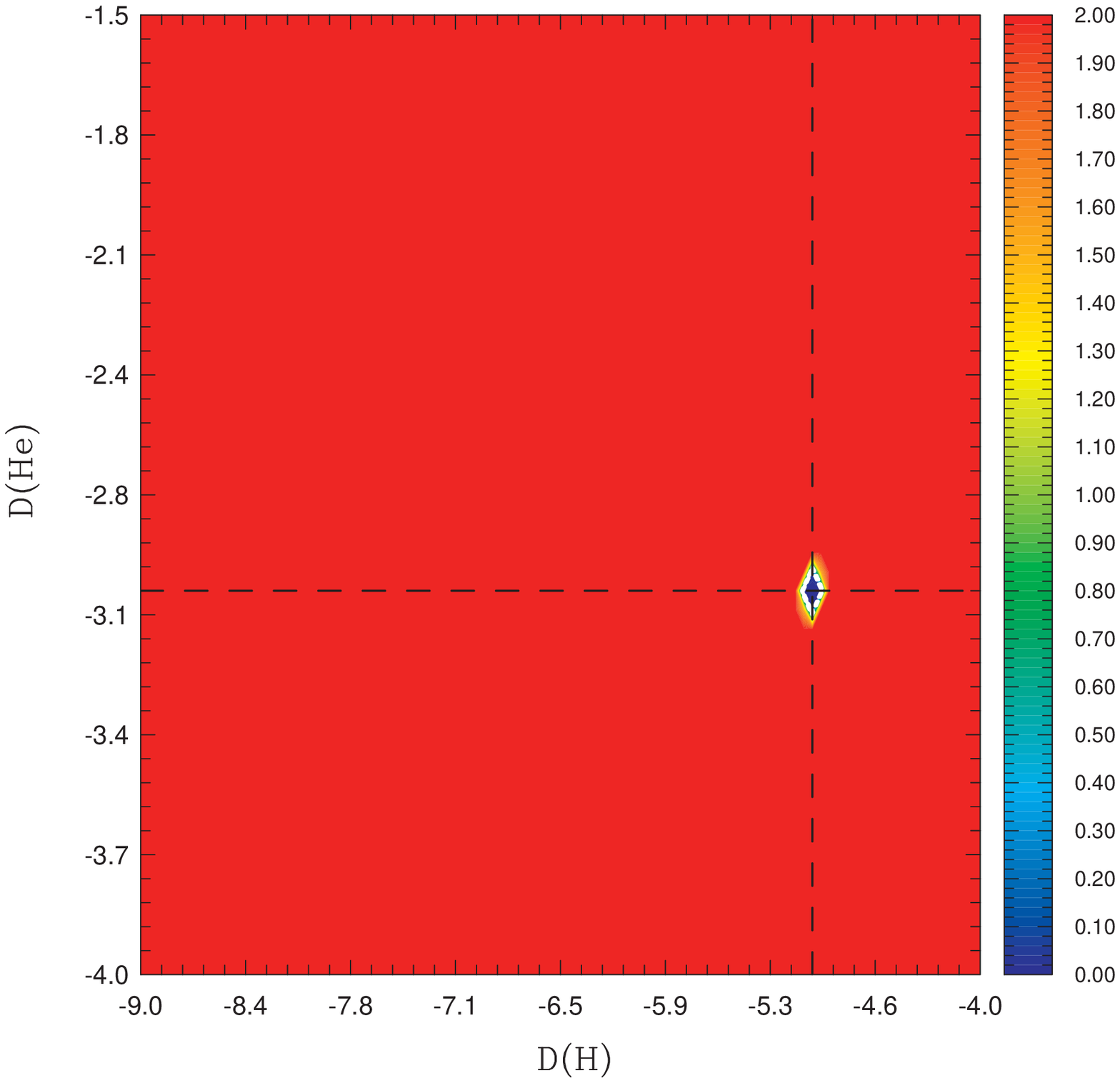} &
    \includegraphics[width=.35\textwidth]{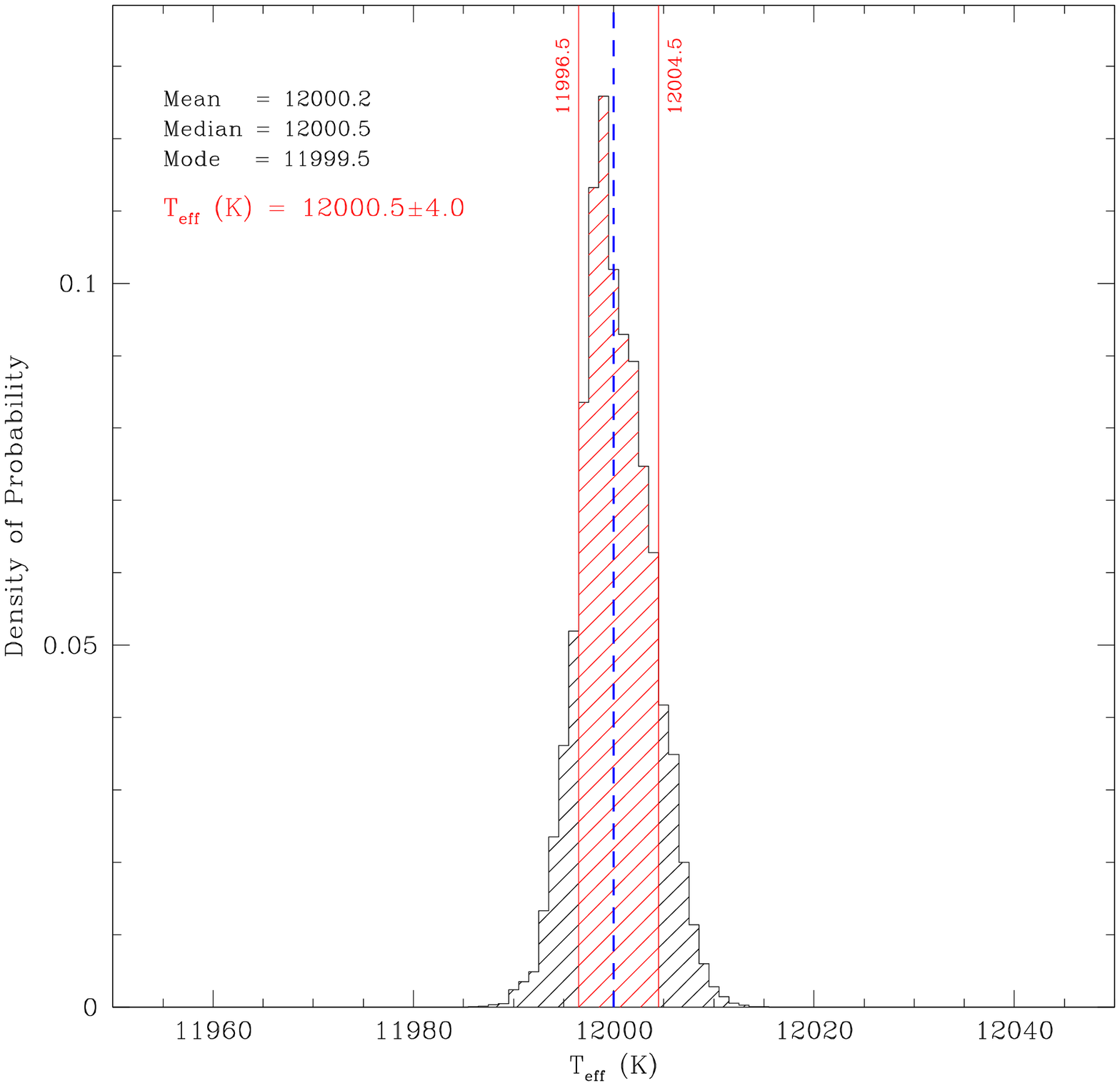} \\
    \includegraphics[width=.35\textwidth]{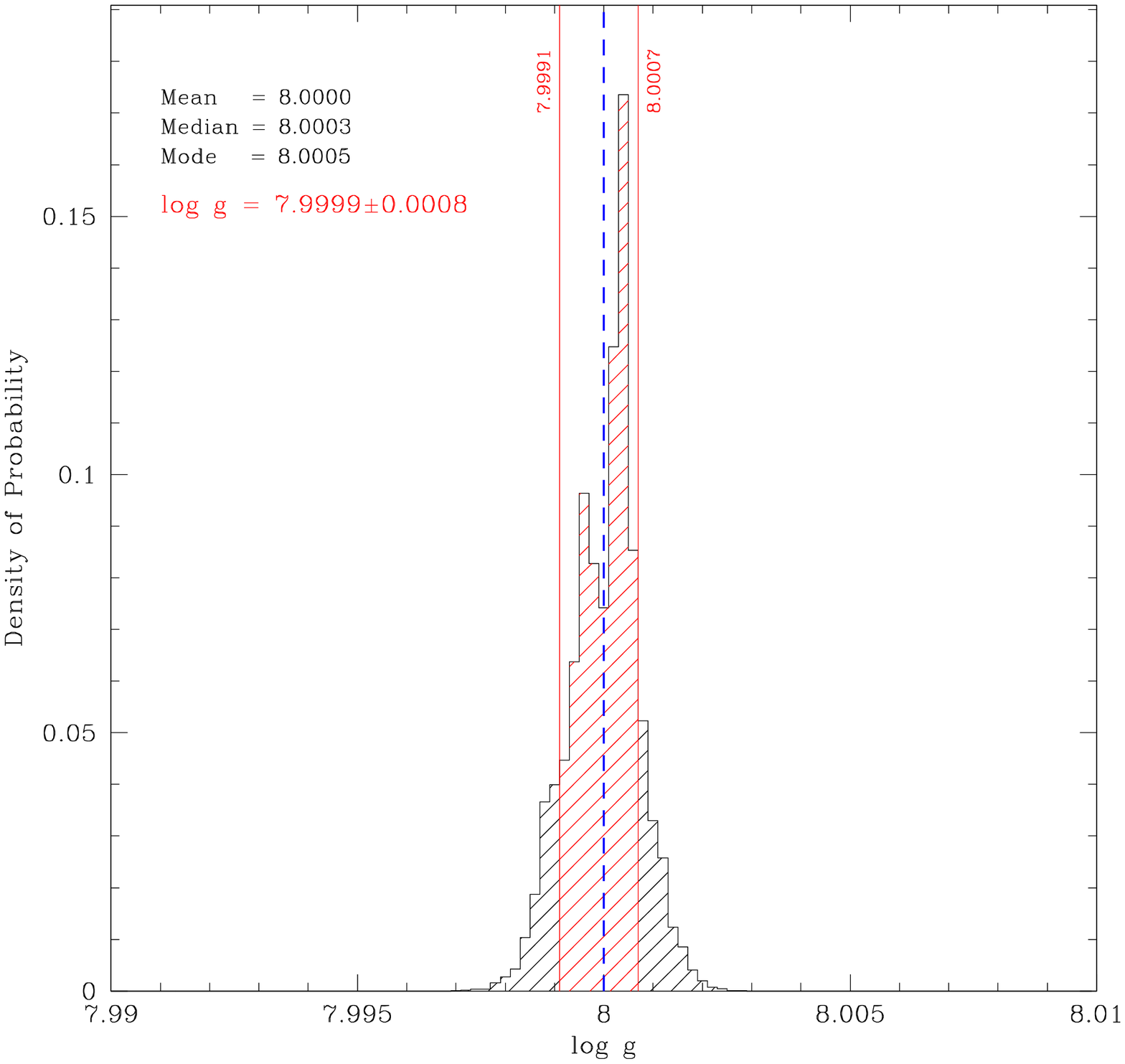}  &
    \includegraphics[width=.35\textwidth]{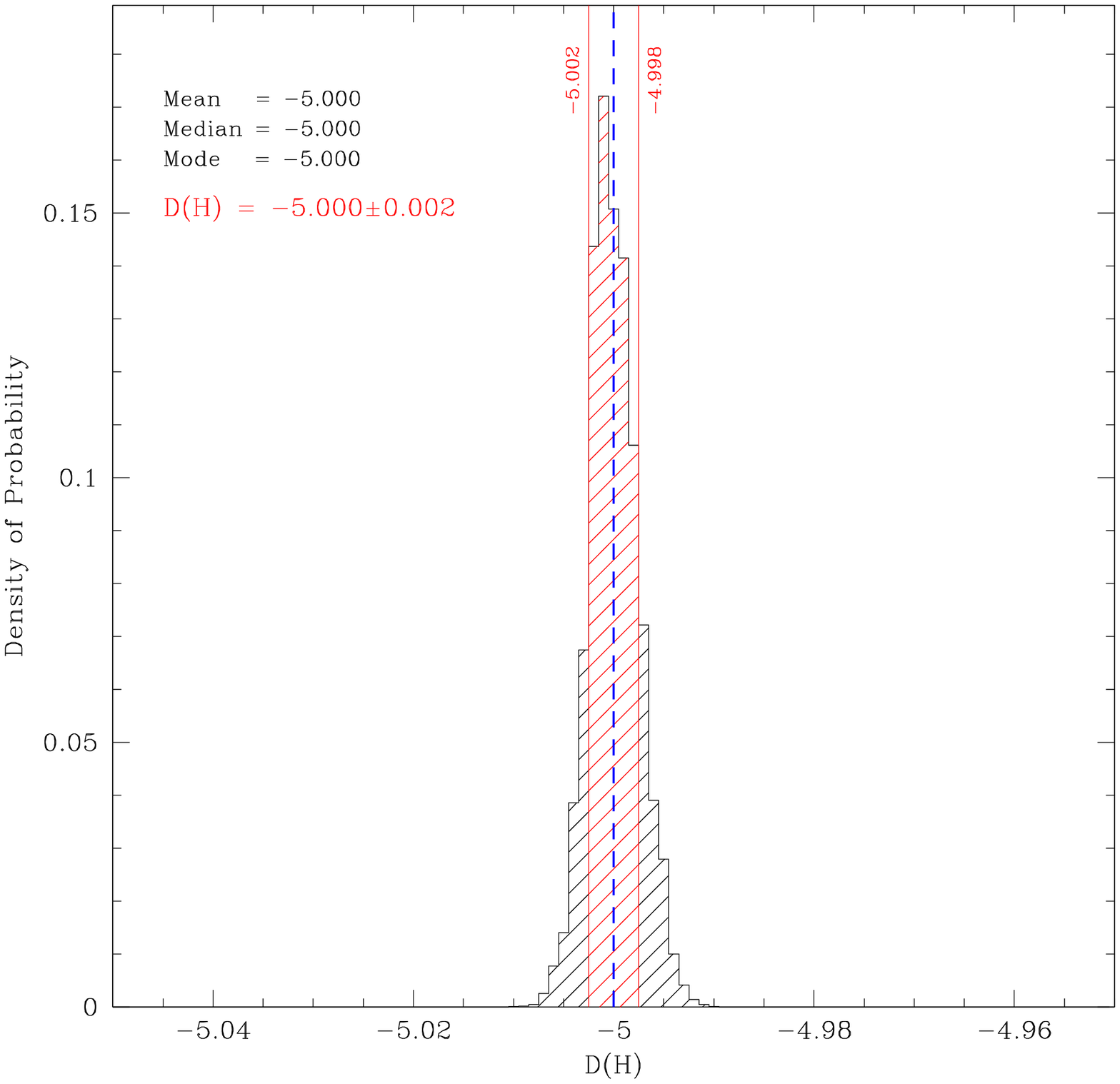} &
    \includegraphics[width=.35\textwidth]{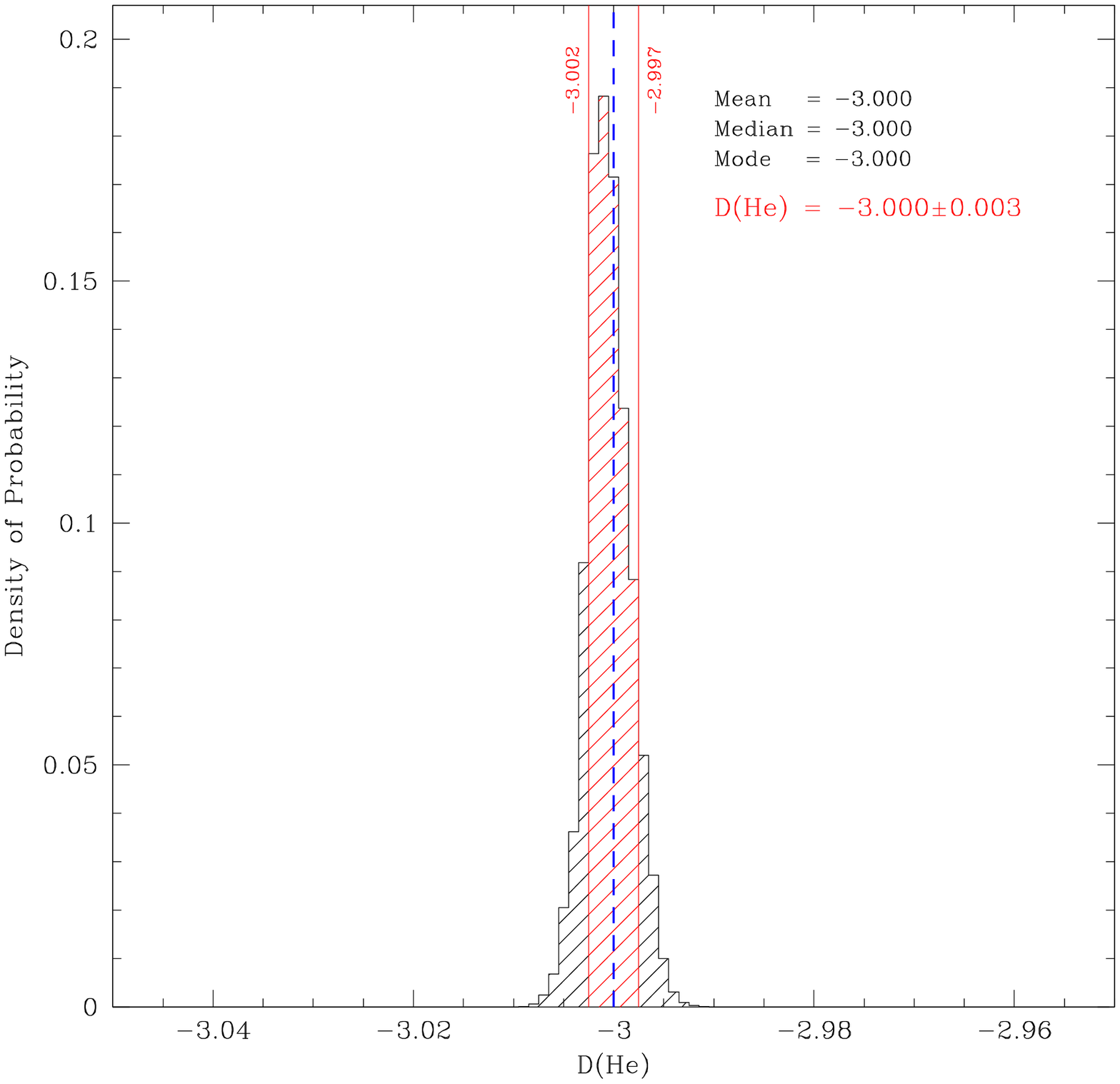} \\
    \includegraphics[width=.35\textwidth]{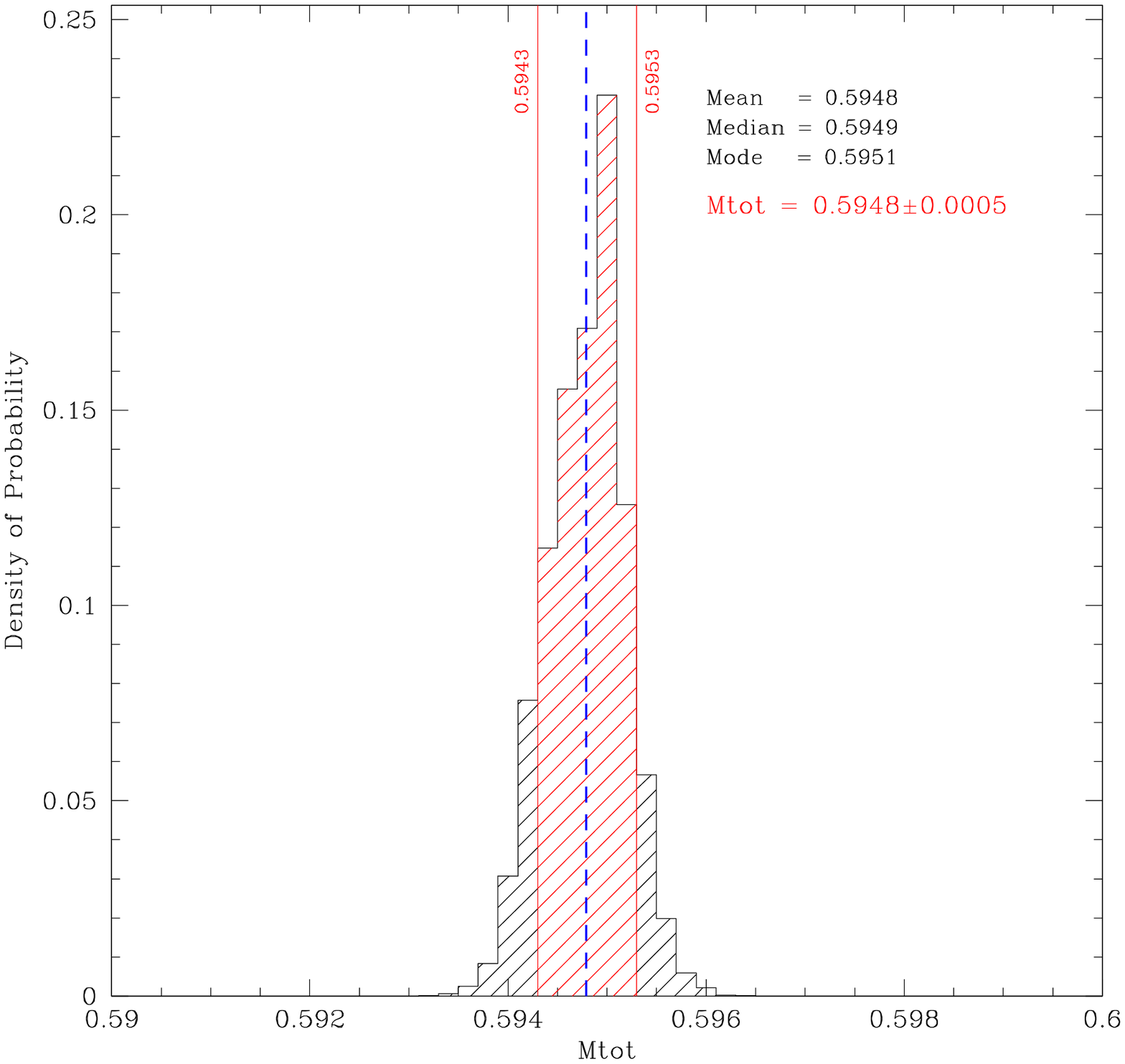} &
    \includegraphics[width=.35\textwidth]{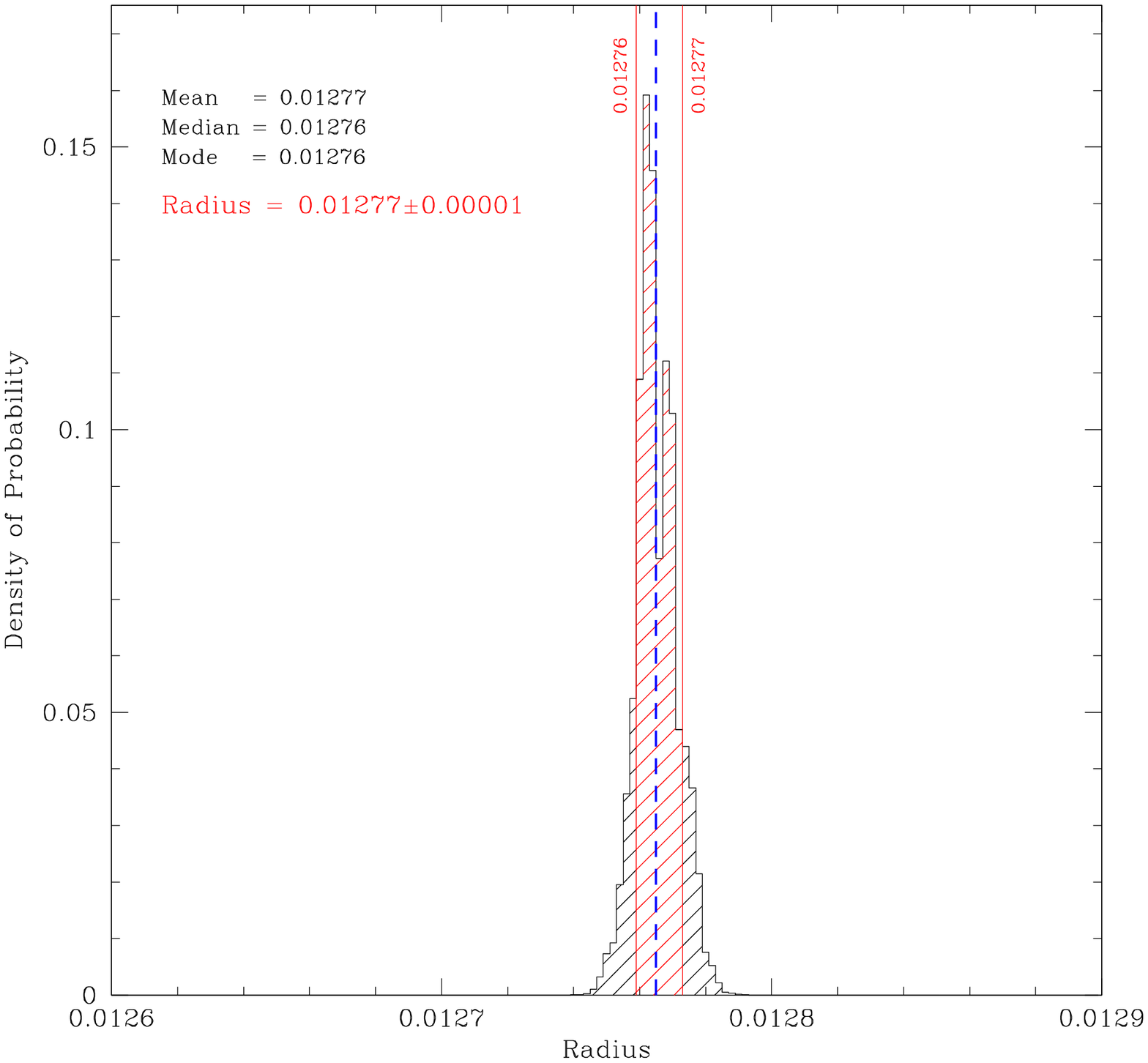} &
    \includegraphics[width=.35\textwidth]{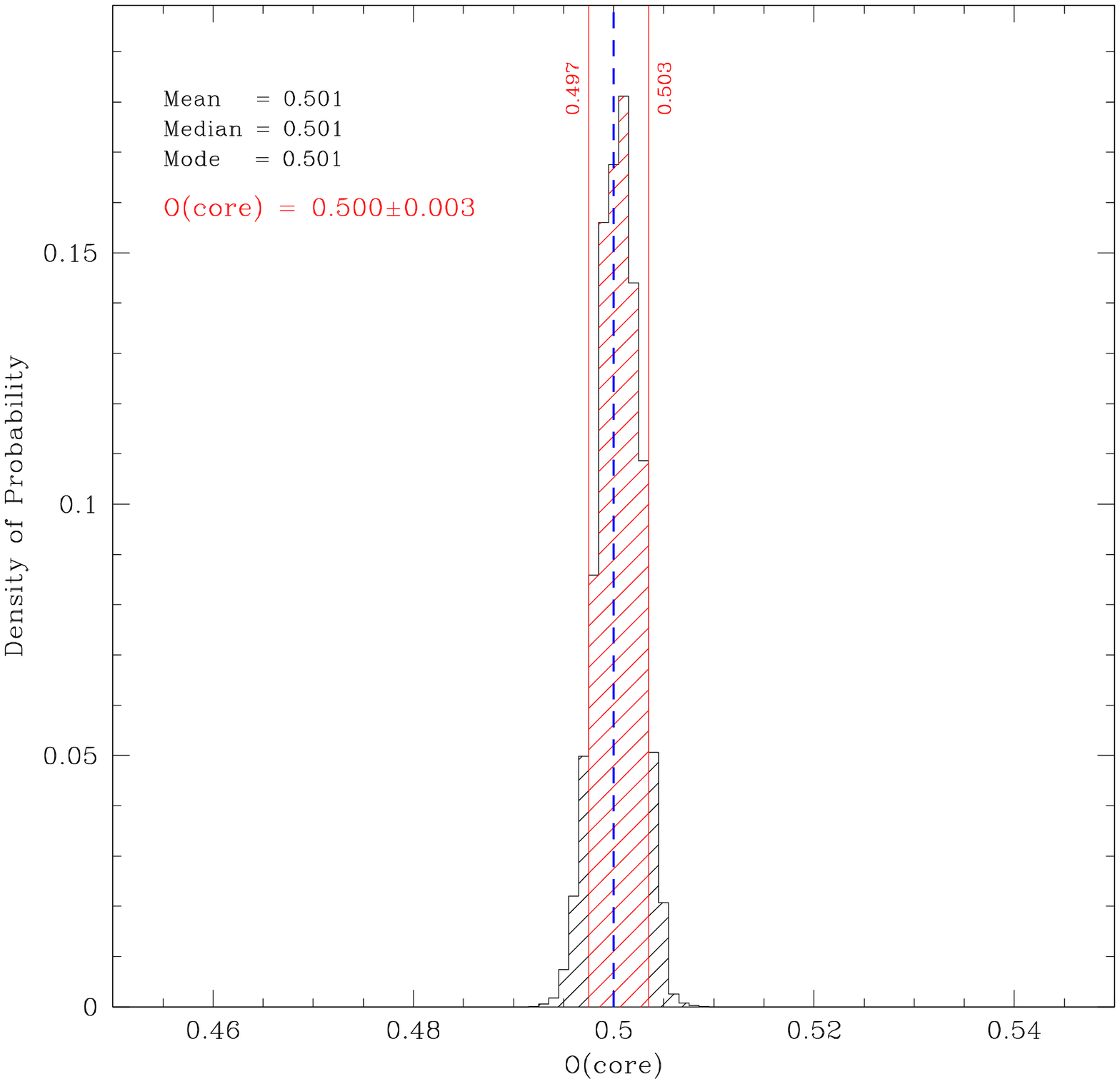}
    \end{tabular}
    \begin{flushright}
Figure~5
\end{flushright}
\end{figure}

\clearpage

\begin{figure}[!h]
\centering
  \begin{tabular}{@{}ccc@{}}
    \includegraphics[width=.35\textwidth]{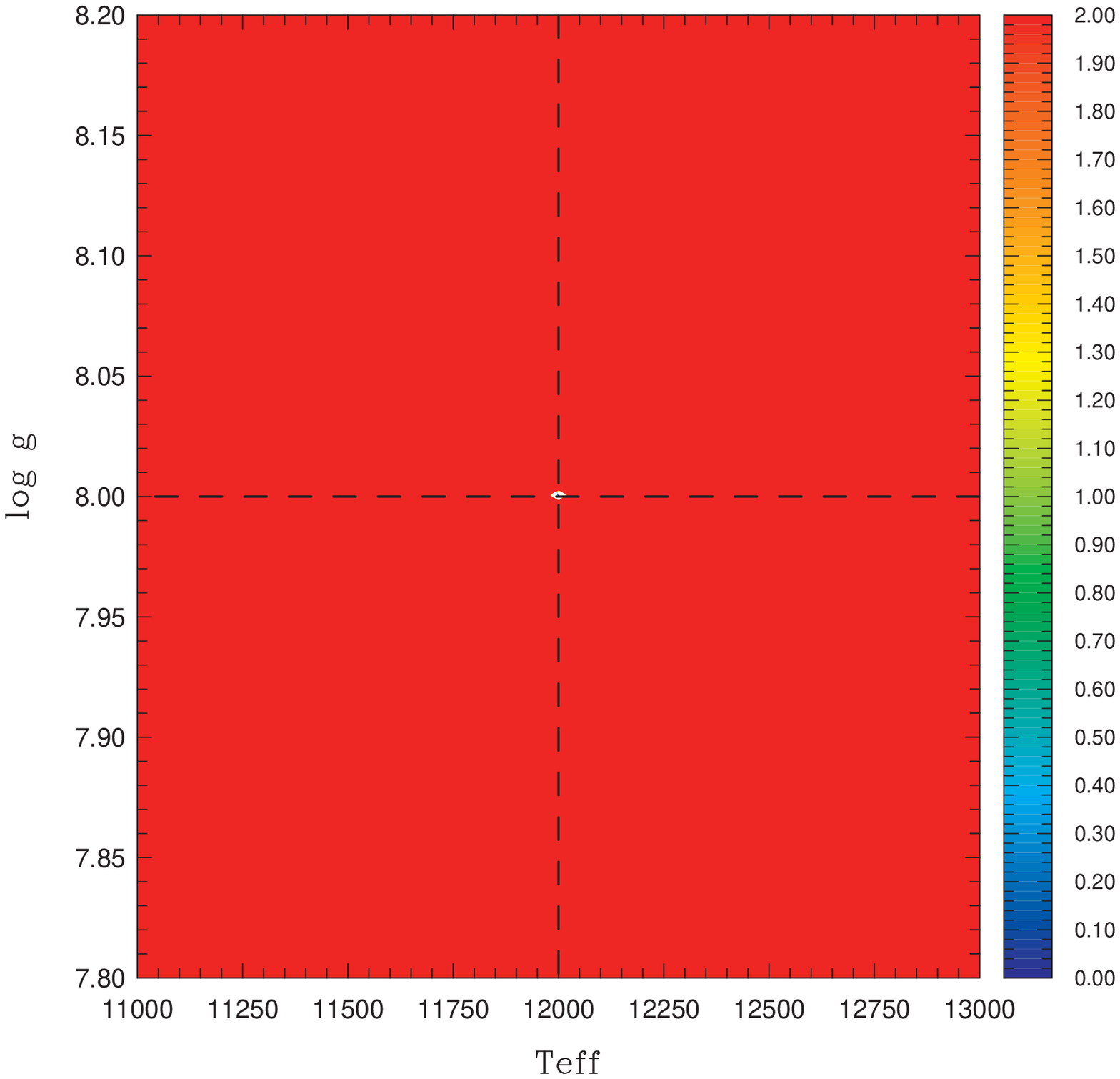} &
    \includegraphics[width=.35\textwidth]{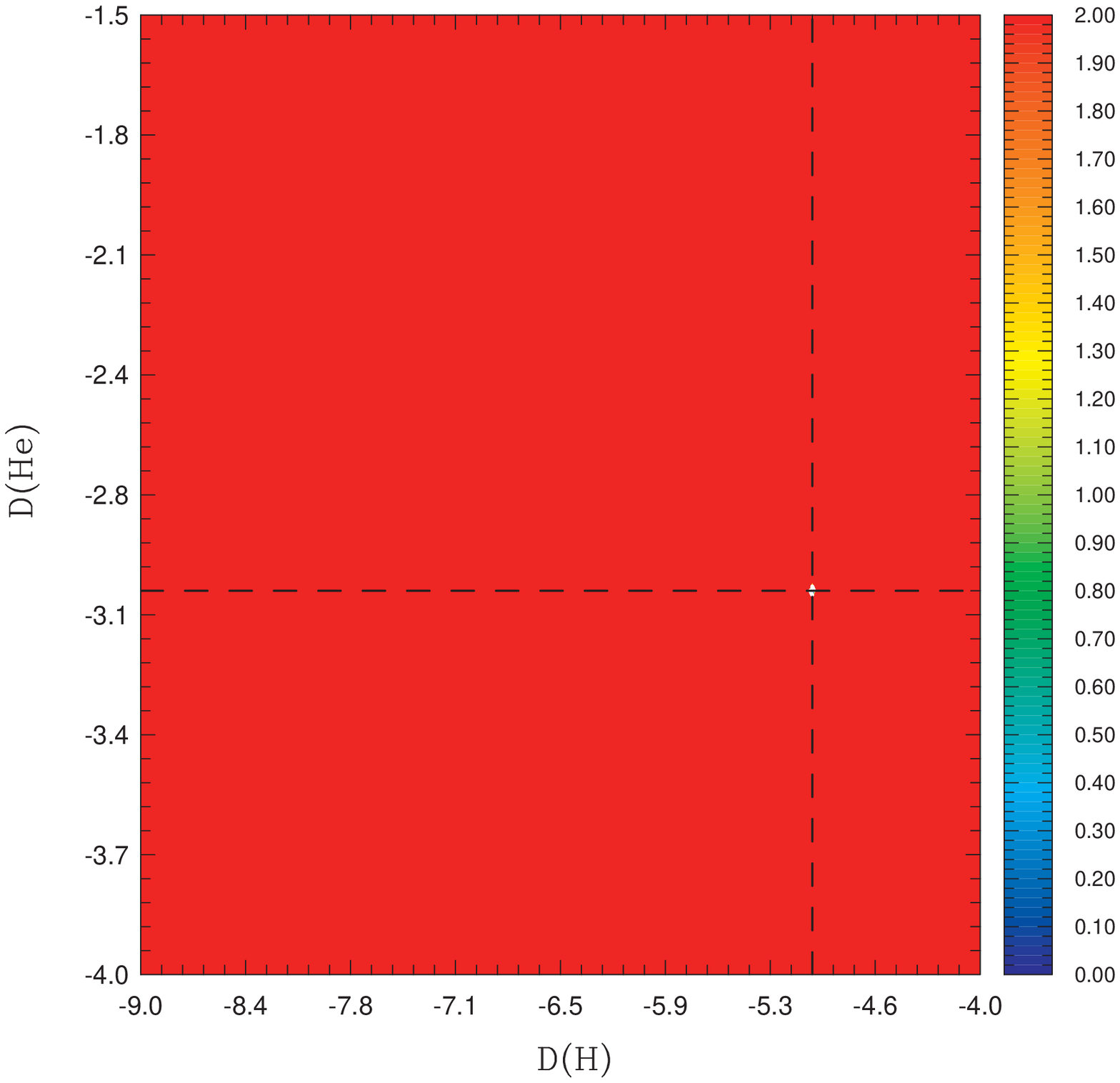} &
    \includegraphics[width=.35\textwidth]{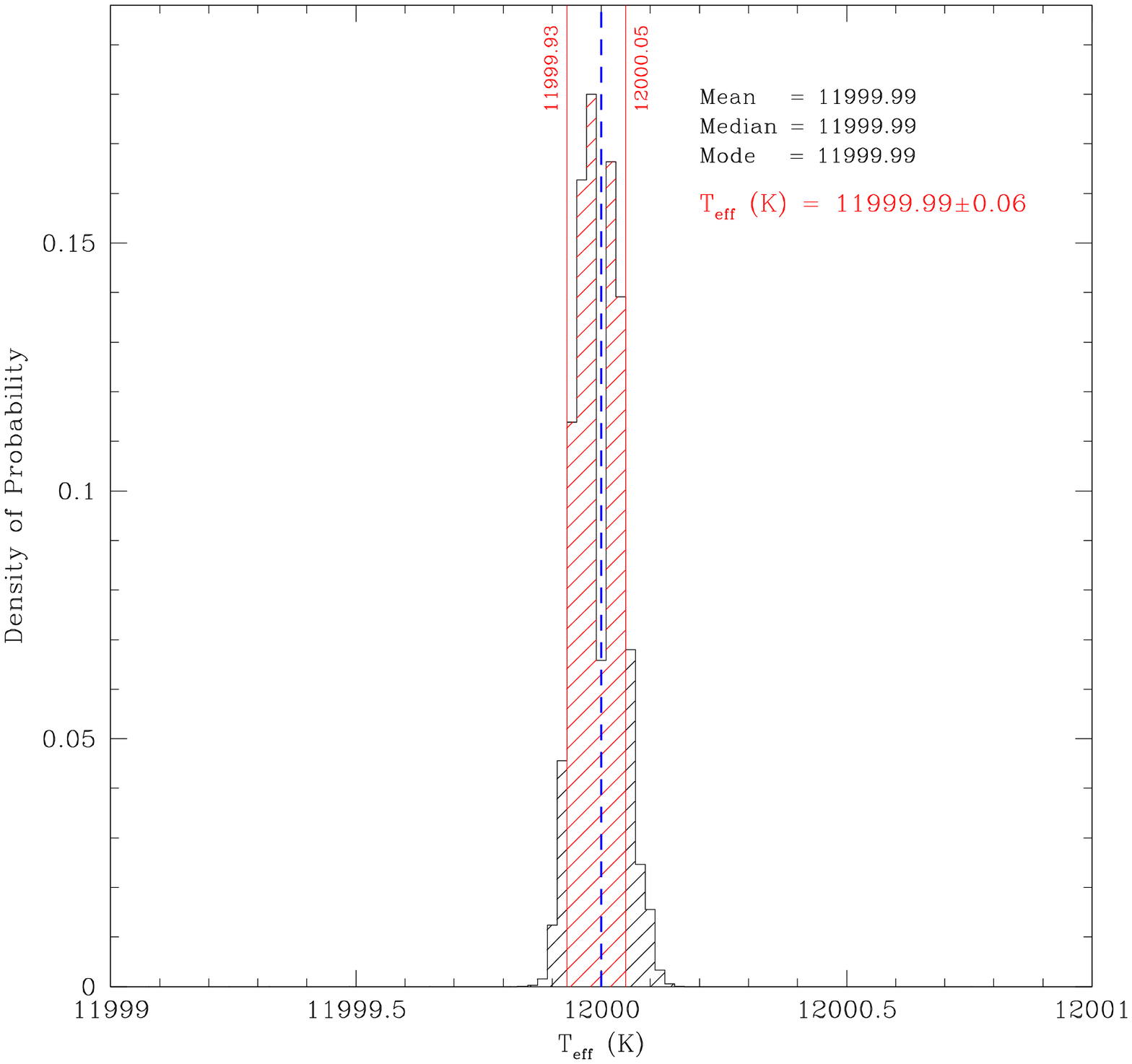} \\
    \includegraphics[width=.35\textwidth]{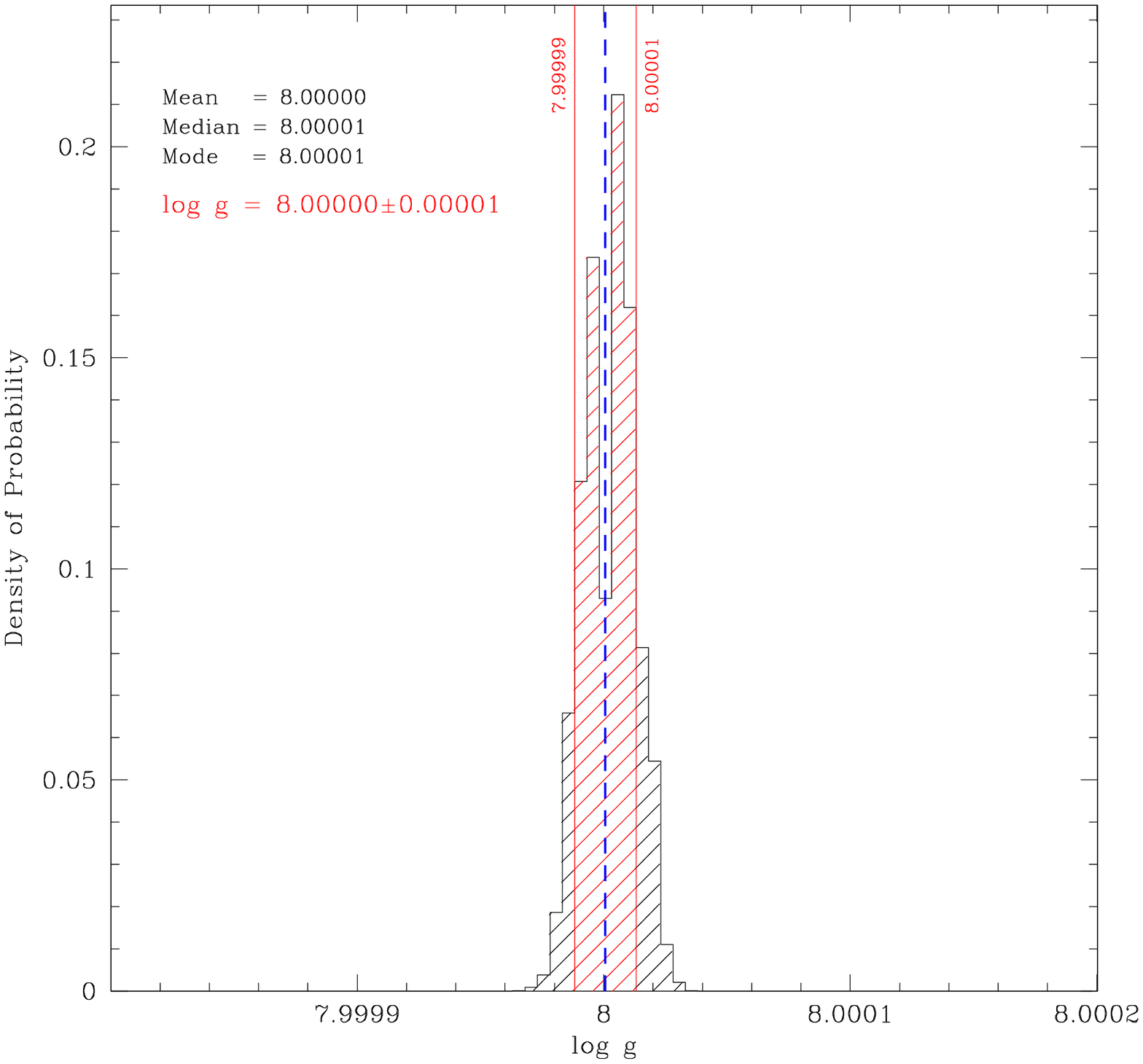}  &
    \includegraphics[width=.35\textwidth]{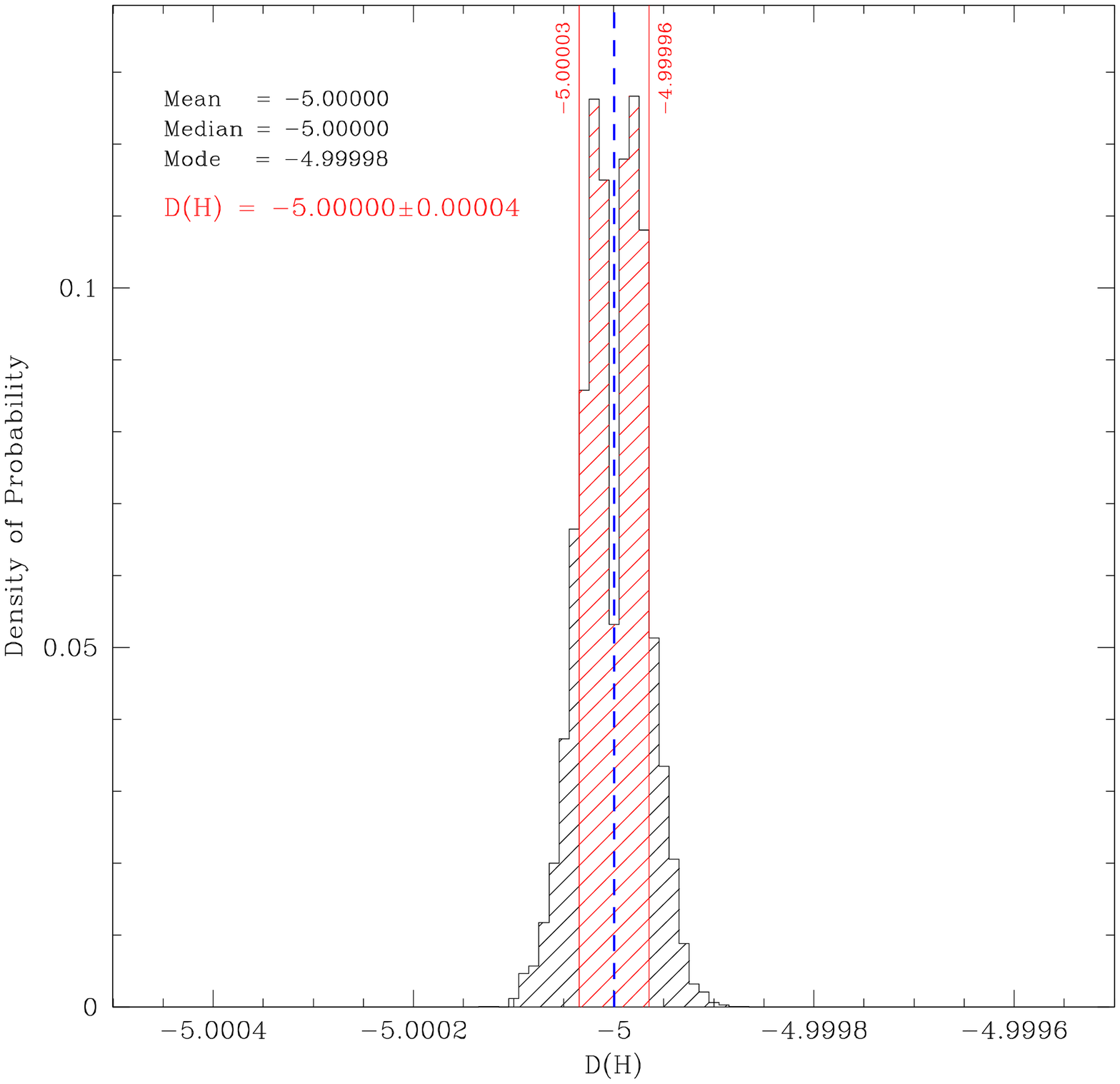} &
    \includegraphics[width=.35\textwidth]{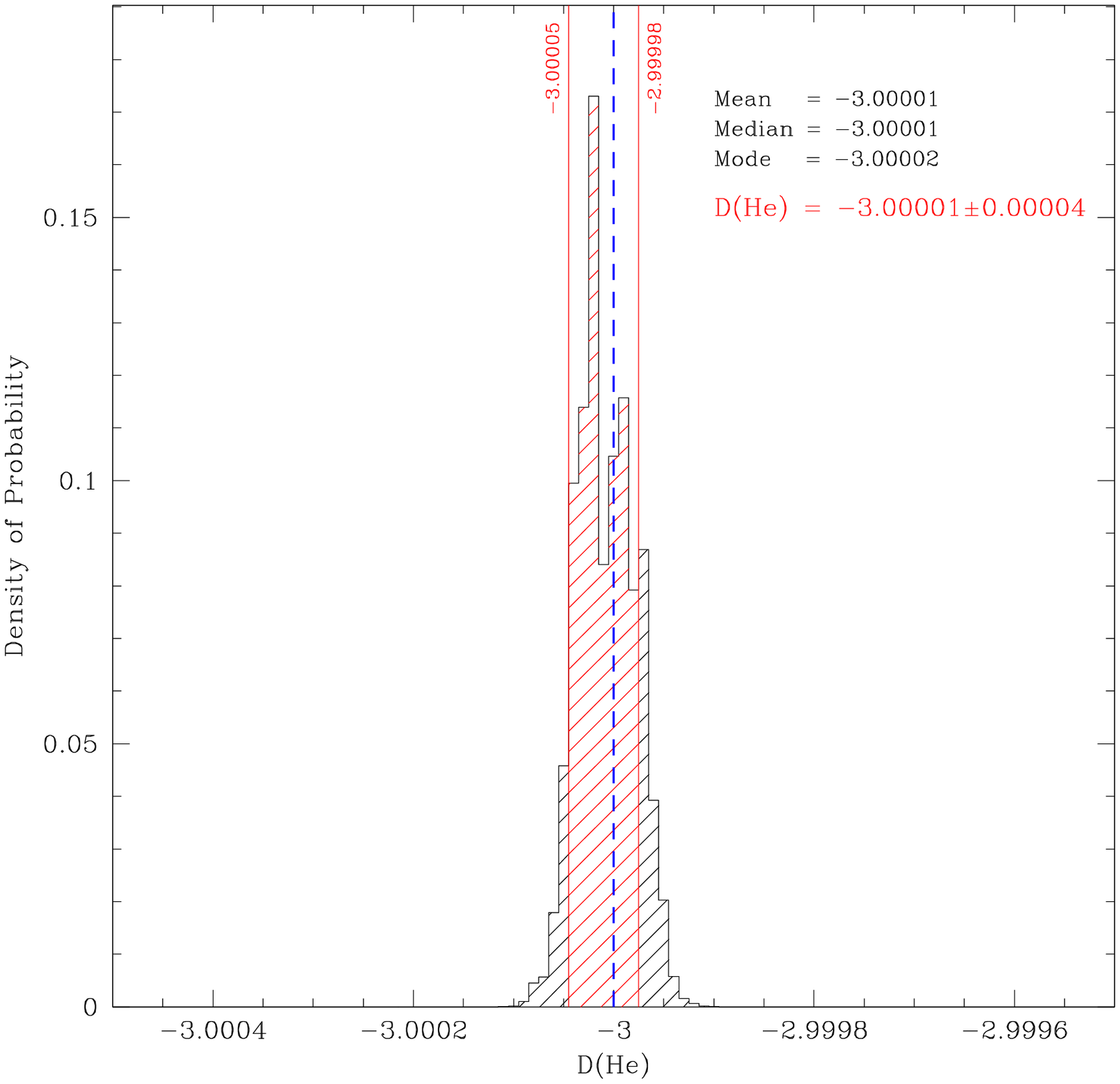} \\
    \includegraphics[width=.35\textwidth]{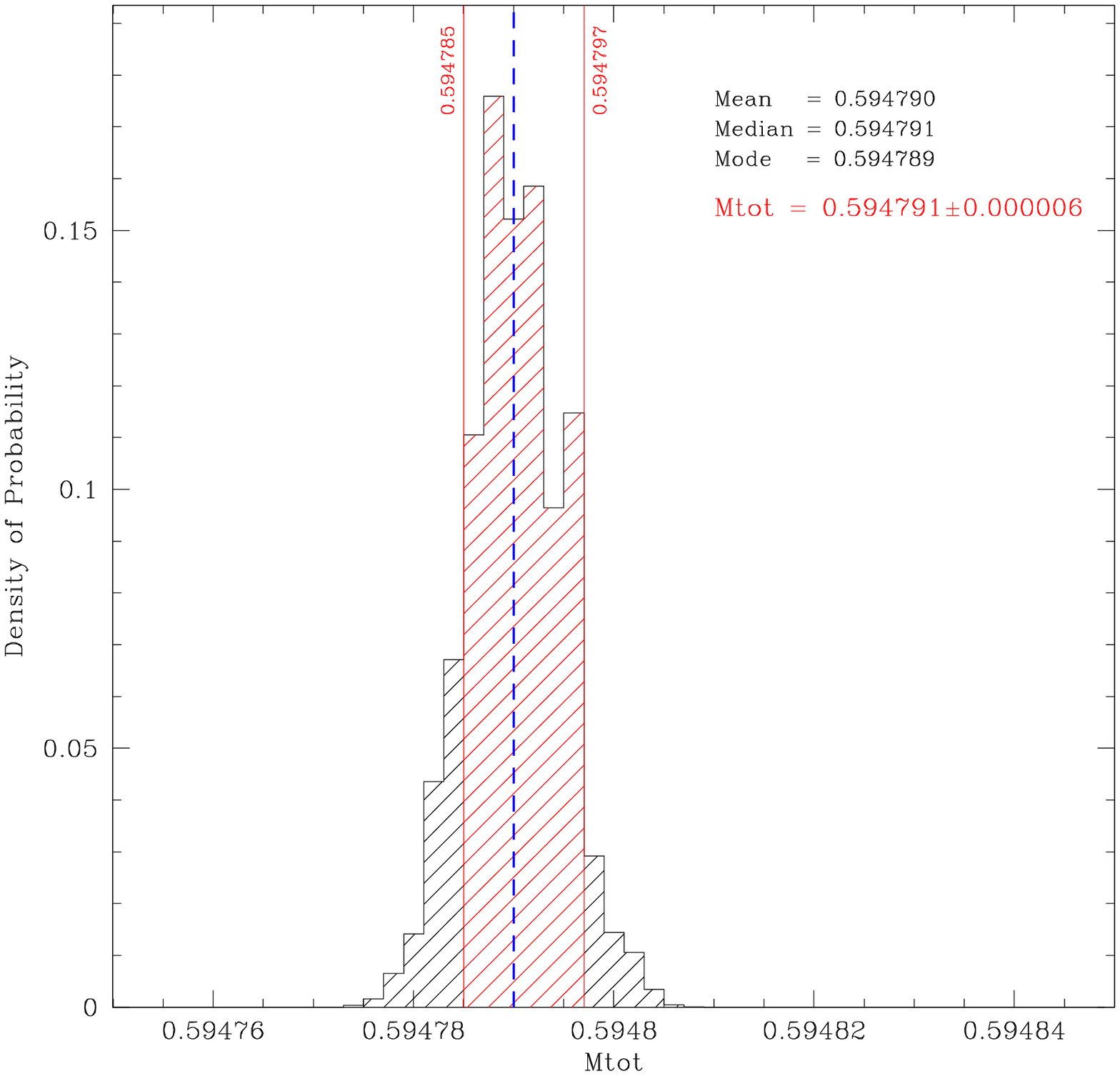} &
    \includegraphics[width=.35\textwidth]{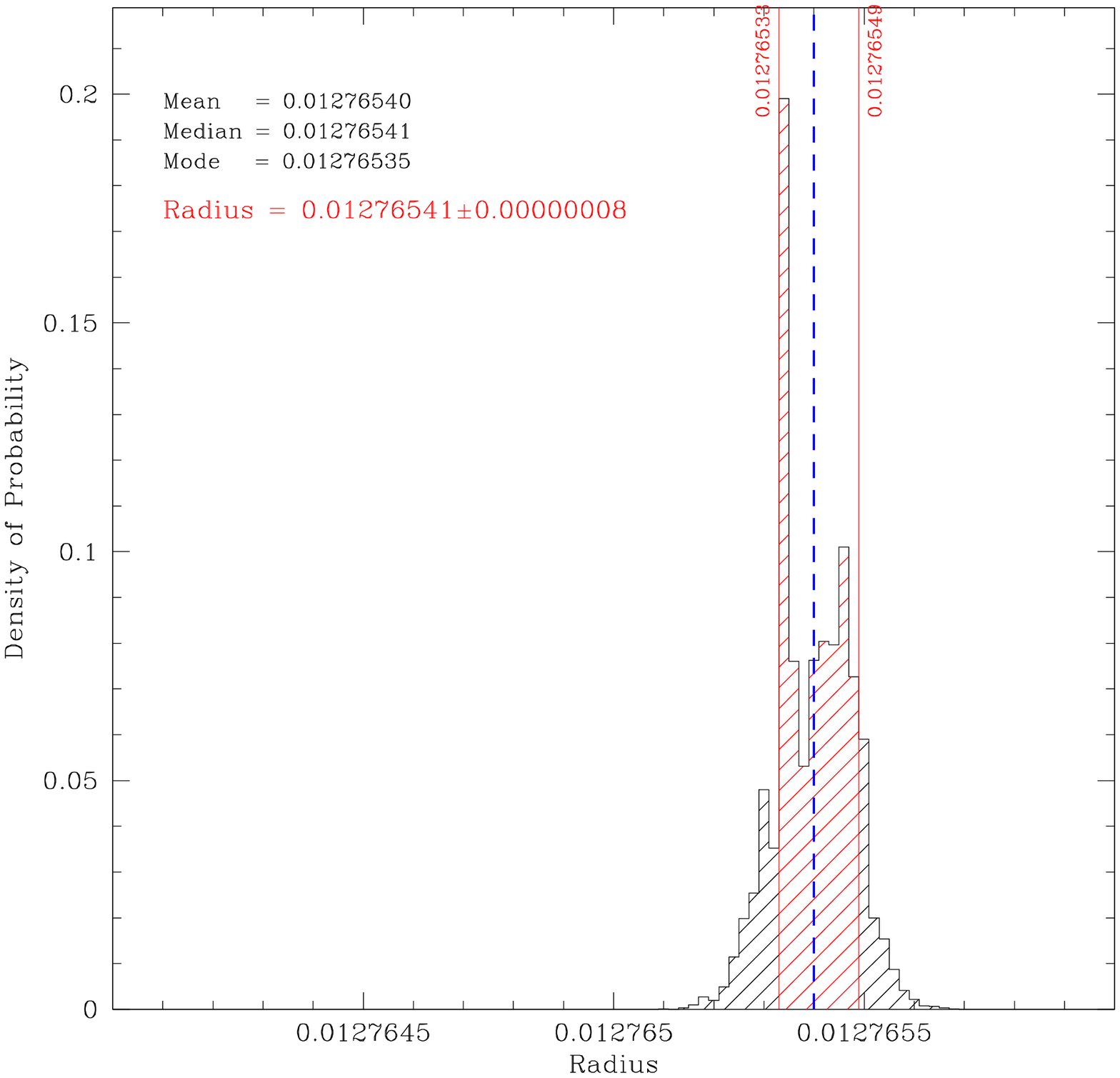} &
    \includegraphics[width=.35\textwidth]{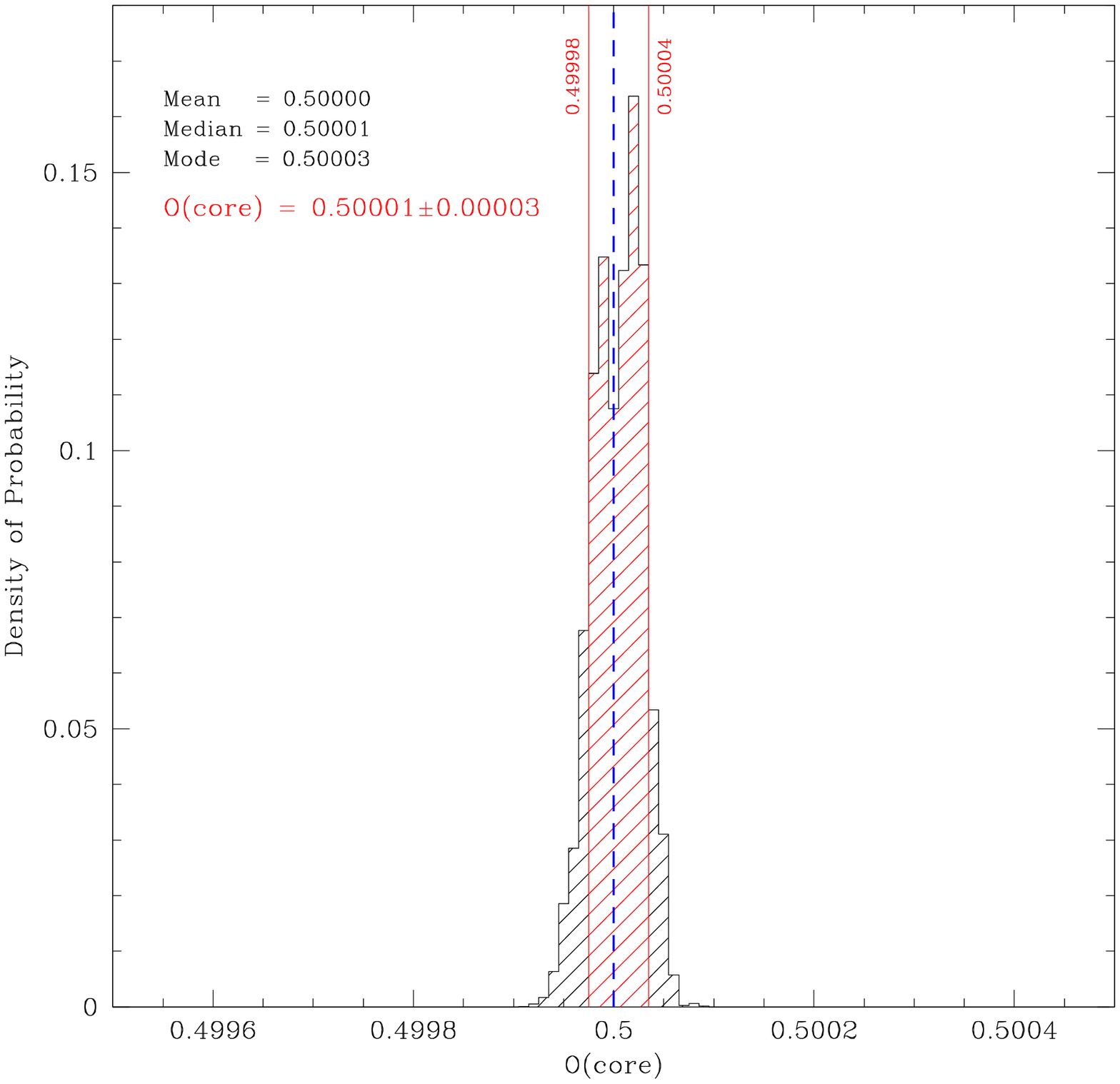}
    \end{tabular}
    \begin{flushright}
Figure~6
\end{flushright}
\end{figure}

\clearpage

\begin{figure}[!h]
\centering
  \begin{tabular}{@{}ccc@{}}
    \includegraphics[width=.35\textwidth]{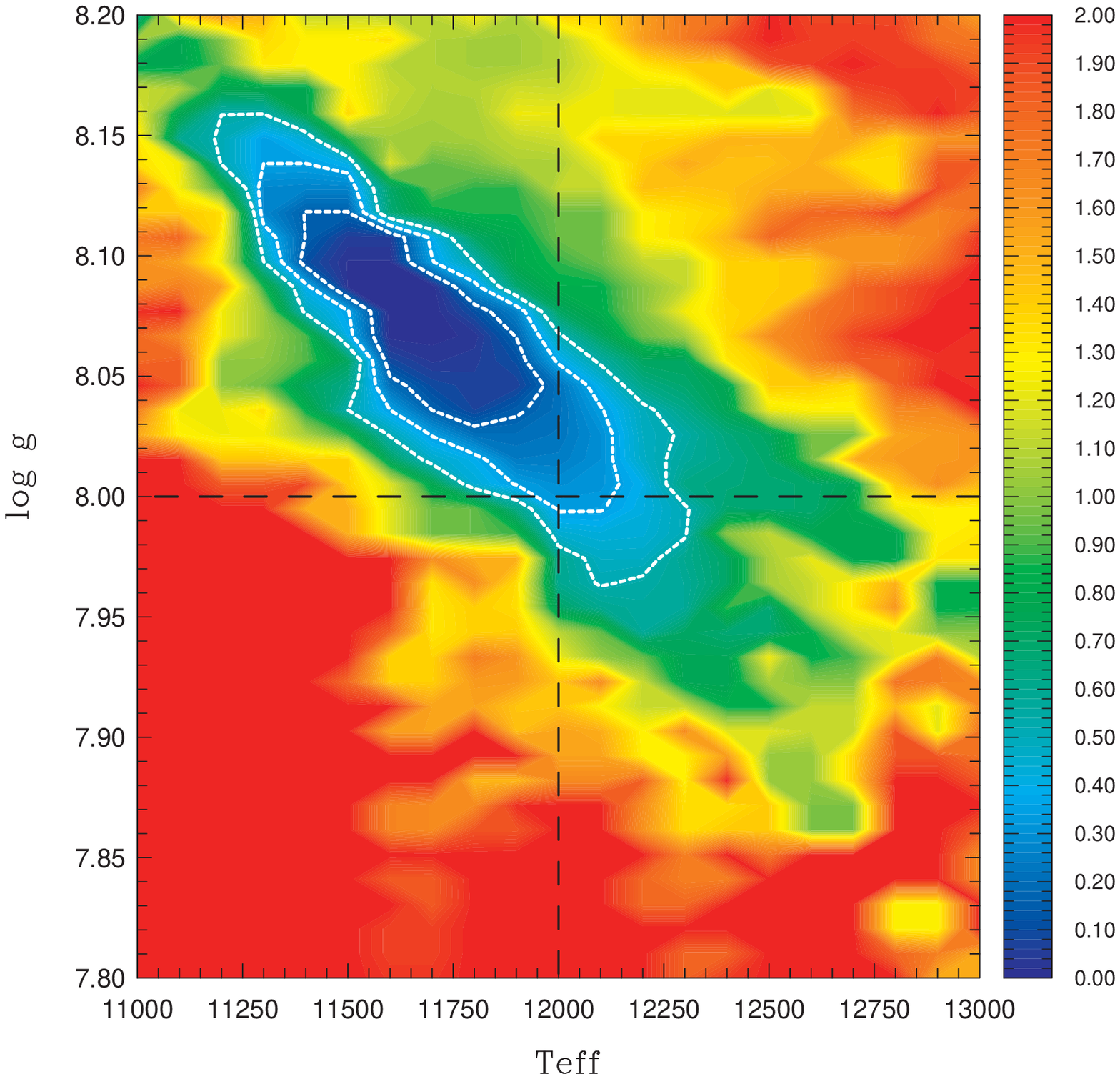} &
    \includegraphics[width=.35\textwidth]{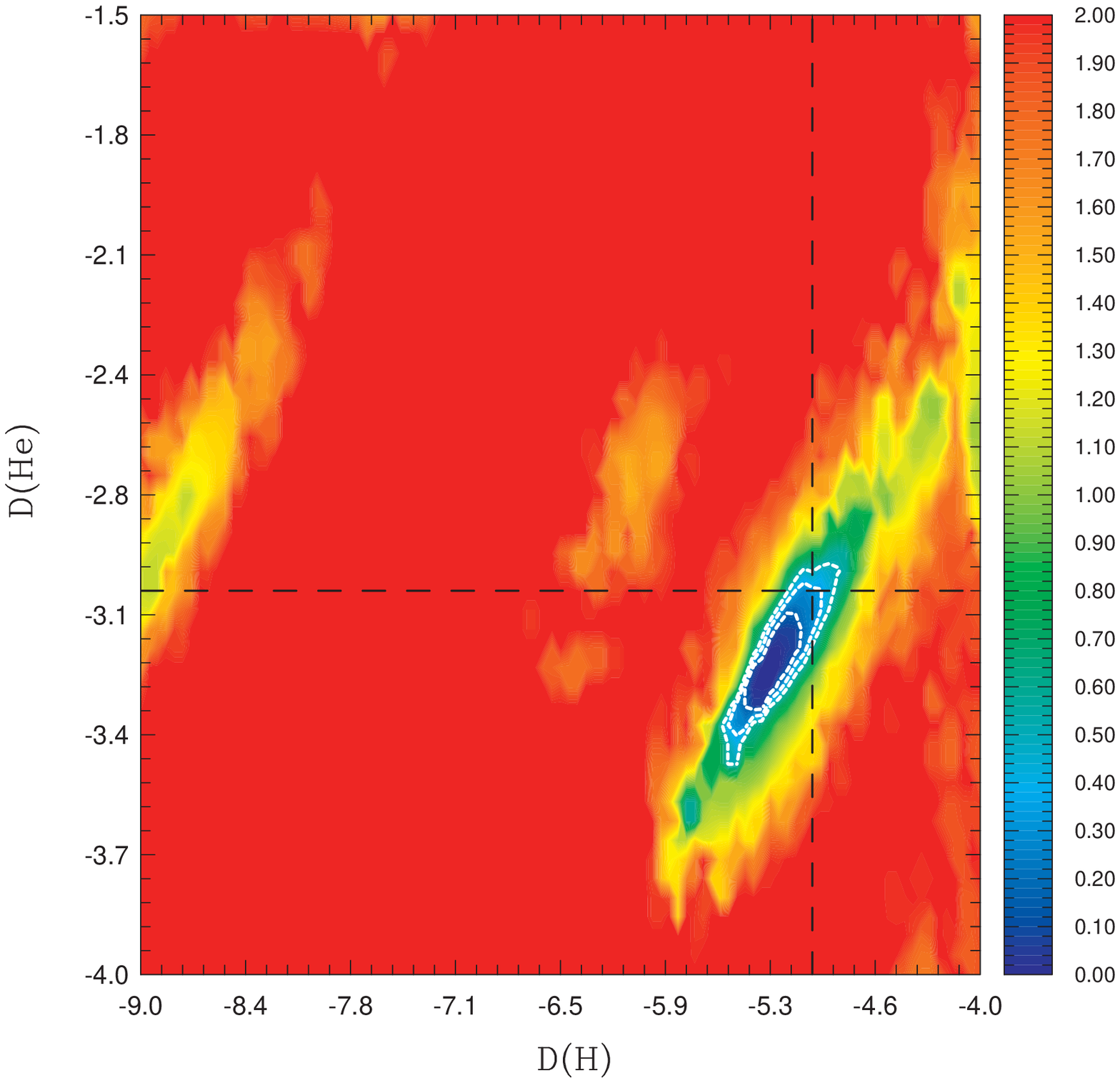} &
    \includegraphics[width=.35\textwidth]{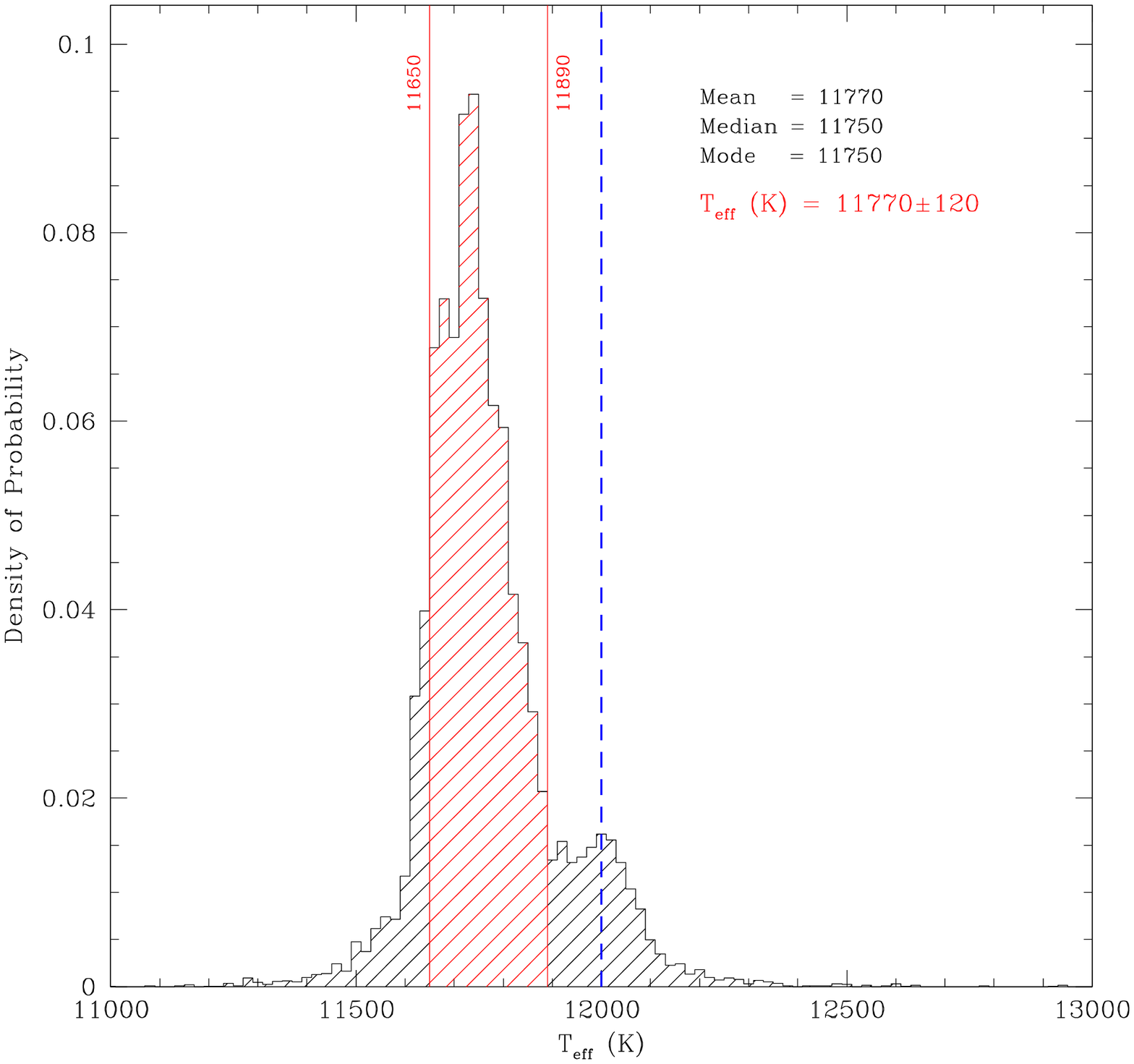} \\
    \includegraphics[width=.35\textwidth]{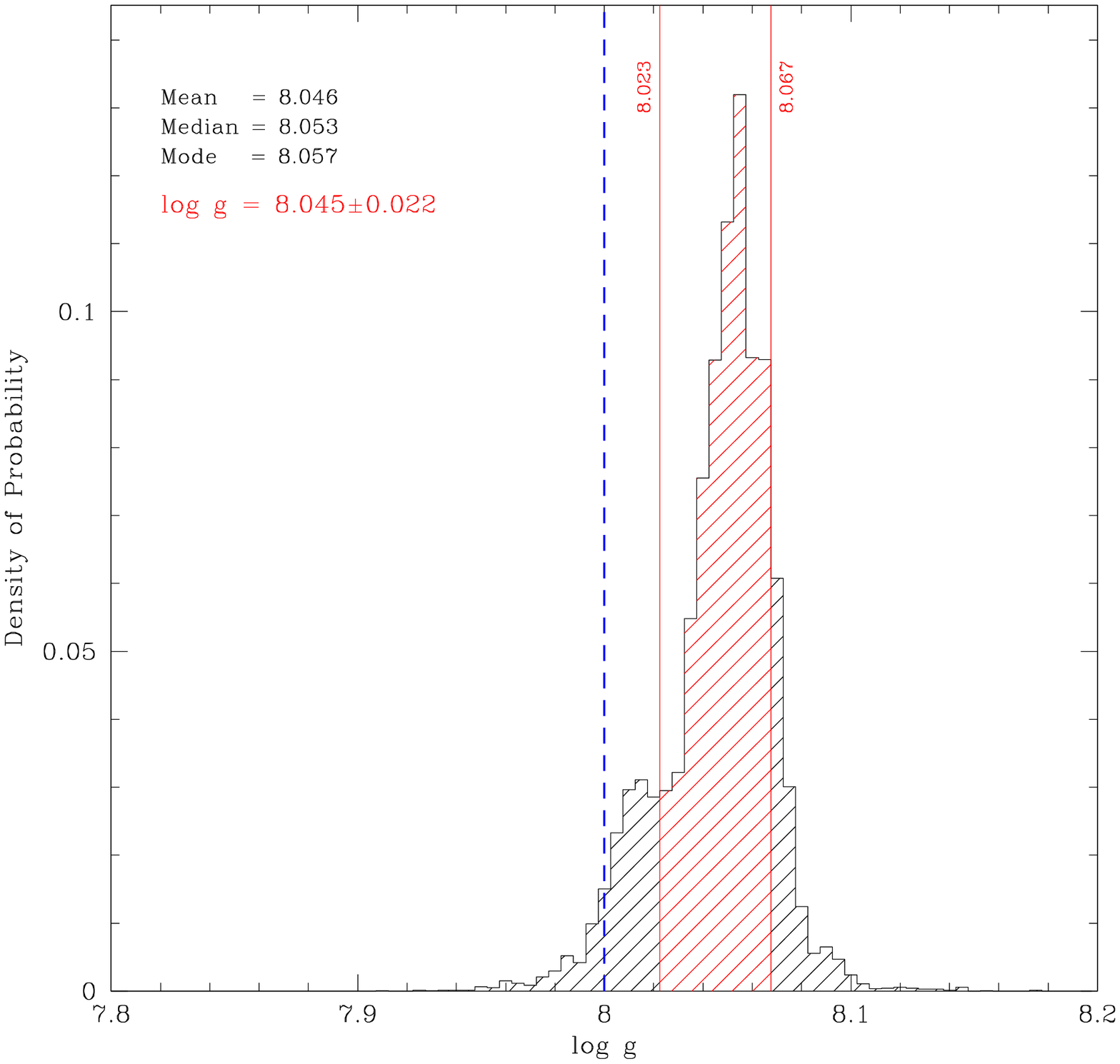}  &
    \includegraphics[width=.35\textwidth]{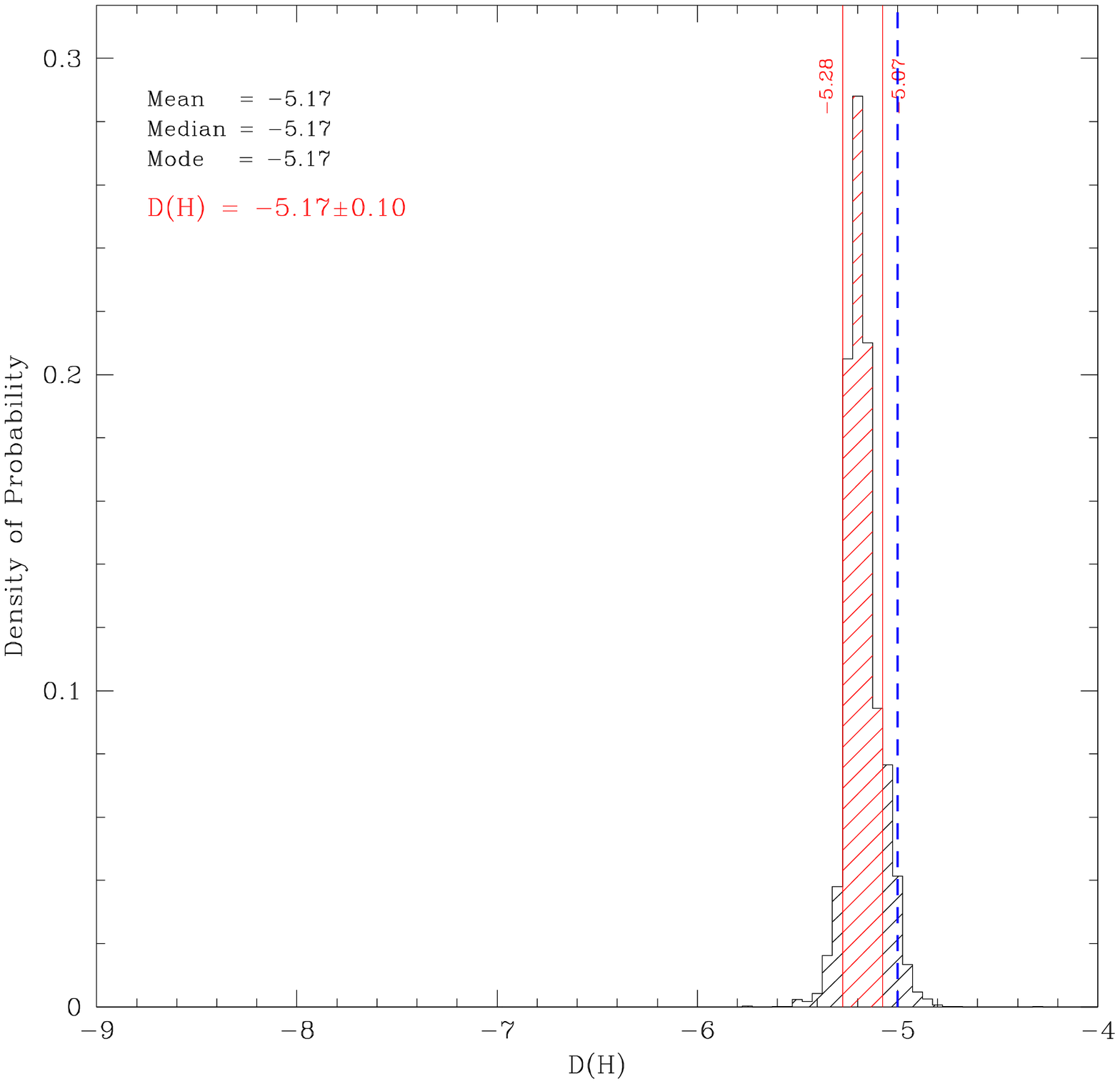} &
    \includegraphics[width=.35\textwidth]{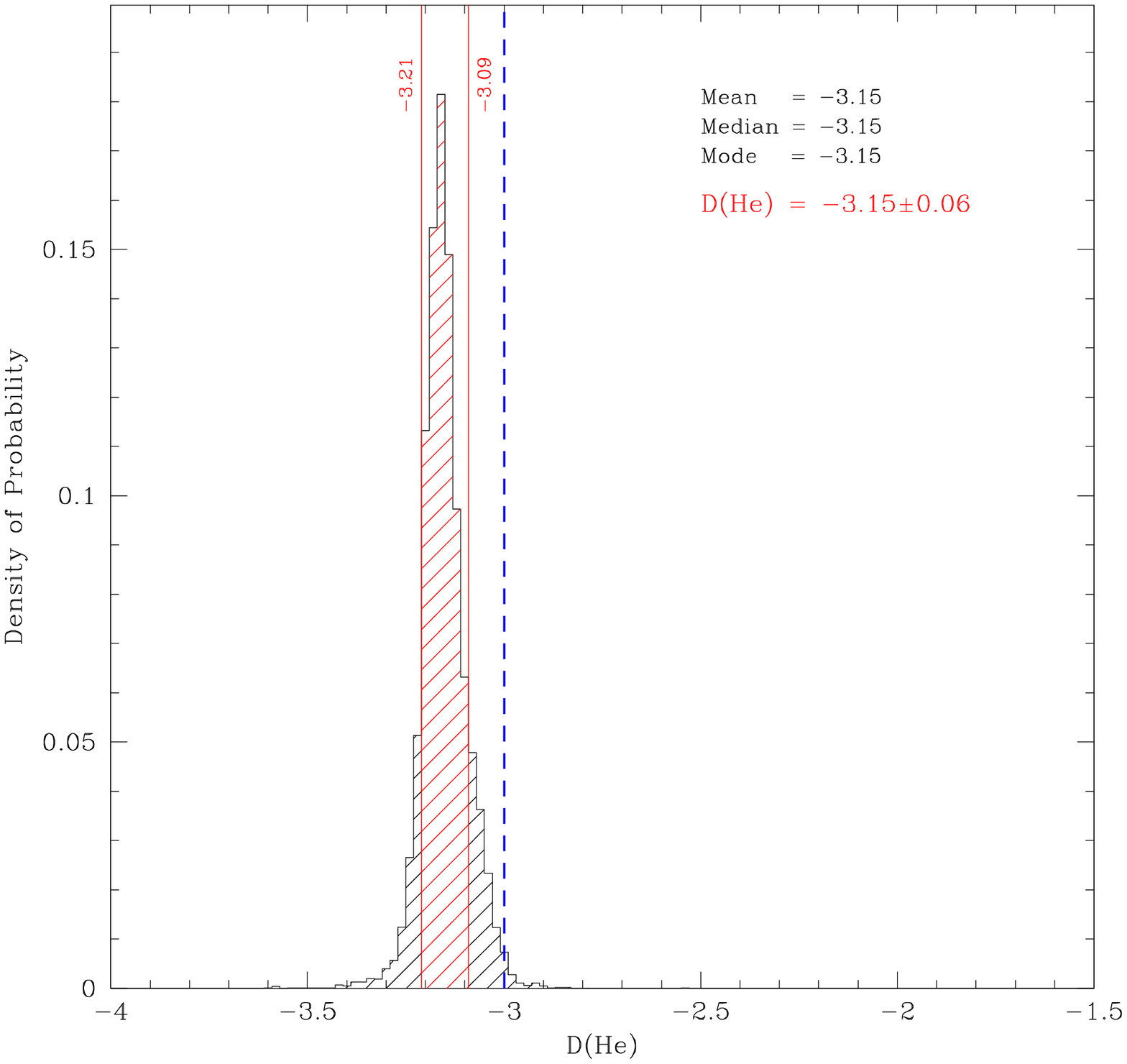} \\
    \includegraphics[width=.35\textwidth]{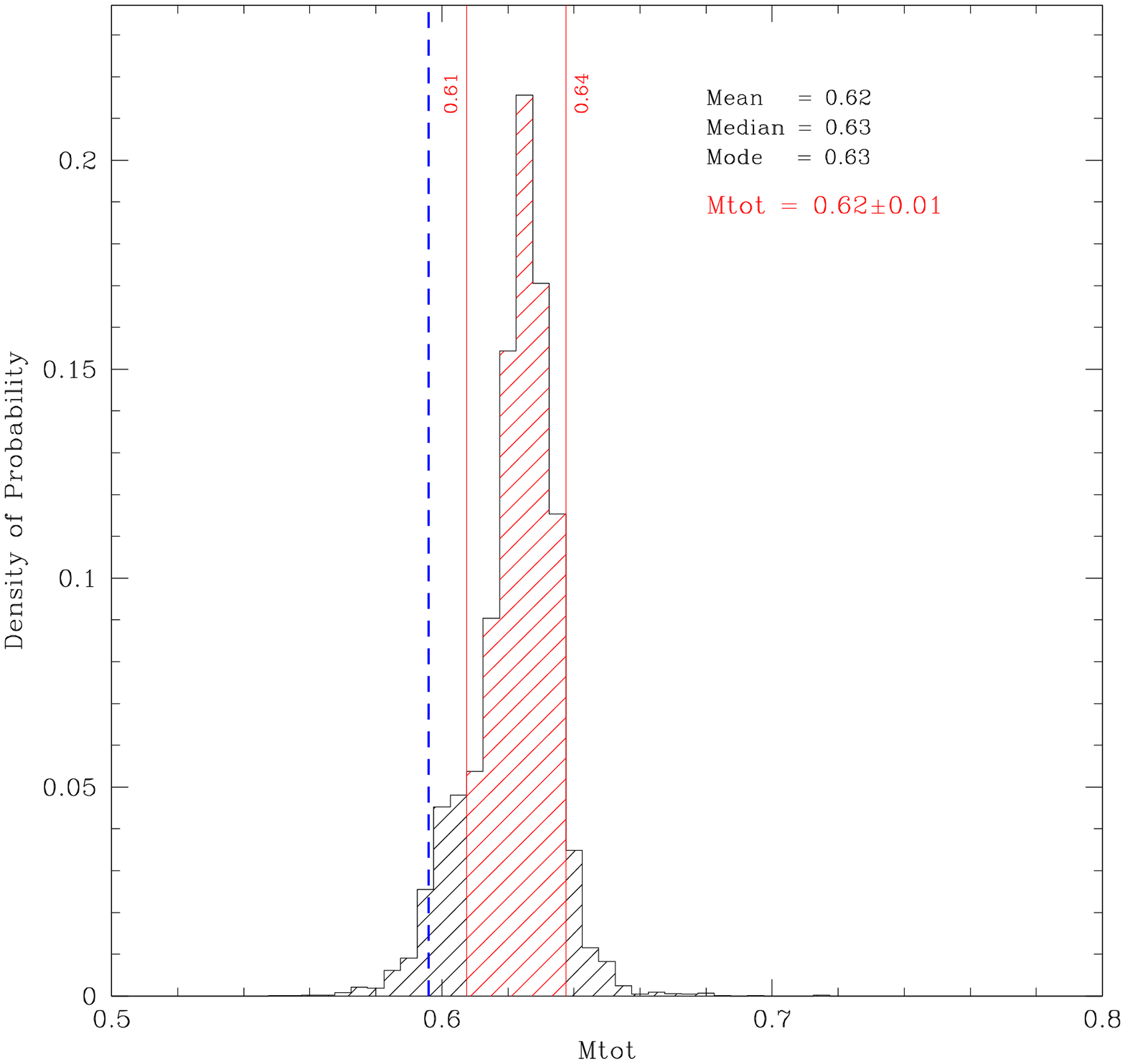} &
    \includegraphics[width=.35\textwidth]{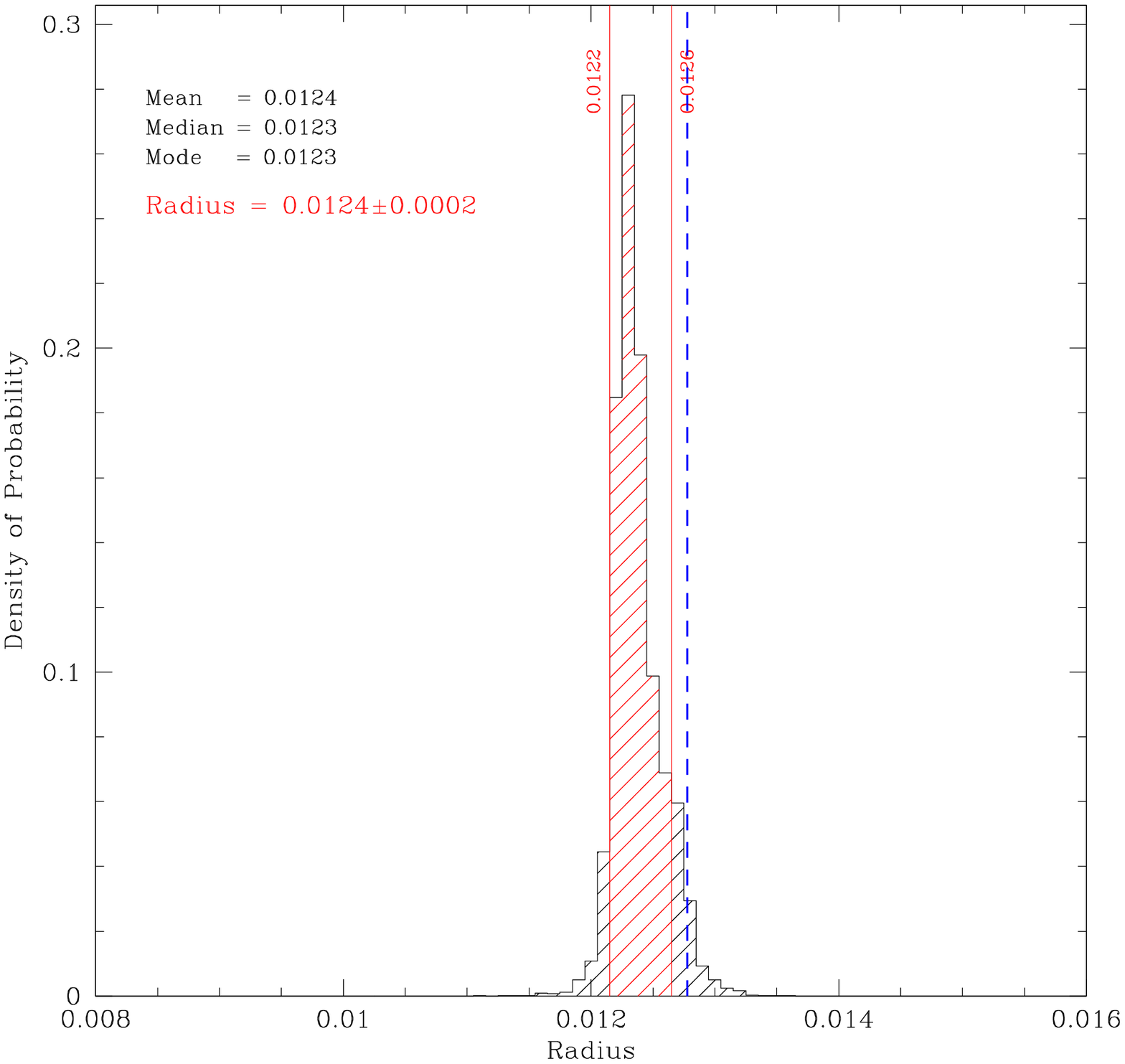} &
    \includegraphics[width=.35\textwidth]{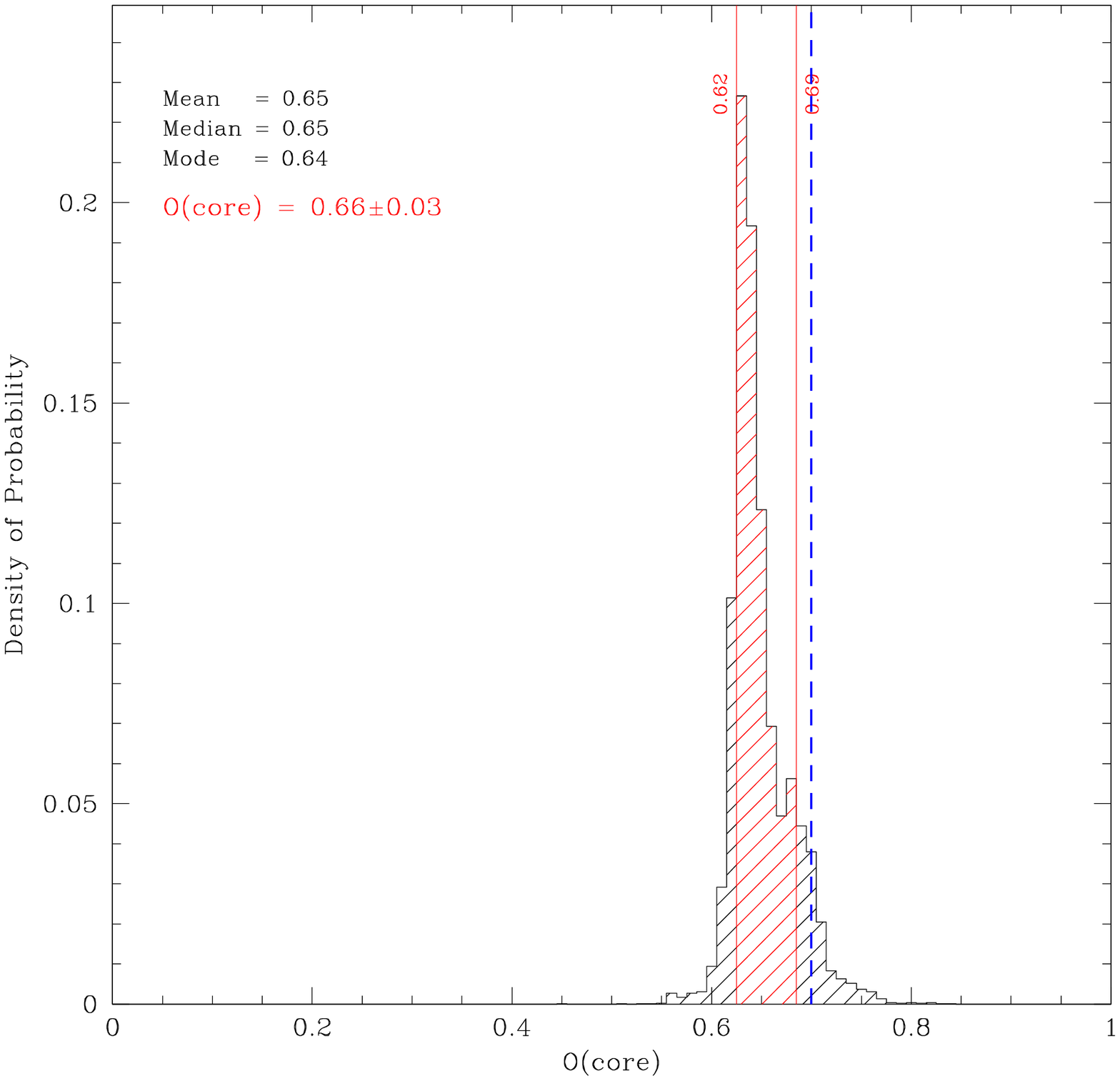} \\
    \end{tabular}
    \begin{flushright}
Figure~7
\end{flushright}
\end{figure}

\clearpage
\addtocounter{figure}{-1}

\begin{figure}[!h]
\centering
  \begin{tabular}{@{}ccc@{}}
    \includegraphics[width=.35\textwidth]{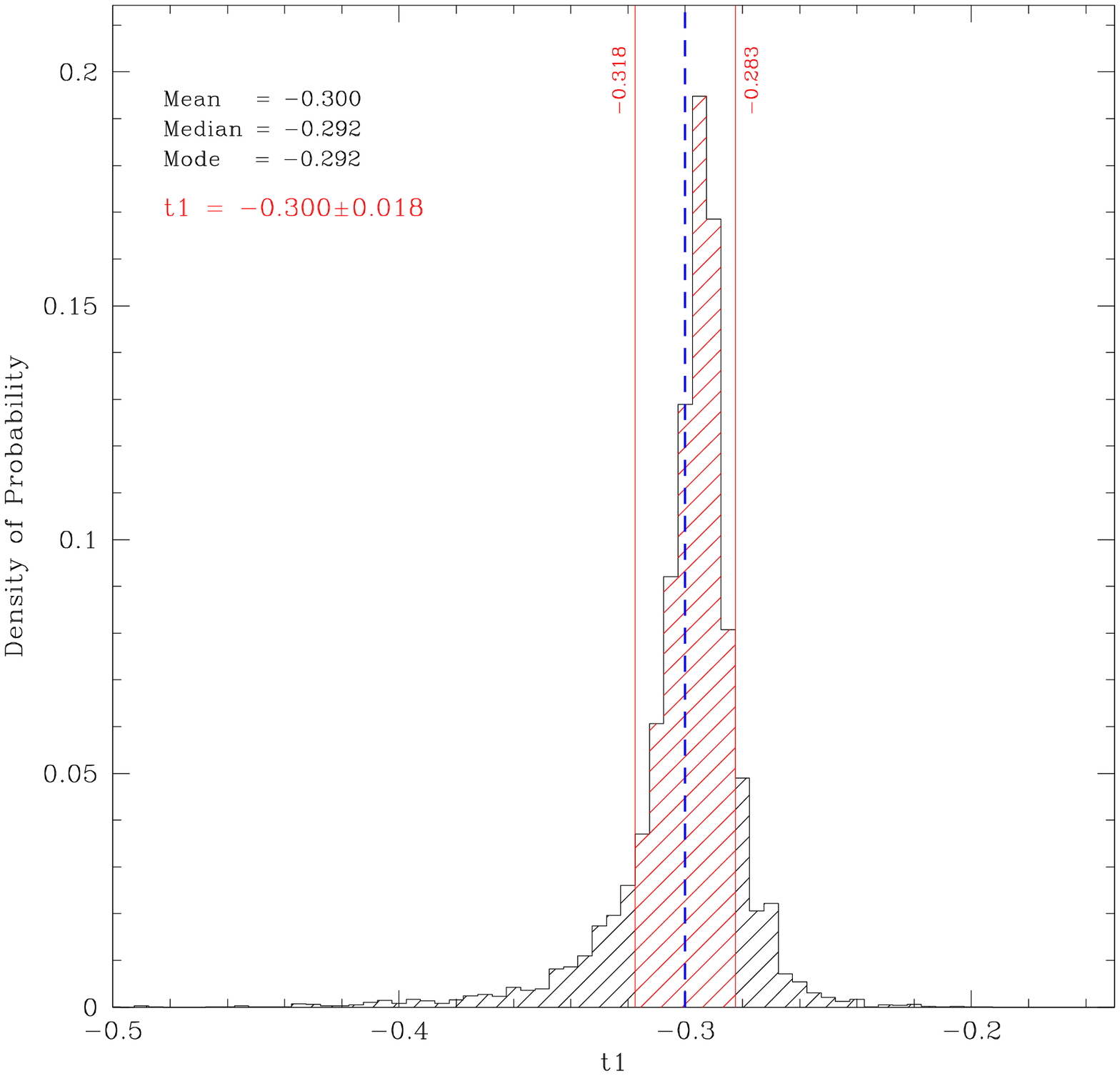} &
    \includegraphics[width=.35\textwidth]{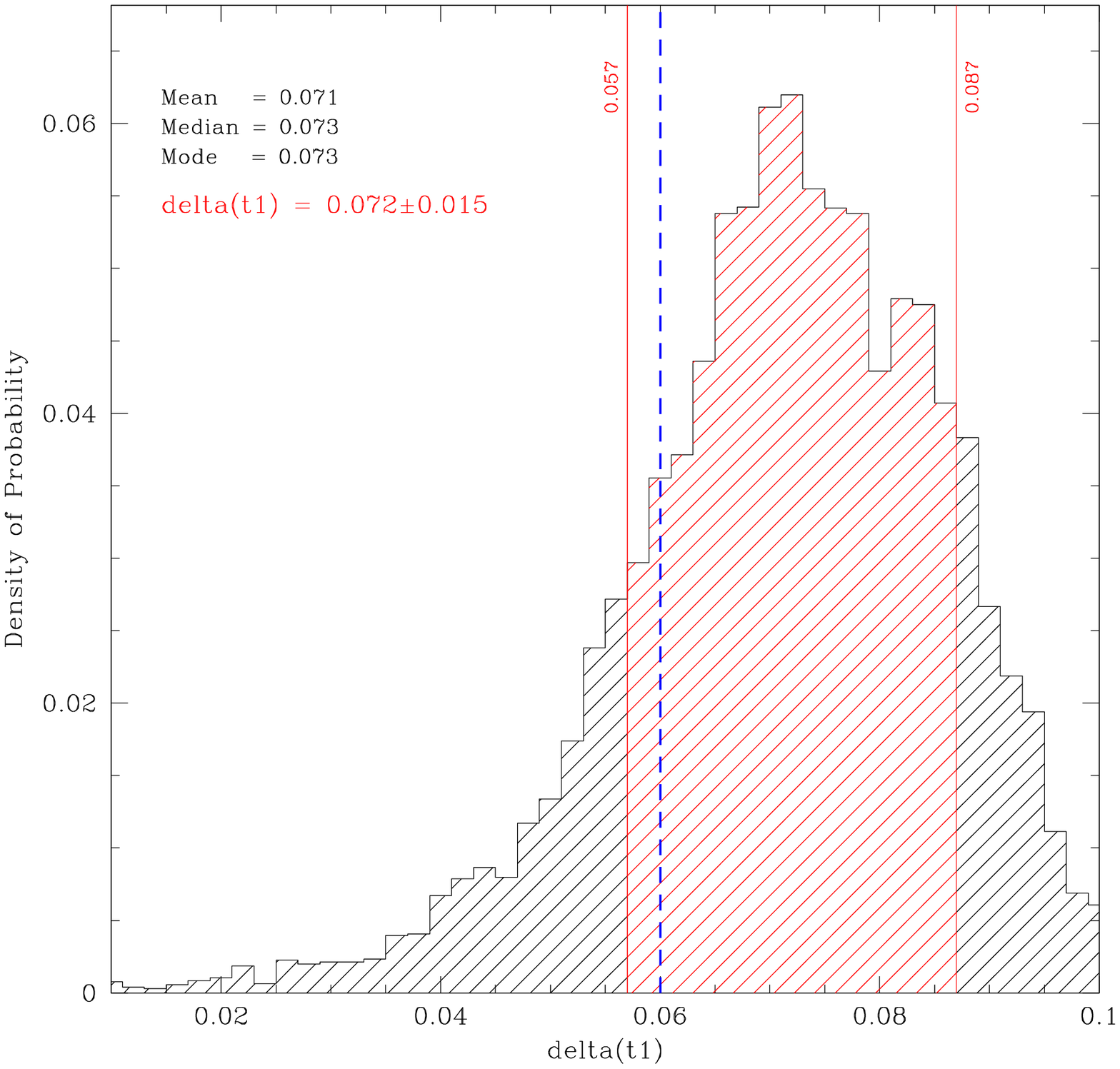}
    \end{tabular}
     \begin{flushright}
Figure~7
\end{flushright}
\end{figure}

\clearpage
\begin{figure}[!h]
\centering
  \begin{tabular}{@{}ccc@{}}
    \includegraphics[width=.35\textwidth]{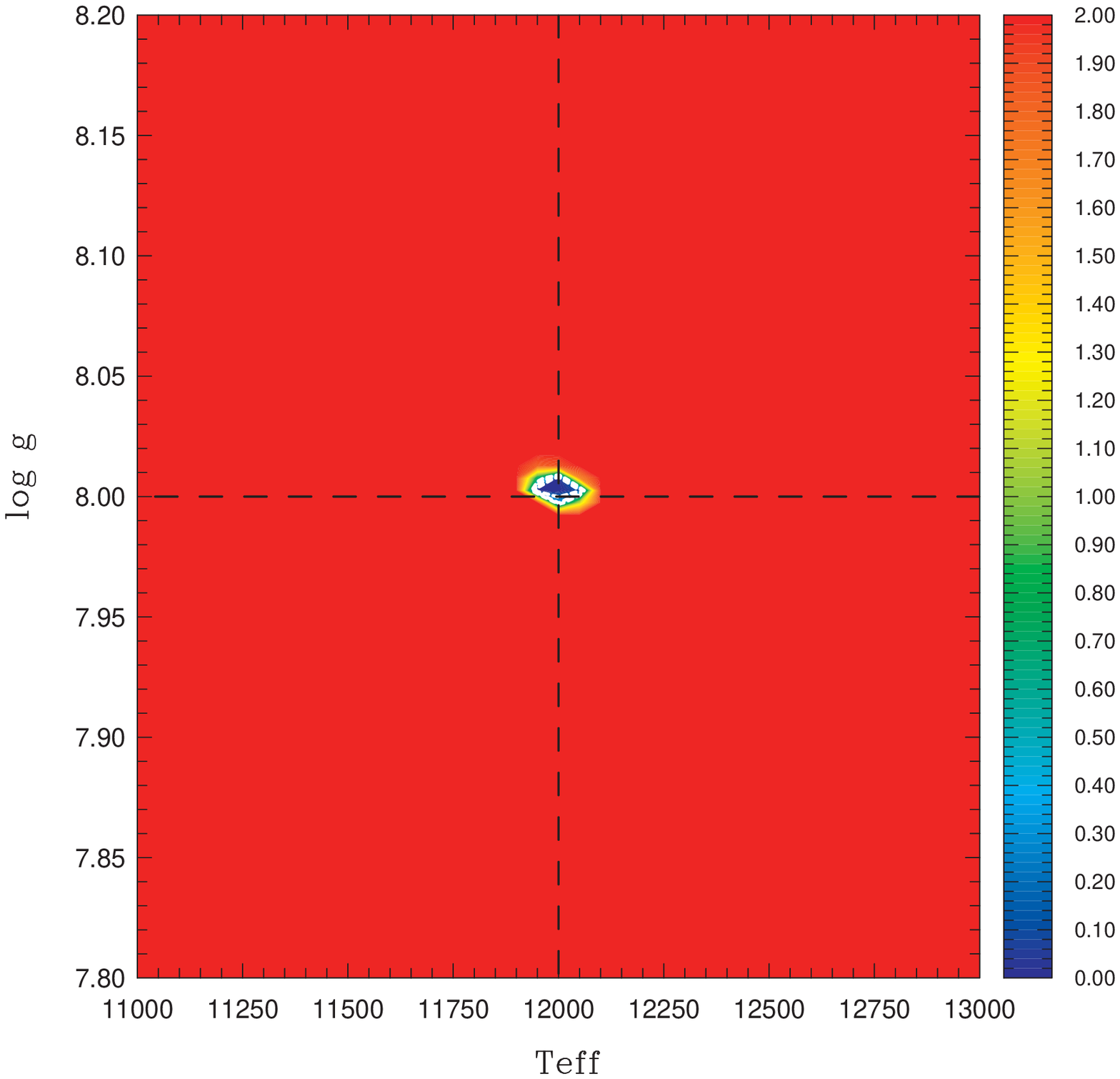} &
    \includegraphics[width=.35\textwidth]{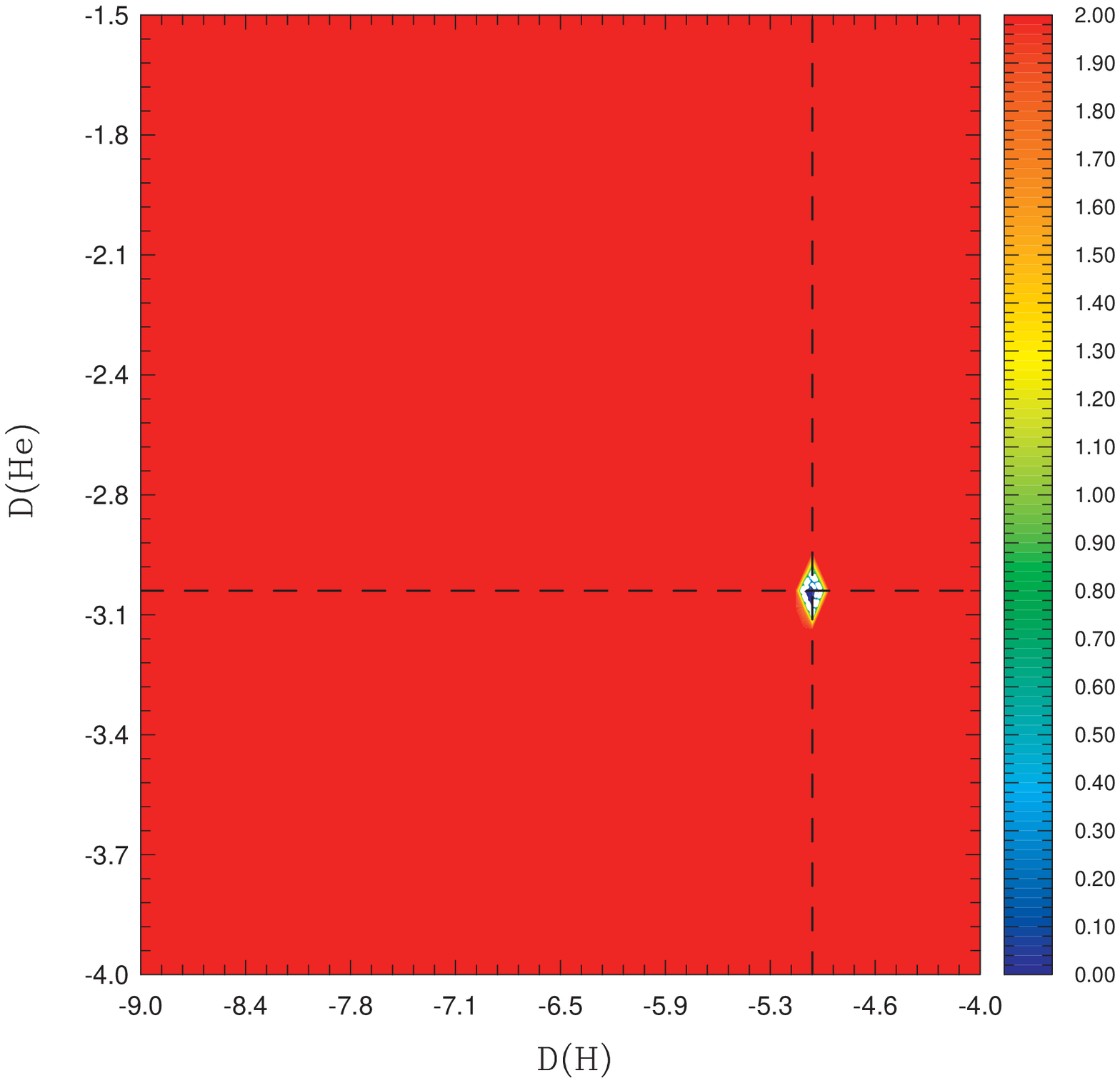} &
    \includegraphics[width=.35\textwidth]{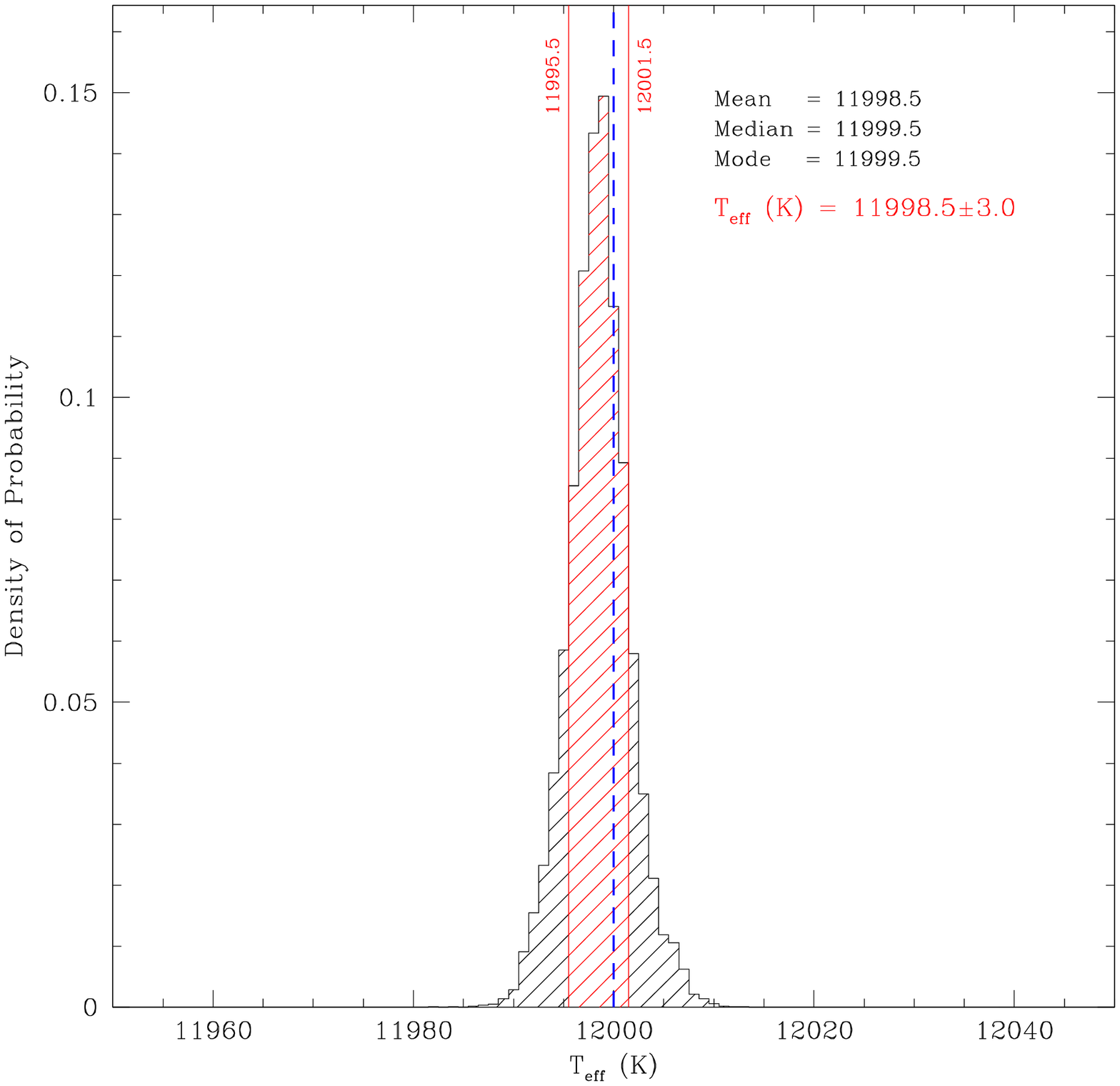} \\
    \includegraphics[width=.35\textwidth]{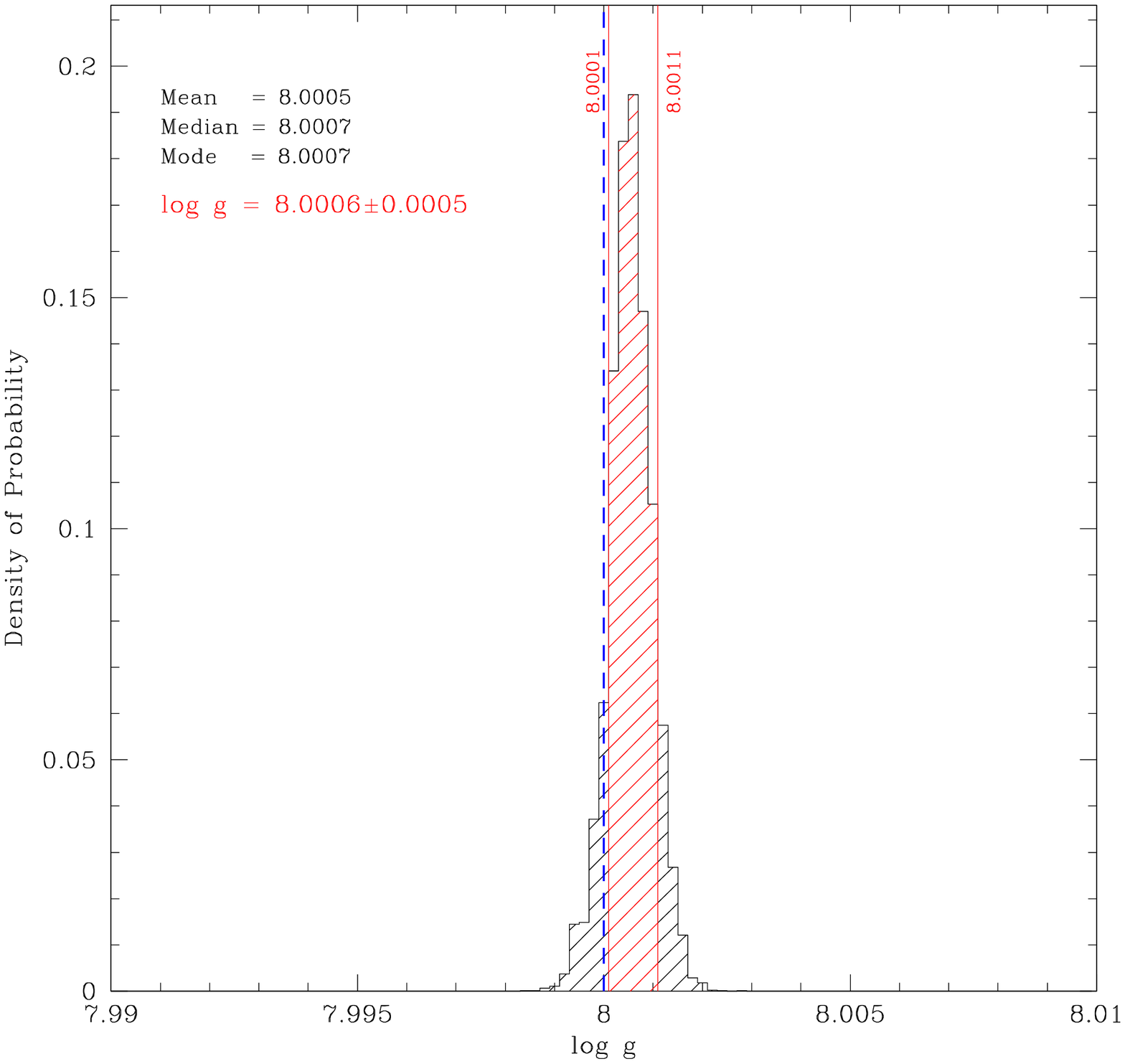}  &
    \includegraphics[width=.35\textwidth]{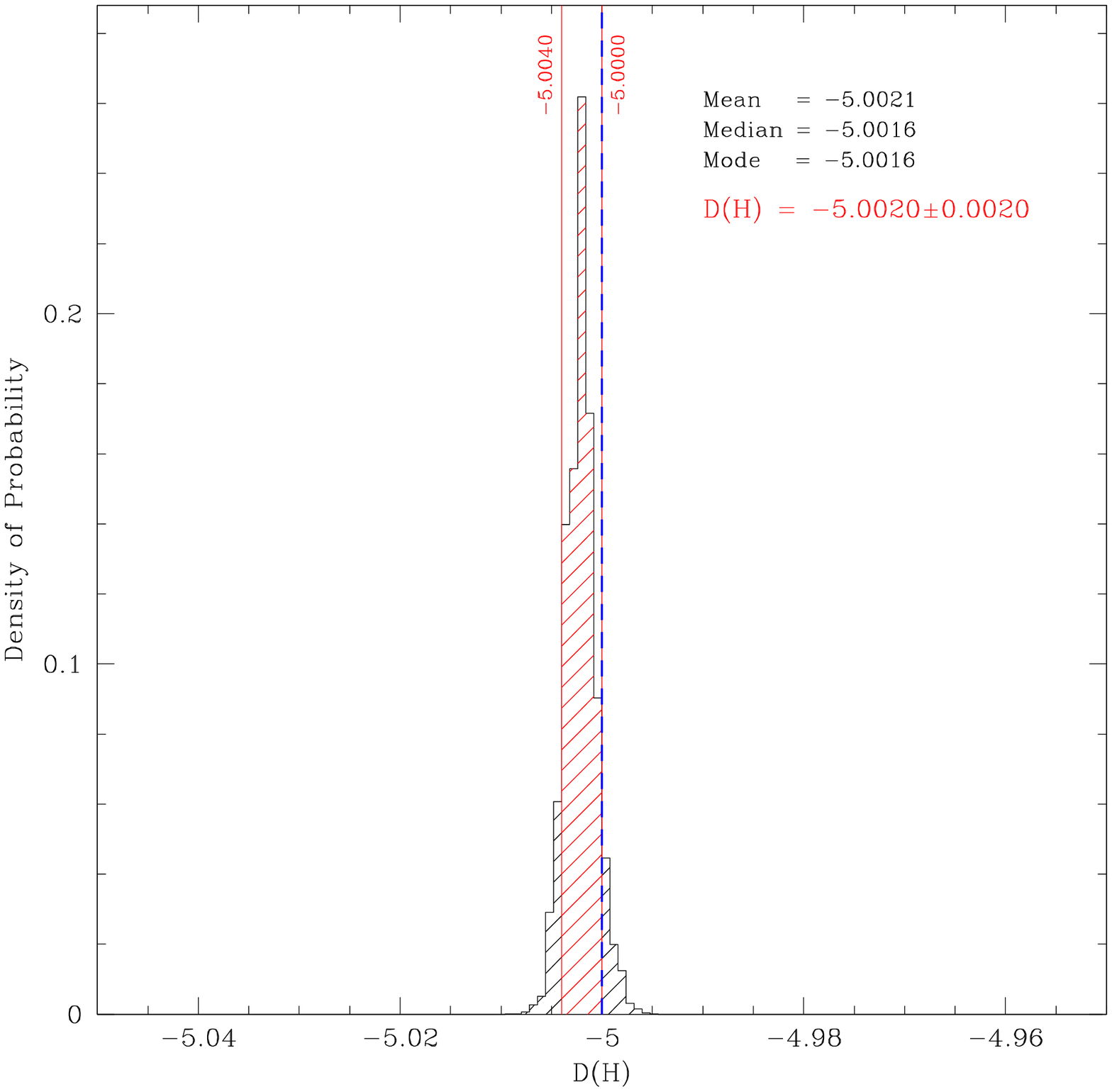} &
    \includegraphics[width=.35\textwidth]{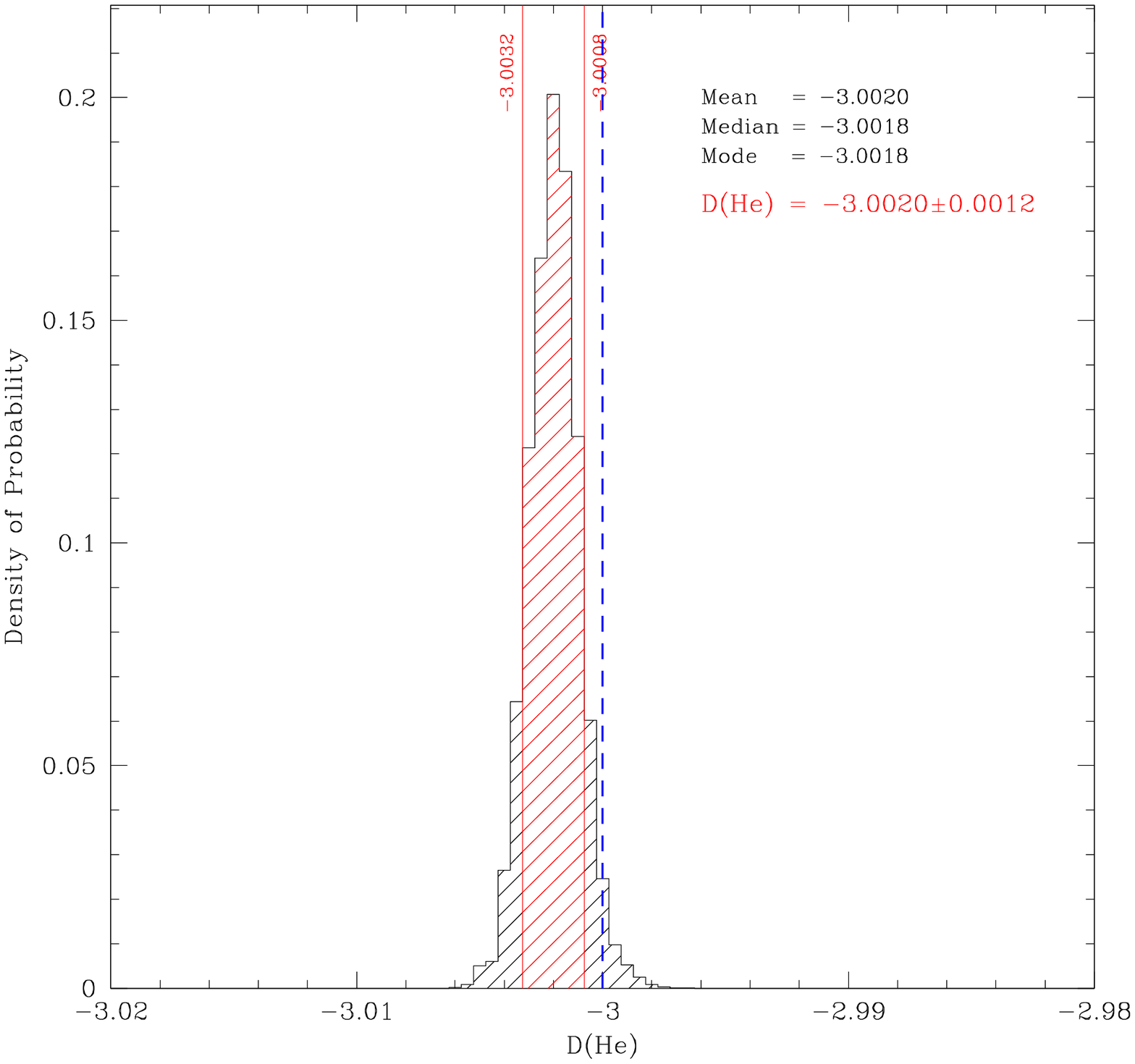} \\
    \includegraphics[width=.35\textwidth]{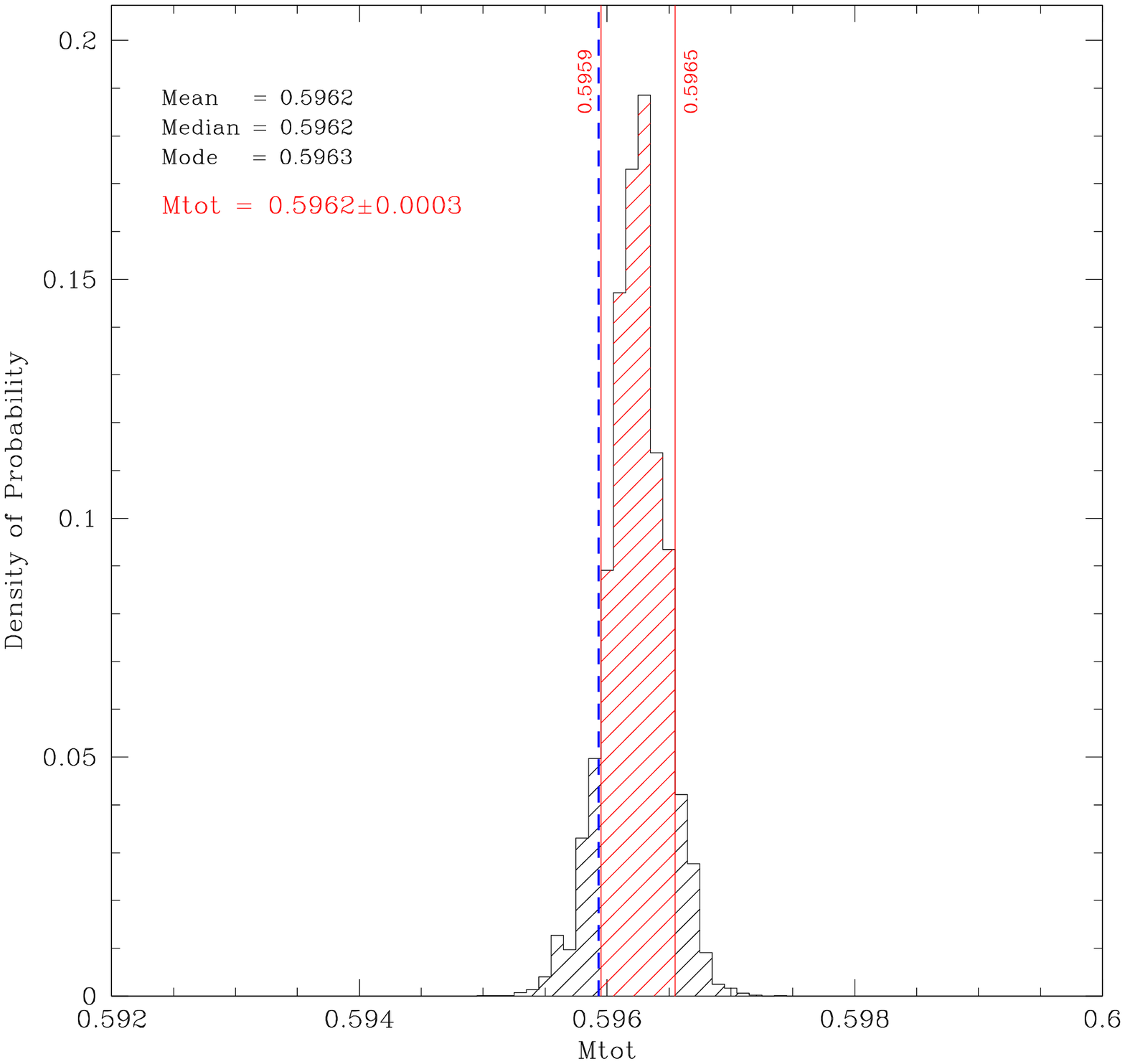} &
    \includegraphics[width=.35\textwidth]{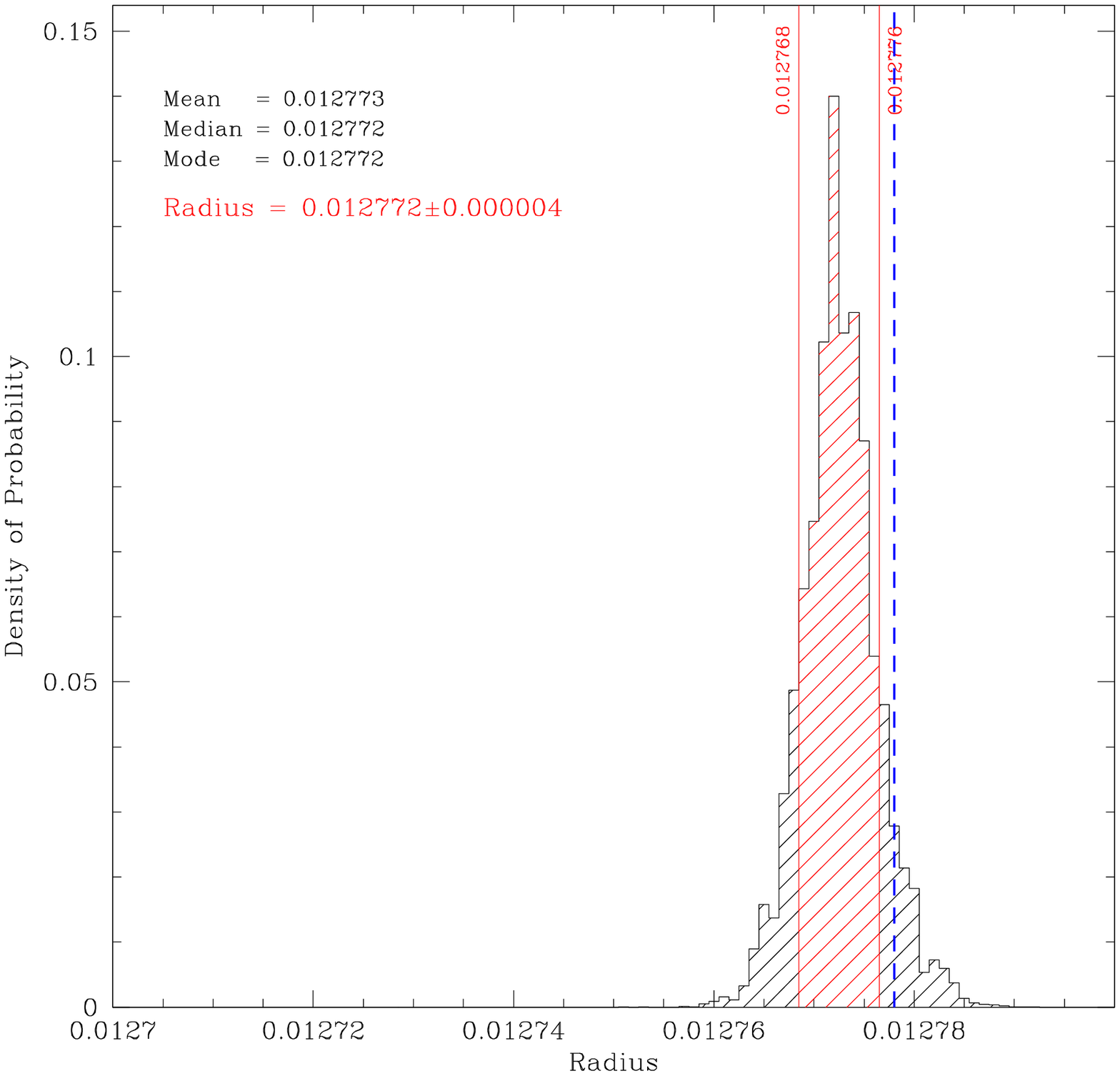} &
    \includegraphics[width=.35\textwidth]{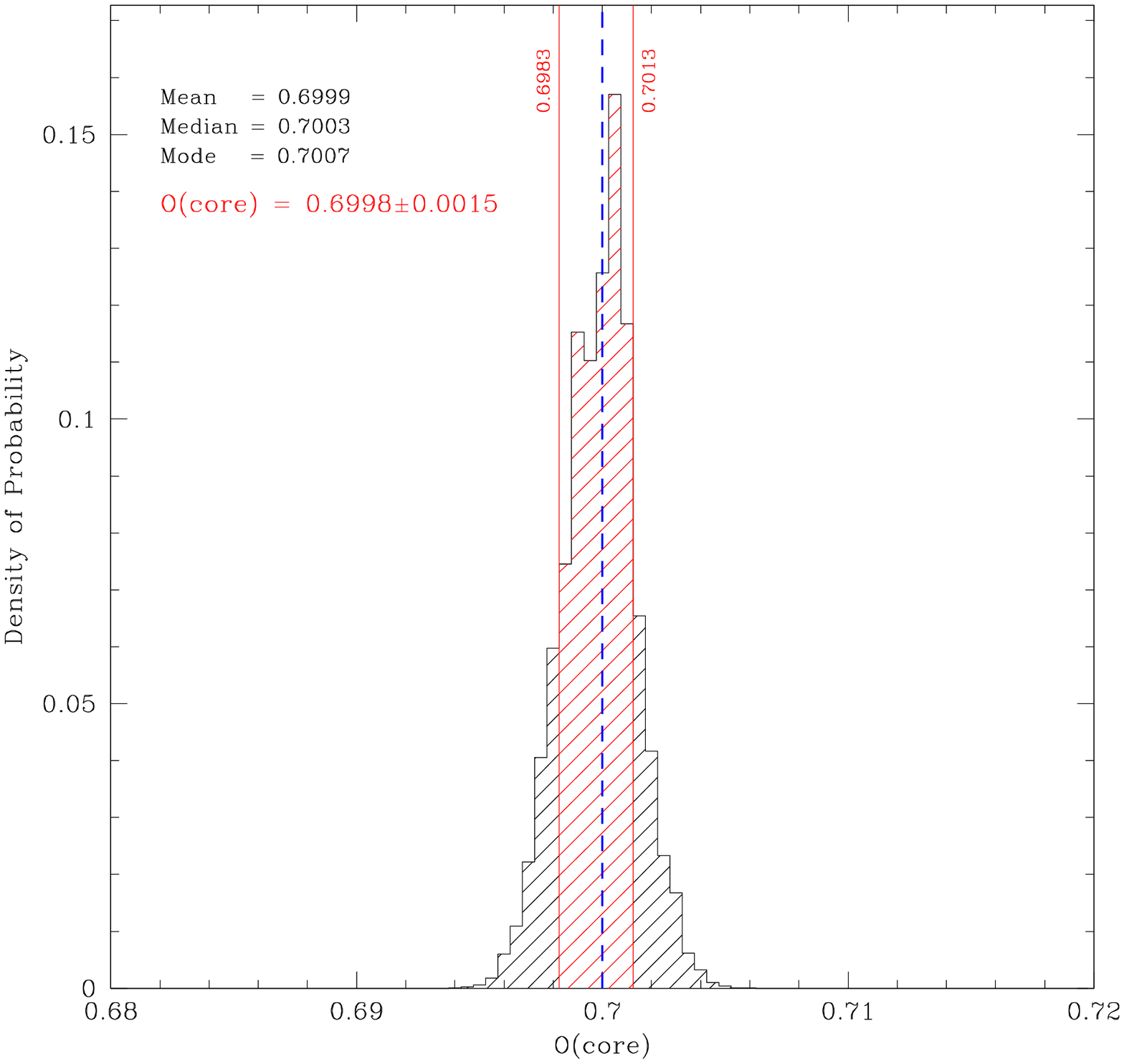} \\
    \end{tabular}
      \begin{flushright}
Figure~8
\end{flushright}
\end{figure}

\clearpage
\addtocounter{figure}{-1}
\begin{figure}[!h]
\centering
  \begin{tabular}{@{}ccc@{}}
    \includegraphics[width=.35\textwidth]{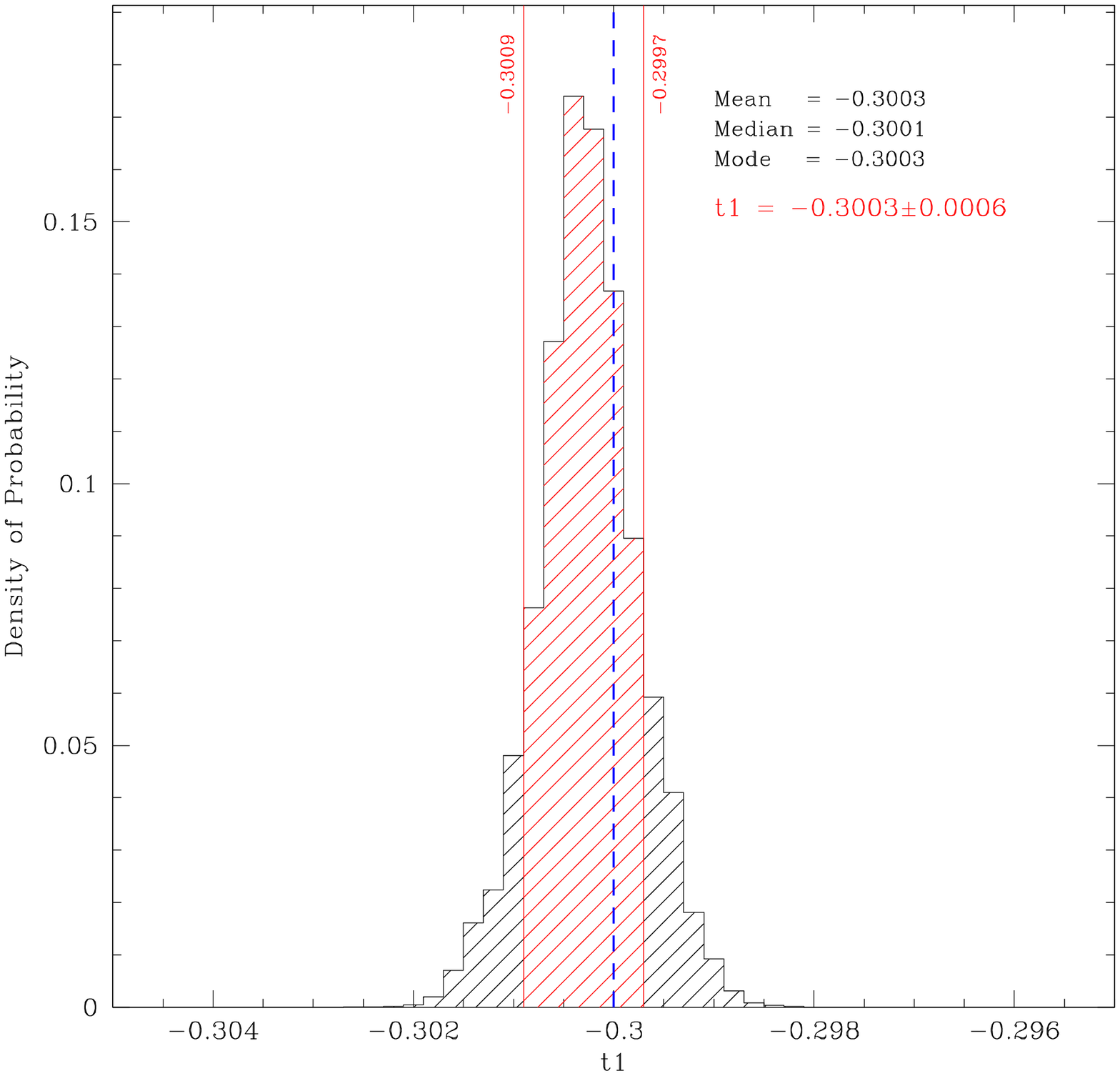} &
    \includegraphics[width=.35\textwidth]{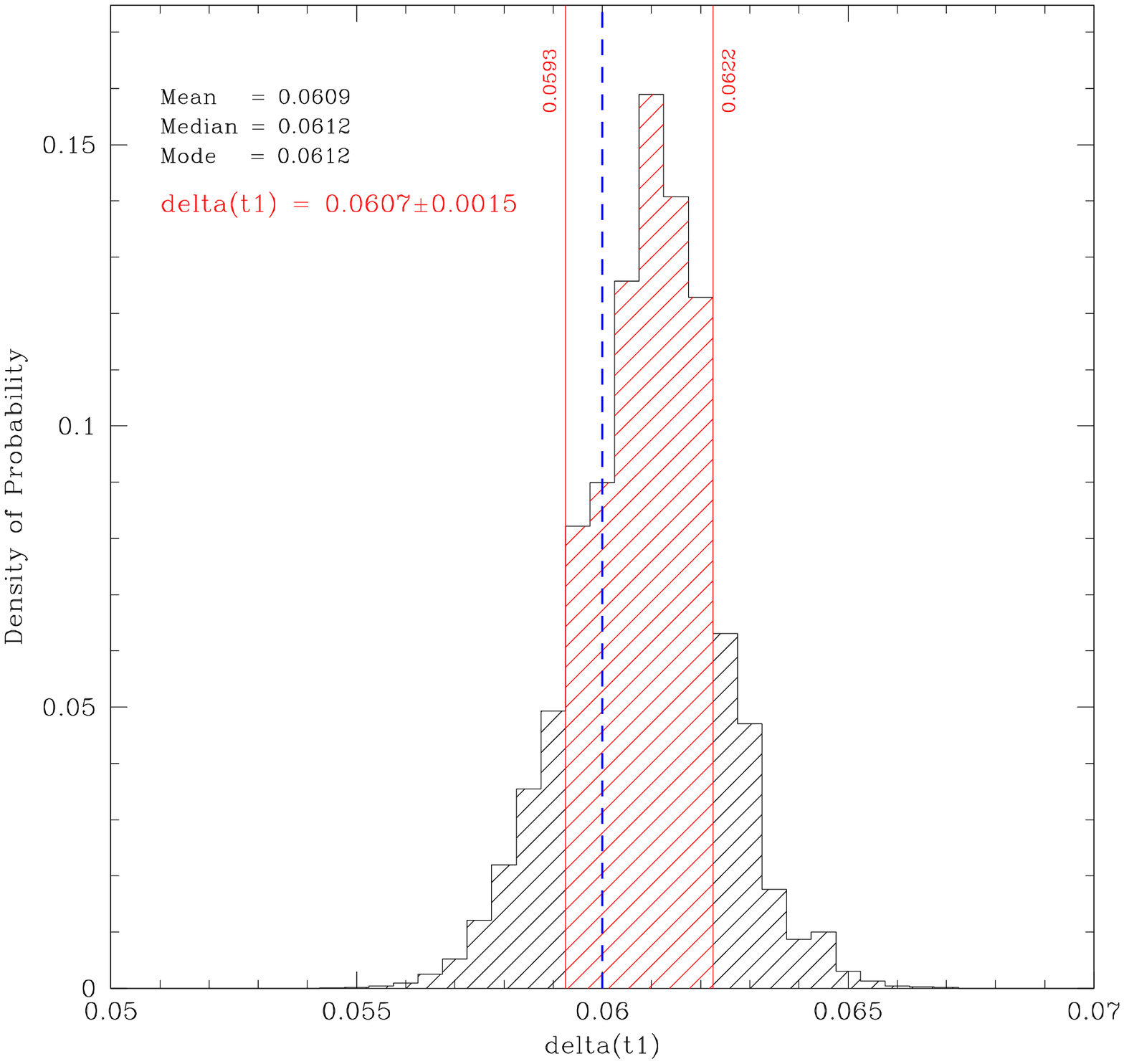}
    \end{tabular}
        \begin{flushright}
Figure~8
\end{flushright}
\end{figure}

\clearpage
\begin{figure}[!h]
\centering
  \begin{tabular}{@{}ccc@{}}
    \includegraphics[width=.35\textwidth]{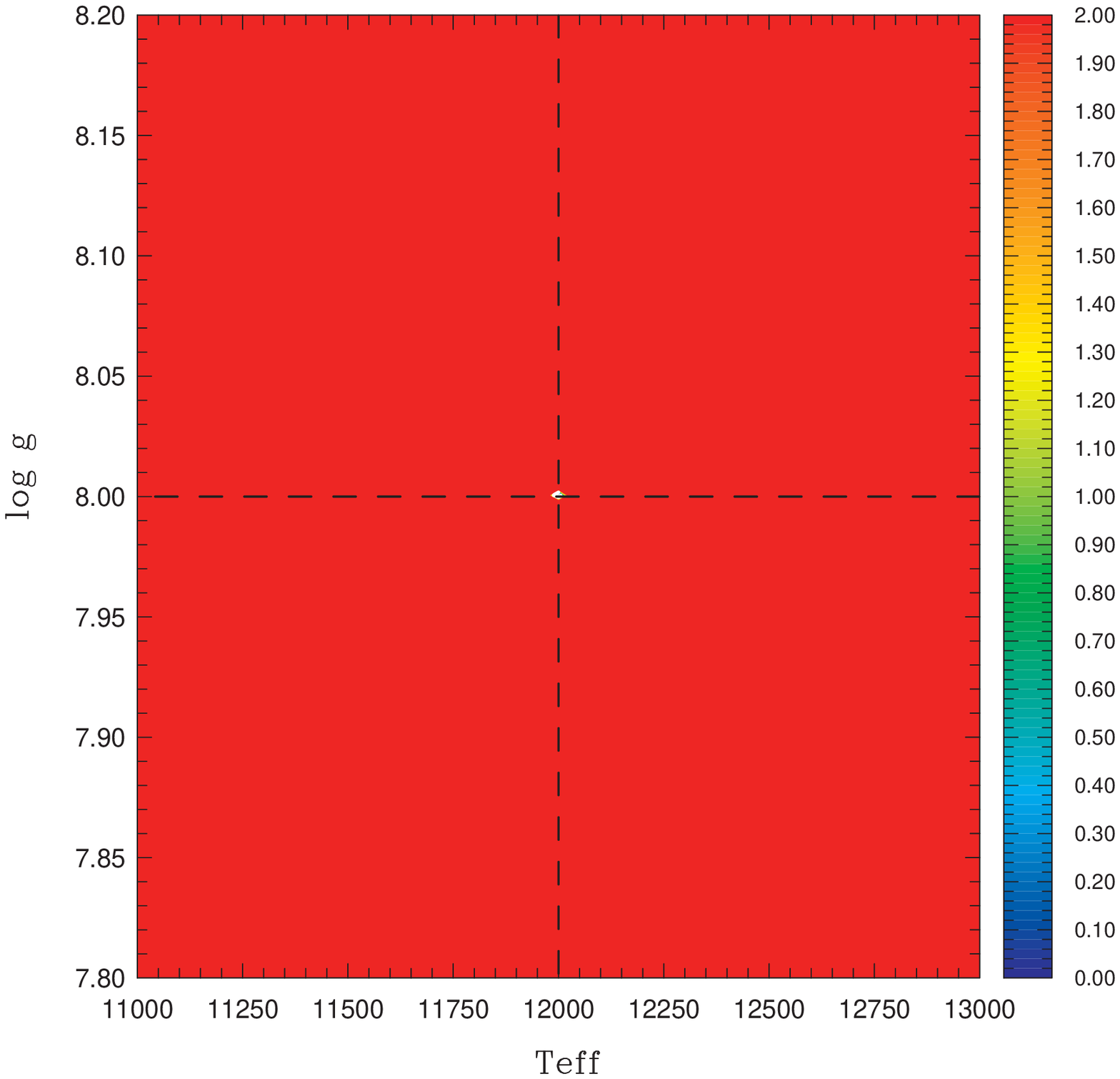} &
    \includegraphics[width=.35\textwidth]{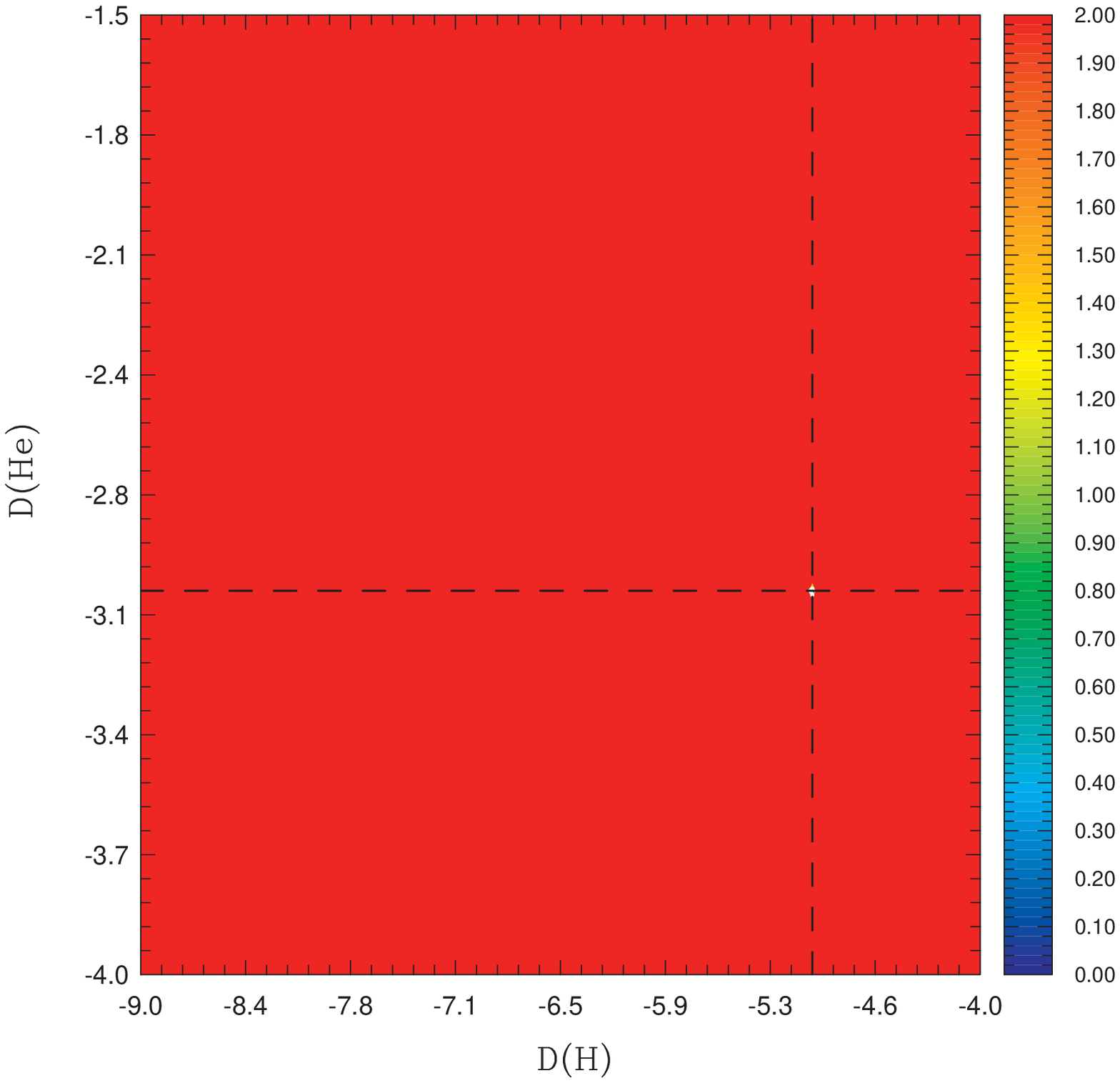} &
    \includegraphics[width=.35\textwidth]{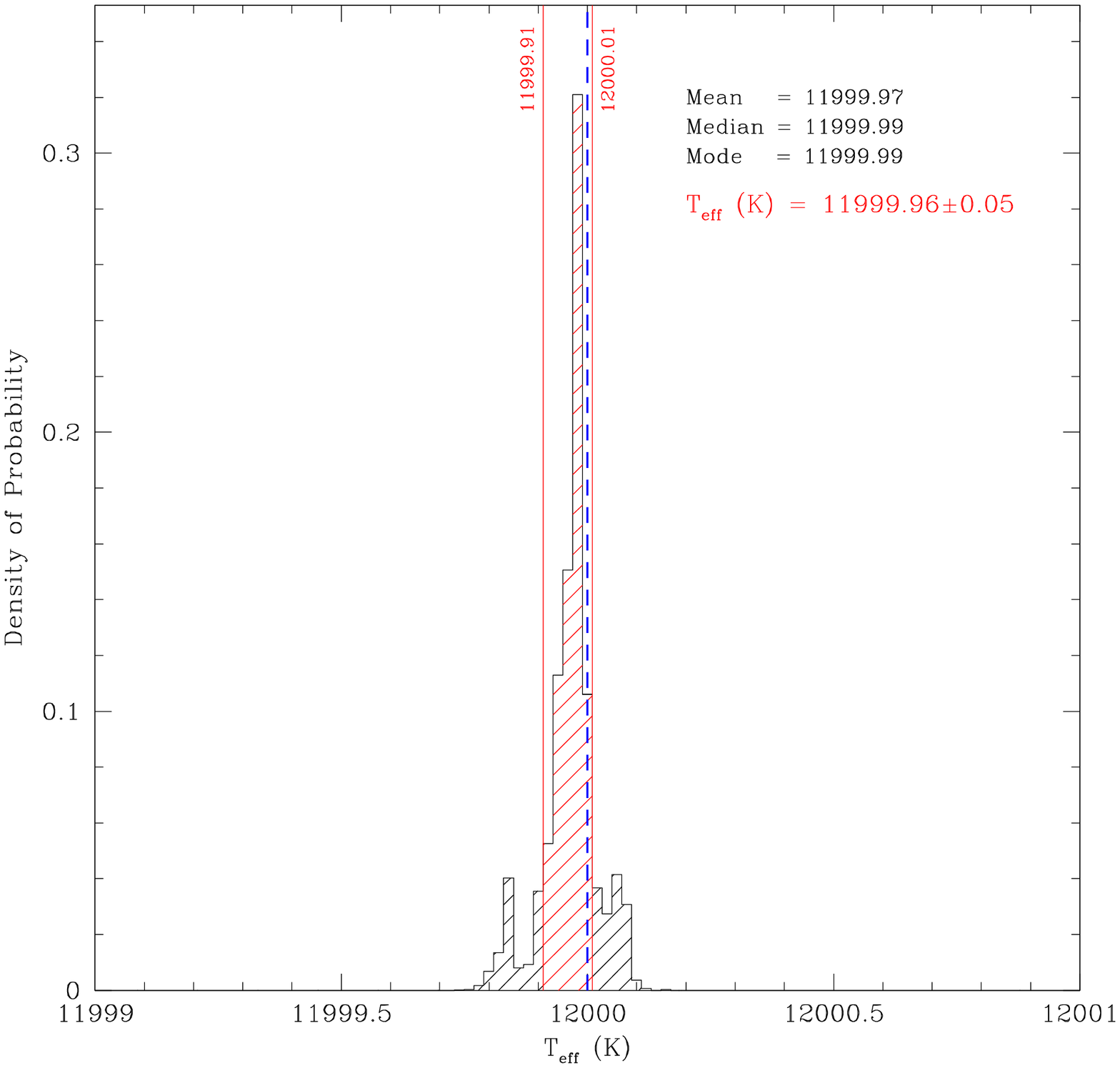} \\
    \includegraphics[width=.35\textwidth]{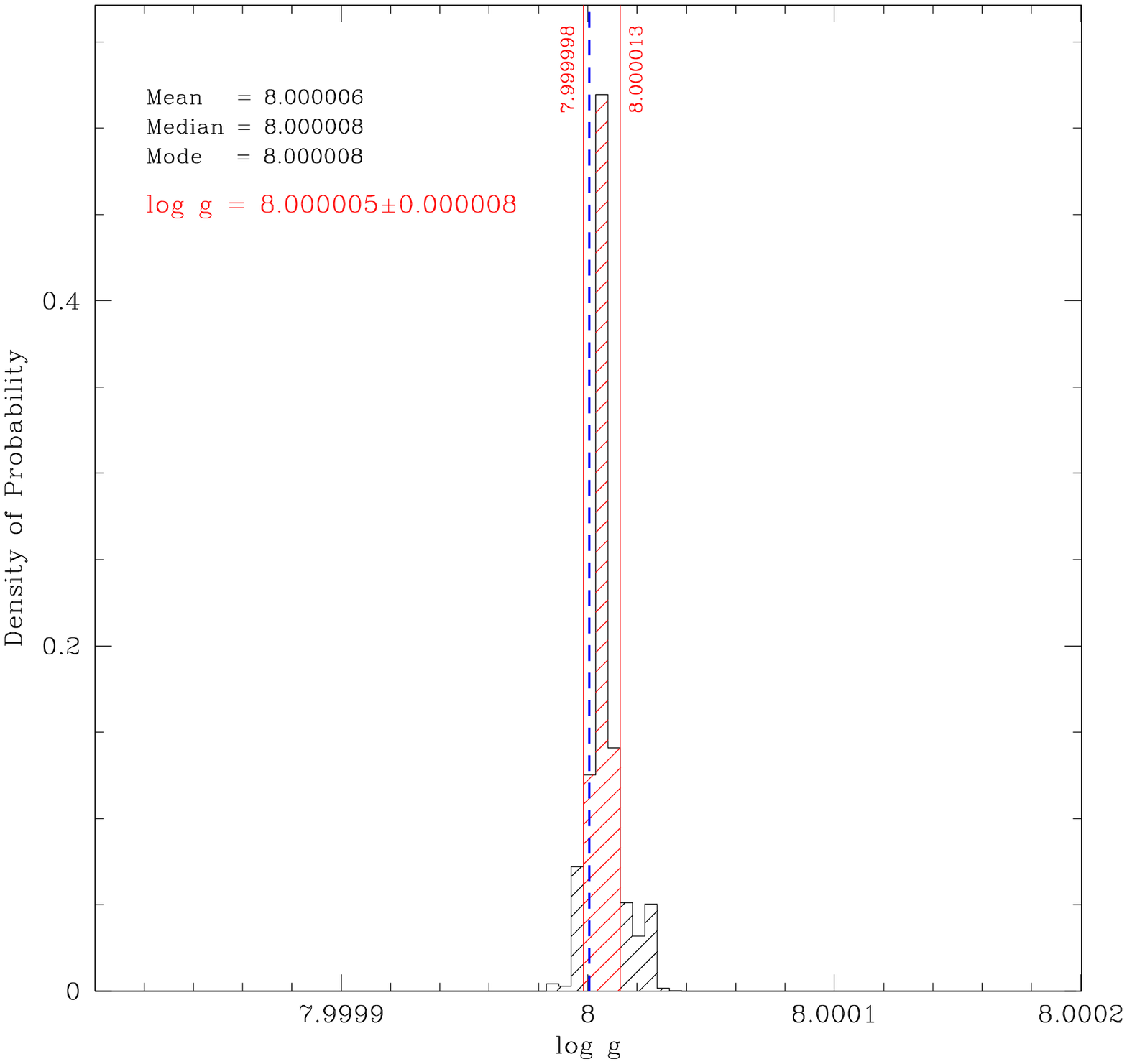}  &
    \includegraphics[width=.35\textwidth]{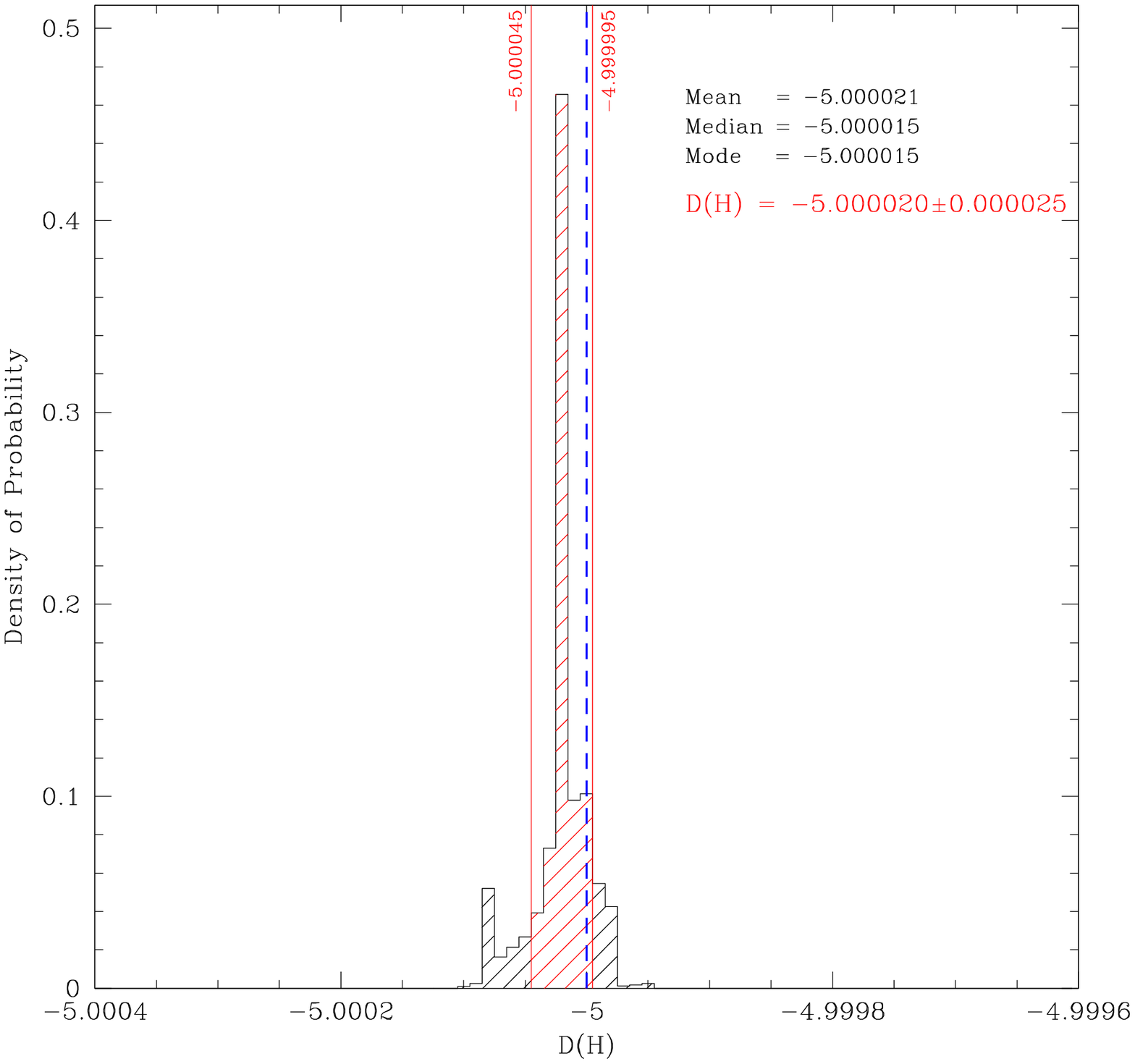} &
    \includegraphics[width=.35\textwidth]{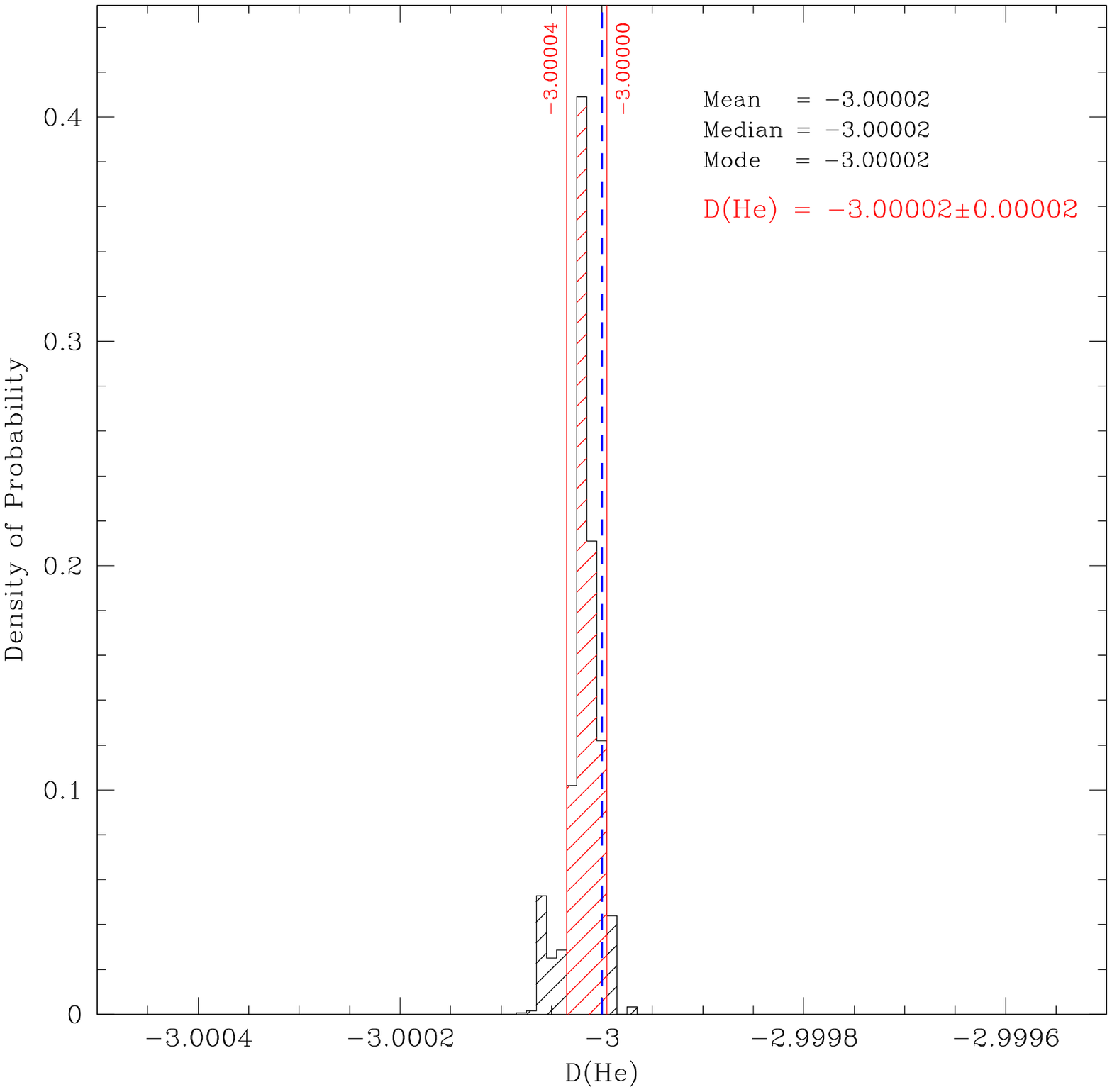} \\
    \includegraphics[width=.35\textwidth]{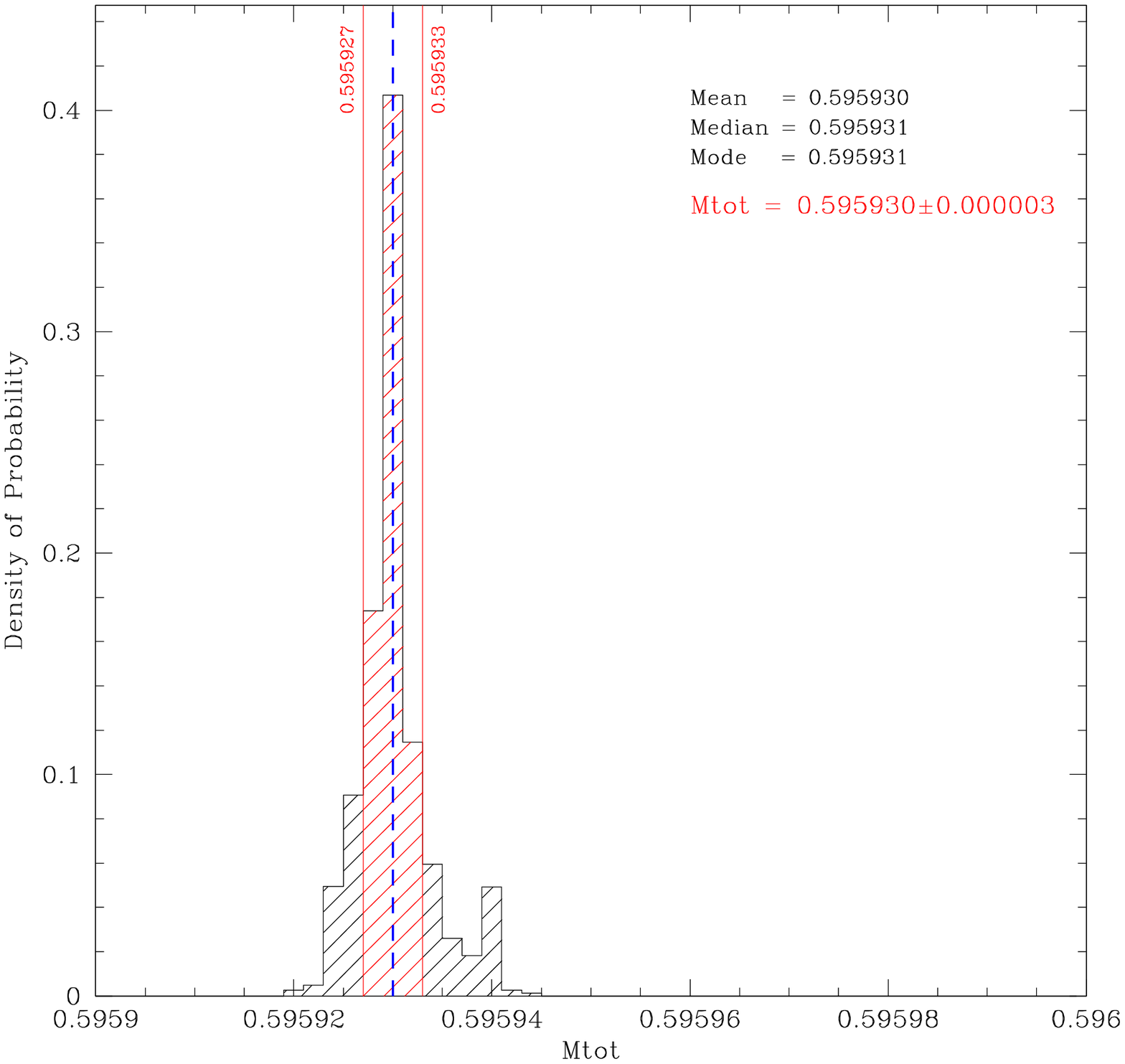} &
    \includegraphics[width=.35\textwidth]{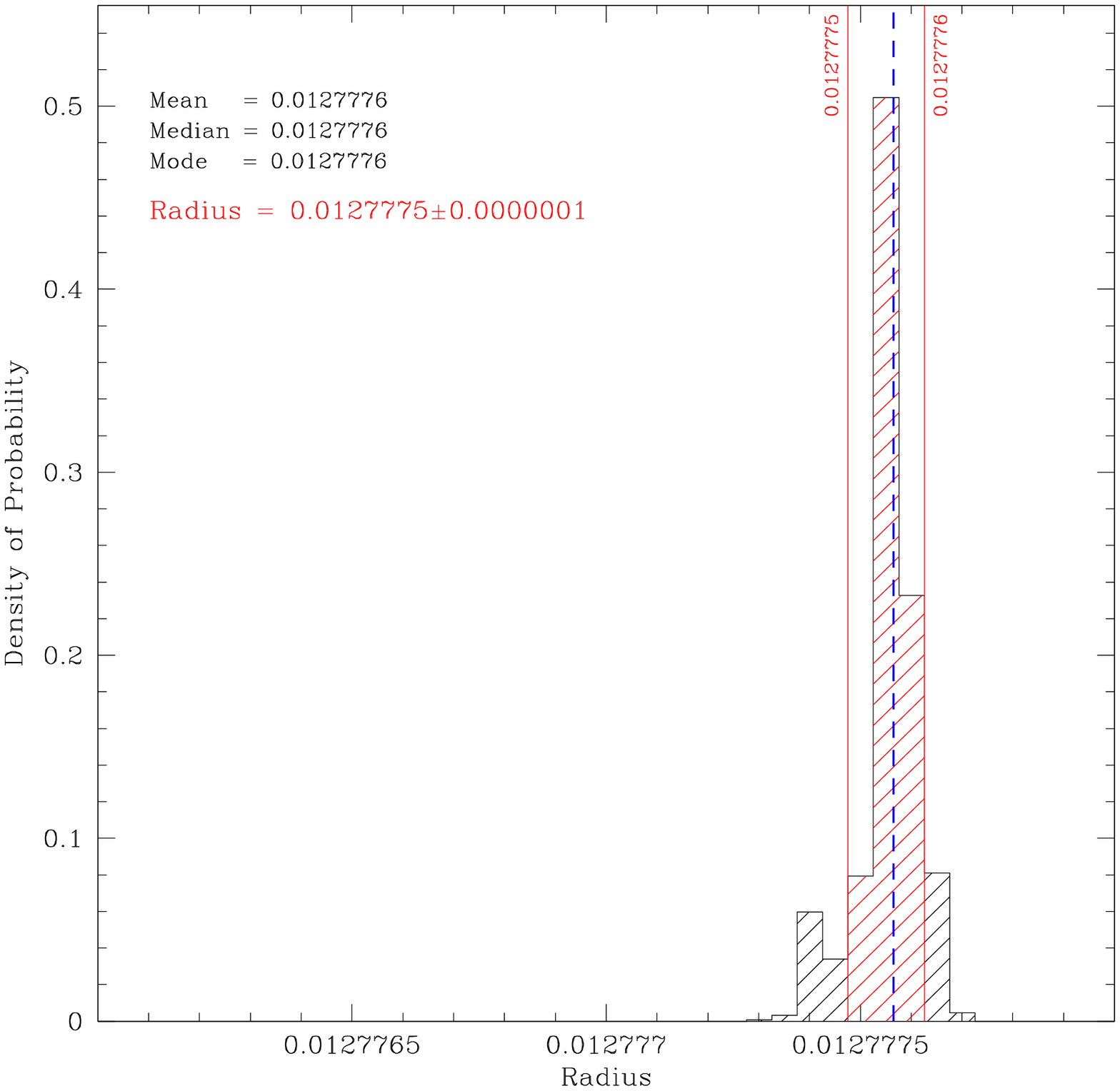} &
    \includegraphics[width=.35\textwidth]{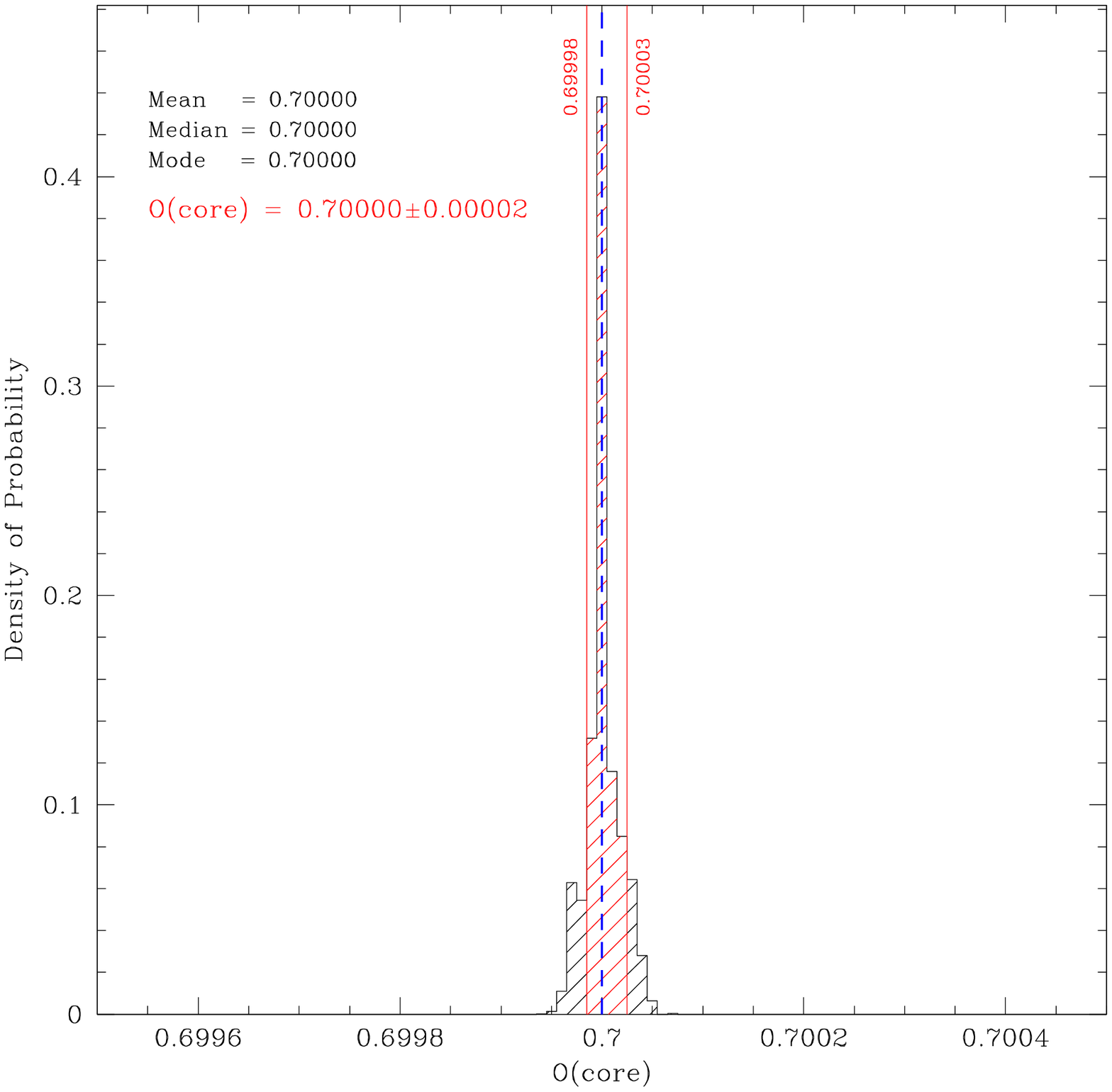} \\
    \end{tabular}
  \begin{flushright}
Figure~9
\end{flushright}
\end{figure}

\clearpage
\addtocounter{figure}{-1}
\begin{figure}[!h]
\centering
  \begin{tabular}{@{}ccc@{}}
    \includegraphics[width=.35\textwidth]{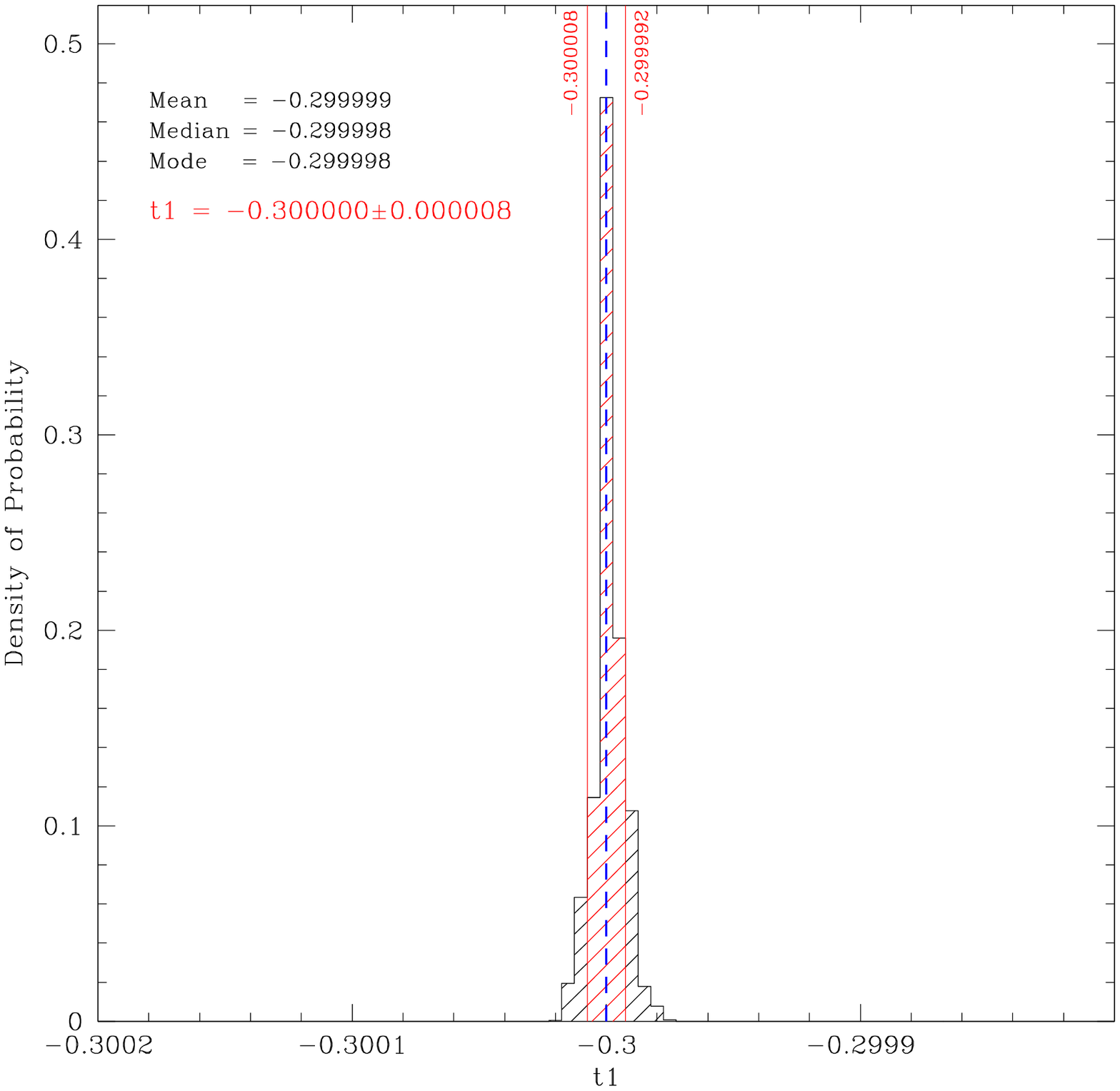} &
    \includegraphics[width=.35\textwidth]{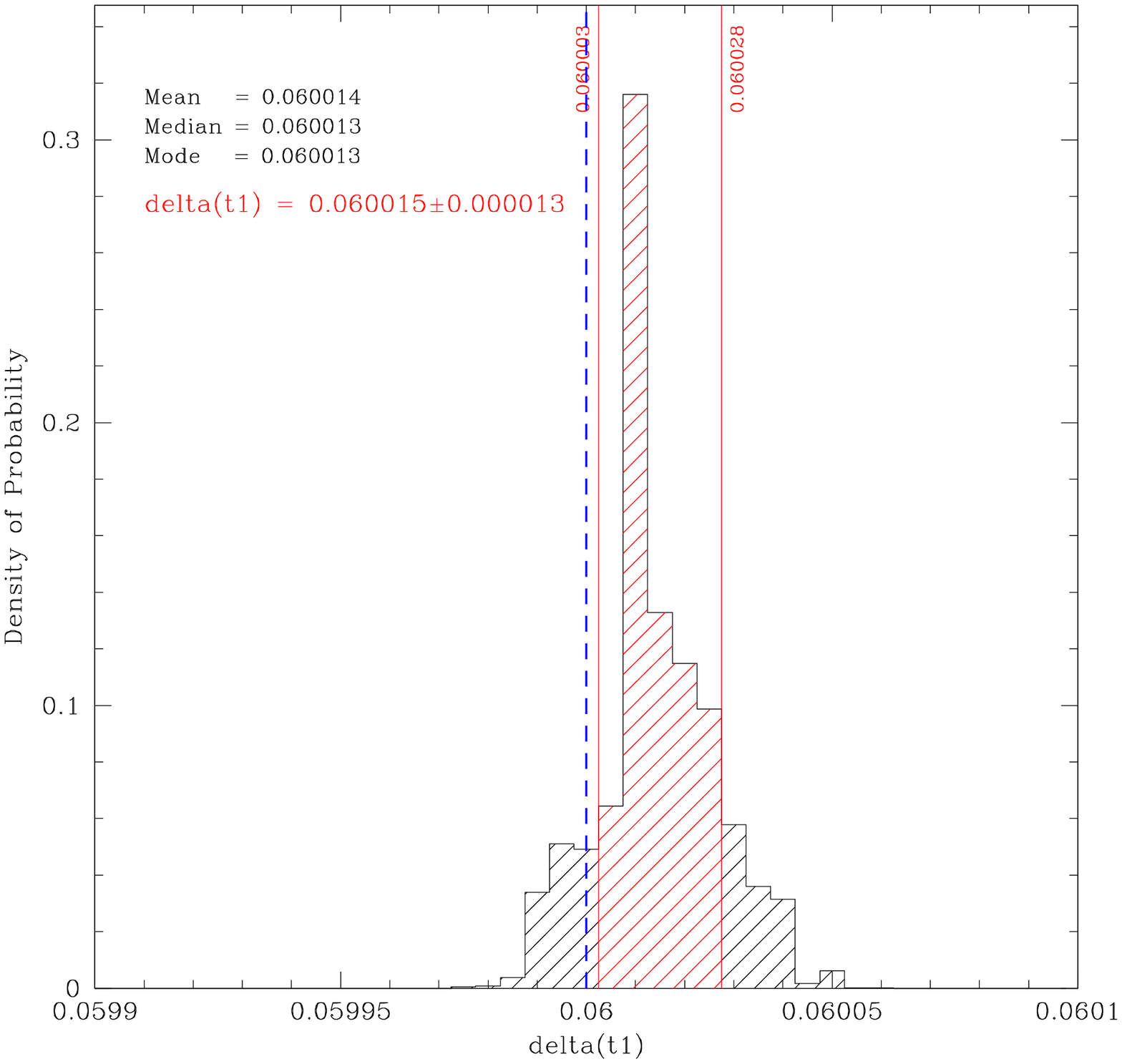}
    \end{tabular}
  \begin{flushright}
Figure~9
\end{flushright}
\end{figure}

\clearpage

\begin{figure}[!h]
\centering
  \begin{tabular}{@{}ccc@{}}
    \includegraphics[width=.35\textwidth]{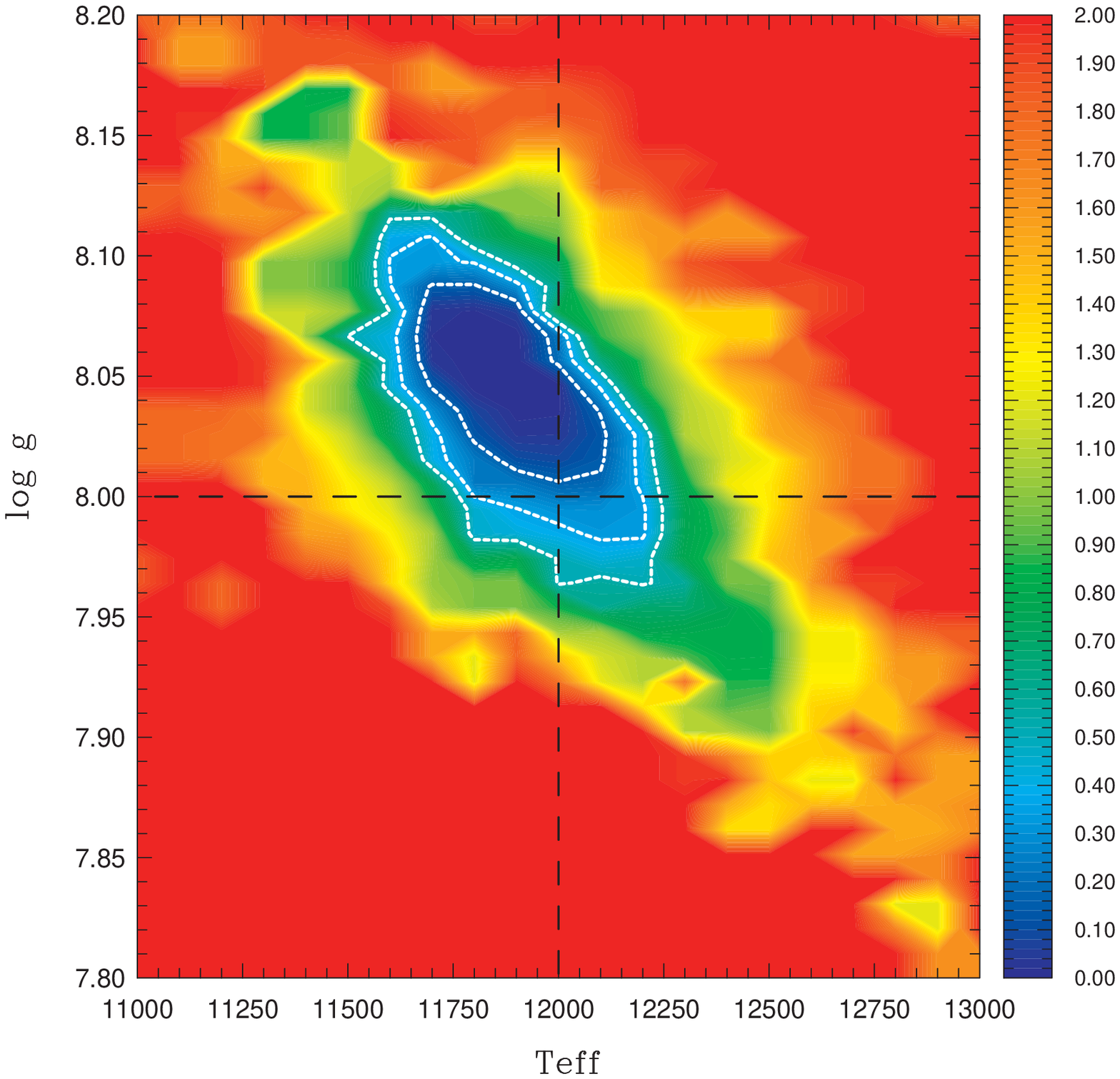} &
    \includegraphics[width=.35\textwidth]{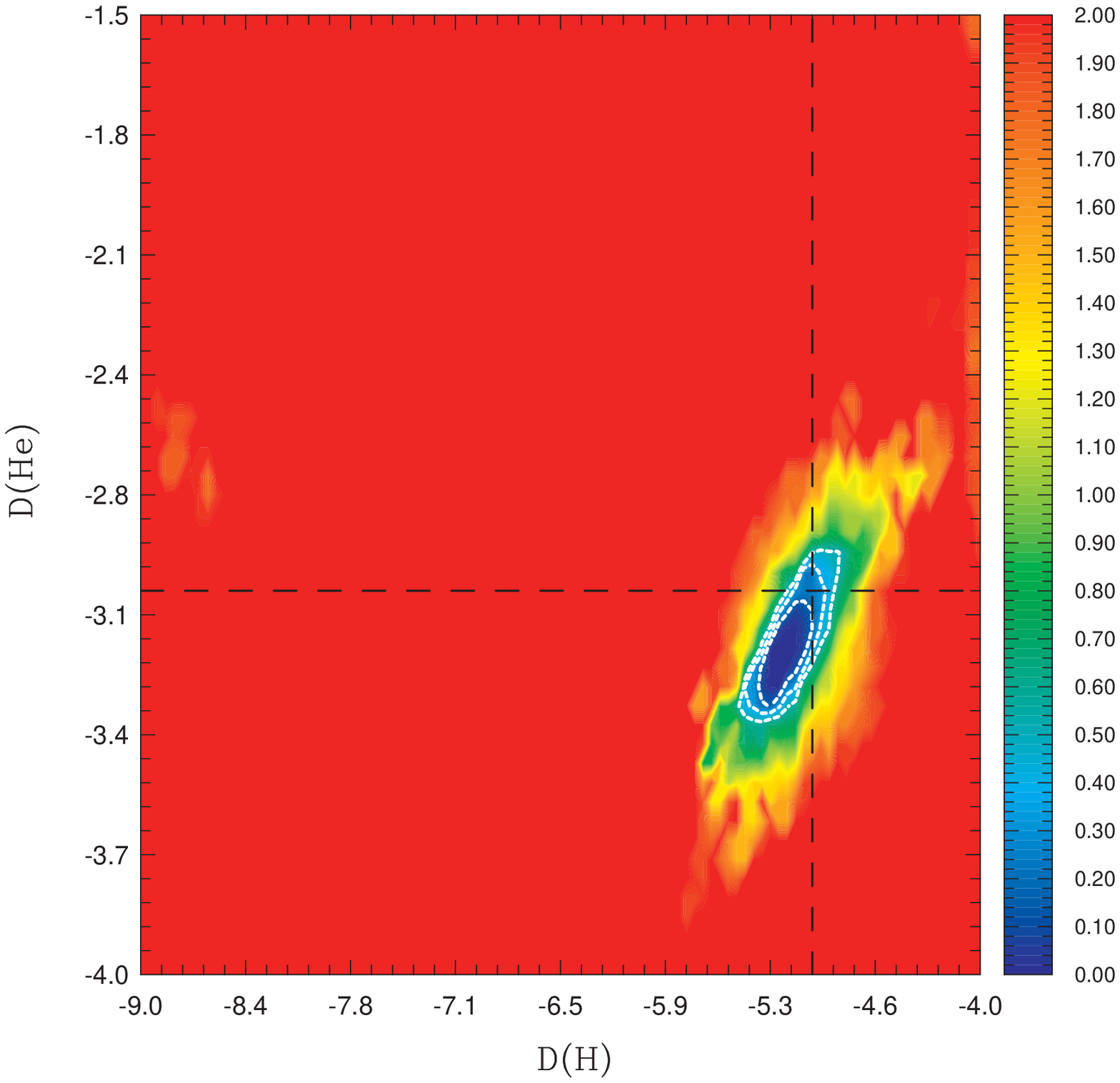} &
    \includegraphics[width=.35\textwidth]{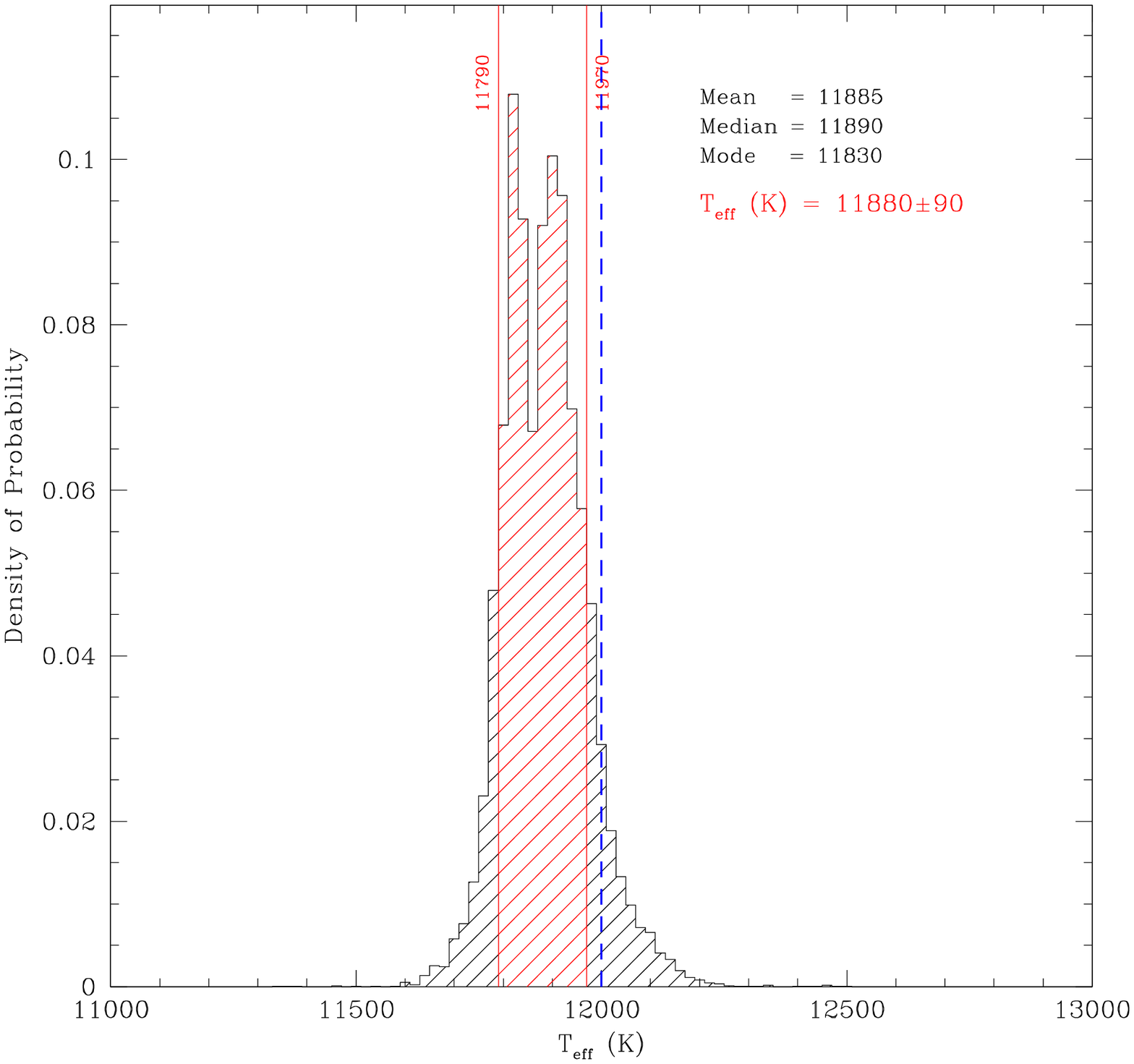} \\
    \includegraphics[width=.35\textwidth]{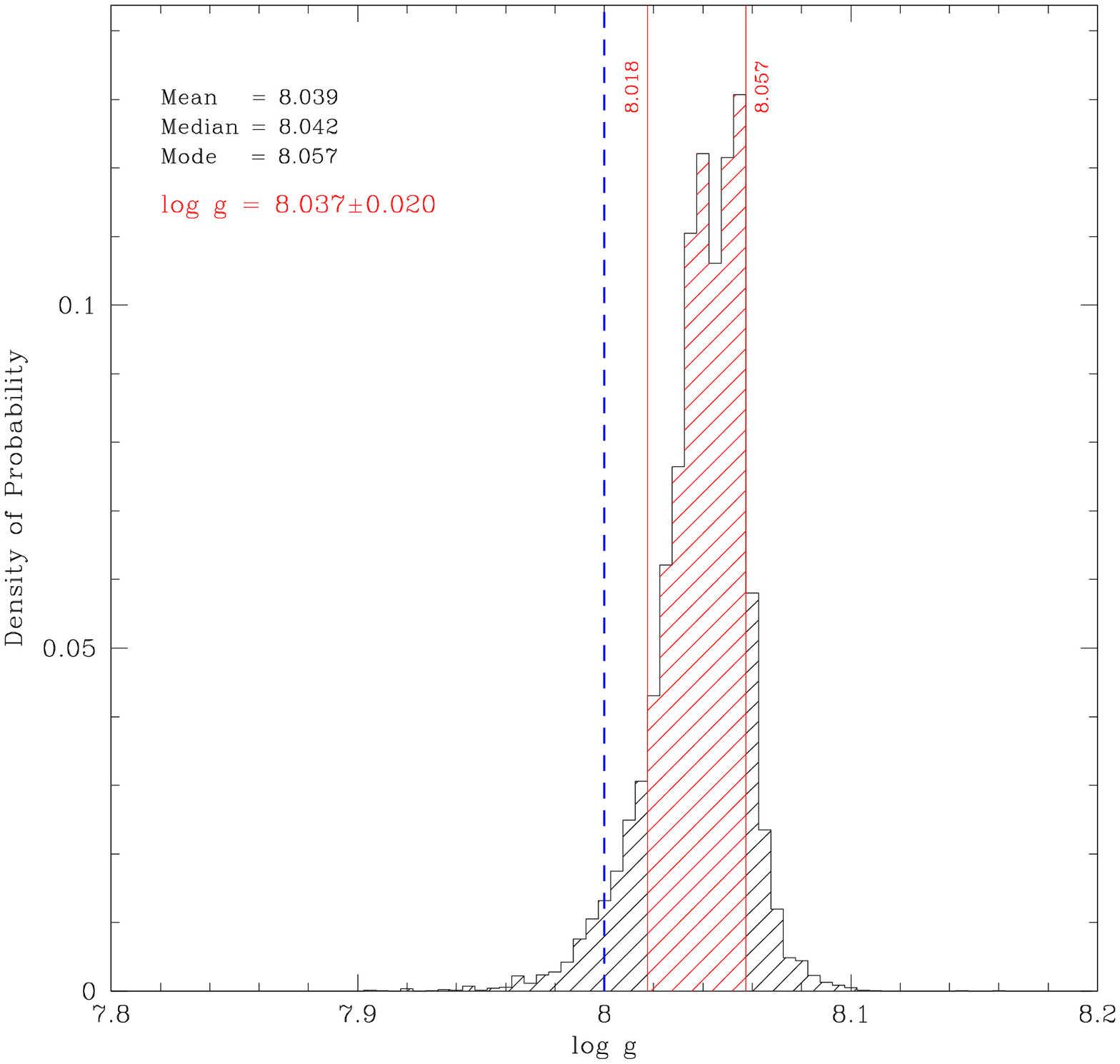}  &
    \includegraphics[width=.35\textwidth]{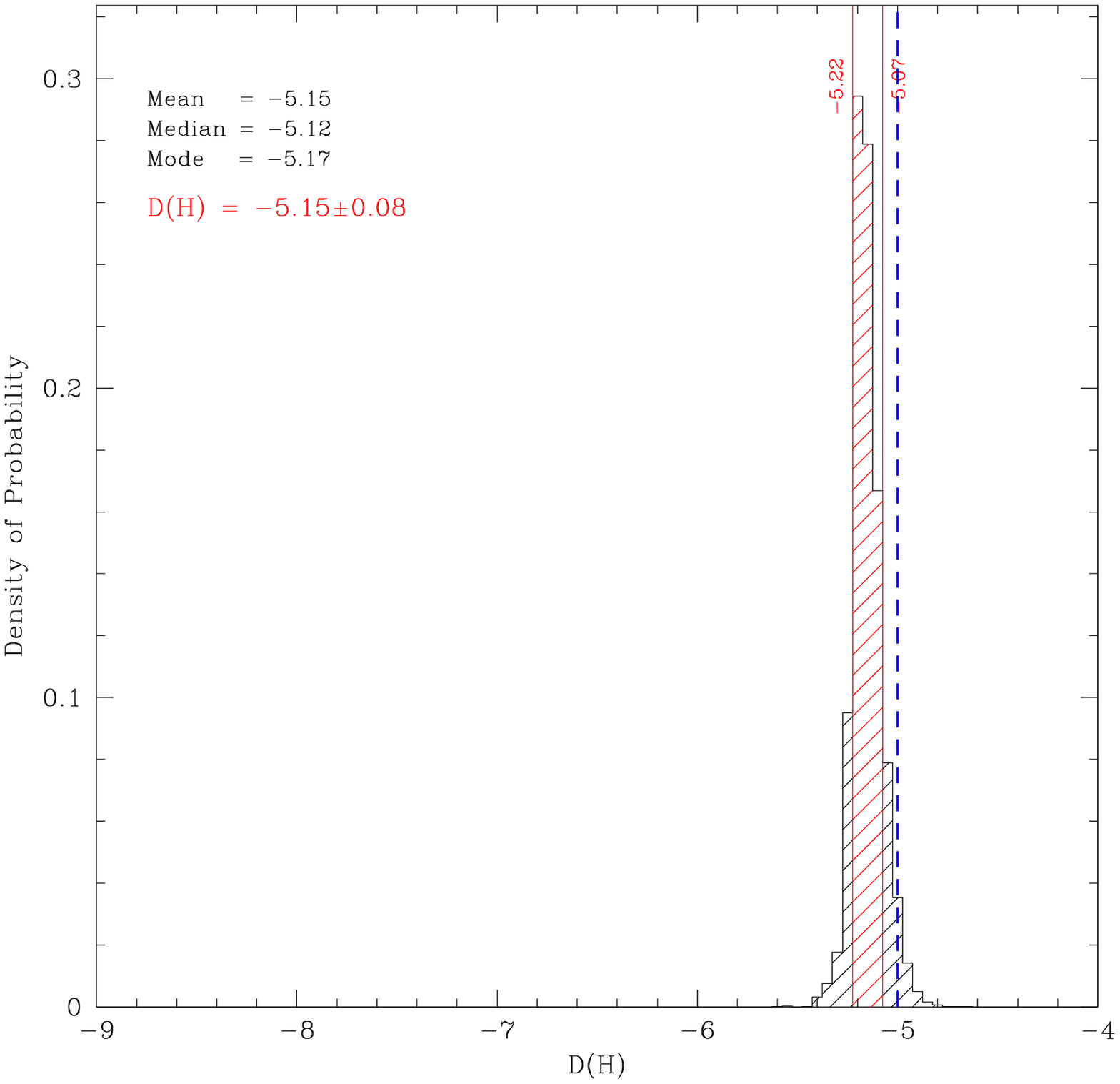} &
    \includegraphics[width=.35\textwidth]{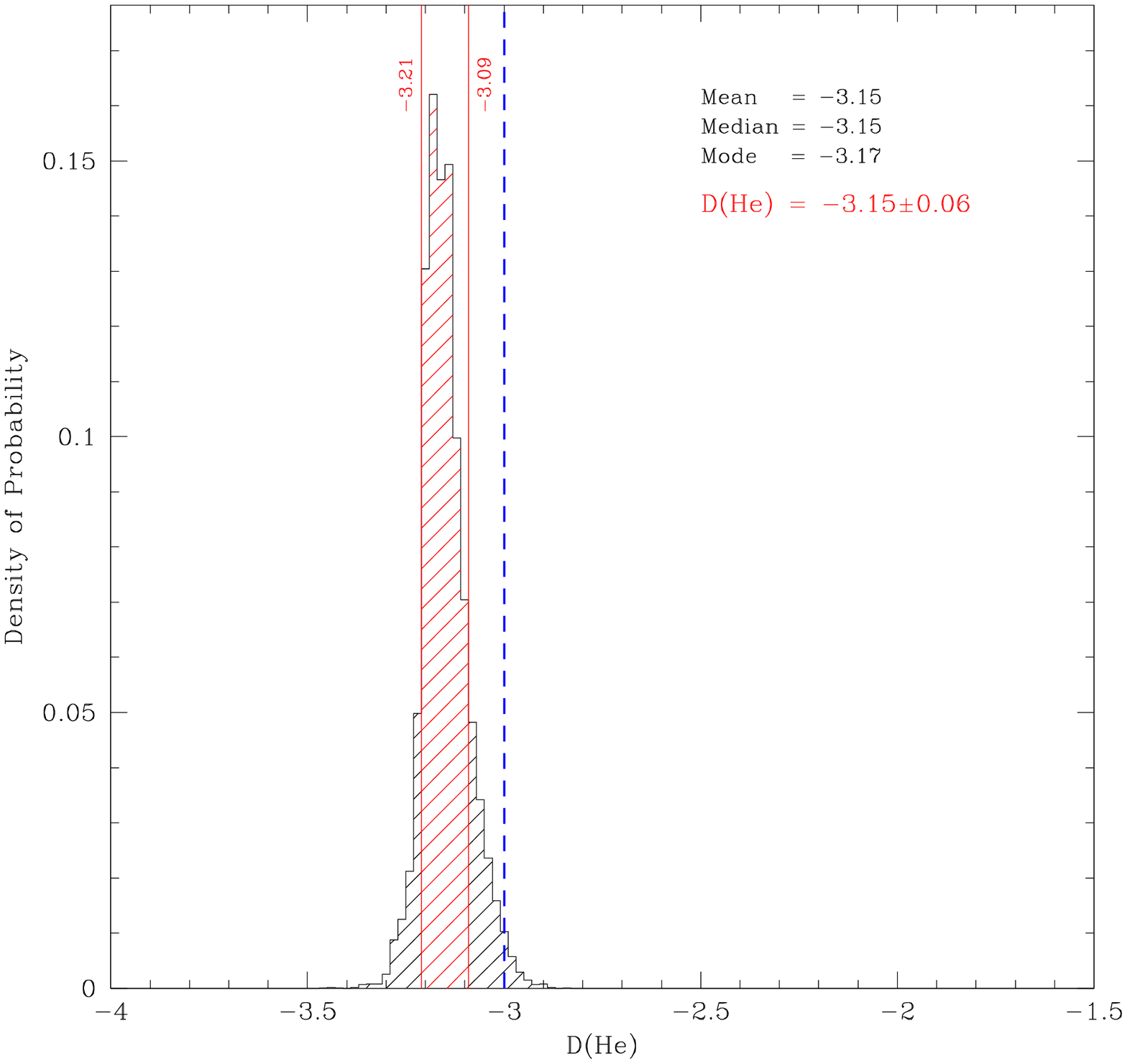} \\
    \includegraphics[width=.35\textwidth]{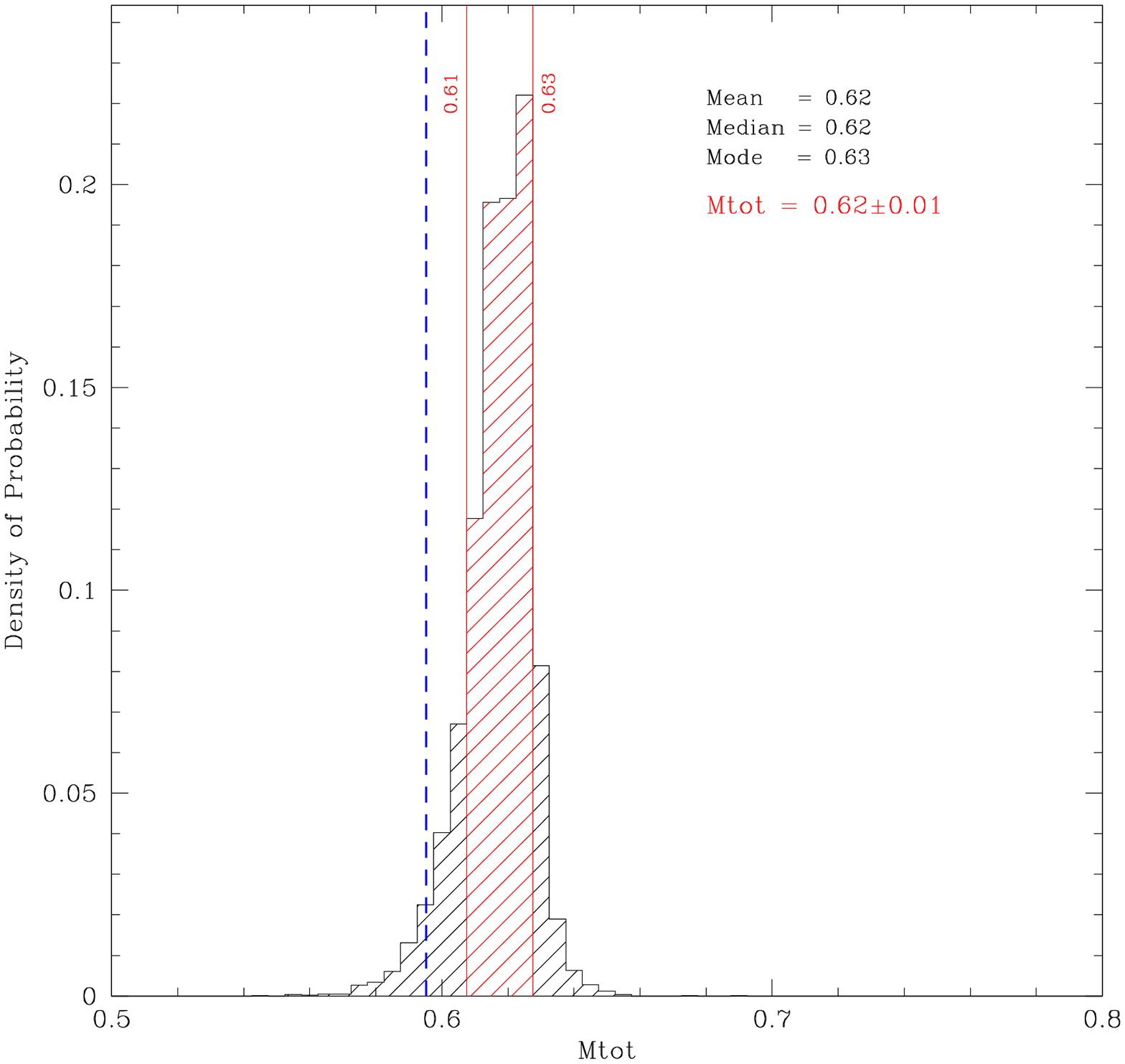} &
    \includegraphics[width=.35\textwidth]{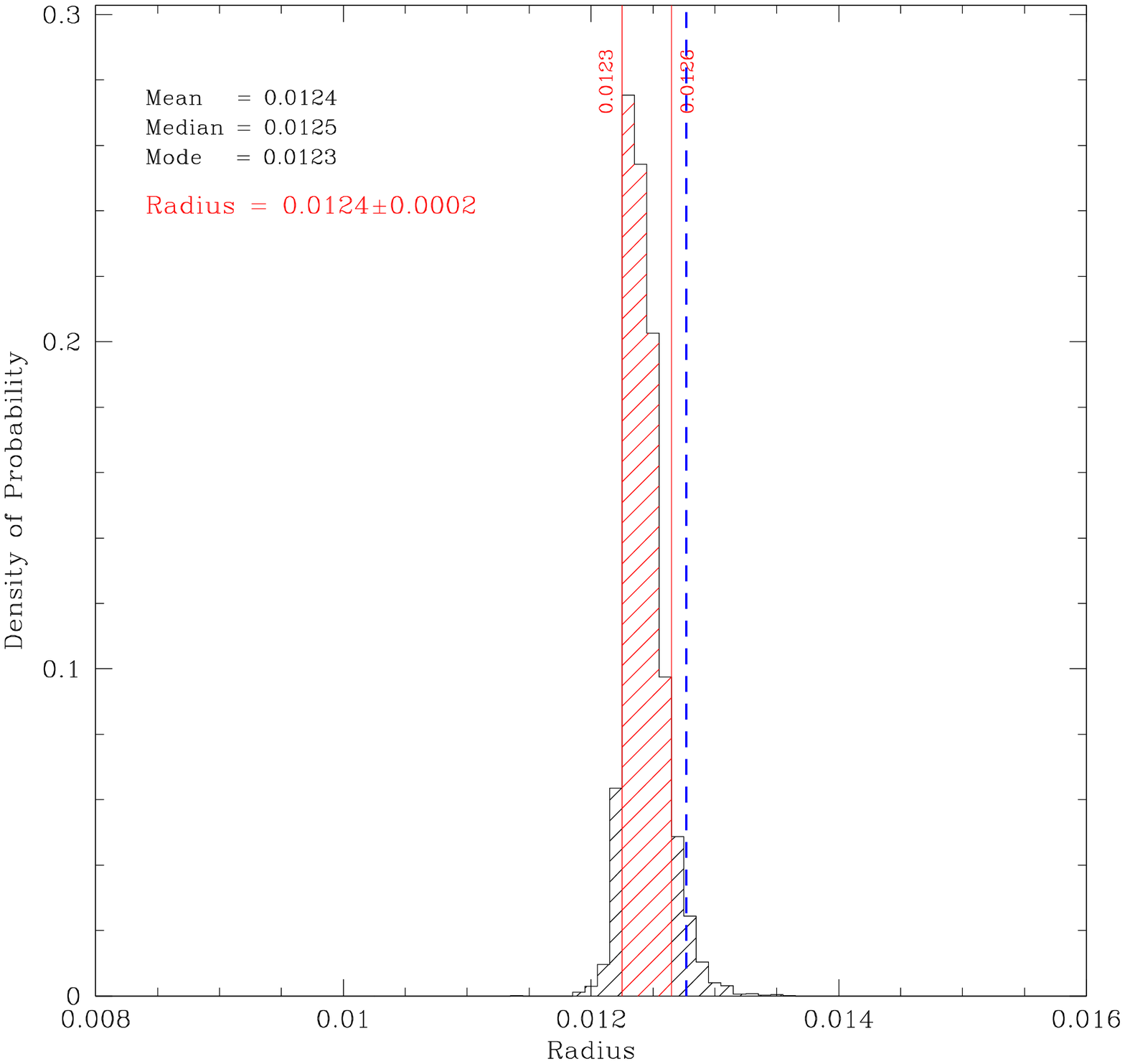} &
    \includegraphics[width=.35\textwidth]{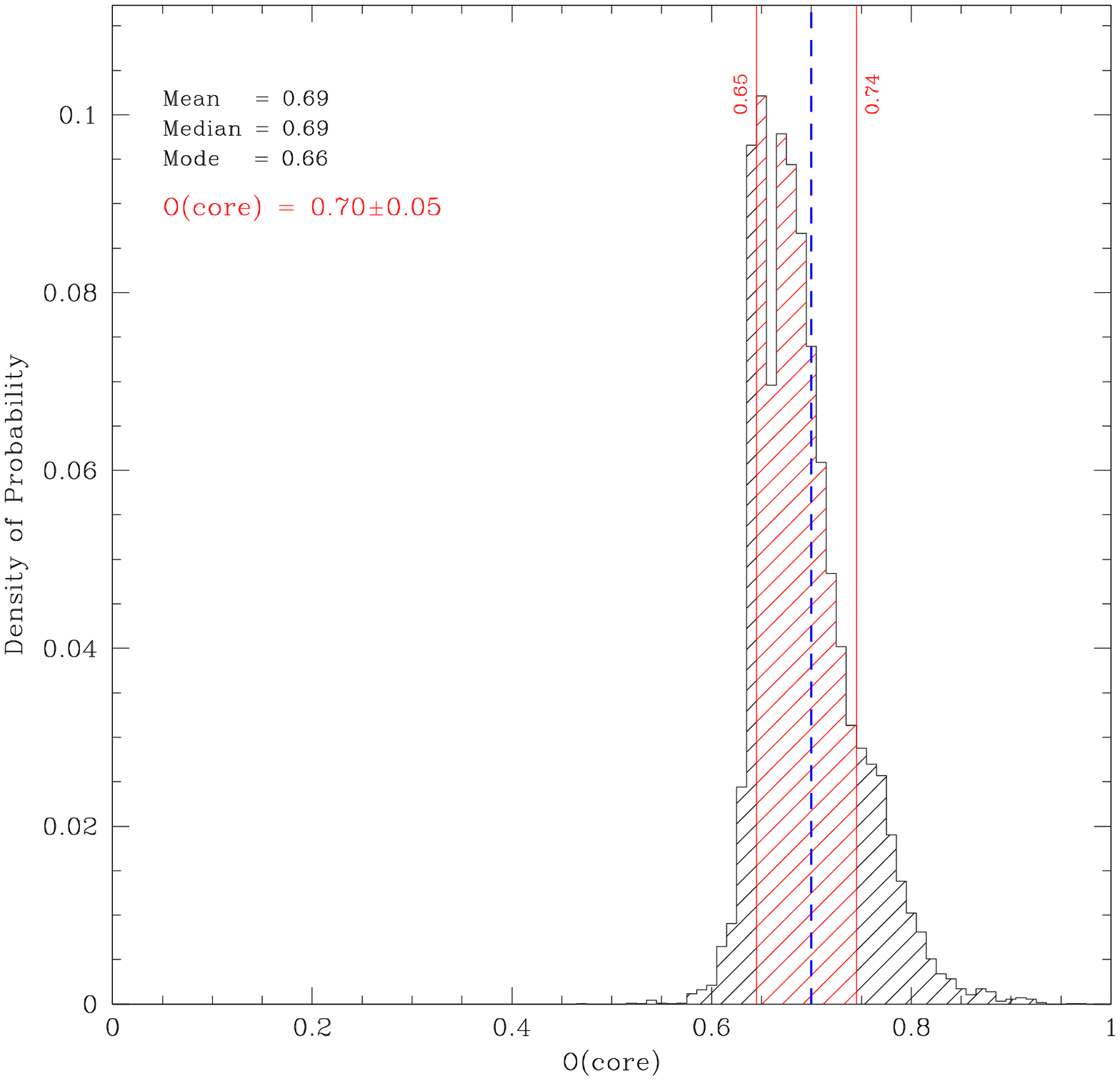} \\
    \end{tabular}
  \begin{flushright}
Figure~10
\end{flushright}
\end{figure}

\clearpage
\addtocounter{figure}{-1}
\begin{figure}[!h]
\centering
  \begin{tabular}{@{}ccc@{}}
    \includegraphics[width=.35\textwidth]{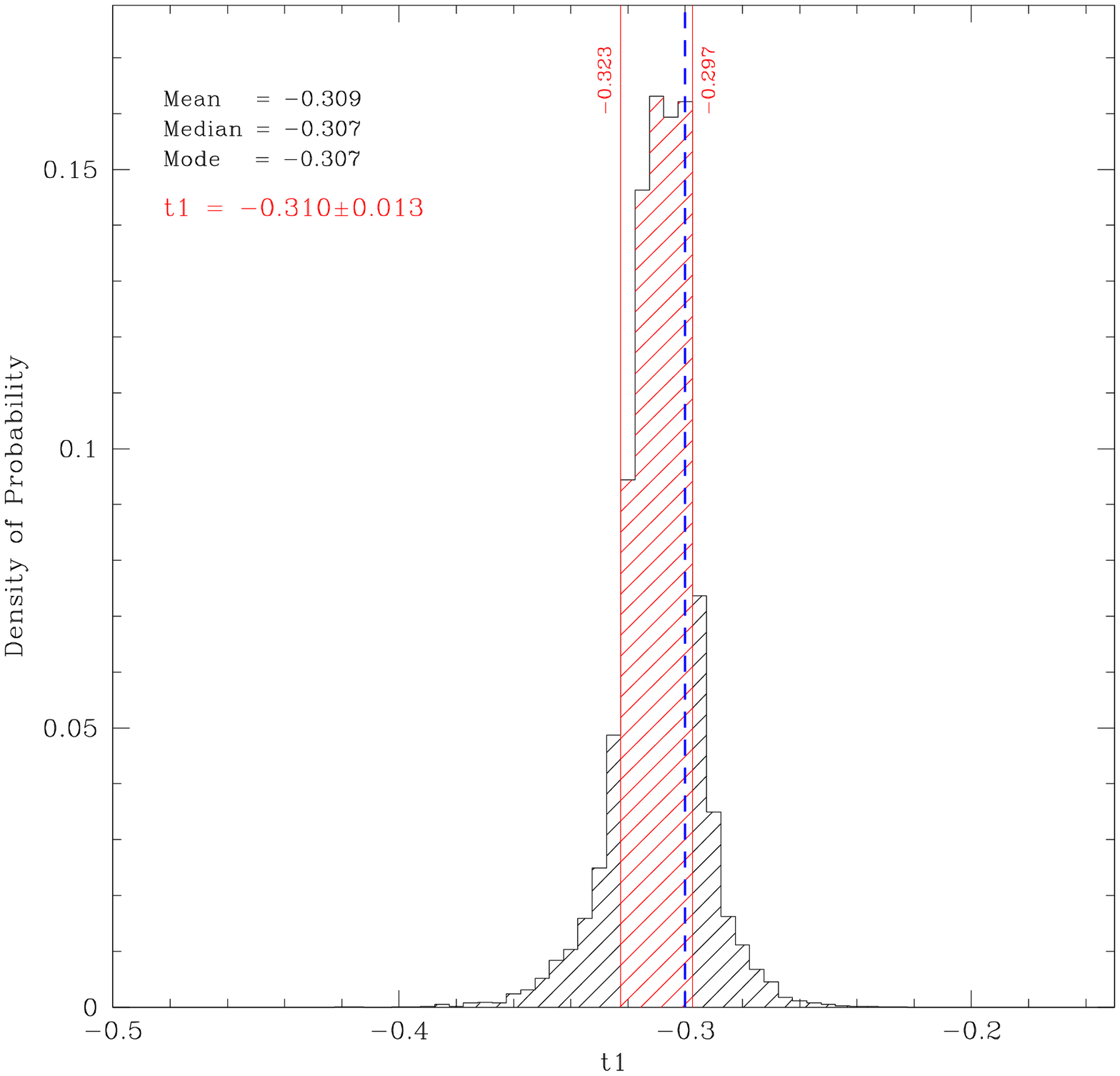} &
    \includegraphics[width=.35\textwidth]{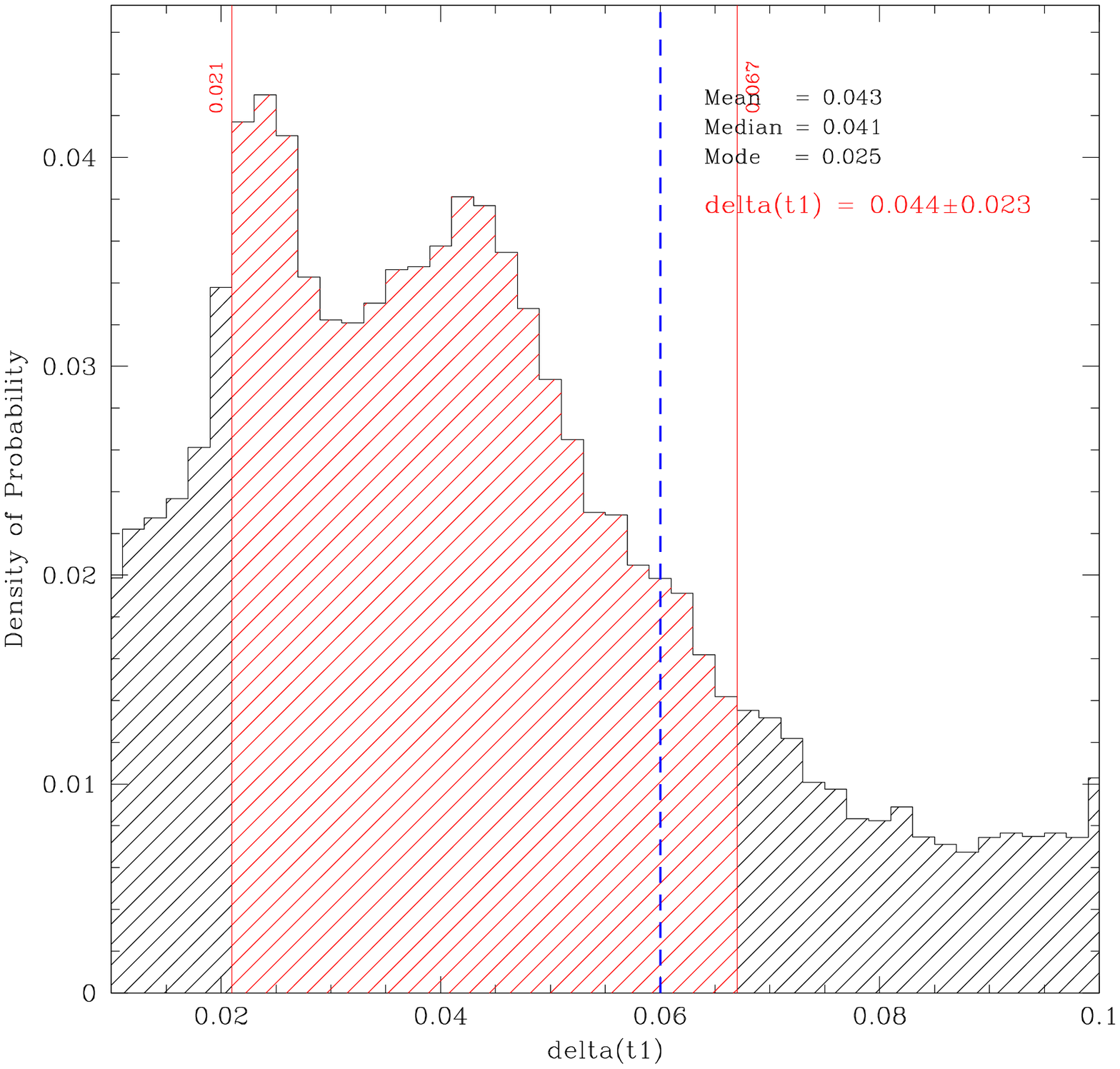} &
    \includegraphics[width=.35\textwidth]{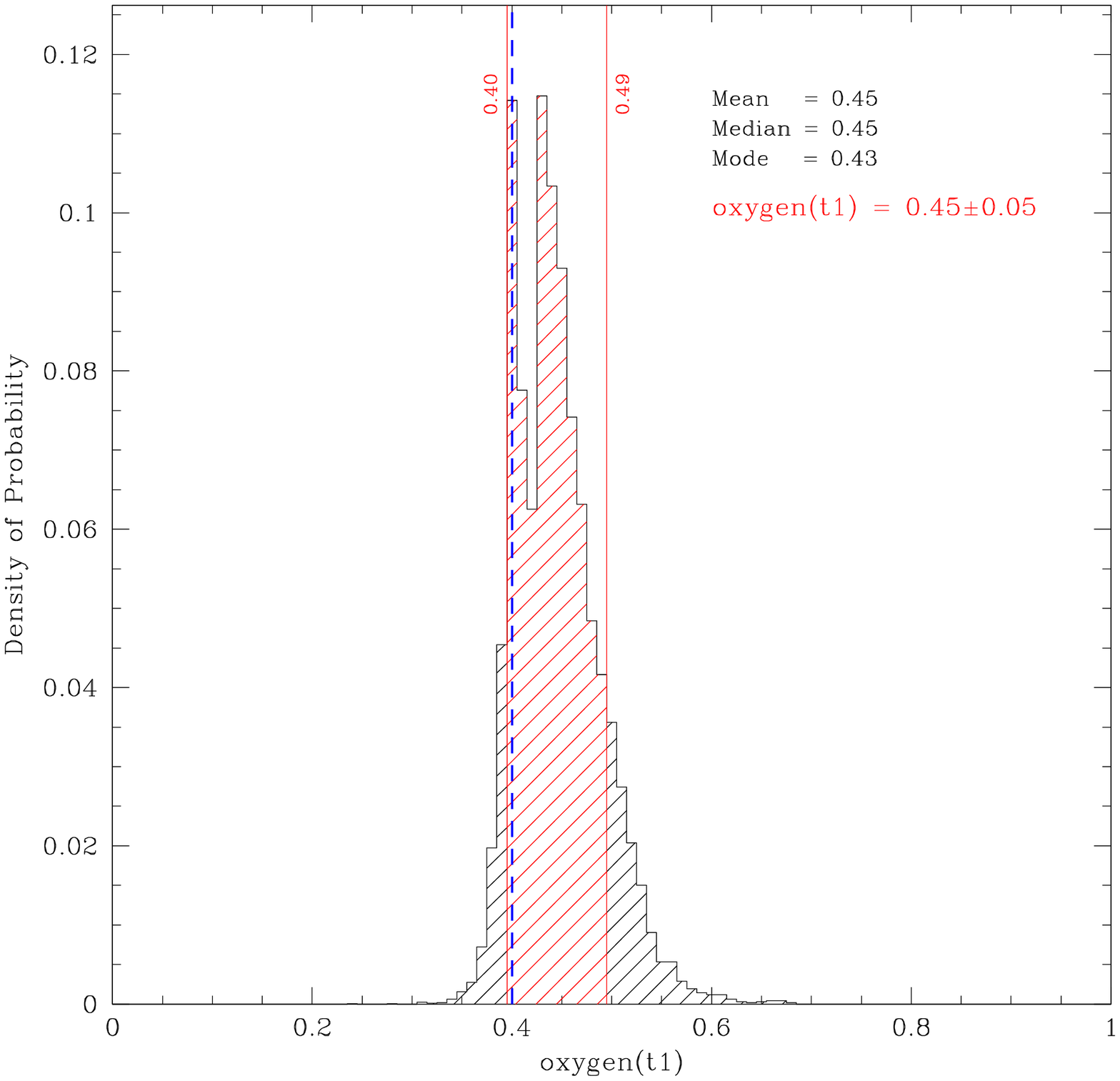} \\
    \includegraphics[width=.35\textwidth]{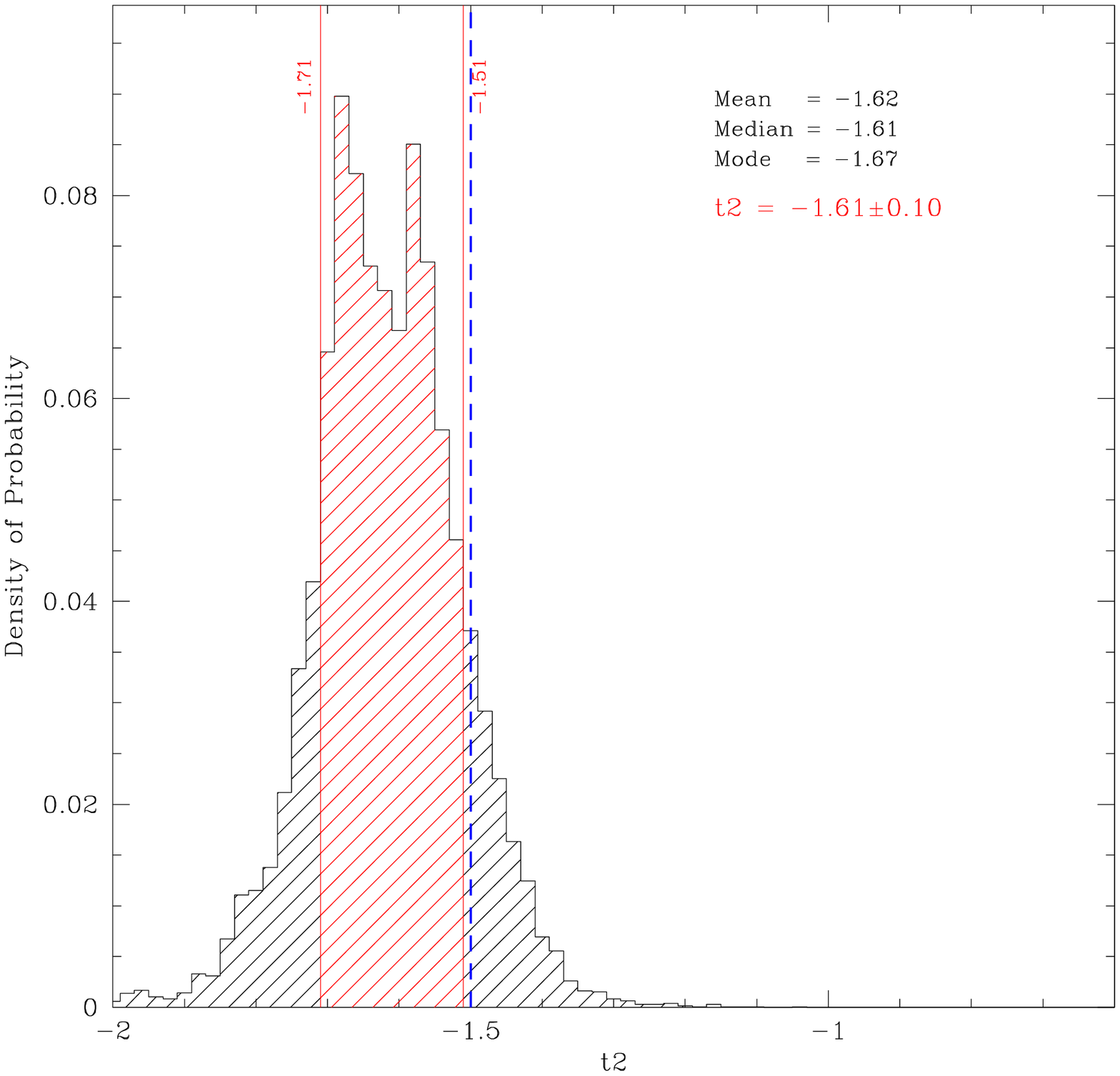} &
    \includegraphics[width=.35\textwidth]{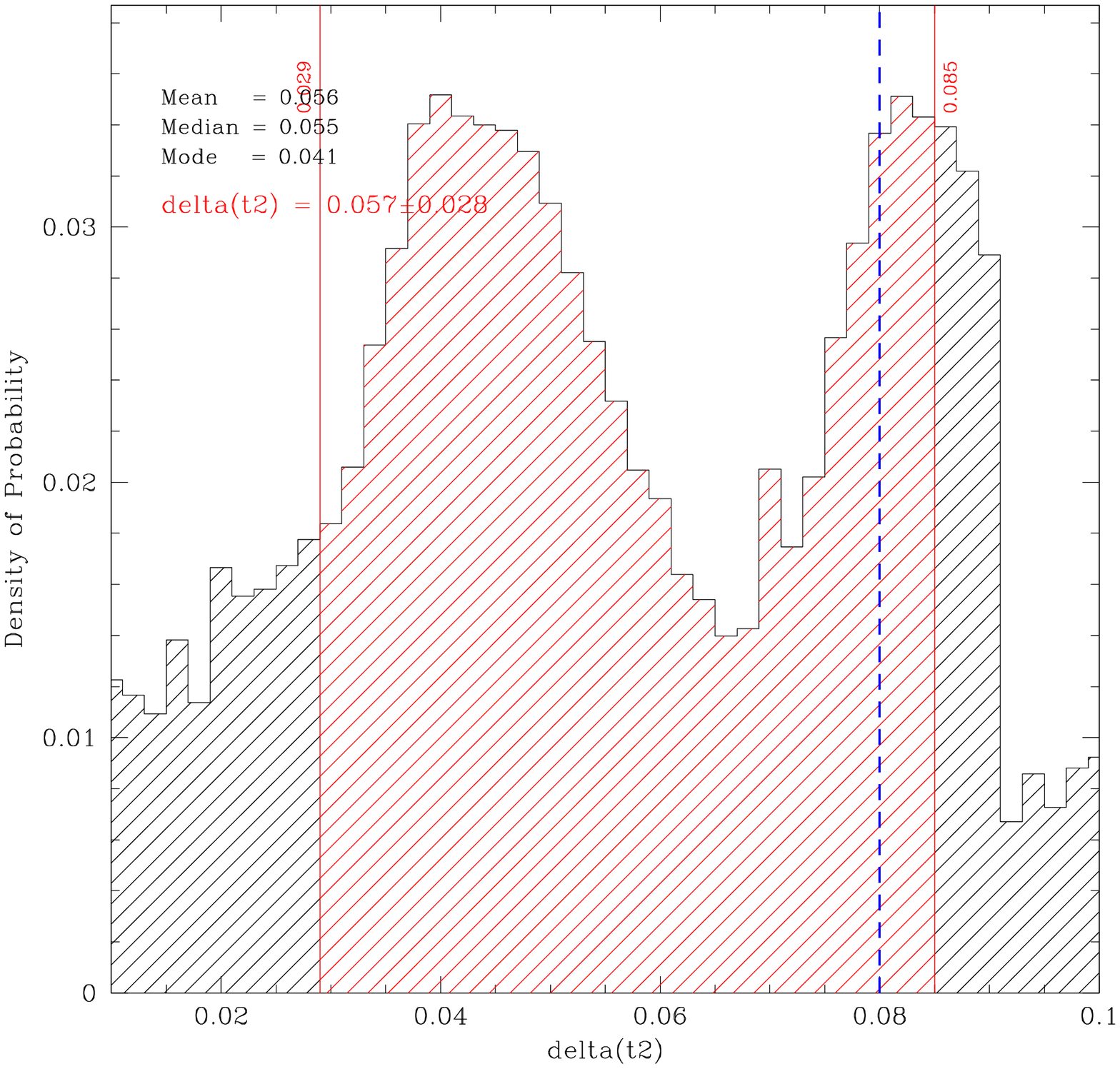} &
    \includegraphics[width=.35\textwidth]{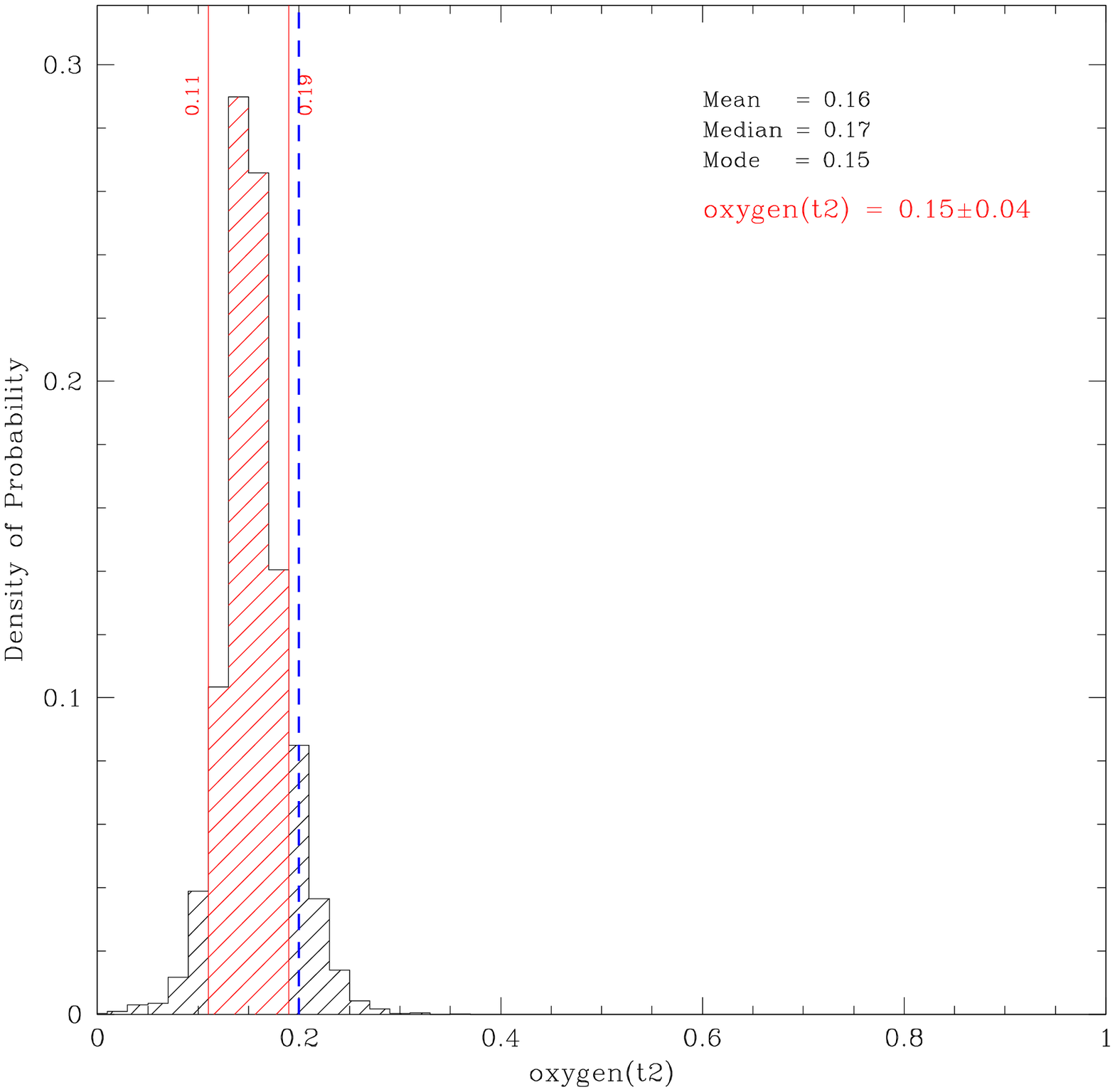}
    \end{tabular}
  \begin{flushright}
Figure~10
\end{flushright}
\end{figure}

\clearpage

\begin{figure}[!h]
\centering
  \begin{tabular}{@{}ccc@{}}
    \includegraphics[width=.35\textwidth]{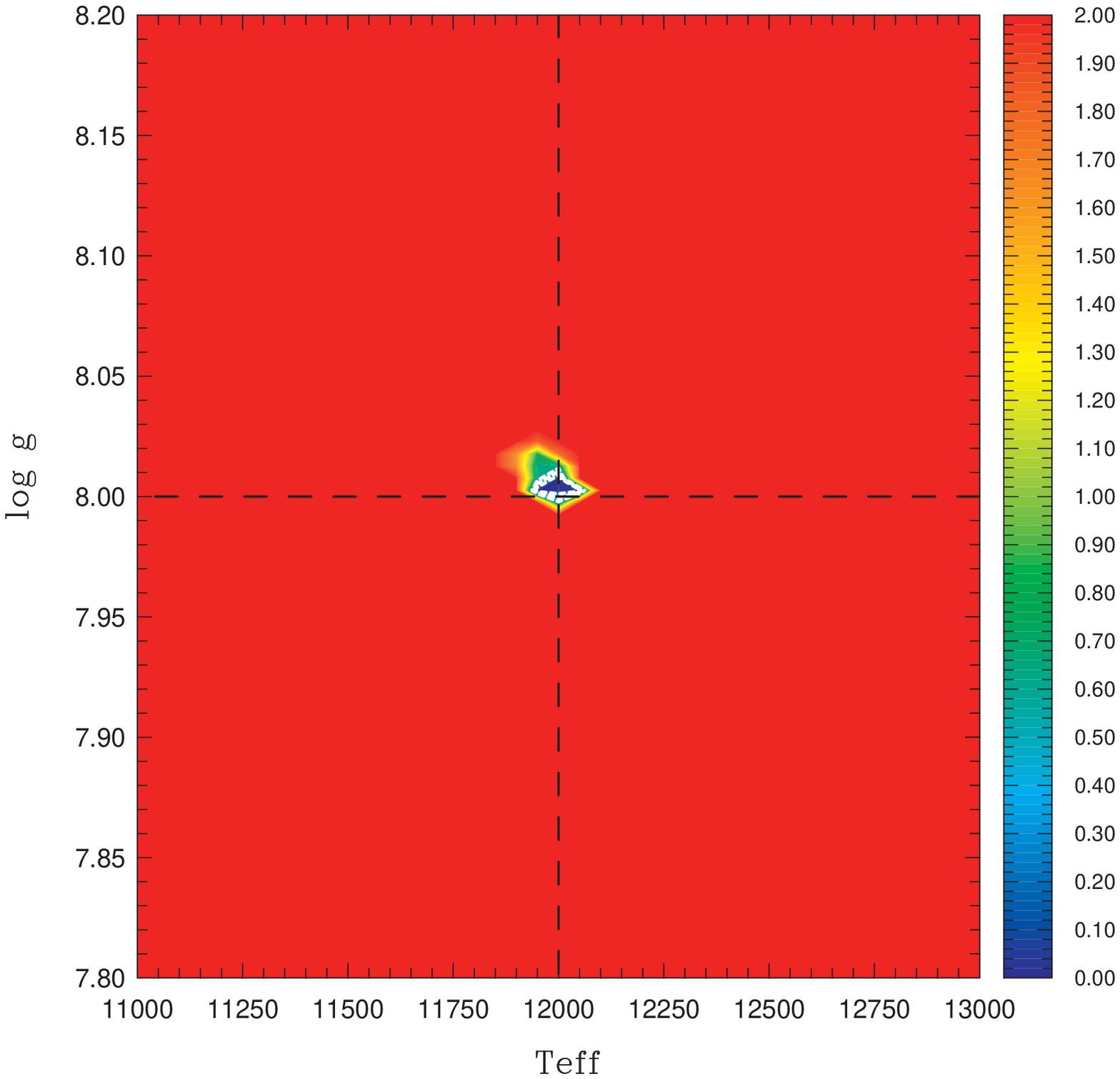} &
    \includegraphics[width=.35\textwidth]{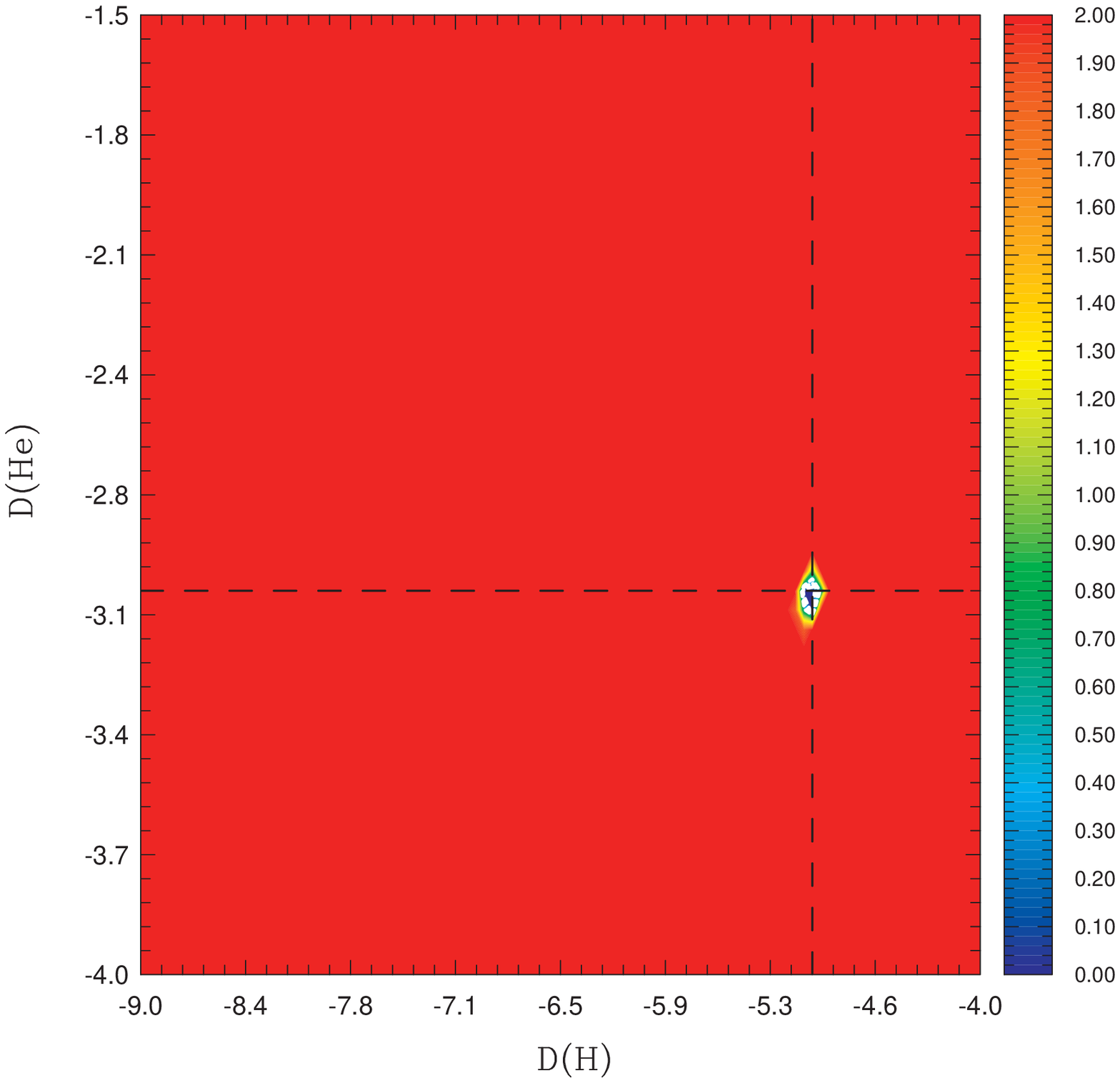} &
    \includegraphics[width=.35\textwidth]{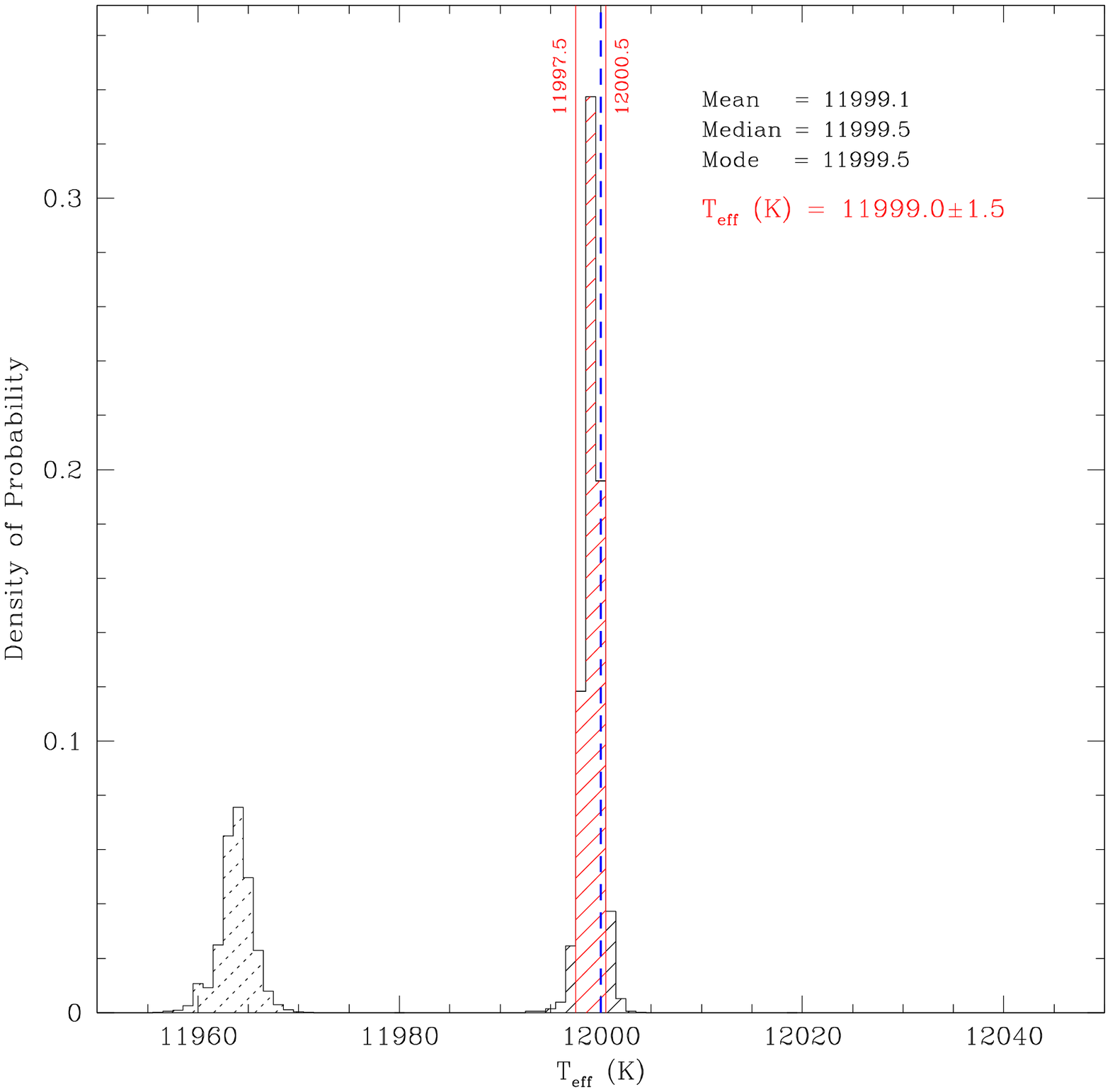} \\
    \includegraphics[width=.35\textwidth]{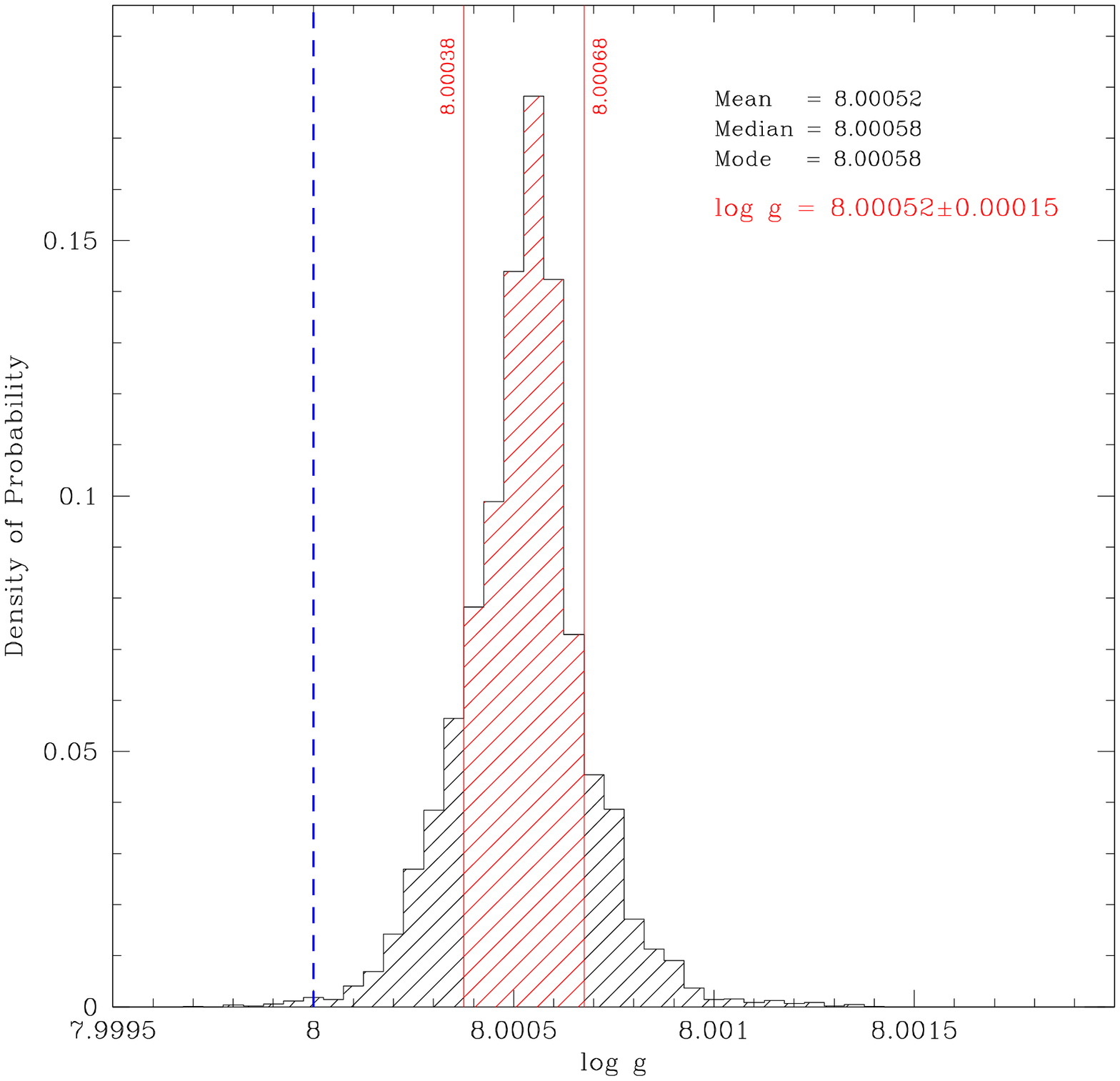}  &
    \includegraphics[width=.35\textwidth]{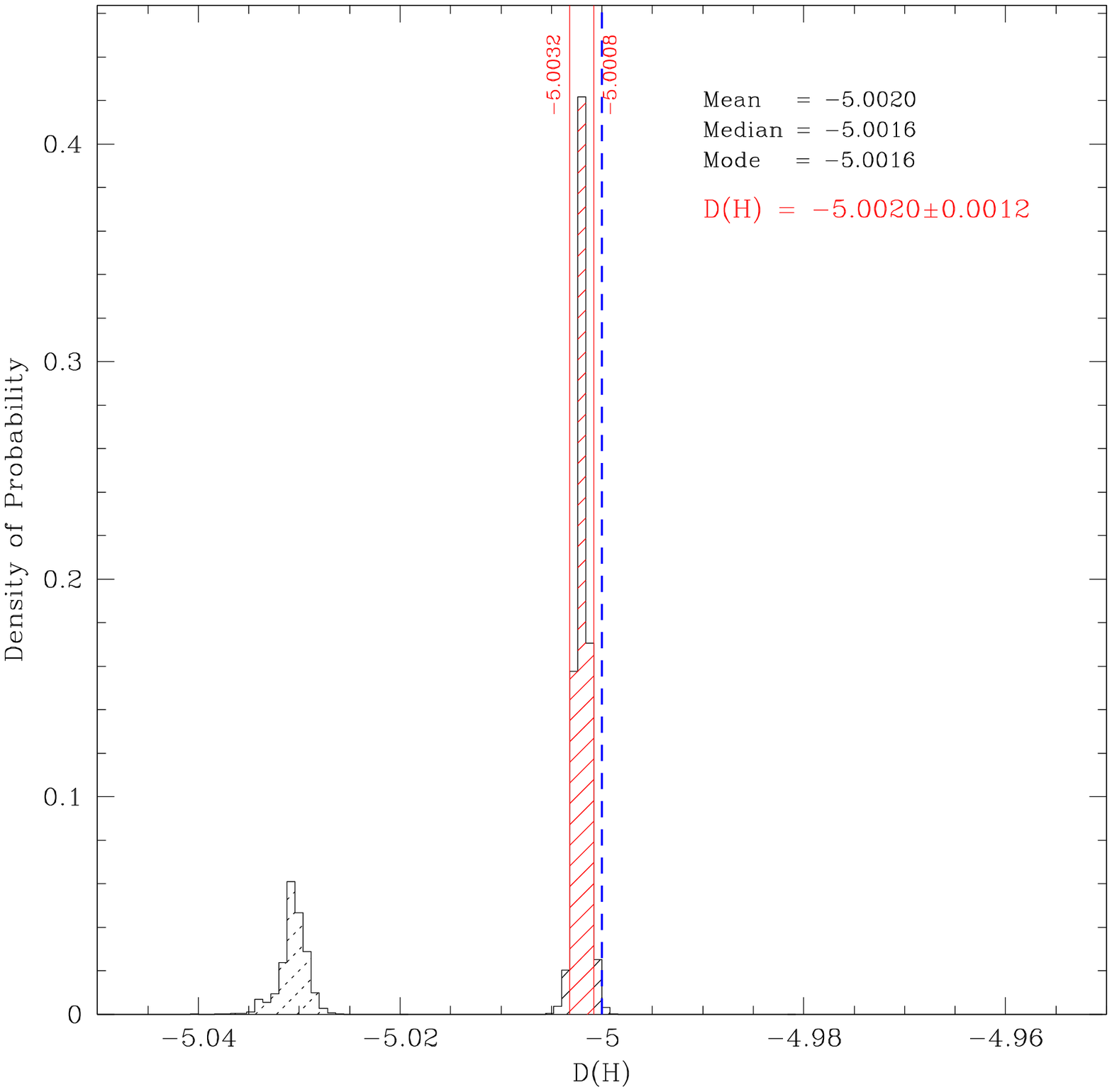} &
    \includegraphics[width=.35\textwidth]{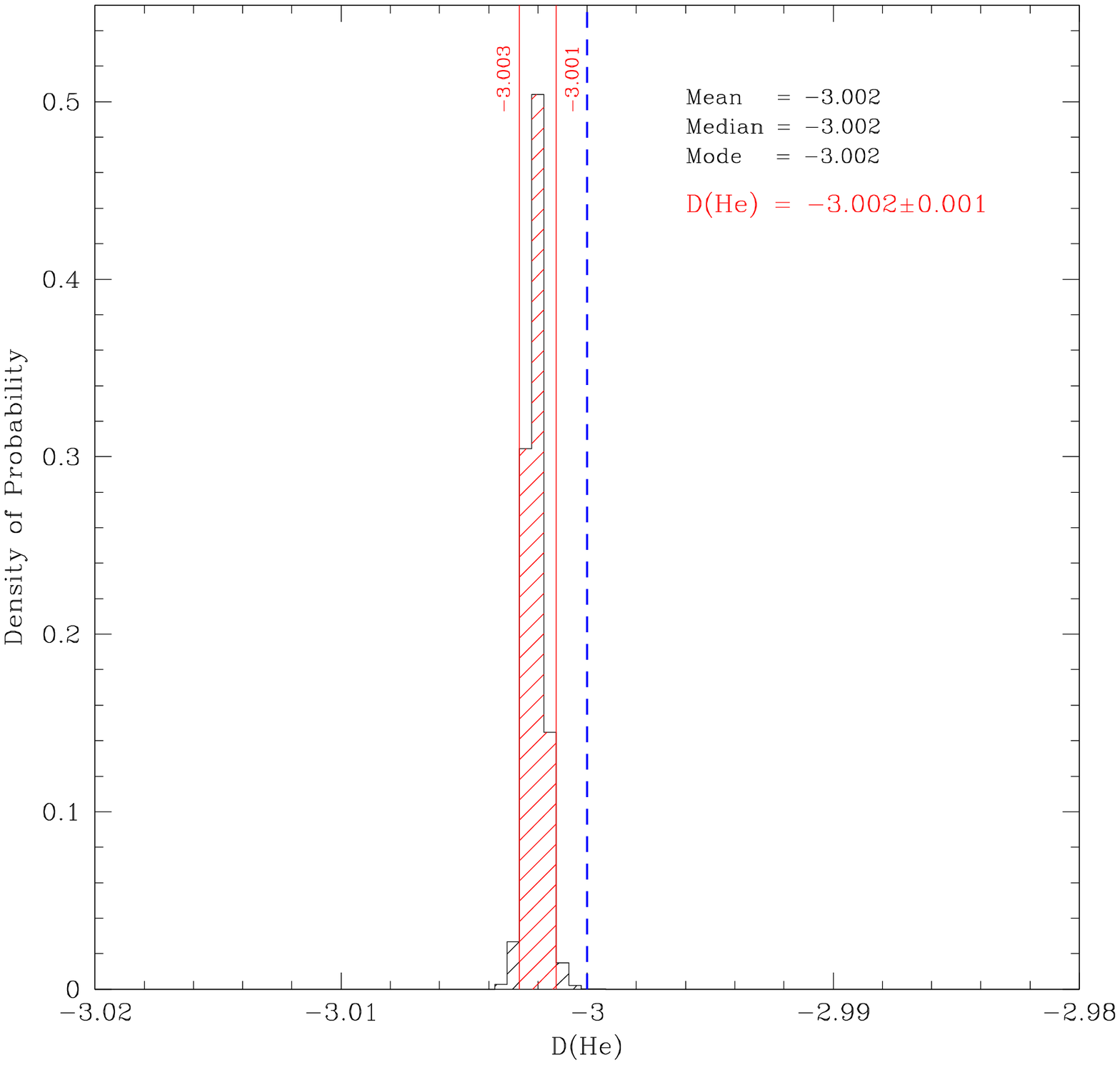} \\
    \includegraphics[width=.35\textwidth]{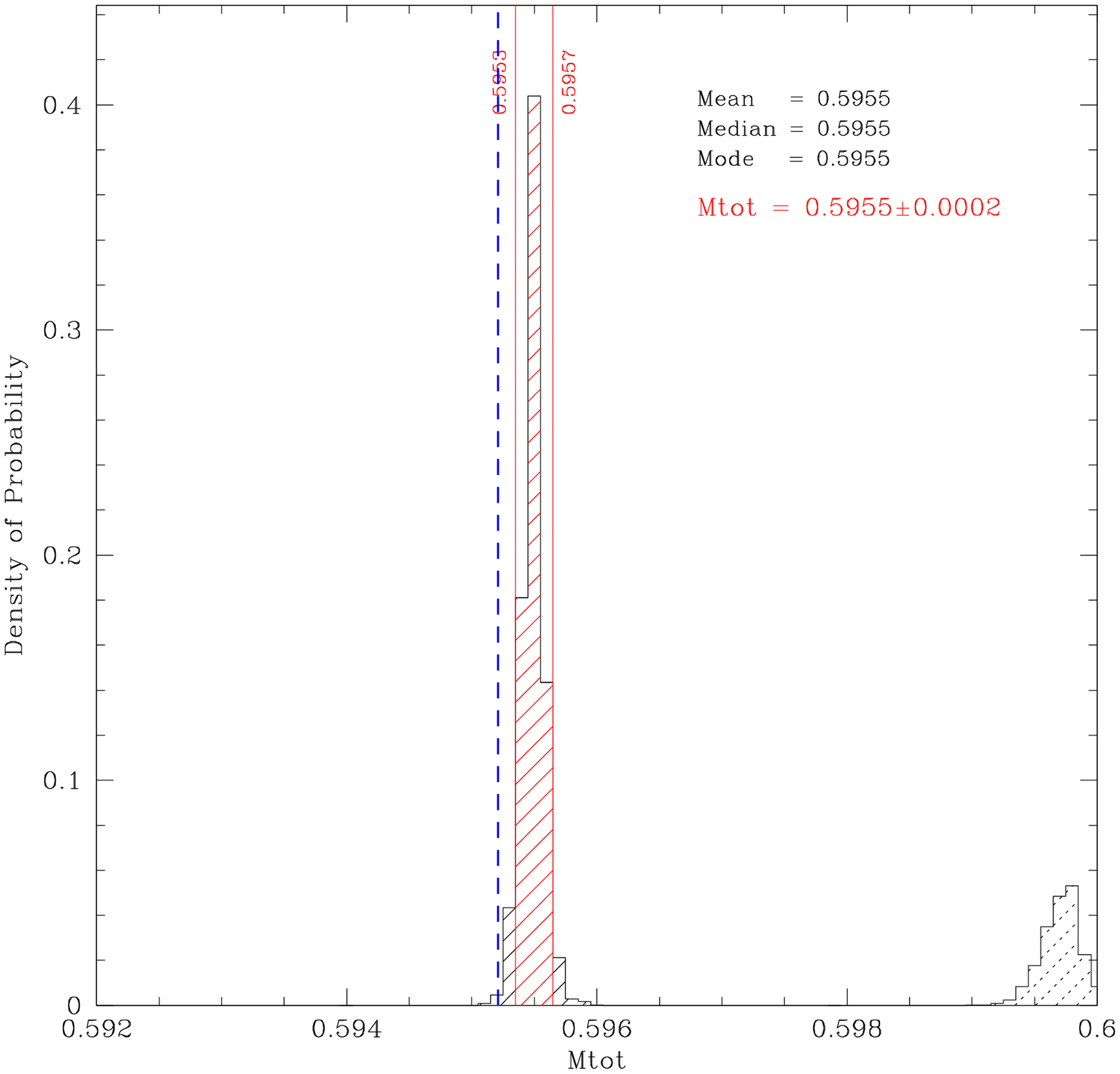} &
    \includegraphics[width=.35\textwidth]{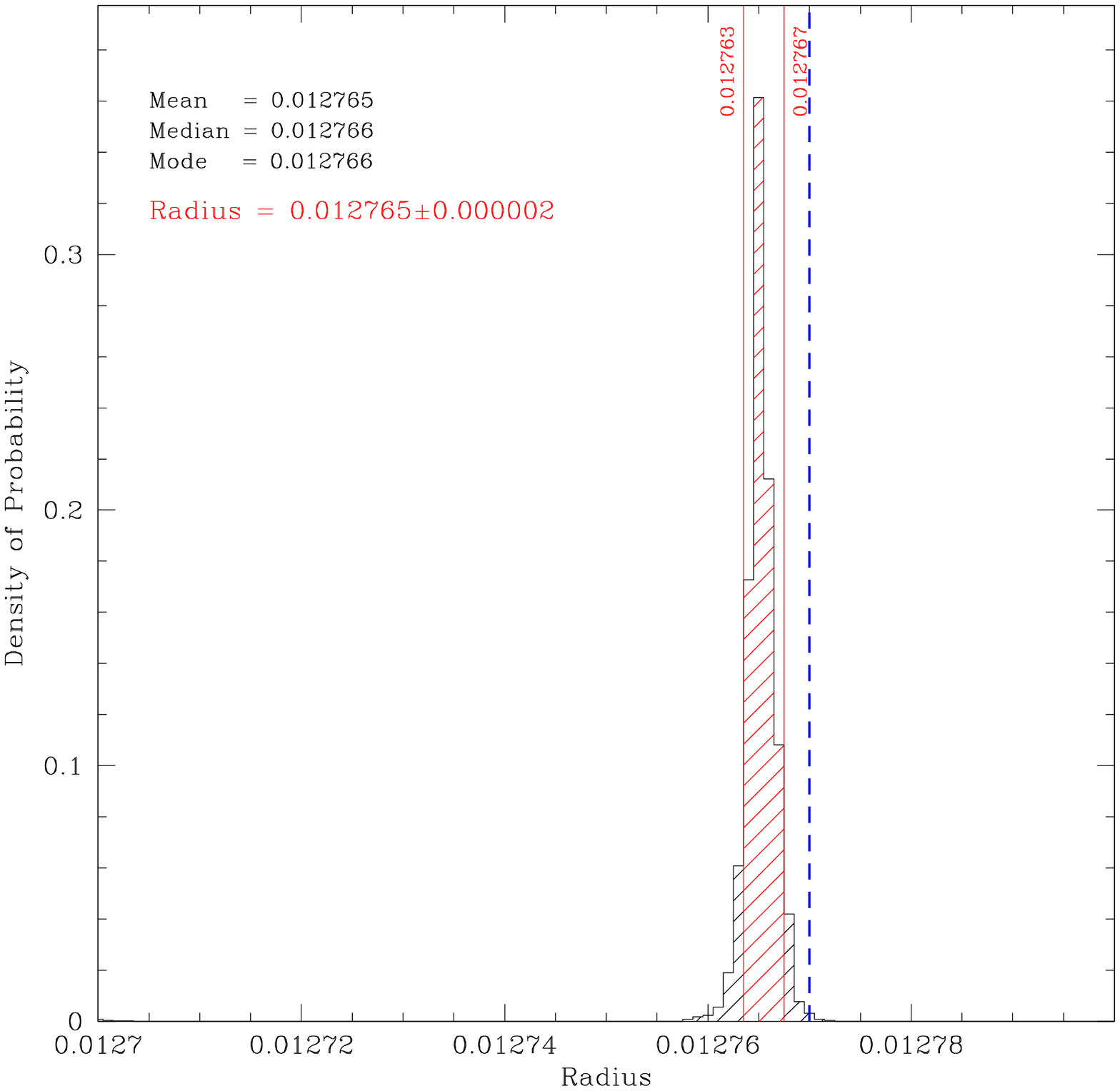} &
    \includegraphics[width=.35\textwidth]{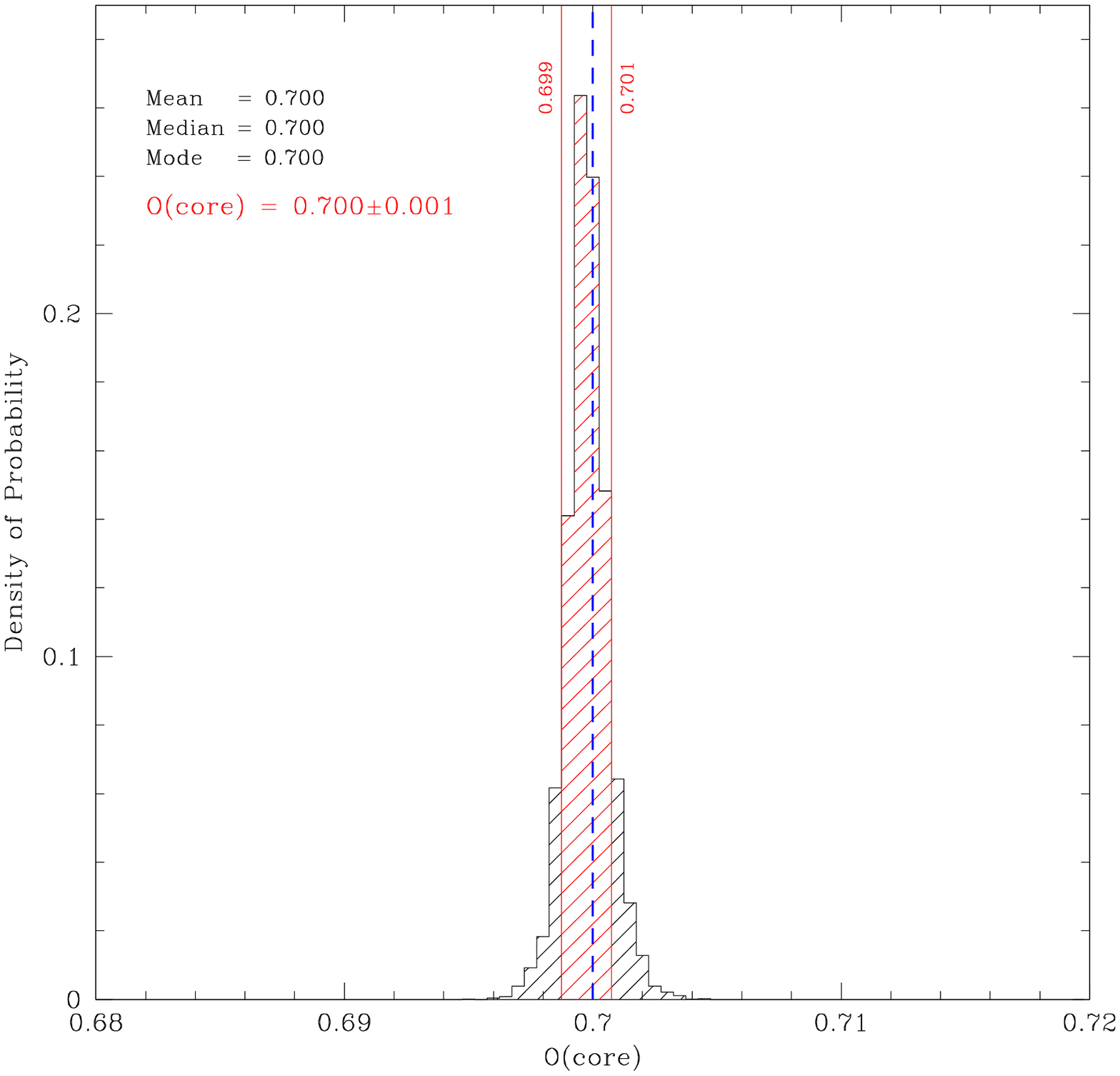} \\
    \end{tabular}
  \begin{flushright}
Figure~11
\end{flushright}
\end{figure}

\clearpage
\addtocounter{figure}{-1}
\begin{figure}[!h]
\centering
  \begin{tabular}{@{}ccc@{}}
    \includegraphics[width=.35\textwidth]{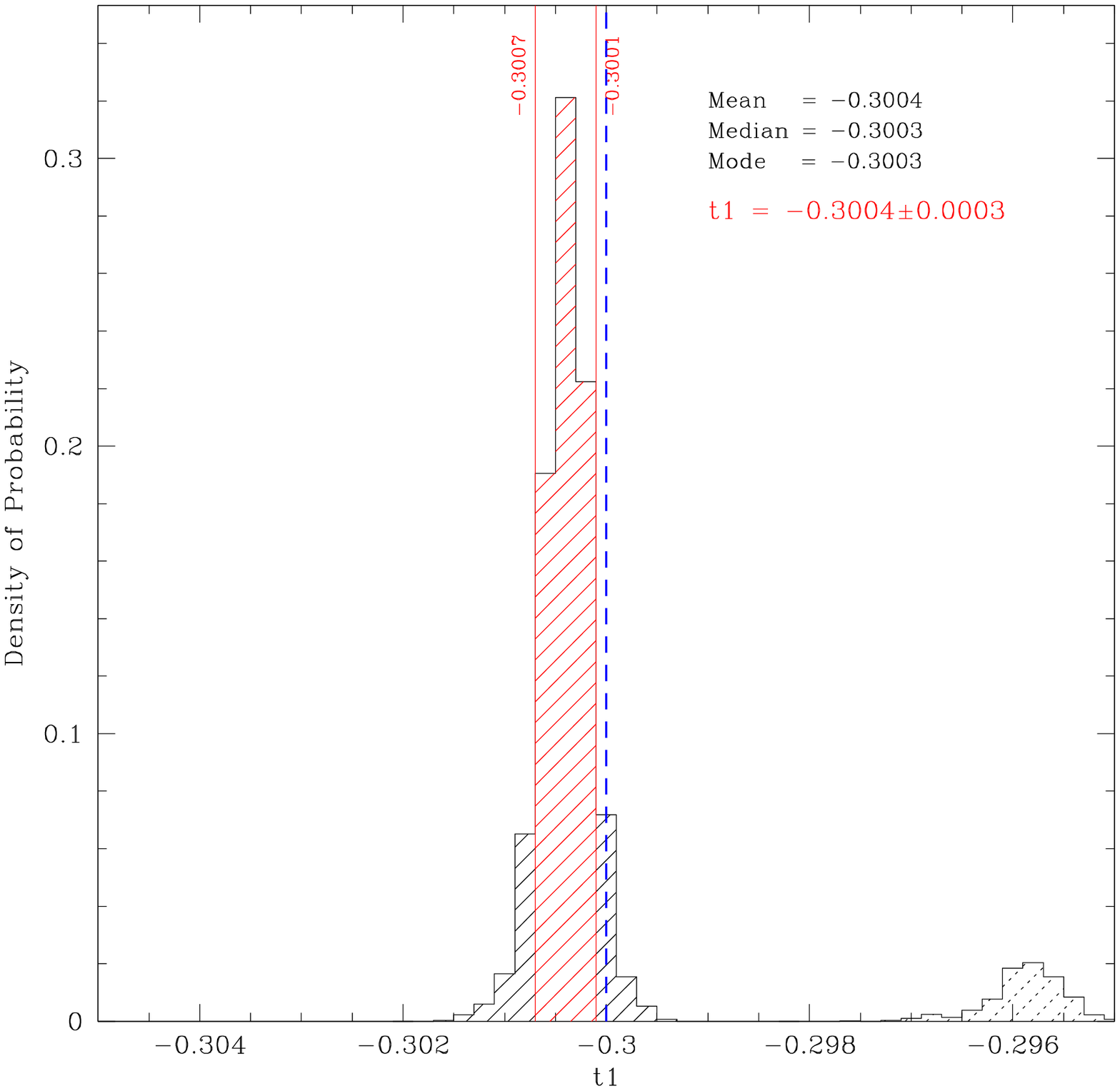} &
    \includegraphics[width=.35\textwidth]{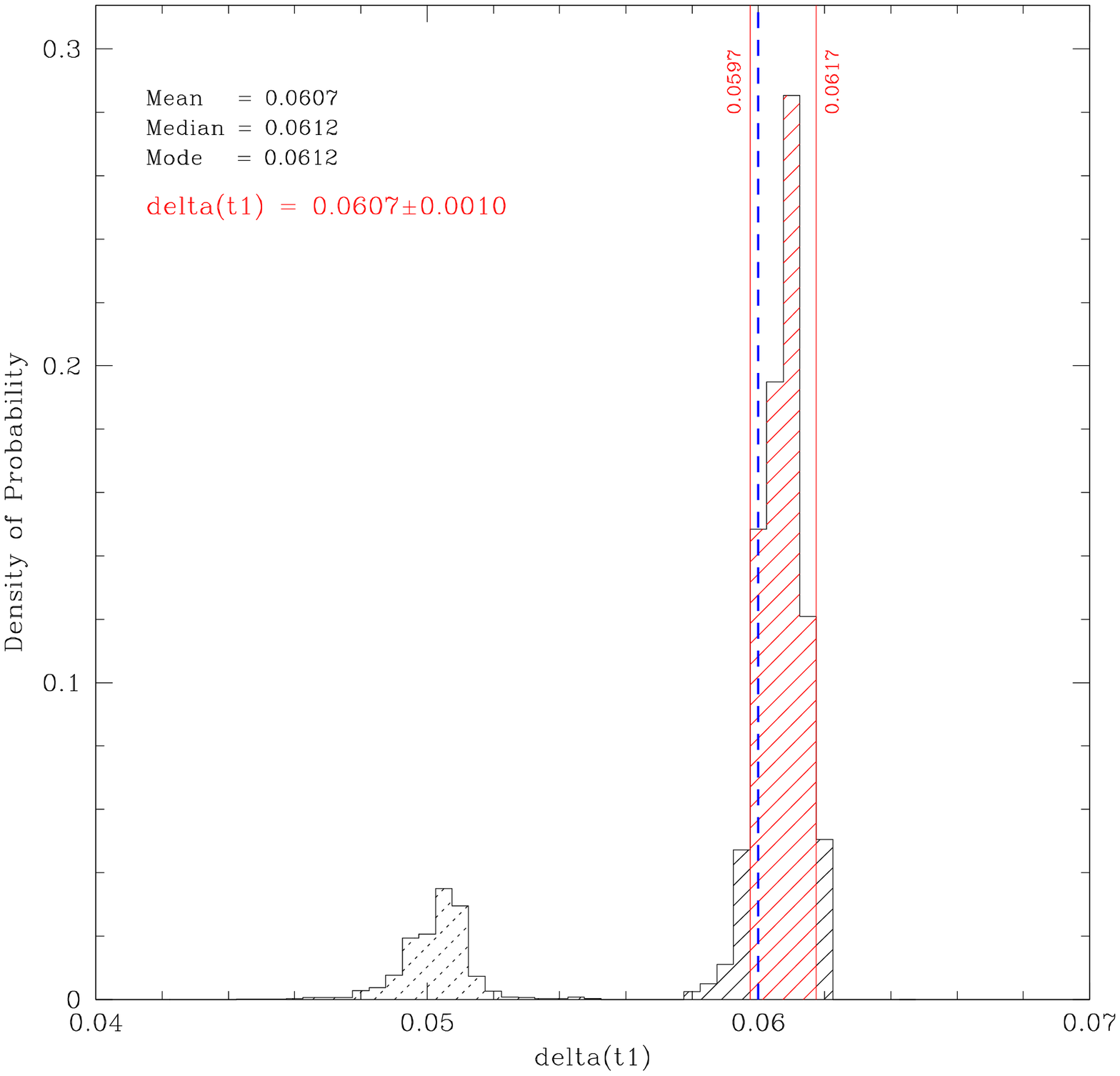} &
    \includegraphics[width=.35\textwidth]{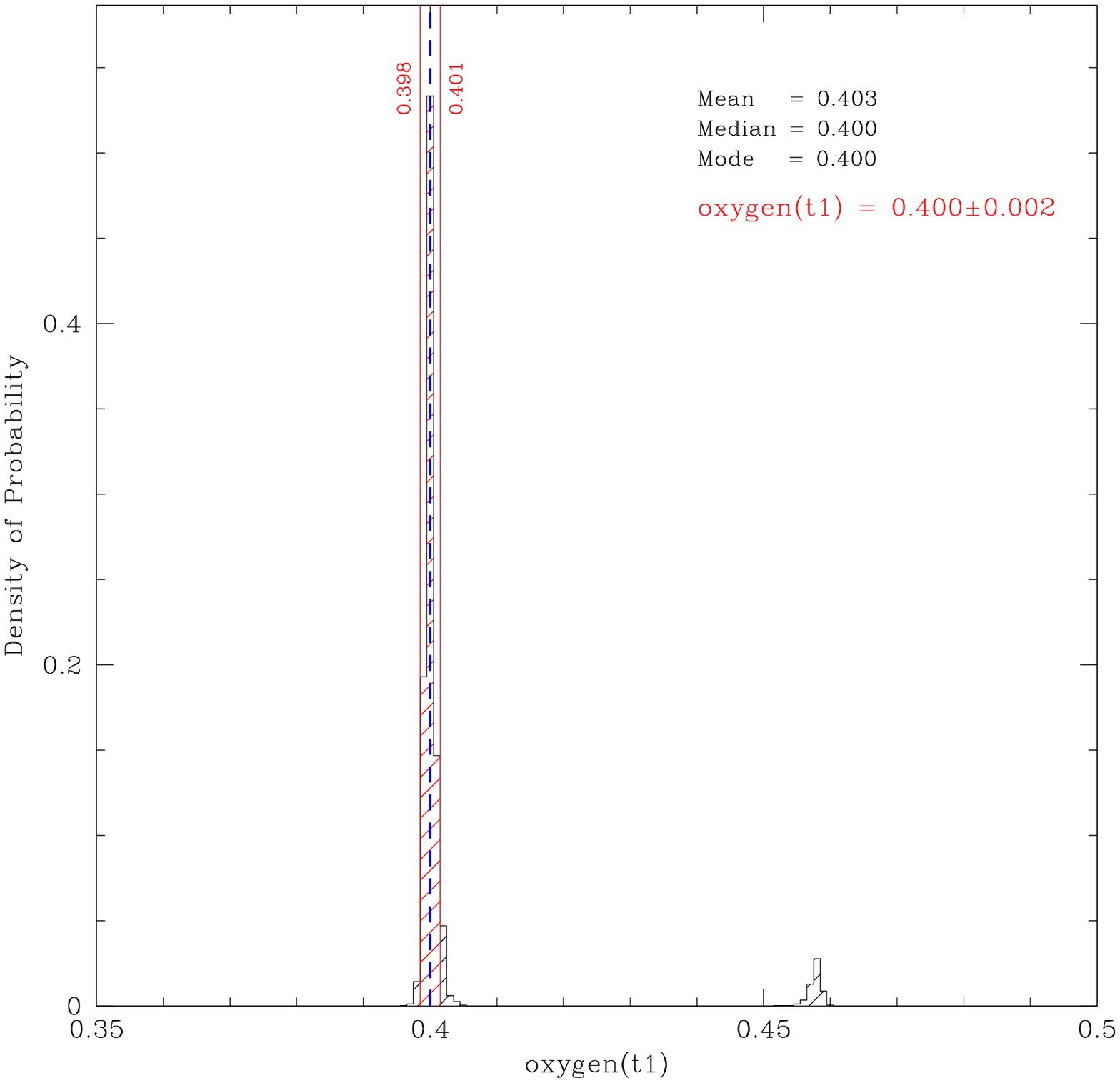} \\
    \includegraphics[width=.35\textwidth]{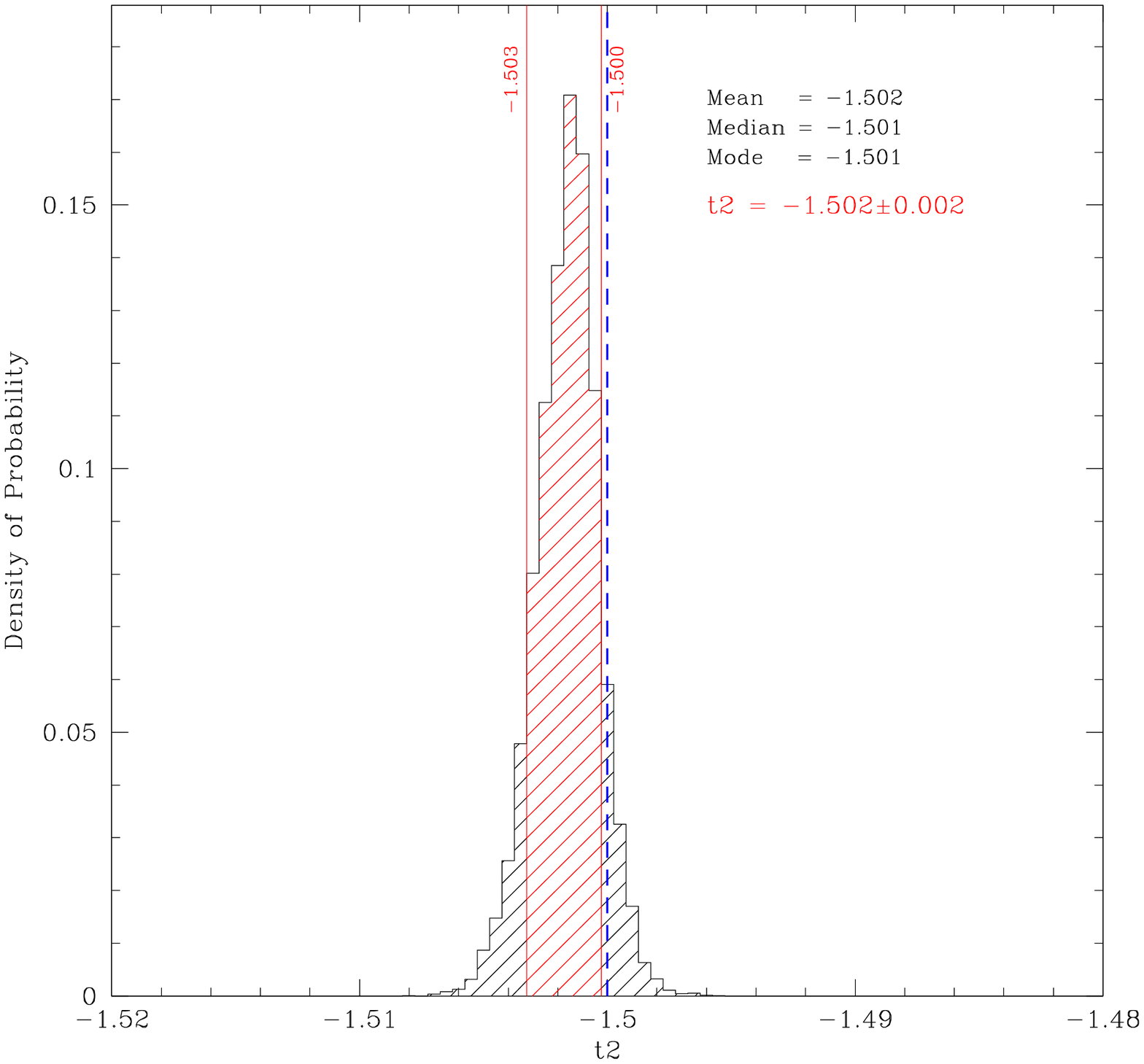} &
    \includegraphics[width=.35\textwidth]{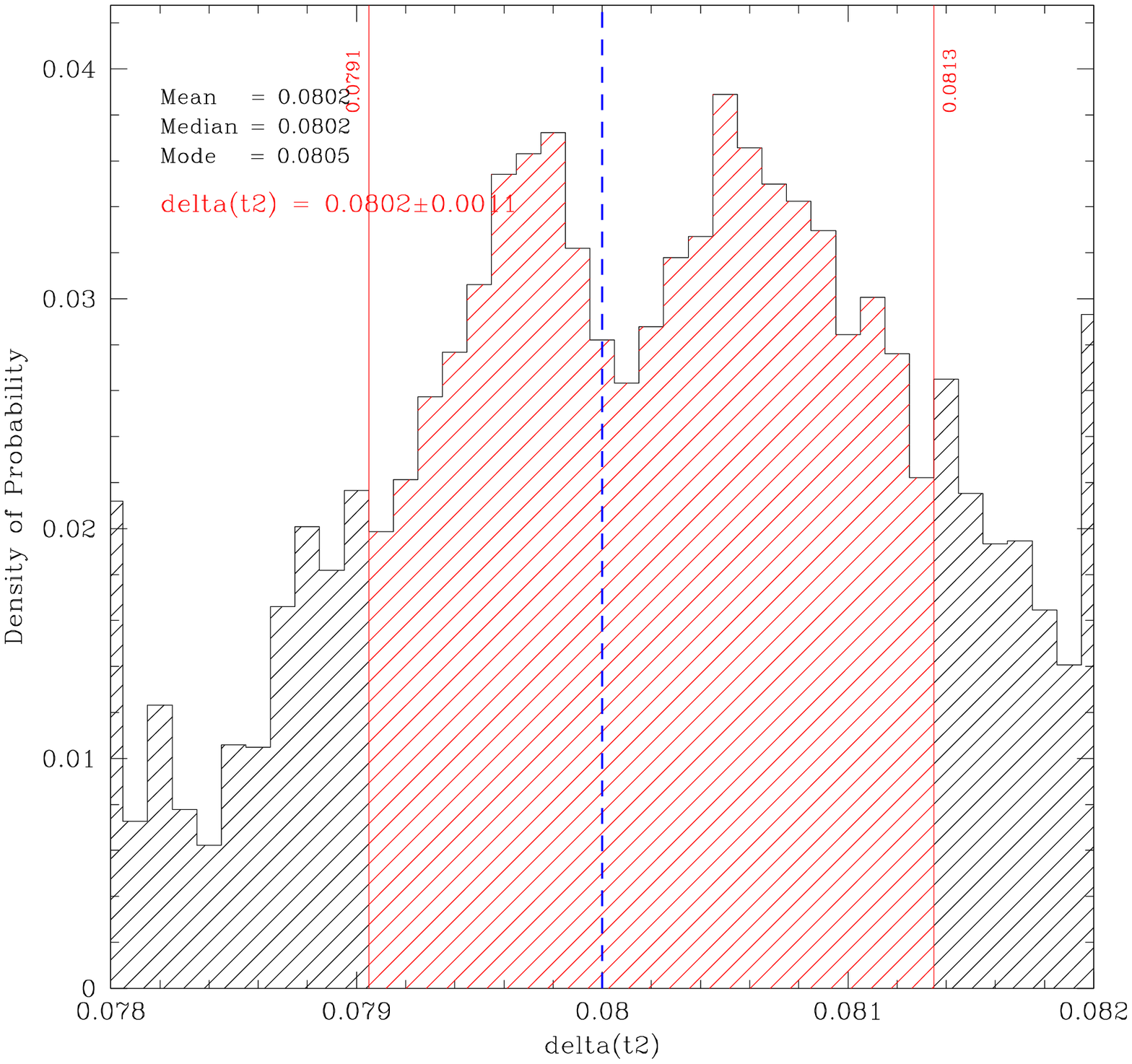} &
    \includegraphics[width=.35\textwidth]{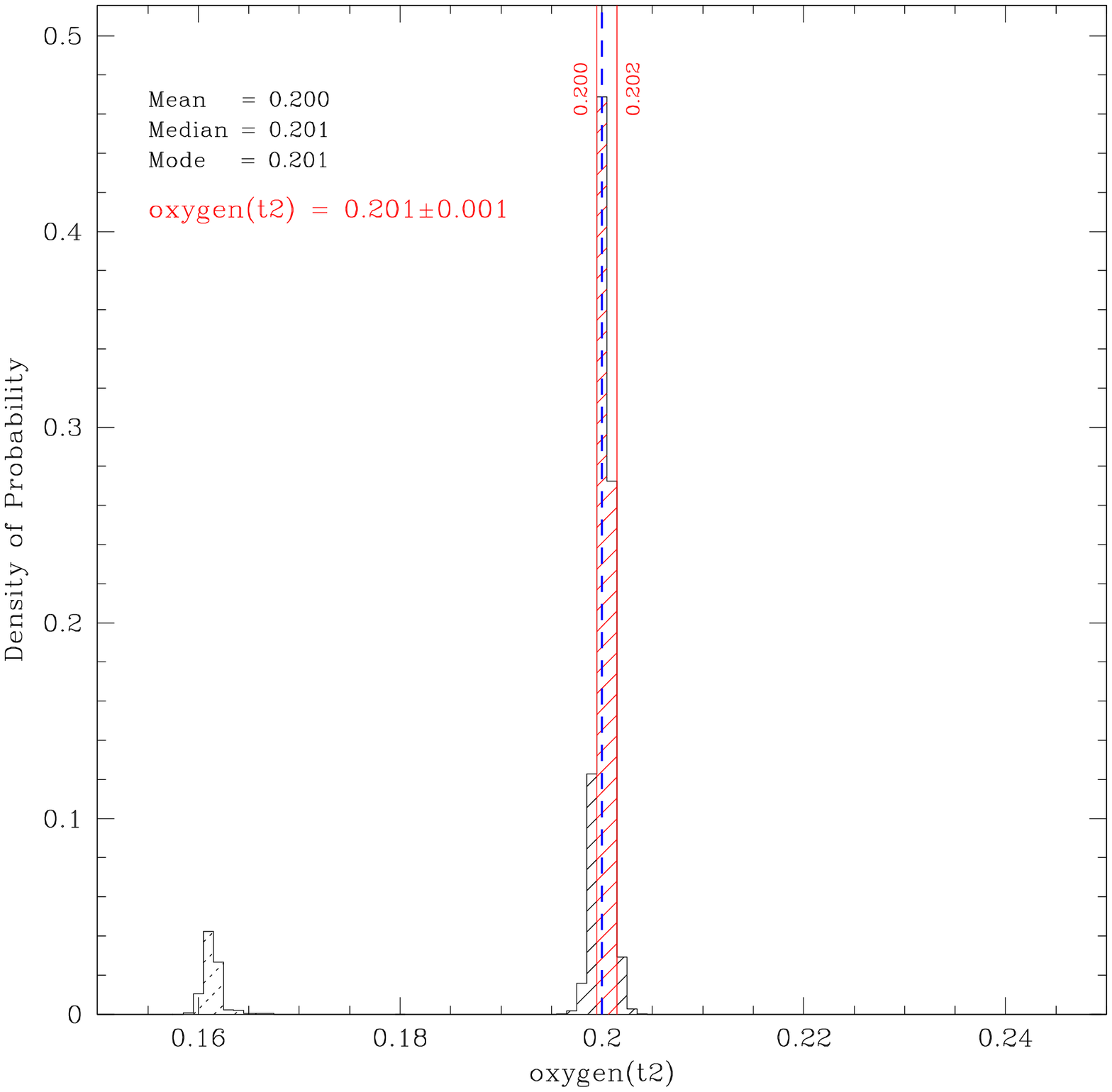}
    \end{tabular}
  \begin{flushright}
Figure~11
\end{flushright}
\end{figure}

\clearpage

\begin{figure}[!h]
\centering
  \begin{tabular}{@{}ccc@{}}
    \includegraphics[width=.35\textwidth]{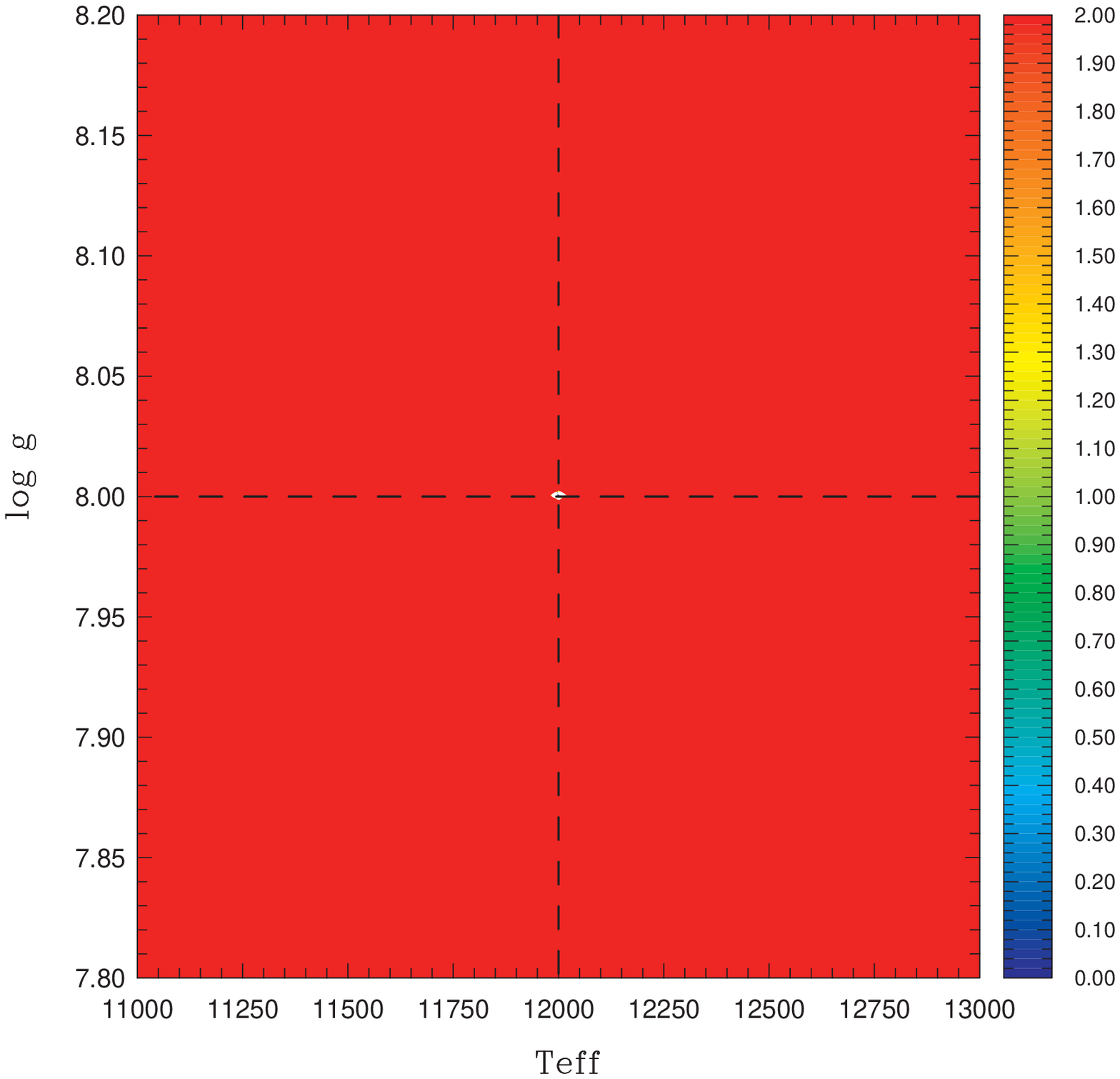} &
    \includegraphics[width=.35\textwidth]{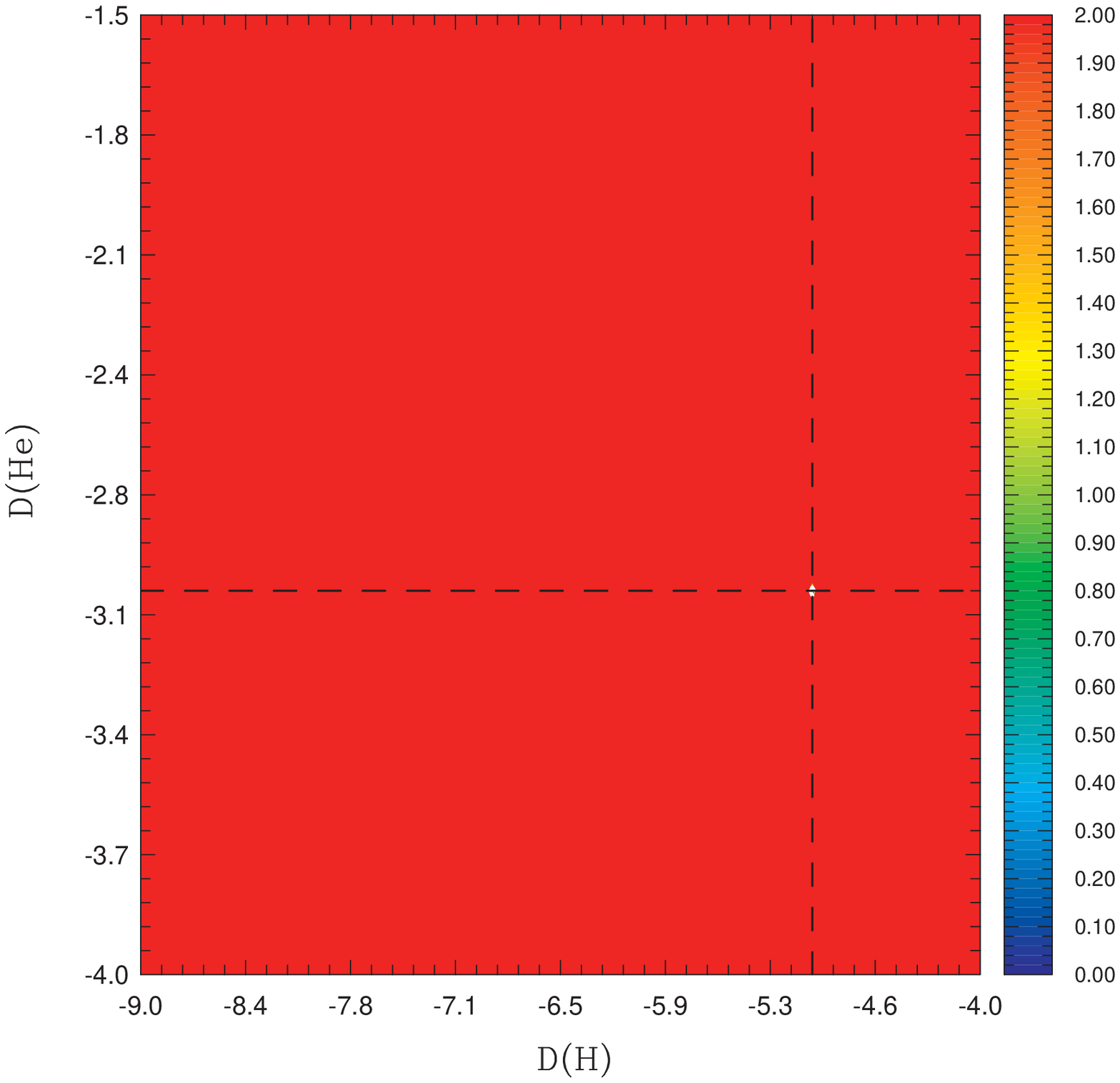} &
    \includegraphics[width=.35\textwidth]{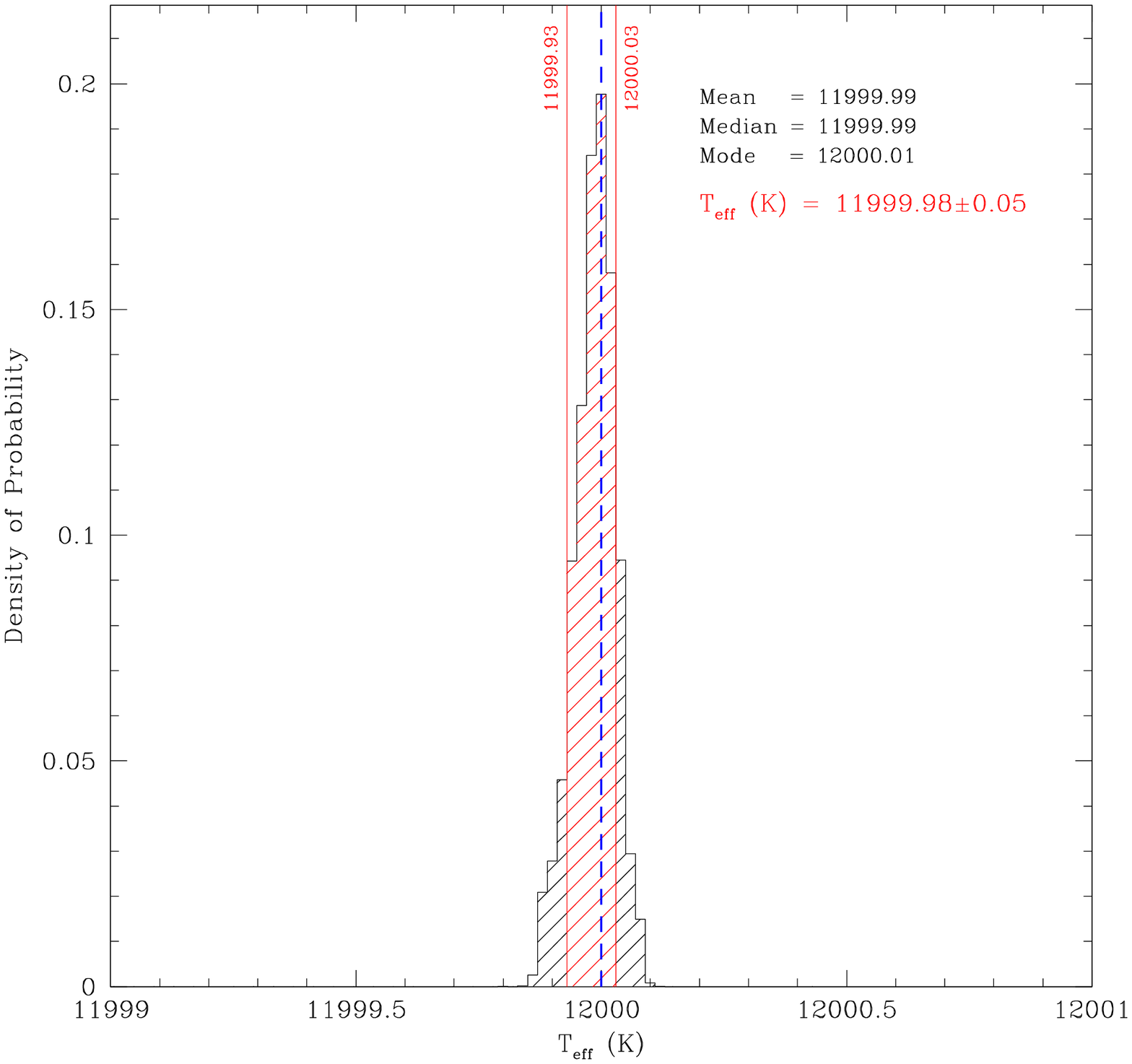} \\
    \includegraphics[width=.35\textwidth]{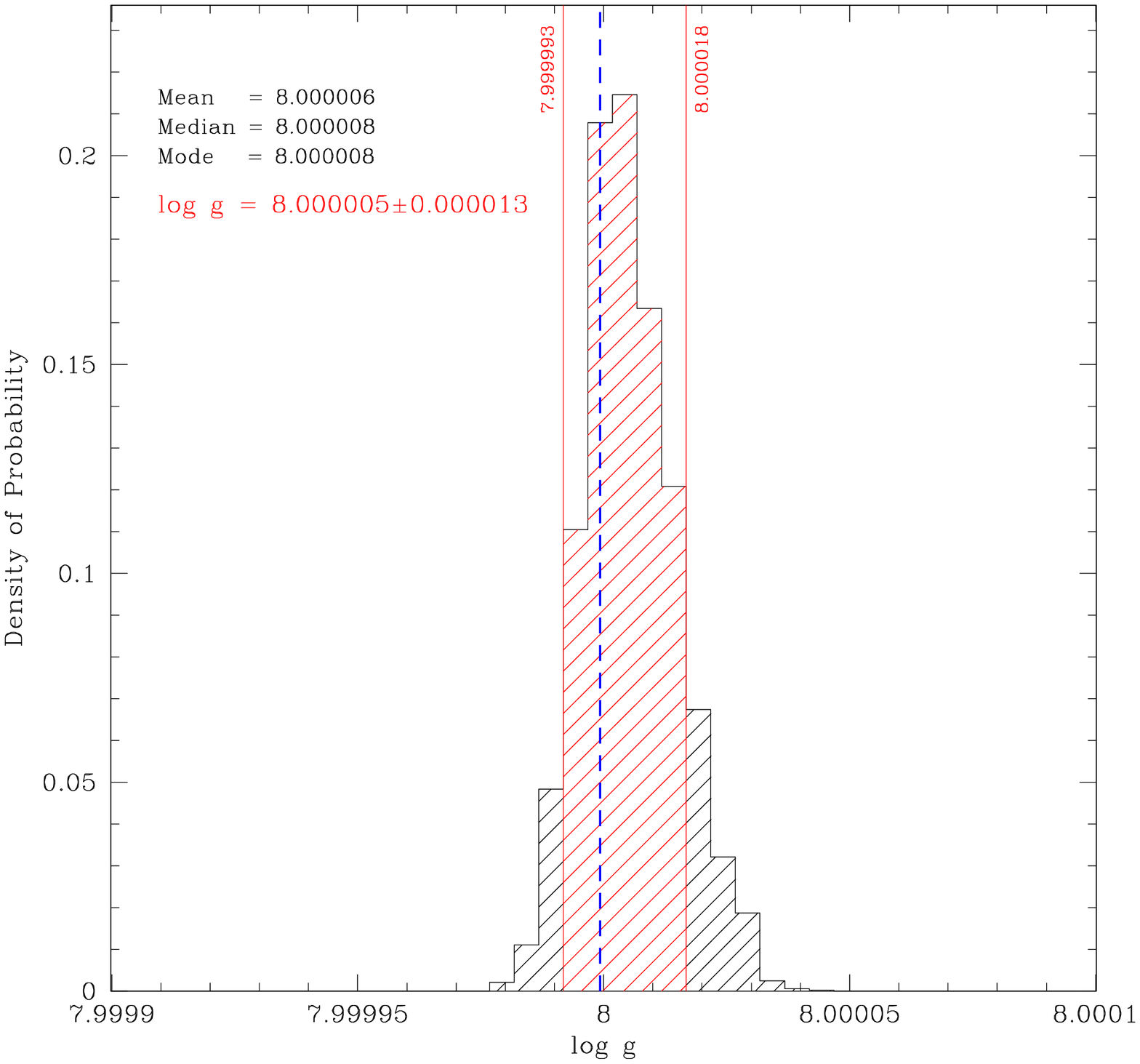}  &
    \includegraphics[width=.35\textwidth]{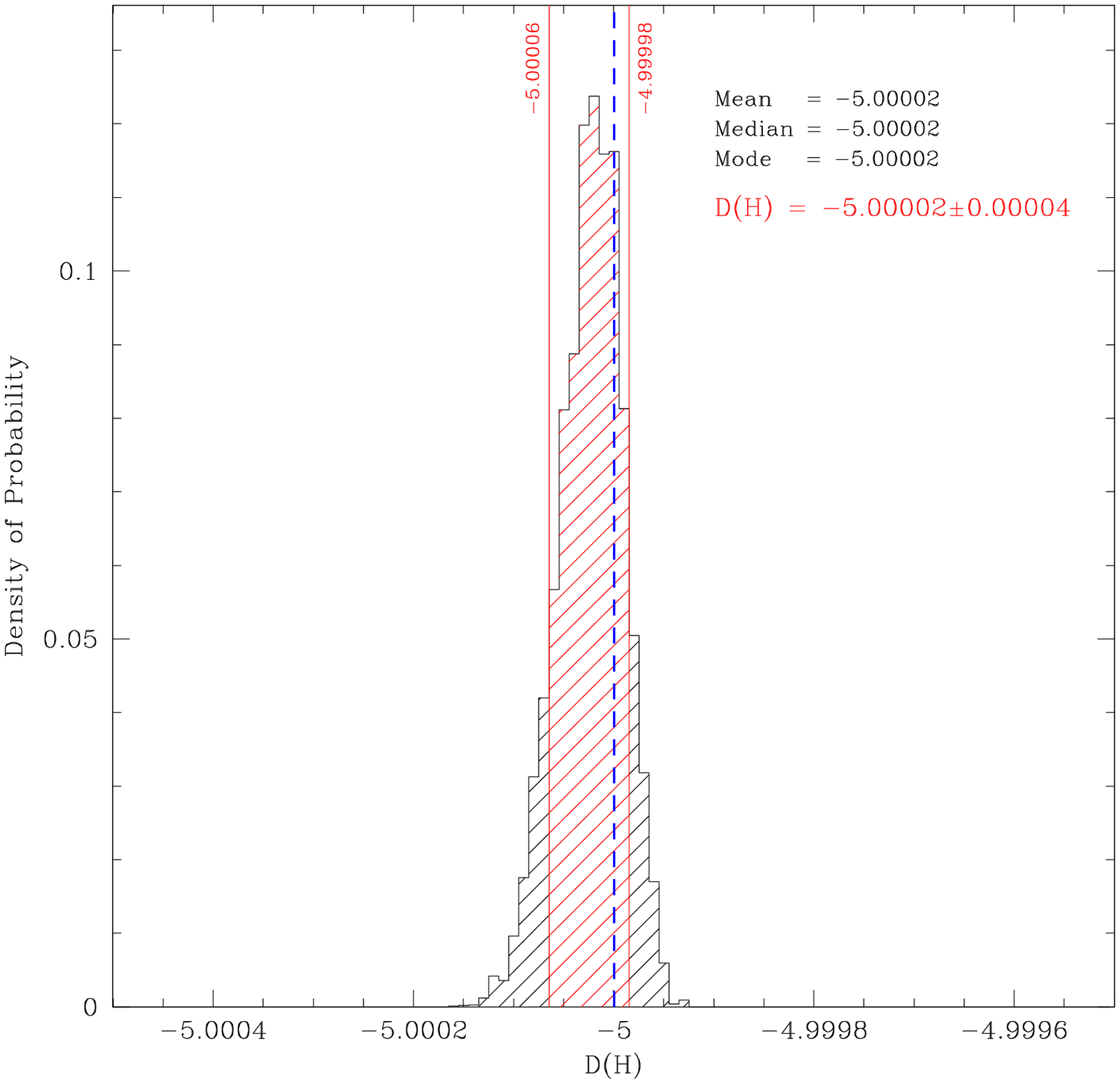} &
    \includegraphics[width=.35\textwidth]{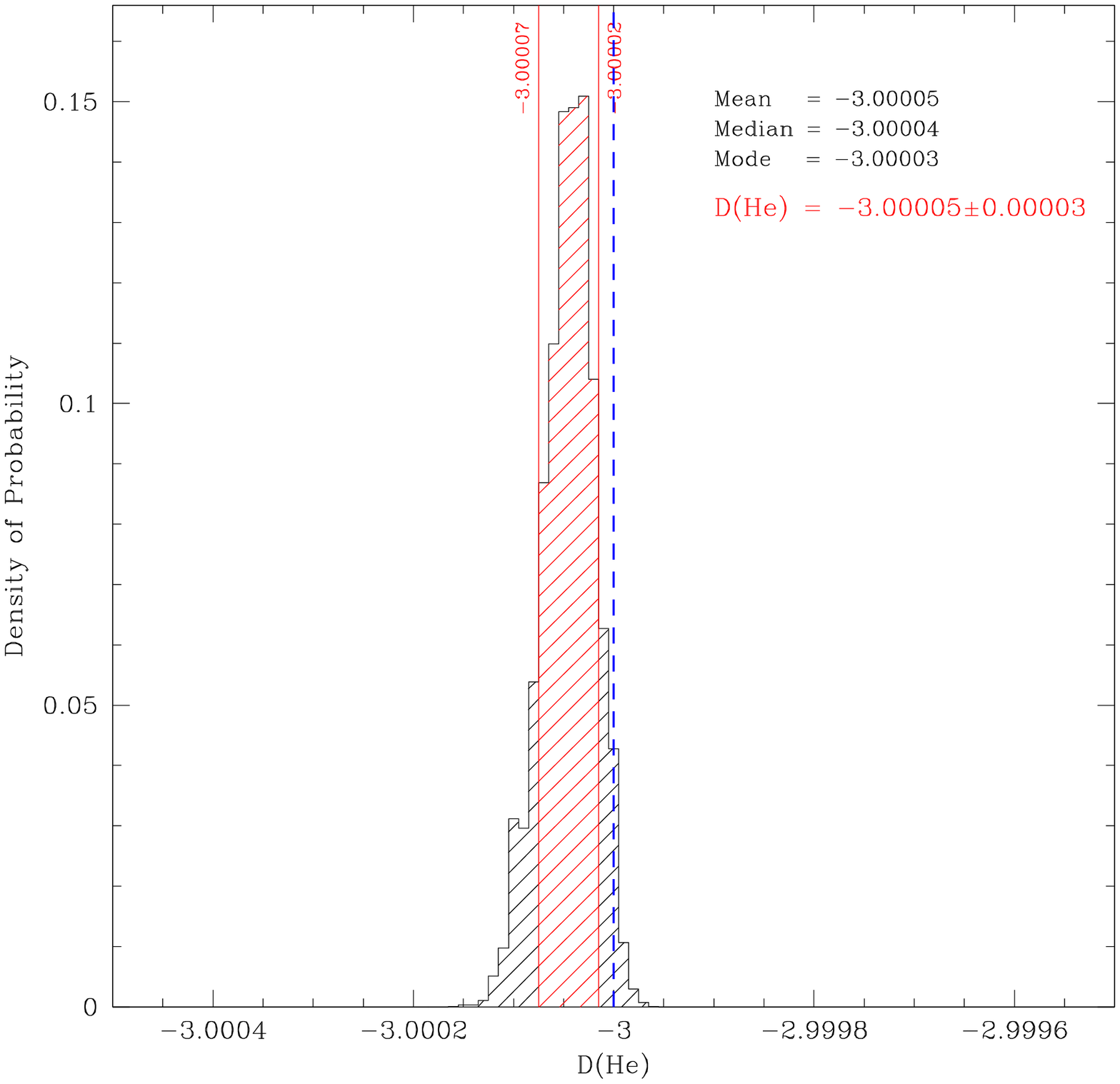} \\
    \includegraphics[width=.35\textwidth]{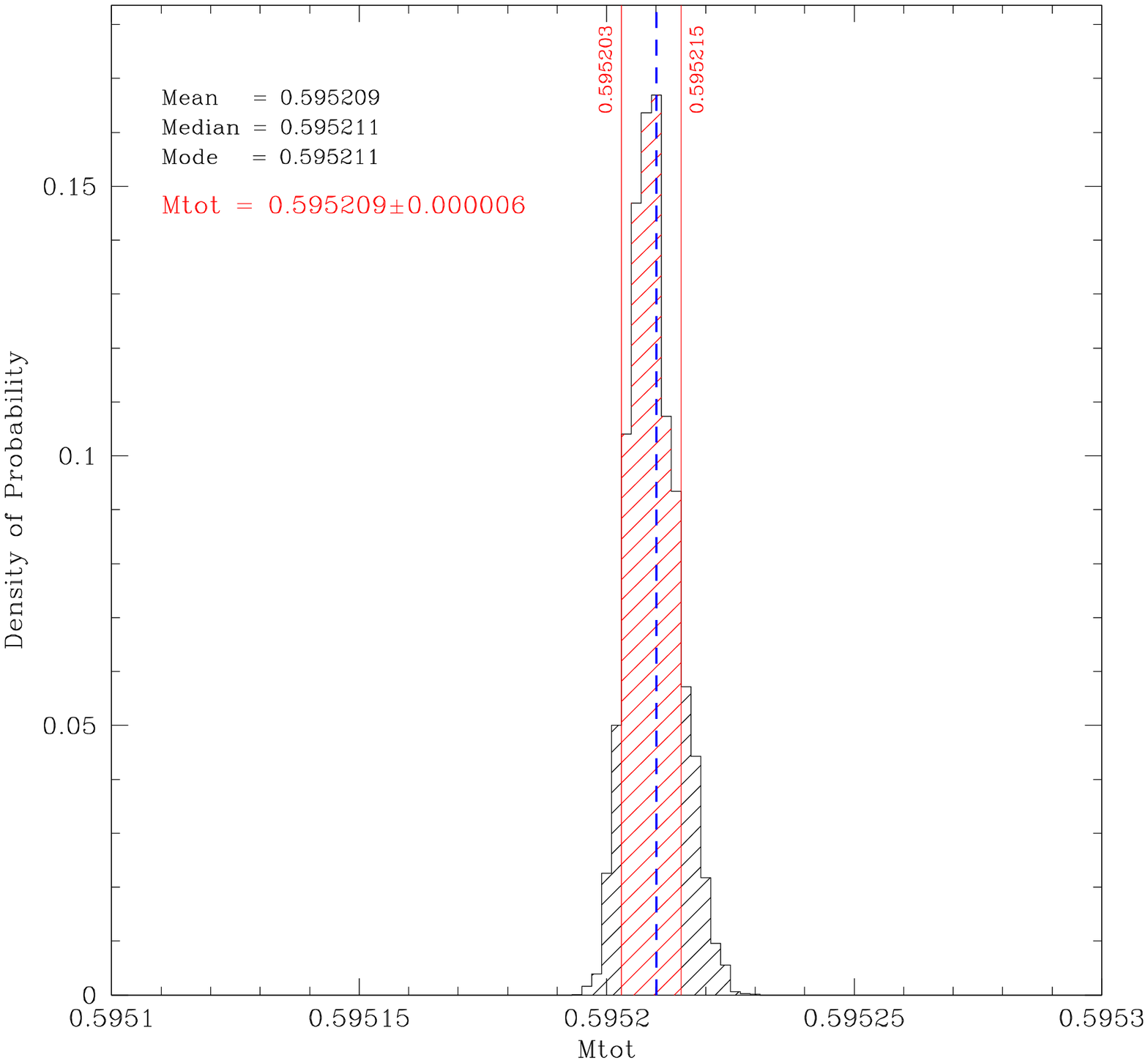} &
    \includegraphics[width=.35\textwidth]{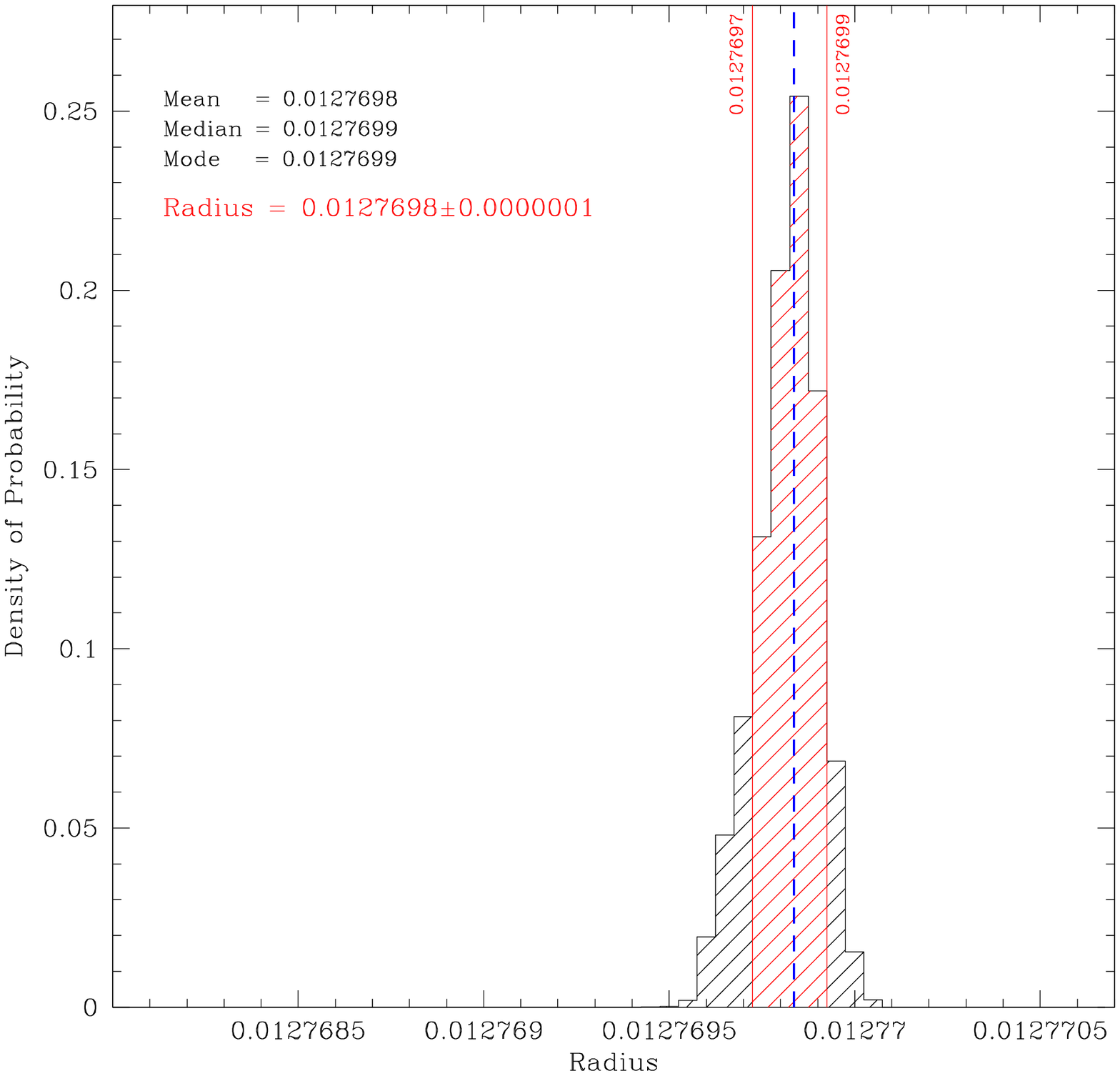} &
    \includegraphics[width=.35\textwidth]{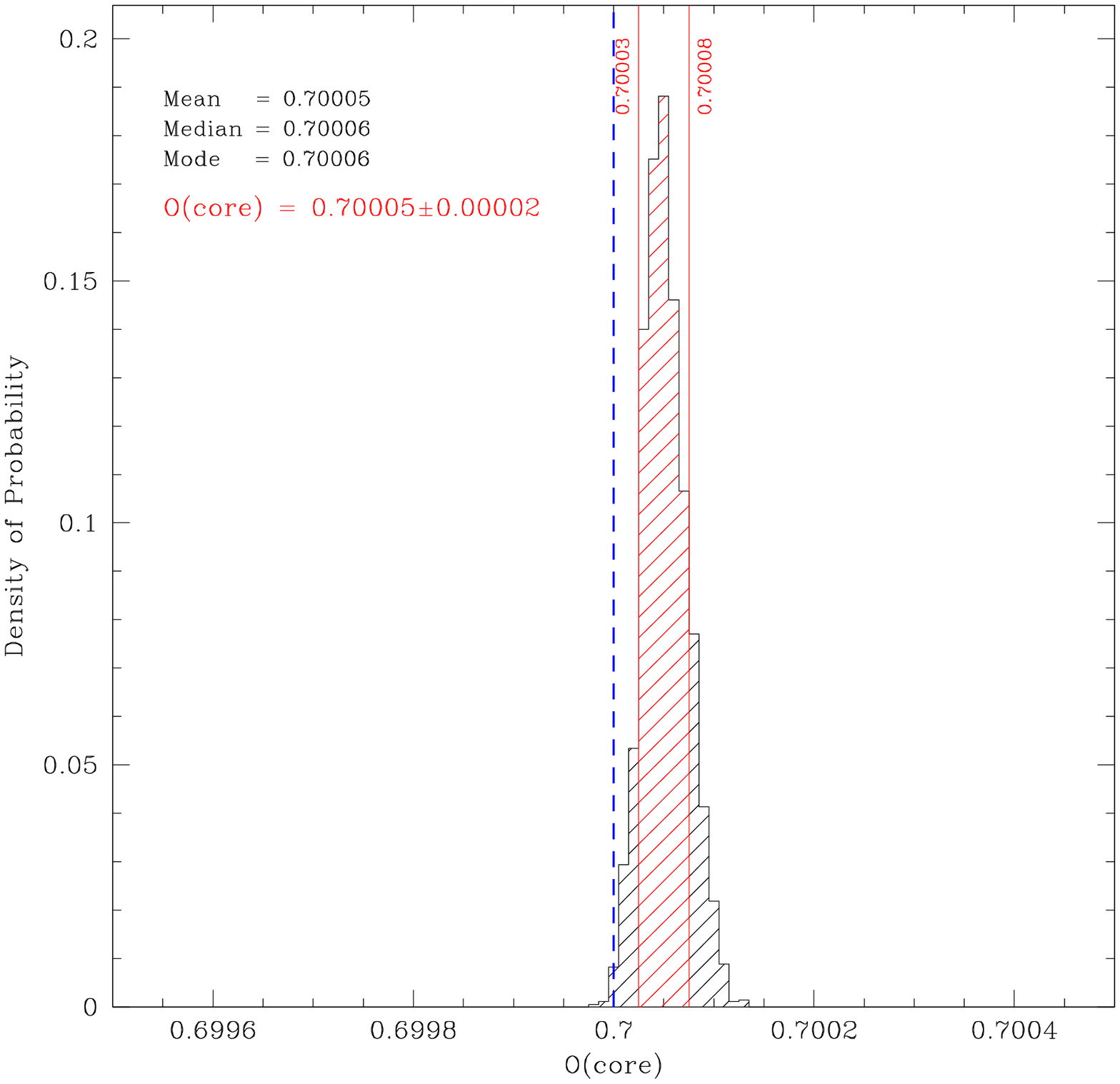} \\
    \end{tabular}
    \begin{flushright}
Figure~12
\end{flushright}
\end{figure}

\clearpage
\addtocounter{figure}{-1}
\begin{figure}[!h]
\centering
  \begin{tabular}{@{}ccc@{}}
    \includegraphics[width=.35\textwidth]{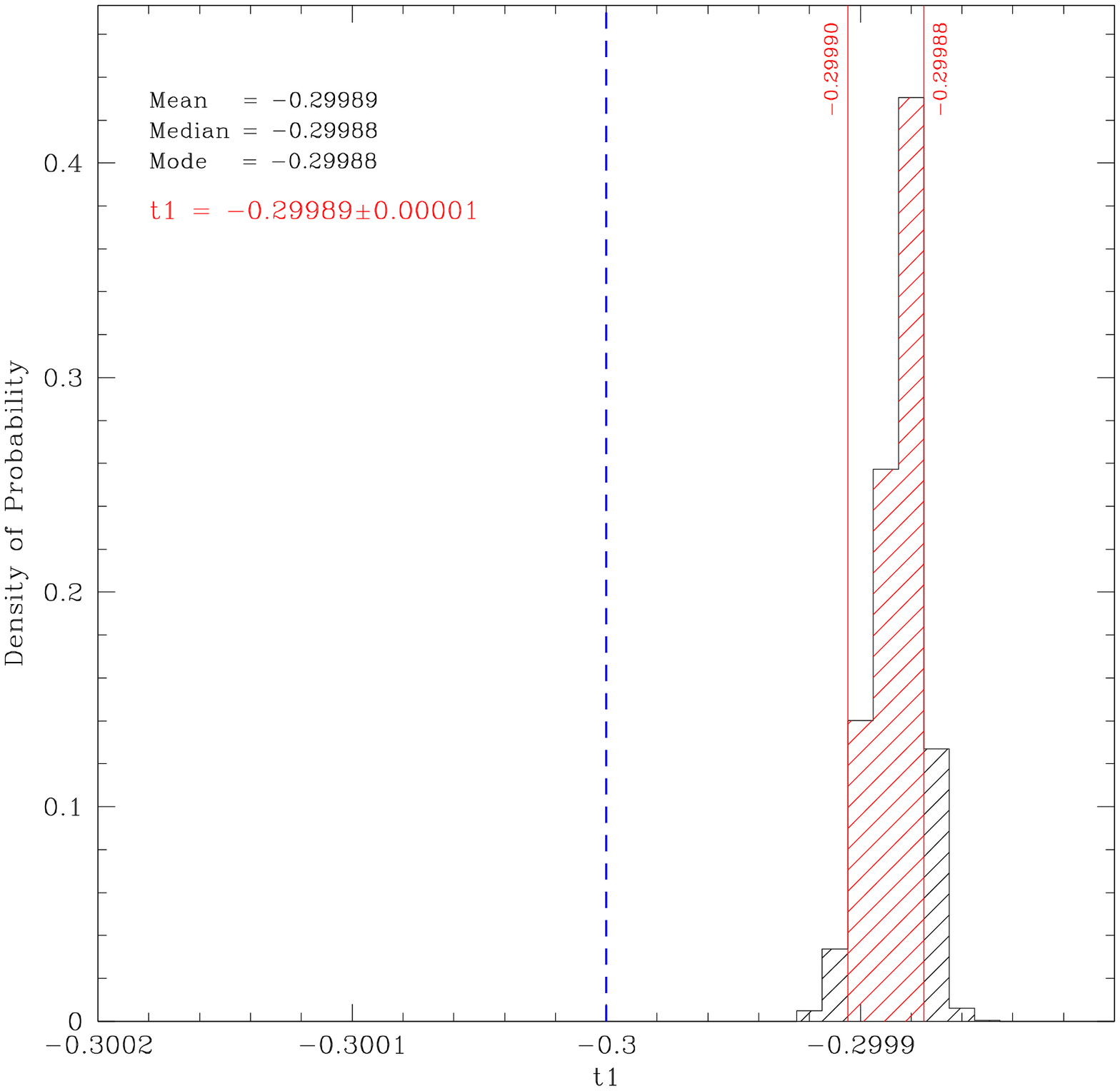} &
    \includegraphics[width=.35\textwidth]{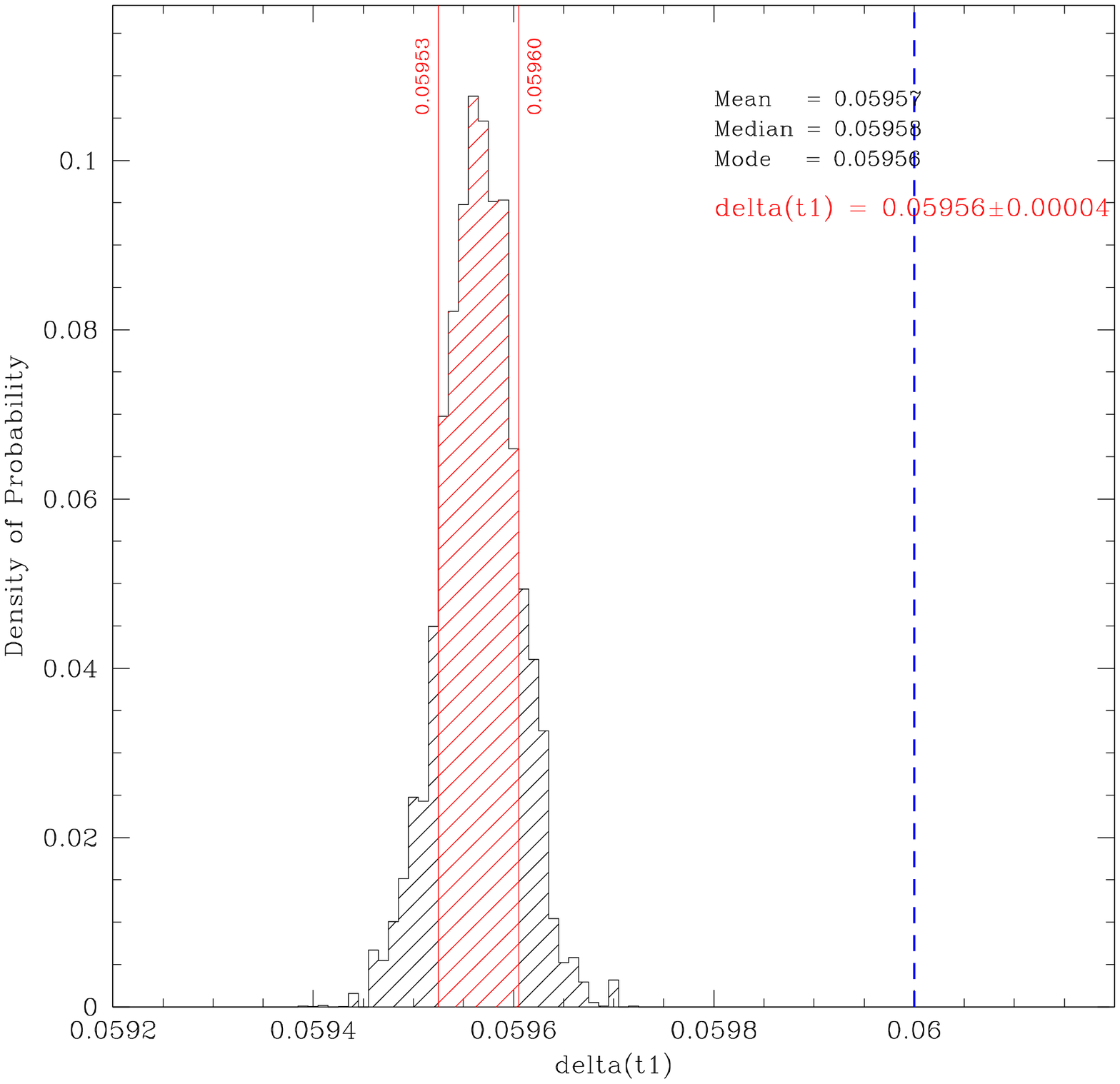} &
    \includegraphics[width=.35\textwidth]{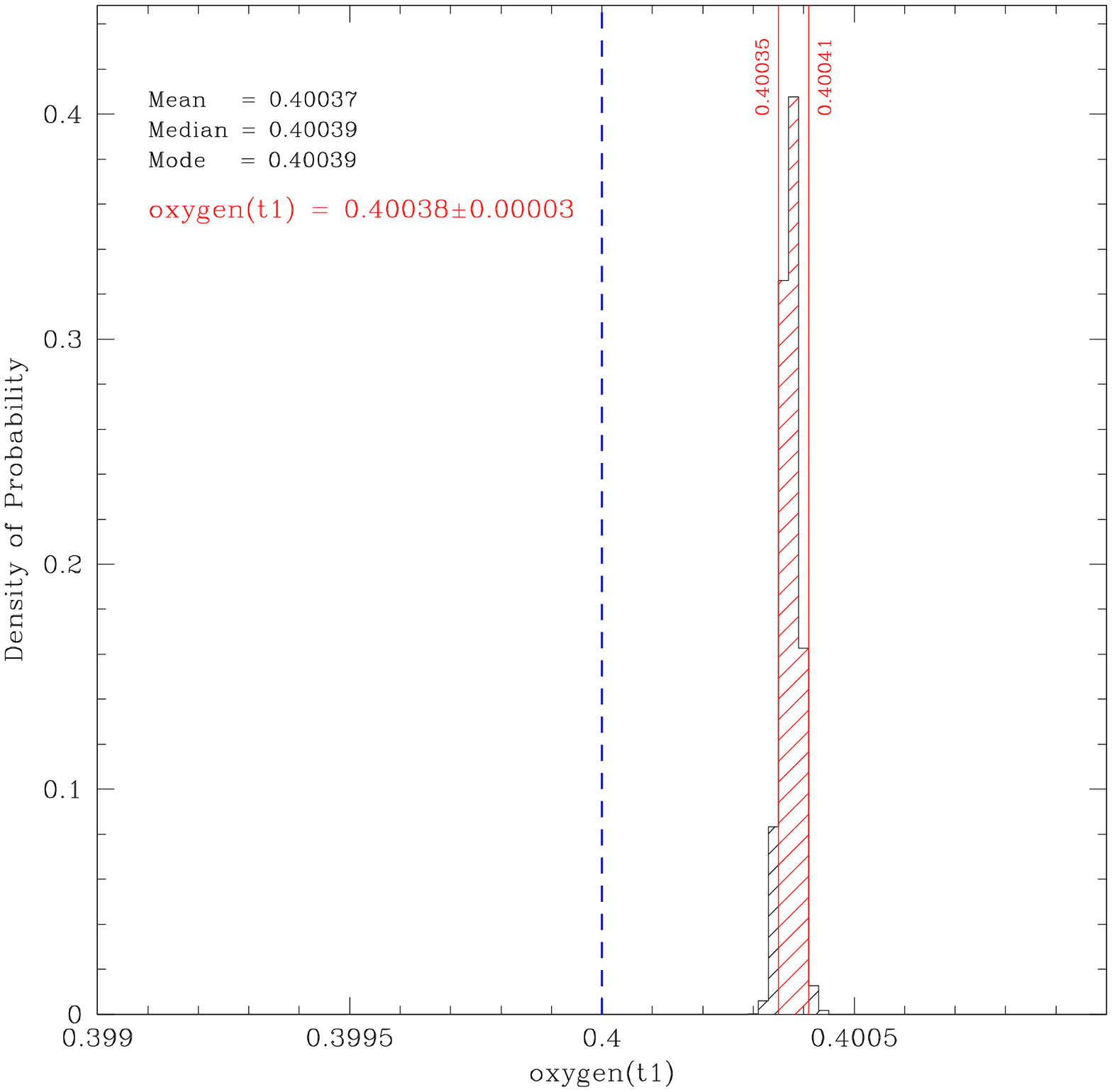} \\
    \includegraphics[width=.35\textwidth]{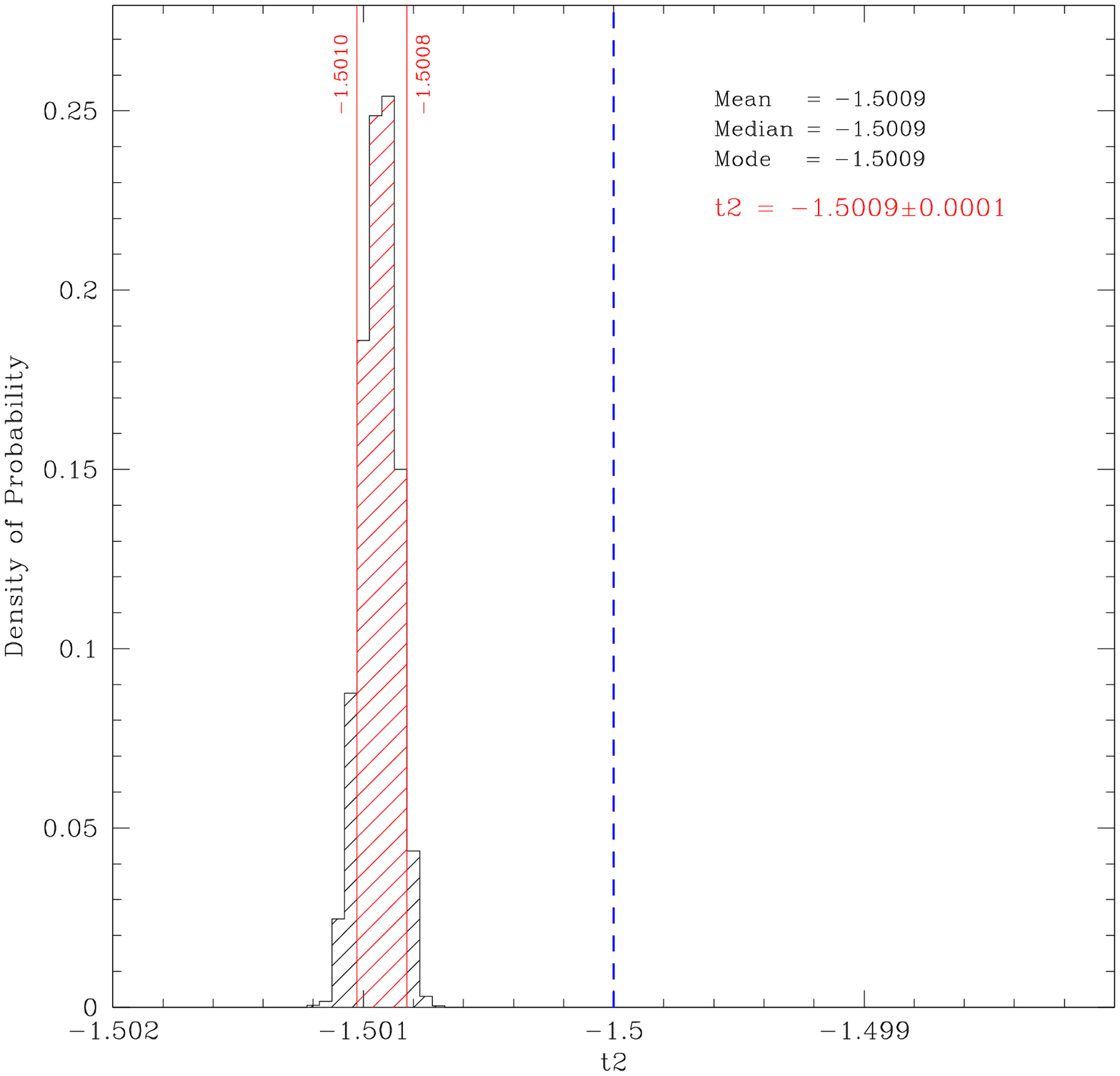} &
    \includegraphics[width=.35\textwidth]{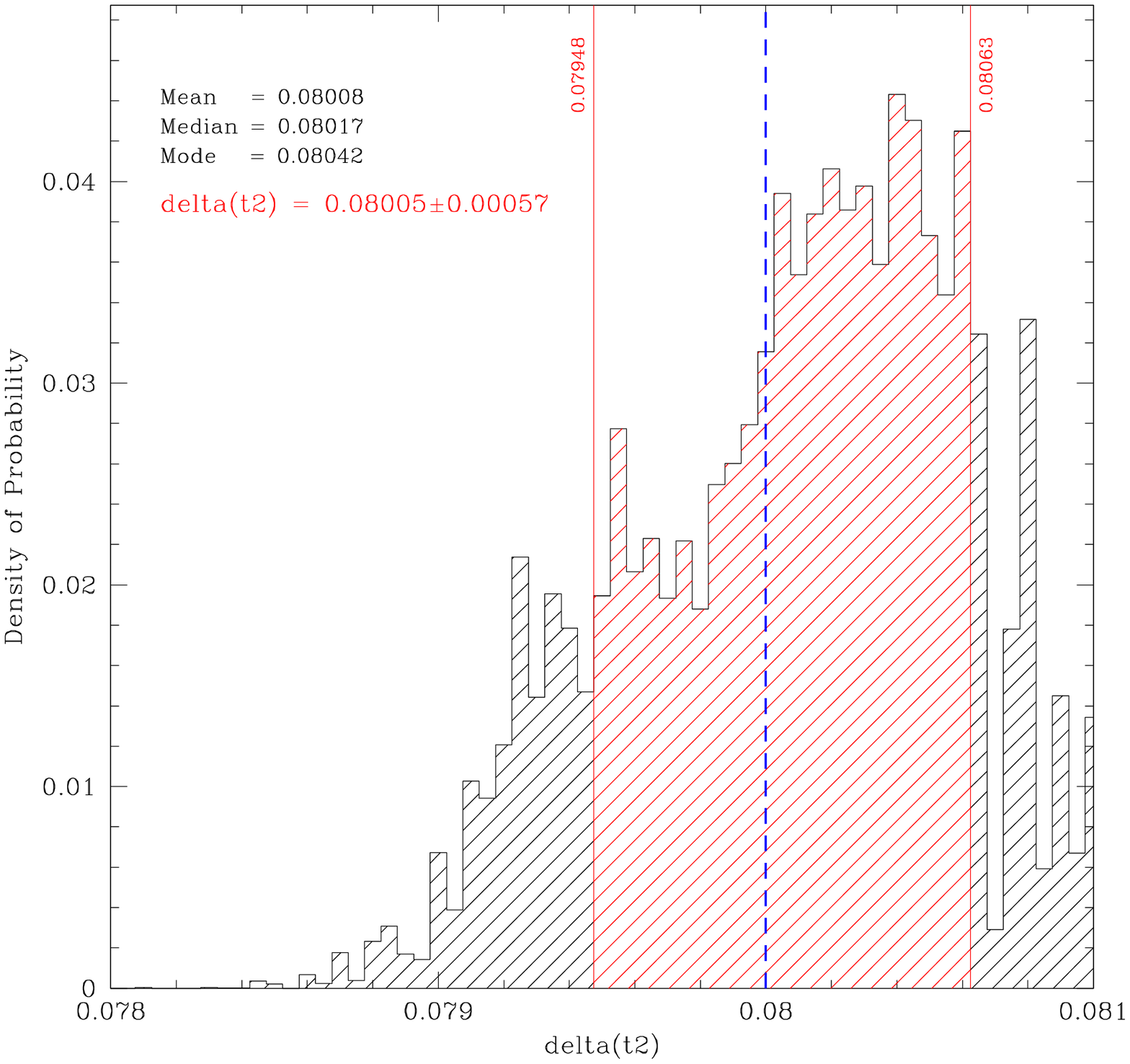} &
    \includegraphics[width=.35\textwidth]{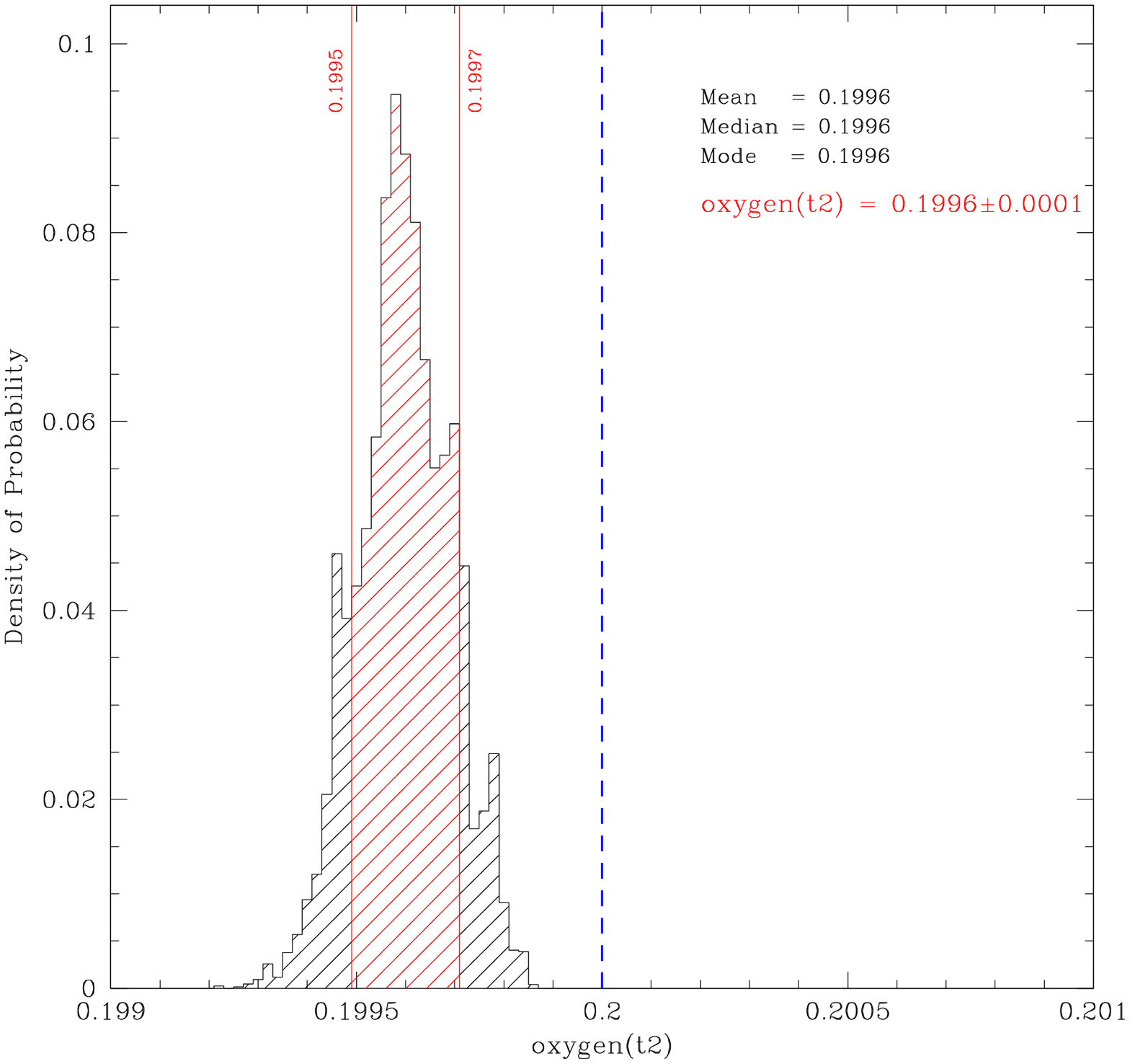}
    \end{tabular}
    \begin{flushright}
Figure~12
\end{flushright}
\end{figure}

\clearpage
\begin{figure}[!h]
 \centering
 \includegraphics[scale=.6]{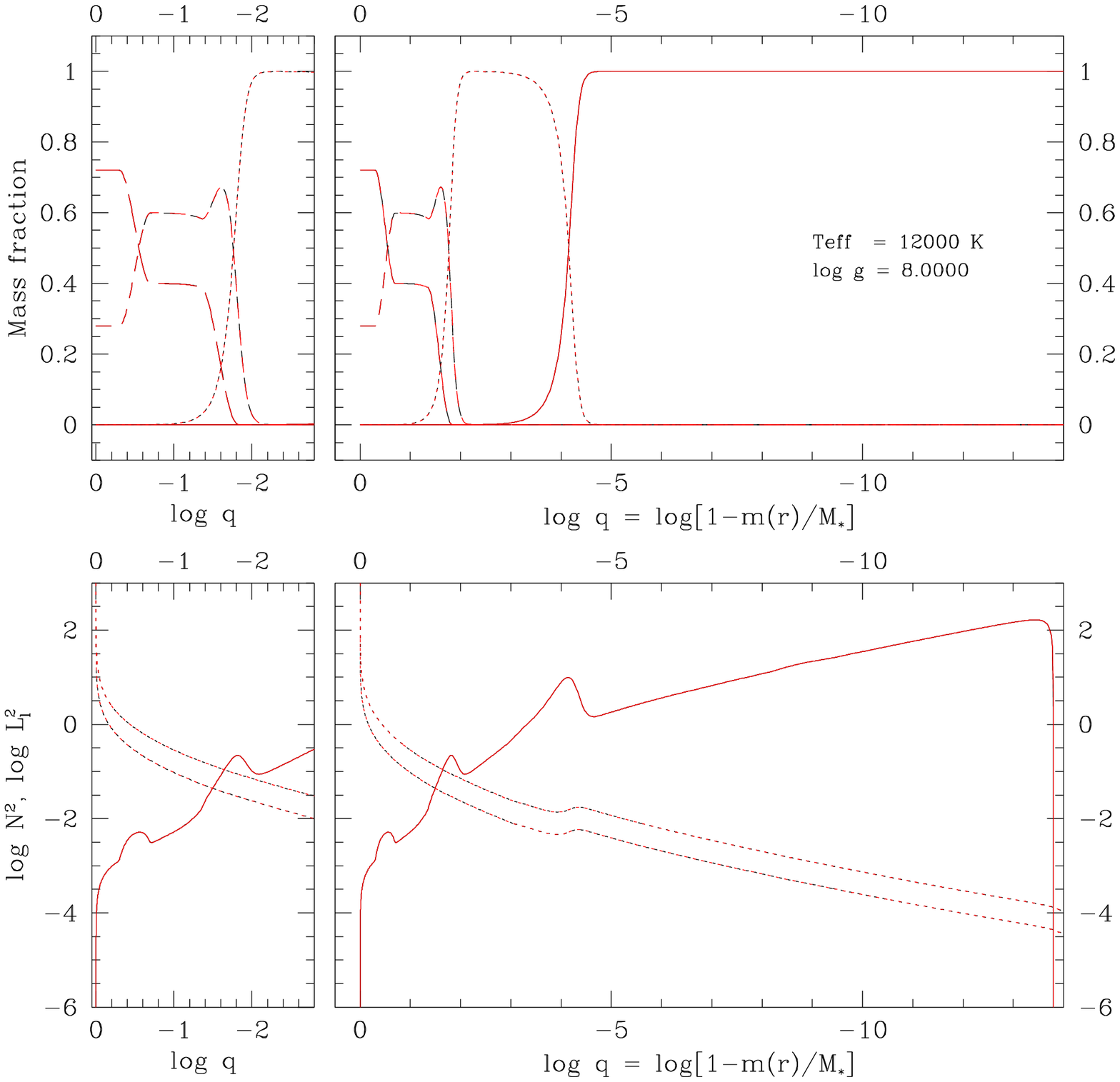}
 \begin{flushright}
Figure~13
\end{flushright}
\end{figure}

\clearpage
\begin{figure}[!h]
 \centering
 \includegraphics[scale=.6]{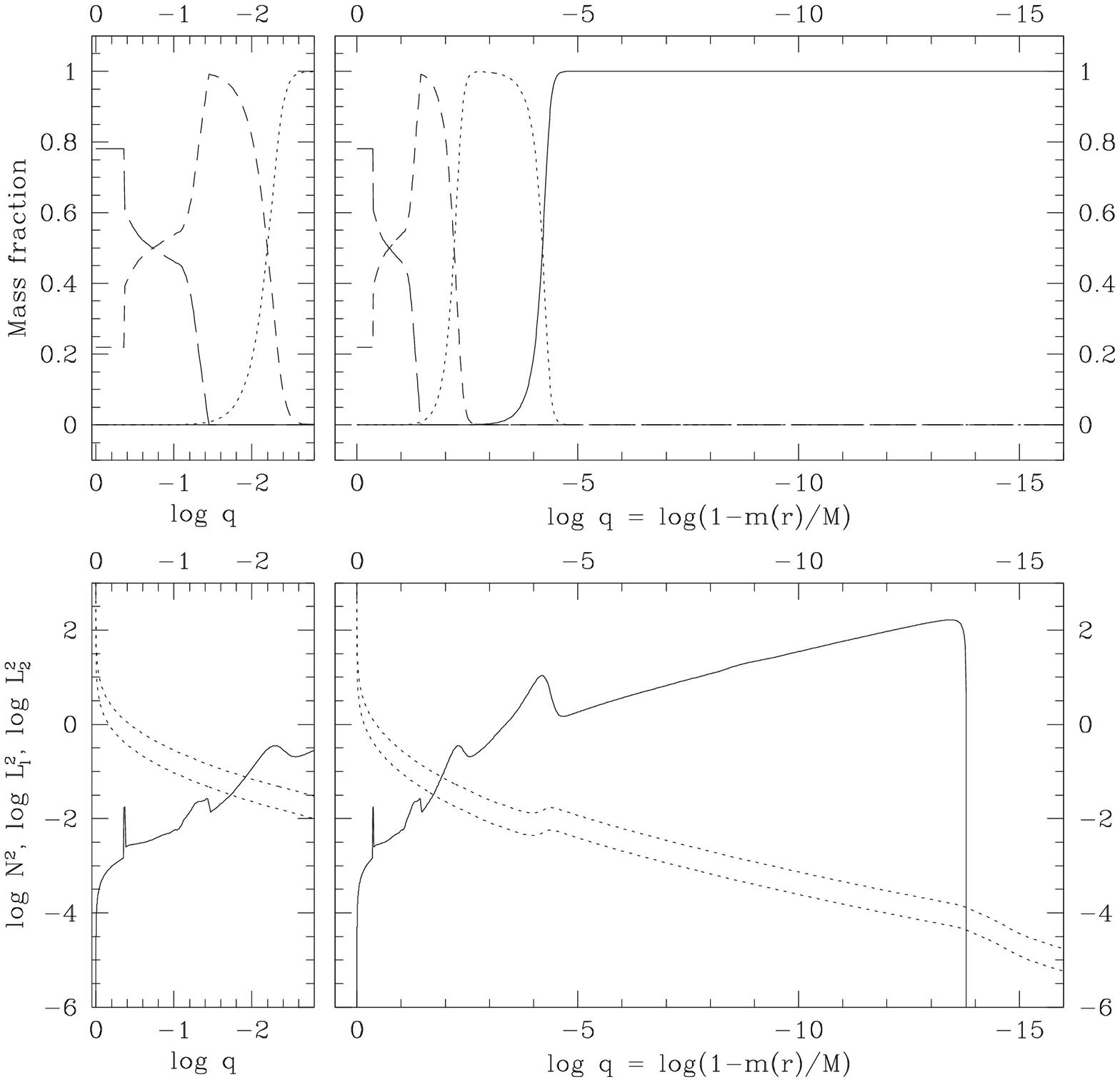}
 \begin{flushright}
Figure~14
\end{flushright}
\end{figure}

\clearpage
\begin{figure}[!h]
 \centering
 \includegraphics[scale=.6]{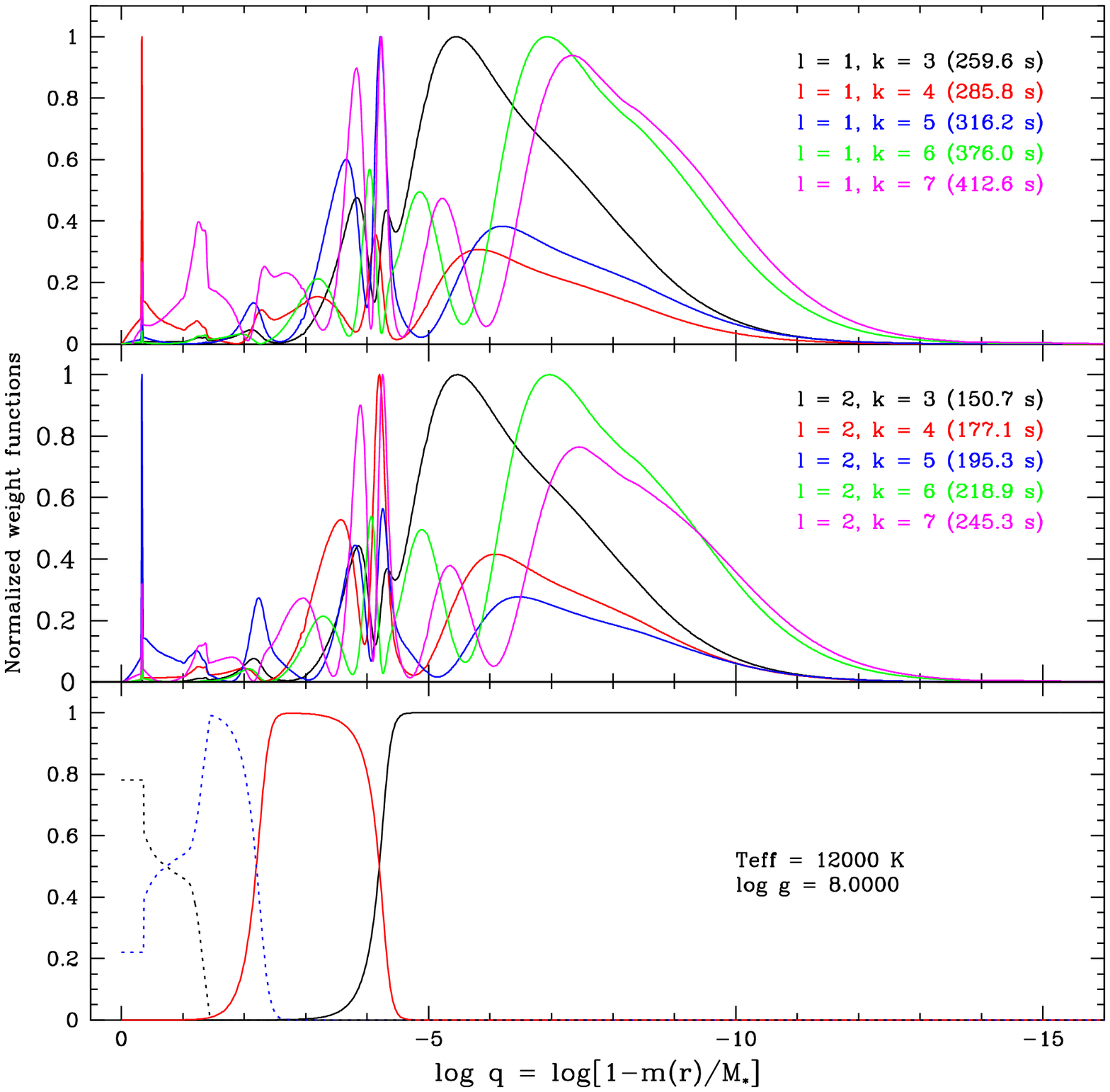}
  \begin{flushright}
Figure~15
\end{flushright}
\end{figure}

\clearpage
\begin{figure}
 \centering
 \includegraphics[scale=.6]{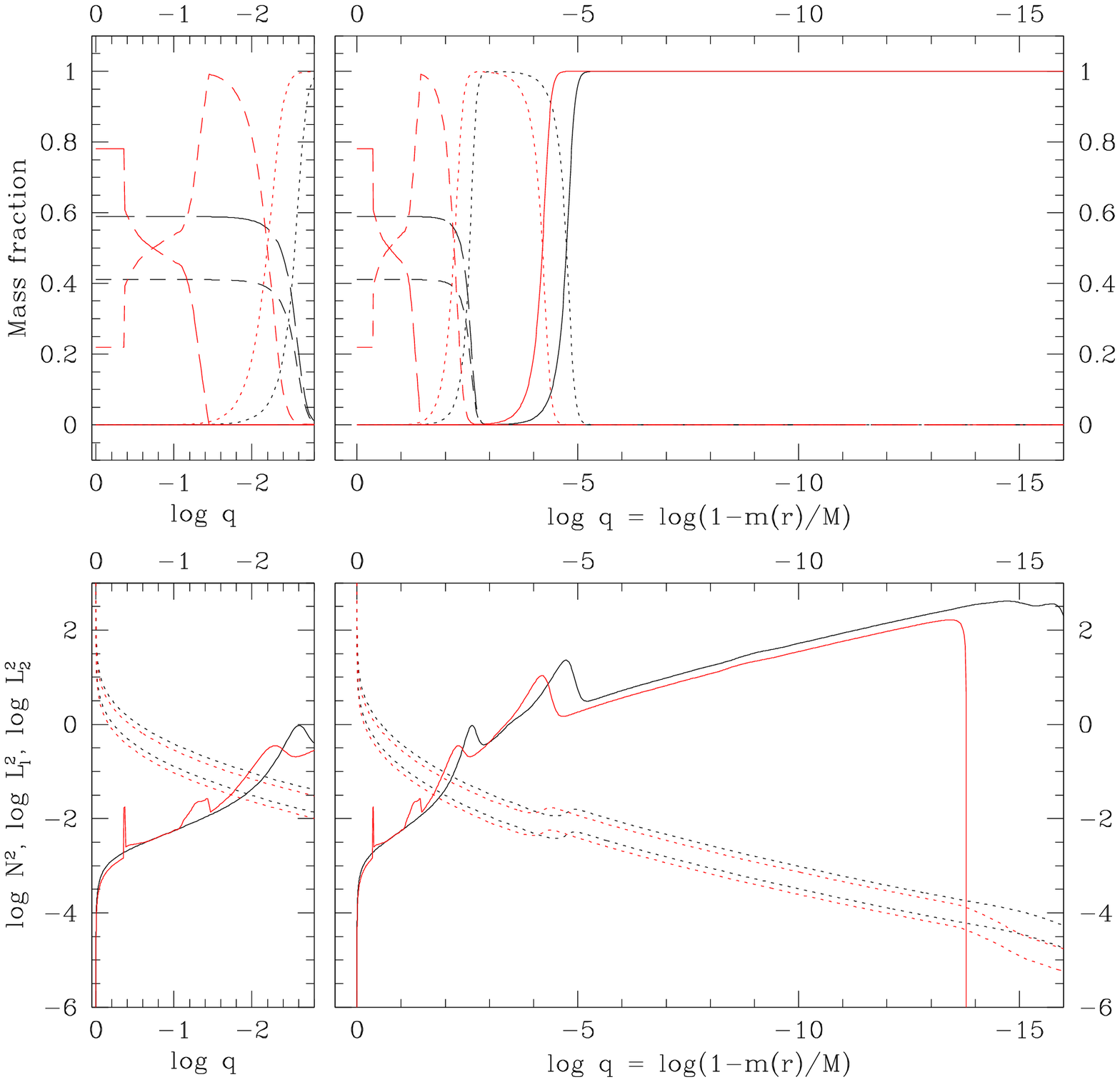}
   \begin{flushright}
Figure~16
\end{flushright}
\end{figure}

\clearpage
\begin{figure}
\centering
\includegraphics[scale=.6]{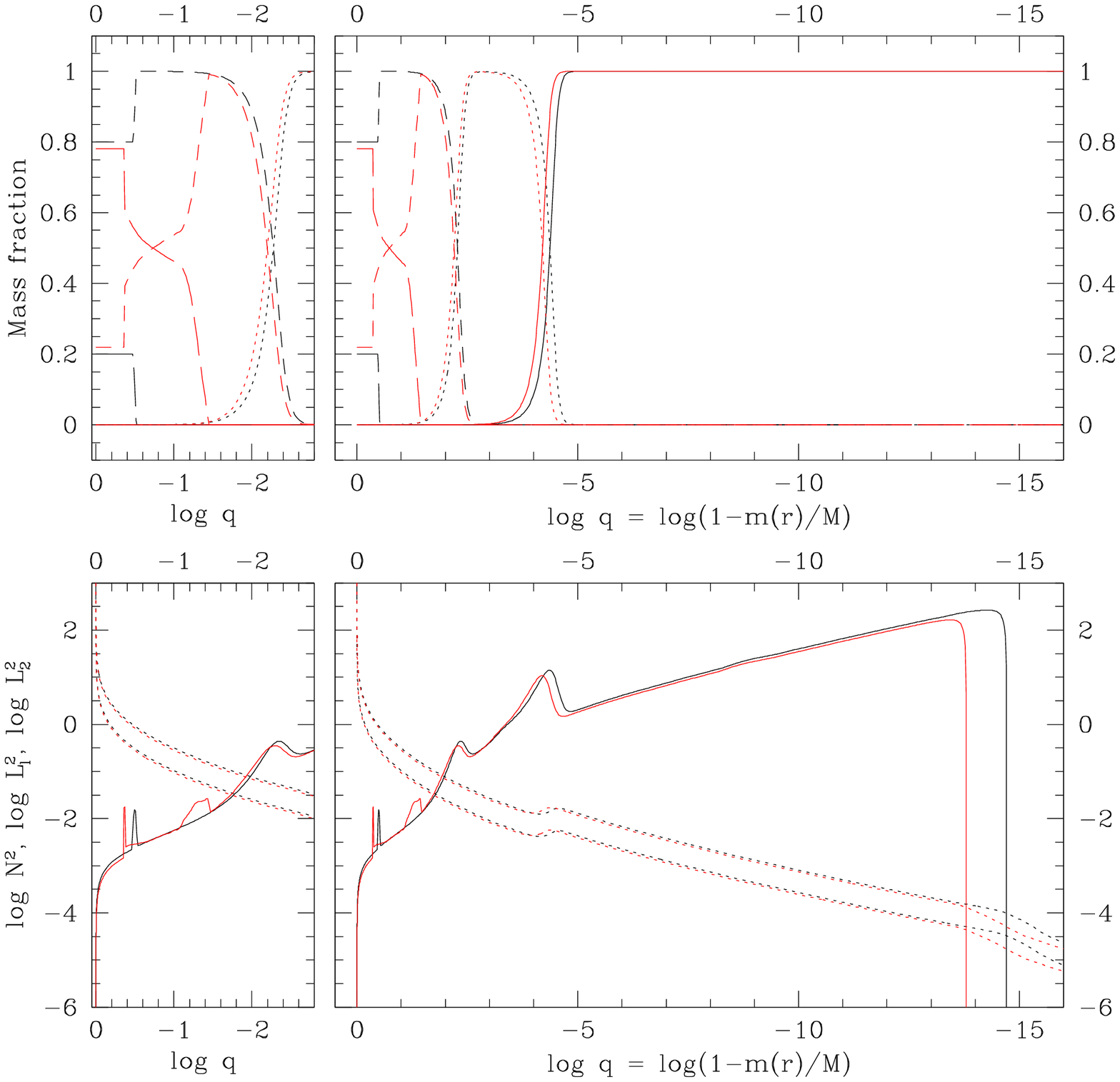}
   \begin{flushright}
Figure~17
\end{flushright}
\end{figure}

\clearpage
\begin{figure}
 \centering
 \includegraphics[scale=.6]{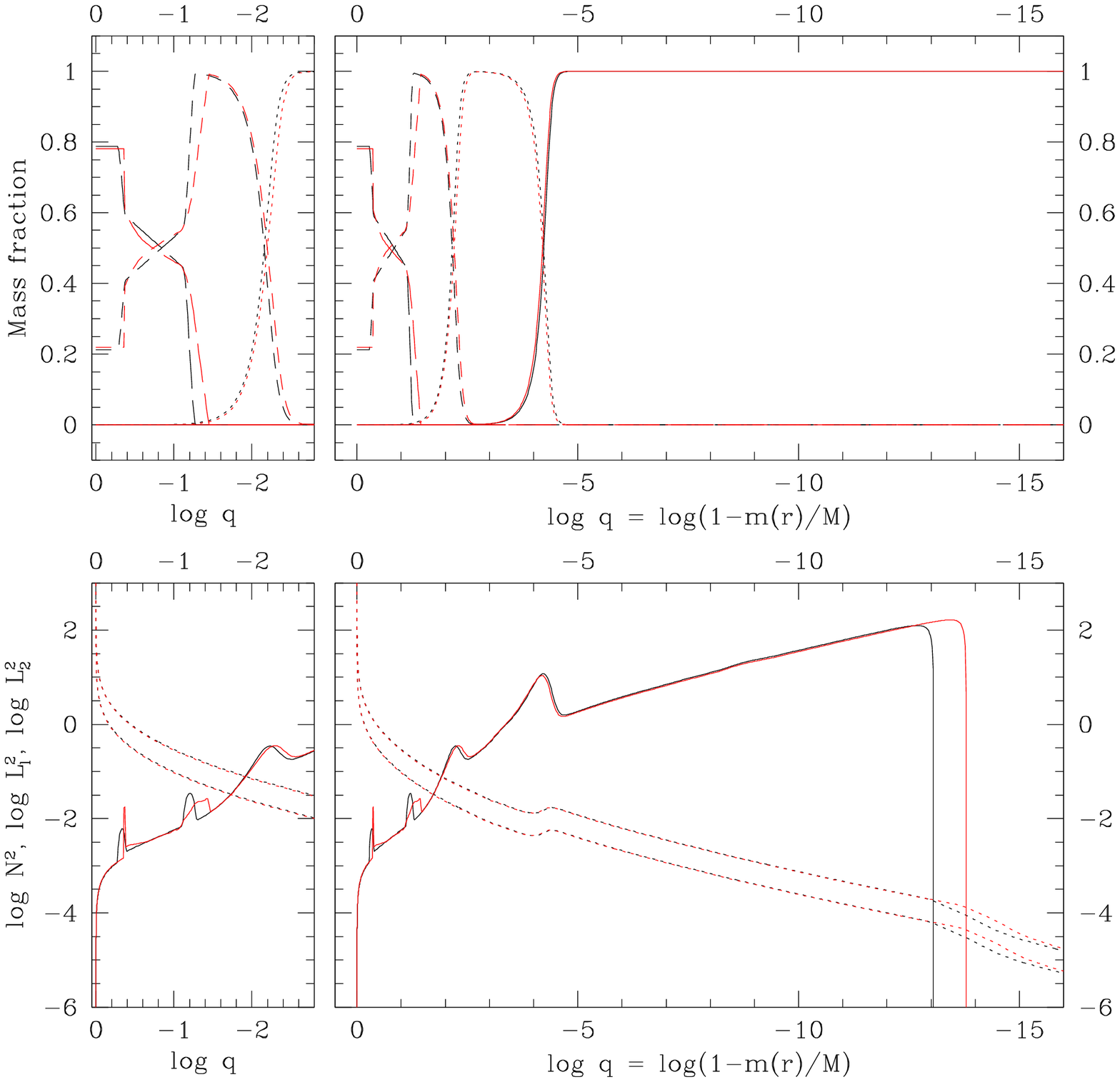}
    \begin{flushright}
Figure~18
\end{flushright}
\end{figure}

\end{document}